\documentclass{article}
\usepackage{arxiv}
\usepackage[utf8]{inputenc}
\usepackage[T1]{fontenc}

\usepackage{amsmath,amssymb,amsfonts,amsthm,mathrsfs,bm,fix-cm}
\usepackage{graphicx,float}
\usepackage[table]{xcolor}
\usepackage{booktabs,array,tabularx,multirow,adjustbox}
\usepackage{tikz}
\usetikzlibrary{positioning,arrows.meta,fit,calc}
\usepackage{enumitem,placeins,microtype,setspace,caption,etoc}
\usepackage[authoryear]{natbib}
\bibpunct{(}{)}{;}{a}{,}{,}
\usepackage{url}
\usepackage[hidelinks,bookmarksnumbered=true]{hyperref}
\providecommand{\doi}[1]{\url{https://doi.org/#1}}

\allowdisplaybreaks[1]
\graphicspath{{Figures/}}
\DeclareGraphicsExtensions{.pdf,.png,.jpg}
\renewcommand{\shorttitle}{\small Probabilistic index models for hierarchical survival outcomes}
\renewcommand{\headeright}{}
\date{}

\theoremstyle{definition}
\newtheorem{theorem}{Theorem}
\newtheorem{lemma}{Lemma}
\newtheorem{proposition}{Proposition}
\newtheorem{condition}{Condition}
\newtheorem{remark}{Remark}

\title{Probabilistic index models for hierarchical survival outcomes}
\author{Xi Fang$^{1,\dagger}$, Zhiqiang Cao$^{2,\dagger}$, Lu Mao$^{3}$, and Fan Li$^{4,*}$ \\[4pt]
  \parbox{0.97\textwidth}{\centering\normalfont\small
    $^{1}$Data Science Institute, Medical College of Wisconsin, Milwaukee, Wisconsin, U.S.A.\\
    $^{2}$Department of Internal Medicine, Yale School of Medicine, New Haven, Connecticut, U.S.A.\\
    $^{3}$Biostatistics and Medical Informatics, University of Wisconsin--Madison, Madison, Wisconsin, U.S.A.\\
    $^{4}$Department of Biostatistics, Yale School of Public Health, New Haven, Connecticut, U.S.A.\\[2pt]
    $^{\dagger}$These authors contributed equally to this work.\\
    $^{*}$Corresponding author, Fan Li (\href{mailto:fan.f.li@yale.edu}{\texttt{fan.f.li@yale.edu}}).}}
\hypersetup{
  pdftitle={Probabilistic index models for hierarchical survival outcomes},
  pdfauthor={Xi Fang, Zhiqiang Cao, Lu Mao, and Fan Li},
  pdfkeywords={Composite endpoints, conditional tie weighting, inverse probability of censoring weighting, probabilistic index model, semi-competing risks, win statistics}
}

\begin{document}
\maketitle

\begin{abstract}
In clinical trials, prioritized composite outcomes allow treatment comparisons to reflect the clinical importance of fatal and nonfatal events. Probabilistic index models relate these comparisons to covariates, but censoring can leave a comparison unresolved. We develop three weighted estimating equations at a prespecified restriction time, with logit, probit, and identity links. Complete history weighting requires both restricted histories to be observed. Resolved comparison weighting also retains comparisons determined before both histories are complete. Conditional tie weighting further uses an observed lower-priority event when higher-priority ties remain uncertain, weighting its regression residual by their conditional probability under a joint model of observed event histories. The construction accommodates a fixed number of nonfatal components and parametric or semiparametric joint models. Under their respective nuisance-model assumptions, the three equations target the same full data regression parameter. We establish consistency and asymptotic normality and derive sandwich estimators accounting for overlapping pairs and nuisance estimation. Across 27 simulation configurations, conditional tie weighting had smaller empirical variance than the other two methods, with reductions of 5.1\%--59.0\% relative to resolved comparison weighting. An HF-ACTION analysis of death and first hospitalization illustrates the additional observed information used by conditional tie weighting. It had smaller estimated variances, while all three methods' treatment intervals included zero.

\end{abstract}

\keywords{Composite endpoints, conditional tie weighting, inverse probability of censoring weighting, probabilistic index model, semi-competing risks, win statistics}

\etocdepthtag.toc{main}

\section{Introduction}
\label{sec:intro}

Composite survival endpoints are commonly used when no single event adequately represents the overall clinical course. In cardiovascular clinical trials, for example, death and hospitalization are both important but are not of equal clinical consequence. Hierarchical composite methods address such differences through a prespecified clinical hierarchy: two individuals are compared on the most important component first, and a lower-priority component is compared only when all higher-priority components are tied. The win ratio is the best-known summary measure arising from this construction \citep{Pocock2012}. Related measures include the win odds \citep{Song2023}, which allocates one half of each tie to both comparison directions, and the win difference, also known as the net benefit \citep{buyse2010generalized}, which contrasts the probabilities of a win and a loss. The pairwise comparison structure also provides a natural route to regression. For example, \citet{Thas2012} introduced probabilistic index models that relate the Mann-Whitney effect measure for non-hierarchical outcomes to the covariate profiles of the two individuals being compared. For hierarchical outcomes, \citet{Mao2021} developed the proportional win fractions model that targets a covariate-specific win ratio under a proportionality assumption, with extensions to accommodate stratified regression \citep{Wang2022} and nonlinear associations \citep{Chakraborty2026}. \citet{Song2023} studied win odds regression as a special case of the probabilistic index regression for non-censored outcomes.

Right censoring is consequential for hierarchical survival outcomes, preventing the pairwise comparison from being fully resolved and rendering standard analysis to depend on the censoring distribution \citep{mao2024defining}. In the two-sample setting, right censoring has been addressed through inverse probability of censoring weighting (IPCW) or through plug-in estimators with estimated survival functions \citep{Dong2020IPCW,Ozenne2021,Peron2021,Cui2025}. For example, \citet{Cui2025} used fixed-horizon weights for lower-priority contributions after observed higher-priority ties, whereas first-priority wins are weighted at the earlier event time. \citet{FangLi2026Tie} proposed a novel conditional tie weighting to replace an unobserved higher-priority tie indicator with its conditional probability to further improve efficiency. In an observational setting, \citet{Huang2026FutureScore} developed a doubly robust win-ratio regression estimator that uses observed event histories to adjust for censoring.
\citet{Wang2026general_wo} introduced a generalized win odds regression for censored hierarchical outcomes and incorporated inverse probability of censoring weights. Their regression model is the special case of PIM with logit link. However, the weighting scheme in \citet{Wang2026general_wo} allows a lower prioriy comparison when censoring prevents determination of a higher-priority comparison. This weighting does not generally give consistent estimator of the full data regression coefficients, even under independent censoring (Appendix \ref{supp:wang-comparison}).

Despite these advances, the general theory of estimation for PIM with censored hierarchical survival outcomes remains incompletely developed. PIM provides a flexible framework for relating prioritized pairwise comparisons to covariates, but extending this framework to right-censored outcomes requires aligning the censoring adjustment with the full-data comparison rule. Even at a fixed restriction time, this alignment does not uniquely determine the censoring weights, raising questions about how alternative weighting schemes can be constructed and affect statistical efficiency. In this paper, we develop the general theory of estimation and inference for PIM with hierarchical survival outcomes at a prespecified restriction time, accommodating general link functions, including the logit, probit, and identity links. Therefore, the regression model in \citet{Wang2026general_wo} can be viewed as a special case under the logit link specification. We then construct three different censoring weighting schemes whose corresponding estimators target the same regression parameter defined in the absence of censoring and establish an ordering of their conditional pairwise score variances when the nuisance quantities are known.
Combining counting process notation with the theory of sparsely correlated estimating equations \citep{Thas2012}, we prove consistency and asymptotic normality and derive sandwich variance estimators accounting for overlapping pairs and estimation of the censoring distribution. Simulations evaluate bias from ignoring censoring, the accuracy of the weighted PIM estimators, and finite-sample efficiency. We apply our developed regression estimators to the HF-ACTION trial and examine how covariate associations with the hierarchical survival outcomes vary across restriction times.

\section{Modeling the probabilistic index with hierarchical survival outcomes}
\label{sec:target}

Let $\{(\bm X_i,\bm Y_i):i=1,\ldots,n\}$ be independent and identically distributed full data observations, where $\bm X_i\in\mathbb R^d$ is a $d$-dimensional baseline covariate vector and $\bm Y_i=(D_i,T_{1i},\ldots,T_{Qi})^\top$ is the ordered composite survival outcome. The fatal event time $T_{0i}=D_i$ is assigned the highest priority, and $T_{qi}$ denotes the time to the $q$th nonfatal event ranked in descending order of clinical importance for $q=1,\ldots,Q$. All pairwise comparisons are made up to a prespecified restriction time $L$, so that the estimand is defined over a fixed horizon. Accordingly, we define the restricted component times $Y_{0i}(L)=D_i\wedge L$ and $Y_{qi}(L)=T_{qi}\wedge D_i\wedge L$ for $q=1,\ldots,Q$, and collect them into $\bm Y_i(L)=\{Y_{0i}(L),\ldots,Y_{Qi}(L)\}^\top$. Each nonfatal time is truncated by $D_i$ and by $L$, reflecting a semi-competing risks structure. For $q=0,\ldots,Q$,  let $\mathcal{E}_{ij}^{(q)}(L)=\left\{Y_{qi}(L)=Y_{qj}(L)\right\}$ denote the event that individuals $i$ and $j$ are tied on component $q$ in the restricted full data. The strict win function is then defined as
\begin{equation}
\mathcal W(\bm Y_i,\bm Y_j)(L)=I\{Y_{0i}(L)>Y_{0j}(L)\}+\sum_{q=1}^{Q}I\left\{Y_{qi}(L)>Y_{qj}(L), \bigcap_{r=0}^{q-1}\mathcal{E}_{ij}^{(r)}(L)\right\},
\label{eq:win-function}
\end{equation}
which means the fatal component is compared first, and the $q$th nonfatal component is compared only when the fatal component and all nonfatal components of higher priority are tied through $L$. The win function $\mathcal W(\bm Y_i,\bm Y_j)(L)$ satisfies three natural conditions. First, it depends on the pair only through the restricted outcomes $\bm Y_i(L)$ and $\bm Y_j(L)$. Second, at most one of $\mathcal W(\bm Y_i,\bm Y_j)(L)$ and $\mathcal W(\bm Y_j,\bm Y_i)(L)$ equals 1. Third, because each nonfatal component time is truncated by death, the win or loss status is determined no later than \(D_i\wedge D_j\wedge L\).

The hierarchical structure of \eqref{eq:win-function} implies that whenever the pair is compared, exactly one component is responsible for the decision. To identify this component, for $q=0,\ldots,Q$, let $\kappa_{ij}^{(q)}(L)=1$ if component $q$ is the first component that distinguishes the pair, and let $\kappa_{ij}^{(q)}(L)=0$ otherwise. The event $\bm Y_i(L)=\bm Y_j(L)$ occurs when no component distinguishes the pair, and such a pair is termed as a \emph{genuine tie} at $L$.
This can occur, for example, when two individuals remain alive and free of every nonfatal event through $L$, in which case all restricted component times are simply equal to $L$. Because the restriction time truncates the comparison, the probability of a genuine tie is typically positive and increases as $L$ decreases. To proceed, we define the probabilistic index response:
\begin{equation}
H_{ij}(L)=\mathcal W(\bm Y_i,\bm Y_j)(L)+\frac{1}{2}I\{\bm Y_i(L)=\bm Y_j(L)\},
\label{eq:pim-response}
\end{equation}
so that a win, a genuine tie, and a loss contribute $1$, $1/2$, and $0$, respectively.
Throughout, $H_{ij}(L)$ in \eqref{eq:pim-response} is the paired response in the restricted full data, defined when both individuals are not right censored up to restriction time $L$.

We consider the probabilistic index model of \citet{Thas2012} for the paired response \eqref{eq:pim-response},
\begin{equation}
E\{H_{ij}(L)\mid\bm X_i^{Y},\bm X_j^{Y}\}=g^{-1}(\bm\beta_L^\top\bm Z_{ij}),
\label{eq:pim-model}
\end{equation}
where \(\bm X_i^{Y} \subseteq \bm{X}_i\), \(\bm X_j^{Y} \subseteq \bm X_j\) are the predictors,  $g$ is a prespecified link function, $\bm Z_{ij}\in\mathbb R^p$ is a known function of $(\bm X_i^Y,\bm X_j^Y)$, and $\bm\beta_L\in\mathbb R^p$ is the regression parameter at restriction time $L$. A canonical choice is the covariate difference $\bm Z_{ij}=\bm X_i^{Y}-\bm X_j^{Y}$. Because exchanging the two individuals reverses the sign of this predictor, the model should respect the reversal identity $H_{ij}(L)+H_{ji}(L)=1$. Hence, the probabilistic index structure imposes a natural constraint on the link function such that $g(1-u)=-g(u)$ for $0<u<1$. This condition is satisfied by the logit and probit link functions. For the identity link, the implied constraint is $g(u)=2u-1$ and therefore $2E\{H_{ij}(L)\mid\bm X_i^{Y},\bm X_j^{Y}\}-1=\bm\beta_L^\top\bm Z_{ij}$, with the linear predictor restricted to $[-1,1]$. Hence, the identity PIM shares the same spirit of a linear probability model. Of note, under a pure covariate difference structure, no free intercept is included because the reversal identity forces the intercept to be zero.

The choice of link function determines the interpretation of the regression coefficients. Under the logit link, $\exp(\bm\beta_L^\top\bm Z_{ij})$ equals
\begin{equation}
\frac{P\{\bm Y_i(L)\succ\bm Y_j(L)\mid\bm X_i^{Y},\bm X_j^{Y}\}+P\{\bm Y_i(L)=\bm Y_j(L)\mid\bm X_i^{Y},\bm X_j^{Y}\}/2}{P\{\bm Y_i(L)\prec\bm Y_j(L)\mid\bm X_i^{Y},\bm X_j^{Y}\}+P\{\bm Y_i(L)=\bm Y_j(L)\mid\bm X_i^{Y},\bm X_j^{Y}\}/2},
\label{eq:conditional-wo}
\end{equation}
which is the conditional win odds at restriction time $L$ \citep{Song2023,Wang2026general_wo}. When ties are absent, \eqref{eq:conditional-wo} is equivalent to the conditional win ratio at restriction time $L$. Under the identity link function, $\bm\beta_L^\top\bm Z_{ij}=E\{\mathcal W(\bm Y_i,\bm Y_j)(L)-\mathcal W(\bm Y_j,\bm Y_i)(L)\mid\bm X_i^{Y},\bm X_j^{Y}\}$ equals the conditional net benefit. Finally, under the probit link function, $E\{H_{ij}(L)\mid\bm X_i^{Y},\bm X_j^{Y}\}=\Phi(\bm\beta_L^\top\bm Z_{ij})$ such that the same probabilistic index is expressed on the standard normal quantile scale. Under the covariate difference specification, each coefficient represents the change in this normal quantile associated with a one unit increase in the corresponding covariate contrast, holding the remaining contrasts fixed. We provide additional discussions around the constraint on link functions and interpretation in Appendix \ref{supp:connections}.

\section{Three censoring weighted estimation schemes}
\label{sec:estimation}

Let $C_i$ denote the censoring time of individual $i$, common to all components. We assume \(P(C_i\ge t\mid\bm Y_i,\bm X_i)=S_C(t\mid\bm X_i^{C})\) for \(t\in[0,L]\), where \(\bm X_i^{C} \subseteq \bm{X}_i\) and $S_C(t\mid\bm x)=P(C_i\ge t\mid\bm X_i^{C}=\bm x)\ge \epsilon$ for $t\in[0,L]$ and some  \(\epsilon>0\). We write $\mu_{ij}(\bm\beta_L)=g^{-1}(\bm\beta_L^\top\bm Z_{ij})$ and \(\bm K_{ij}(\bm\beta_L)=\frac{\partial\mu_{ij}(\bm\beta_L)}{\partial\bm\beta_L} V^{-1}\{\mu_{ij}(\bm\beta_L)\}\), where \(V(\bullet)\) is a working variance function satisfying \(V(u)=V(1-u)\) (Appendix Lemma \ref{supp:lem:score-symmetry}). Let $\mathcal{I}_n=\{(i,j)\mid 1\le i<j\le n\}$, and without censoring, $\bm\beta_L$ is estimated from the full-data estimating equations of \citet{Thas2012}:
\begin{equation}
\frac{2}{n(n-1)}\sum_{(i,j)\in \mathcal{I}_n}\bm K_{ij}(\bm\beta_L)\{H_{ij}(L)-\mu_{ij}(\bm\beta_L)\}=\bm0.
\label{eq:full-data-ee}
\end{equation}
Its population root $\bm\beta_L^0$ solves $E[\bm K_{ij}(\bm\beta_L)\{H_{ij}(L)-\mu_{ij}(\bm\beta_L)\}]=\bm0$, which is the true regression parameter under model \eqref{eq:pim-model}.
Under right censoring, the full-data response \(H_{ij}(L)\) may not be fully observed and we instead observe \(\widetilde H_{ij}(L)\in\{1,0,1/2\}\), indicating observed win, loss, and tie. Since the observed tie is either a genuine tie at restriction time $L$ or a censoring-induced tie, \(\widetilde H_{ij}(L)\) is not necessarily equal to \(H_{ij}(L)\) and requires adjustment.

\subsection{Complete history weighting}
\label{sec:complete-history}

We define the complete history indicator for each pair as \(R_{ij}^{\mathrm{CHW}}(L)=I\{C_i\ge Y_{0i}(L),C_j\ge Y_{0j}(L)\}\). That is, $R_{ij}^{\mathrm{CHW}}(L)=1$ if and only if both individuals remain uncensored until death or restriction time $L$, whichever occurs first. Conditional on the event times and covariates, the pairwise censoring survival function evaluated at the minimum of death and $L$ is \(\pi_{ij}^{\mathrm{CHW}}(L)=S_C\left(Y_{0i}(L)\mid\bm X_i^{C}\right)S_C\left(Y_{0j}(L)\mid\bm X_j^{C}\right)\). We prove in Appendix \ref{supp:chw-rcw-mean} that
\begin{align*}
E\left[\left.\frac{R_{ij}^{\mathrm{CHW}}(L)}{\pi_{ij}^{\mathrm{CHW}}(L)}
\{\widetilde{H}_{ij}(L)-\mu_{ij}(\bm\beta_L)\} \right|\bm Y_i(L),\bm Y_j(L),\bm X_i,\bm X_j\right]=H_{ij}(L)-\mu_{ij}(\bm\beta_L).
\end{align*}
Therefore, writing $\bm U_{ij}^{\mathrm{CHW}}(\bm\beta_L)=\bm K_{ij}(\bm\beta_L)\frac{R_{ij}^{\mathrm{CHW}}(L)}{\pi_{ij}^{\mathrm{CHW}}(L)}\left\{\widetilde H_{ij}(L)-\mu_{ij}(\bm\beta_L)\right\}$ as the weighted score, the weighted estimating equations $\bm\Psi_n^{\mathrm{CHW}}(\bm\beta_L)=\frac{2}{n(n-1)}\sum_{(i,j)\in \mathcal{I}_n}\bm U_{ij}^{\mathrm{CHW}}(\bm\beta_L)$ have the same expectation as the full-data estimating equations
in \eqref{eq:full-data-ee}, for every $\bm\beta_L$.
Since $S_C$ is unknown in practice, we can estimate it using the Cox model $\lambda_i^C(t)=\lambda_0^C(t)\exp(\bm\gamma^\top\bm X_i^{C})$ with the Breslow estimator of the baseline cumulative hazard \(\Lambda_0^C(t)=\int_0^t\lambda_0^C(s)\,ds\).
The final estimator $\widehat{\bm\beta}_L^{\mathrm{CHW}}$ solves the weighted estimating equations $\widehat{\bm\Psi}_n^{\mathrm{CHW}}(\bm\beta_L)=\bm0$.

\begin{figure}[!htbp]
\centering
\includegraphics[width=0.99\textwidth,height=0.70\textheight,keepaspectratio]{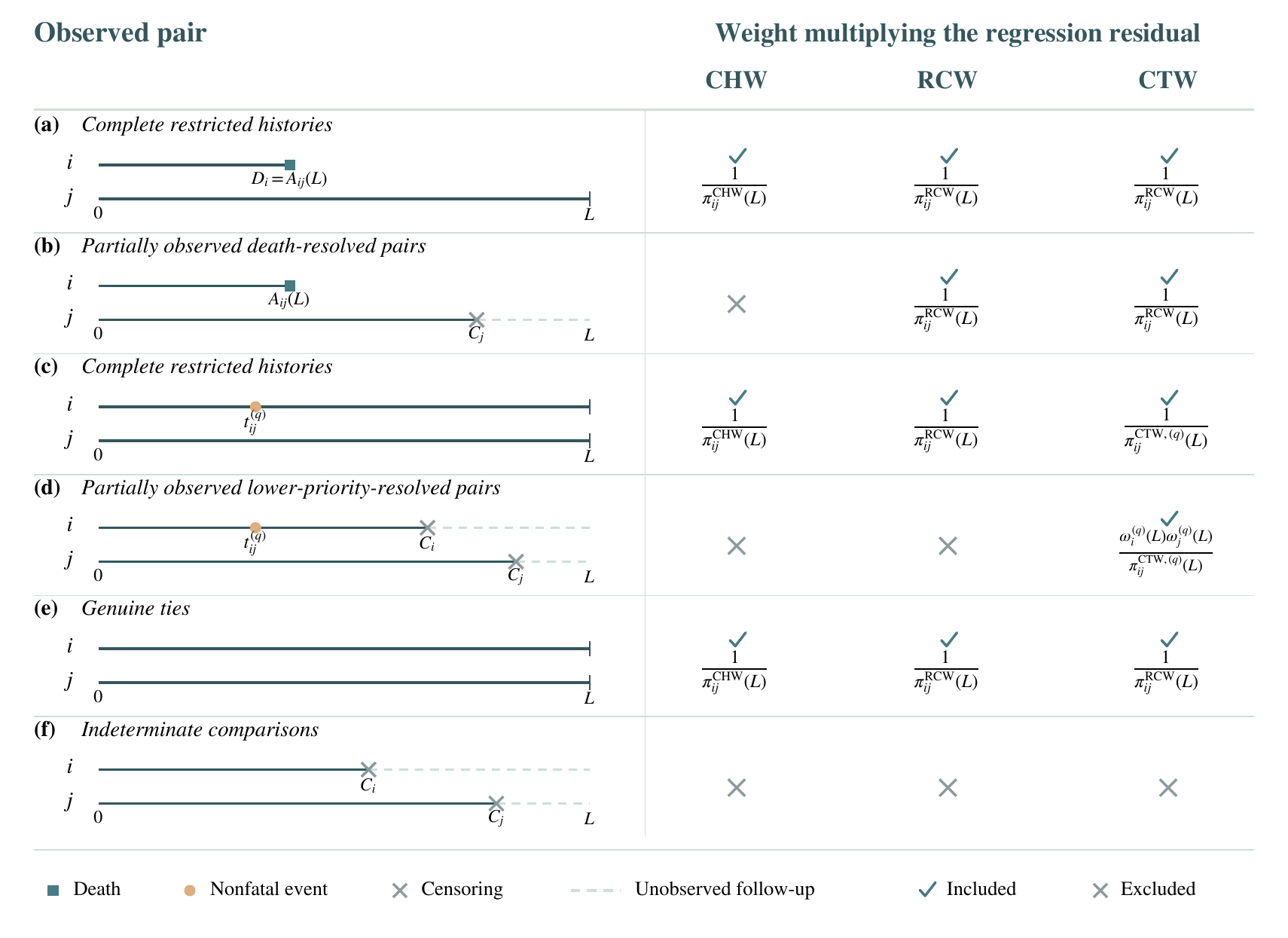}
\caption{Observed pair categories and weights for complete history weighting (CHW), resolved comparison weighting (RCW), and conditional tie weighting (CTW), with death prioritized over \(Q\) nonfatal components. In (a) and (c), both restricted histories are complete, and the comparison is determined by death and a lower-priority component, respectively. In (b), death resolves the comparison before the other individual is censored. In (d), suppose individual \(i\) is censored alive at \(c<L\), with \(C_i=c\). The nonfatal ordering is observed, but the required higher-priority ties are not fully observed. CTW retains this contribution when \(\omega_i^{(q)}(L)\omega_j^{(q)}(L)>0\). Panel (e) shows a genuine tie with response \(1/2\). Panel (f) includes all remaining indeterminate comparisons, including observed lower-priority orderings whose required tie probability is zero. A checkmark indicates inclusion and a cross indicates exclusion. In (c) and (e), \(\pi_{ij}^{\mathrm{CHW}}(L)=\pi_{ij}^{\mathrm{RCW}}(L)=S_C(L\mid\bm X_i^C)S_C(L\mid\bm X_j^C)\).}

\label{fig:weight-information-progression}
\end{figure}

Despite its simplicity, complete history weighting (CHW) does not effectively use all available information. It requires each restricted history to be observed through $Y_{0i}(L)$ and $Y_{0j}(L)$, even when the pairwise comparison is resolved earlier. To be more specific, if one individual dies before \(L\) while the other is alive and under observation, the survivor wins. However, if the survivor is later censored before $L$, such a pair corresponds to $R_{ij}^{\mathrm{CHW}}(L)=0$ and will be excluded. We refer to such a pair as \emph{partially observed death-resolved pair}; see Figure \ref{fig:weight-information-progression}(b) for such an example pair.
Under heavy censoring, therefore, CHW tends to exclude more pairs for which a win or loss is already fully determined.

\subsection{Resolved comparison weighting}
\label{sec:estimating-equation}

Unlike CHW, resolved comparison weighting (RCW) additionally leverages all partially observed death-resolved pairs by requiring both individuals to be observed only through the common determination time, defined as
\begin{equation*}
A_{ij}(L)=\max_{0\le r\le Q}\{Y_{ri}(L)\wedge Y_{rj}(L)\}=Y_{0i}(L)\wedge Y_{0j}(L)=D_i\wedge D_j\wedge L.
\end{equation*}
For each component \(r \in \{0,\ldots, Q\} \), \(Y_{ri}(L)\le Y_{0i}(L)=D_i\wedge L\), and similarly for individual \(j\). Thus, \(Y_{ri}(L)\wedge Y_{rj}(L)\le Y_{0i}(L)\wedge Y_{0j}(L)\), and the maximum equals the term corresponding to \(r=0\). With continuous death times, fatal comparisons can be resolved before \(L\), whereas nonfatal comparisons and genuine ties require both individuals to be followed through \(L\). Hence, $A_{ij}(L)$ is referred to as the common determination time for both individuals.
Let \(R_{ij}^{\mathrm{RCW}}(L)=I\{C_i\ge A_{ij}(L),C_j\ge A_{ij}(L)\}\) be the resolved comparison indicator and \(\pi_{ij}^{\mathrm{RCW}}(L)=S_C\left(A_{ij}(L)\mid\bm X_i^{C}\right) S_C\left(A_{ij}(L)\mid\bm X_j^{C}\right)\) be the product of the conditional survival functions evaluated at the common determination time. Under covariate-dependent censoring, we can show that \(E\{R_{ij}^{\mathrm{RCW}}(L)\mid\bm Y_i,\bm Y_j,\bm X_i^{C},\bm X_j^{C}\}=\pi_{ij}^{\mathrm{RCW}}(L)\).
It follows that \(R_{ij}^{\mathrm{RCW}}(L)\ge R_{ij}^{\mathrm{CHW}}(L)\), with strict inequality when an observed death resolves the comparison and the other individual is subsequently censored before \(L\),
as in Figure \ref{fig:weight-information-progression}(b). In other words, \(R_{ij}^{\mathrm{RCW}}(L)=1> R_{ij}^{\mathrm{CHW}}(L)=0\) only for partially observed death-resolved pairs.

Weighting the regression residual of each resolved pair by \(1/\pi_{ij}^{\mathrm{RCW}}(L)\) gives the RCW pair contribution to the estimating equations
\begin{equation}
\bm U_{ij}^{\mathrm{RCW}}(\bm\beta_L)=\bm K_{ij}(\bm\beta_L)\frac{R_{ij}^{\mathrm{RCW}}(L)}{\pi_{ij}^{\mathrm{RCW}}(L)}\{\widetilde H_{ij}(L)-\mu_{ij}(\bm\beta_L)\},
\label{eq:ipcw-kernel}
\end{equation}
with zero contribution when \(R_{ij}^{\mathrm{RCW}}(L)=0\). Define
\(\bm\Psi_n^{\mathrm{RCW}}(\bm\beta_L)=\frac{2}{n(n-1)}\sum_{(i,j)\in \mathcal{I}_n}\bm U_{ij}^{\mathrm{RCW}}(\bm\beta_L)\)
and let \(\widehat{\bm\beta}_L^{\mathrm{RCW}}\) solve the weighted estimating equations \(\widehat{\bm\Psi}_n^{\mathrm{RCW}}(\bm\beta_L)=\bm0\).
Under covariate-dependent censoring, we prove in Appendix \ref{supp:unbiased} that
$$E\{\bm U_{ij}^{\mathrm{RCW}}(\bm\beta_L)\mid\bm Y_i(L),\bm Y_j(L),\bm X_i,\bm X_j\}=\bm K_{ij}(\bm\beta_L)\{H_{ij}(L)-\mu_{ij}(\bm\beta_L)\}.$$
Thus, CHW and RCW correspond to the same full-data estimating equations for every \(\bm\beta_L\).

While CHW shares the same spirit of the conventional inverse-probability-weighting scheme in the missing outcome context, RCW carries a similar principle to the inverse-probability-weight used to estimate marginal win probabilities in \citet{Dong2020IPCW} and \citet{Cui2025}, with some differences. For fatal comparisons, RCW agrees with \citet{Dong2020IPCW} in evaluating both censoring survival functions at the earlier observed death time. For nonfatal comparisons, \citet{Dong2020IPCW} evaluate these functions at the earlier observed nonfatal event time that establishes the comparison, whereas RCW instead evaluates them at restriction time \(L\), since the required higher priority ties should be confirmed through \(L\). This feature then agrees with the observation requirements and censoring evaluation times in the weighting scheme of \citet{Cui2025} developed for the two-sample comparison context. RCW incorporates these observation requirements into the PIM and therefore includes the \citet{Cui2025} estimator as a special case under a trivial regression model without covariates. Despite using more information than CHW, RCW does not leverage an observed nonfatal comparison unless all higher priority components are confirmed to be tied. As a notable example, if both individuals are subsequently censored before \(L\), it remains unknown whether \(D_i\wedge D_j>L\), and because the fatal comparison is indeterminate, the non-fatal comparisons that could have been resolved before \(L\) are excluded by RCW, or \(R_{ij}^{\mathrm{RCW}}(L)=0\). We refer to such excluded pairs by RCW as \emph{partially observed lower-priority-resolved pairs}; see Figure \ref{fig:weight-information-progression}(d) for a schematic illustration of such a pair. With higher censoring proportion, therefore, RCW tends to exclude more pairs for which a win or loss could have already been determined by lower-priority components that occur prior to censoring, and this exclusion can lead to suboptimal efficiency in the final regression estimator.

\FloatBarrier
\subsection{Conditional tie weighting}
\label{sec:ctw}

To further incorporate partially observed lower-priority-resolved pairs, we generalize the conditional tie weighting (CTW), originally developed in \citet{FangLi2026Tie} for two-sample comparisons, to the regression context. For a win or loss on component \(q\) to determine the hierarchical comparison, all higher-priority components must be genuinely tied, meaning that \(\bigcap_{r=0}^{q-1}\mathcal E_{ij}^{(r)}(L)\) holds based on the full data. Since censoring may leave these component comparisons indeterminate, CTW weights the associated estimating function by the conditional probability that these genuine ties hold. These probabilities are derived from a joint survival model for the event histories, accommodating any fixed $Q$ nonfatal components. Specifically, we set \(T_{qi} = \infty\) if event \(q\) does not occur before death,  and define
\begin{align*}
    \mathcal{O}_i =& \left\{\bm X_i,\ D_i\wedge C_i,\ I(D_i\le C_i),\ \left\{T_{qi}\wedge D_i\wedge C_i,\ I(T_{qi}<D_i\wedge C_i)\right\}_{q=1}^{Q}\right\},\\
    \mathcal{O}_i(L) =& \left\{\bm X_i,\ D_i\wedge C_i\wedge L,\ I(D_i\le C_i\wedge L),\ \left\{T_{qi}\wedge D_i\wedge C_i\wedge L,\ I(T_{qi}<D_i\wedge C_i\wedge L)\right\}_{q=1}^{Q}\right\},
\end{align*}
as the observed data for individual \(i\) and data available through \(L\), respectively.
We first define the following conditional survival probabilities indexed by the non-fatal components:
\begin{align}
\omega_i^{(1)}(L)&=P\left(D_i>L\mid \mathcal O_i(L)\right),\nonumber\\
\omega_i^{(2)}(L)&=P\left(D_i>L,\ T_{1i}>L\mid \mathcal O_i(L)\right),\nonumber\\
&\ldots\nonumber\\
\omega_i^{(Q)}(L)&=P\left(D_i>L,\ T_{1i}>L,\ldots,\ T_{Q-1,i}>L \mid \mathcal O_i(L)\right).\label{eq:ctw-q}
\end{align}
For component $q=1,\ldots,Q$, $\omega_i^{(q)}(L)=0$ if death or a higher-priority nonfatal event is observed by \(L\), and $\omega_i^{(q)}(L)=1$ if individual \(i\) is alive through \(L\) without experiencing any higher-priority nonfatal events.
Under the conditions in Appendix \ref{supp:ctw-identification}, \(\omega_i^{(q)}(L)\omega_j^{(q)}(L)\) is the conditional probability of the required higher-priority ties given \(\mathcal O_i(L)\) and \(\mathcal O_j(L)\).

For \(q=0,\ldots,Q\),  we define \(R_{ij}^{\mathrm{CTW},(q)}(L)=1\) if one individual has an observed event on component \(q\) before \(L\), while the other is alive, uncensored and has not experienced that event on component $q$. Otherwise, set \(R_{ij}^{\mathrm{CTW},(q)}(L)=0\). When \(R_{ij}^{\mathrm{CTW},(q)}(L)=1\), we further denote the corresponding event time by \(t_{ij}^{(q)}\) and thus \(\widetilde H_{ij}^{(q)}(L)=1\) if the event occurs in individual \(j\), and \(\widetilde H_{ij}^{(q)}(L)=0\) otherwise. For \(q\ge1\), it follows that pairs with \(R_{ij}^{\mathrm{CTW},(q)}(L)=1\) include partially observed lower-priority-resolved pairs whose higher-priority comparisons remain indeterminate. Next, we define \(\widetilde H_{ij}^{(T)}(L)=1/2\) as the observed genuine-tie response, and \(R_{ij}^{\mathrm{CTW},T}(L)=1\) if both individuals are followed through \(L\) without death or any nonfatal event, and \(R_{ij}^{\mathrm{CTW},T}(L)=0\) otherwise. When \(R_{ij}^{\mathrm{CTW},T}(L)=1\), the pair has an observed genuine tie on all \(Q+1\) components, with observed response \(\widetilde H_{ij}(L)=1/2\) (Appendix \ref{supp:ctw-observation}). When comparison can be resolved based on death, \(t_{ij}^{(0)}=A_{ij}(L)\), whereas an observed genuine tie requires \(A_{ij}(L)=L\) with censoring probability \(\pi_{ij}^{\mathrm{RCW}}(L)\). For a nonfatal comparison on component \(q\), the joint censoring survival function is instead evaluated at \(t_{ij}^{(q)}\), as \(\pi_{ij}^{\mathrm{CTW},(q)}(L) =S_C(t_{ij}^{(q)}\mid\bm X_i^C)S_C(t_{ij}^{(q)}\mid\bm X_j^C)\) for \(q=1,\ldots,Q\). The CTW pair contribution to the estimating equations can then be written as
\begin{align}
\bm U_{ij}^{\mathrm{CTW}}(\bm\beta_L)
=\bm K_{ij}(\bm\beta_L)\bigg[
&\frac{R_{ij}^{\mathrm{CTW},(0)}(L)}{\pi_{ij}^{\mathrm{RCW}}(L)}
\{\widetilde H_{ij}^{(0)}(L)-\mu_{ij}(\bm\beta_L)\}\nonumber\\
&+\sum_{q=1}^{Q}
\frac{R_{ij}^{\mathrm{CTW},(q)}(L)\omega_i^{(q)}(L)\omega_j^{(q)}(L)}
{\pi_{ij}^{\mathrm{CTW},(q)}(L)}
\{\widetilde H_{ij}^{(q)}(L)-\mu_{ij}(\bm\beta_L)\}\nonumber\\
&+\frac{R_{ij}^{\mathrm{CTW},T}(L)}{\pi_{ij}^{\mathrm{RCW}}(L)}
\{\widetilde{H}_{ij}^{(T)}(L)-\mu_{ij}(\bm\beta_L)\}\bigg].
\label{eq:ctw-score}
\end{align}

For \(R_{ij}^{\mathrm{RCW}}(L)=1\), $H_{ij}(L)=\sum_{q=0}^Q \widetilde H_{ij}^{(q)}(L)+\widetilde H_{ij}^{(T)}(L)$. A notable feature of CTW is assigning separate weights to the response at each component, and the weights for $q=1,\ldots,Q$ leverage the conditional tie probabilities for all higher-priority components.
The first response \(\widetilde{H}_{ij}^{(0)}(L)\), with \(R_{ij}^{\mathrm{CTW},(0)}(L)=1\), applies when an observed death determines the pair's win or loss. Since death has the highest priority, the RCW scheme can be already applied. The second term accounts for nonfatal comparisons, which determine the pair's win or loss only when all higher-priority components are genuinely tied. For \(q=1,\ldots,Q\), \(R_{ij}^{\mathrm{CTW},(q)}(L)\) identifies an observed comparison on component \(q\), while \(\omega_i^{(q)}(L)\omega_j^{(q)}(L)\) gives the conditional probability of the required higher-priority ties given \(\mathcal O_i(L)\) and \(\mathcal O_j(L)\). CTW multiplies the complete regression residual \(\widetilde H_{ij}^{(q)}(L)-\mu_{ij}(\bm\beta_L)\) by this probability, while inverse censoring weighting at \(t_{ij}^{(q)}\) accounts for observation of the nonfatal comparison. This allows partially observed lower-priority-resolved pairs excluded by RCW to contribute when the higher-priority comparisons remain indeterminate. The third term, with \(R_{ij}^{\mathrm{CTW},T}(L)=1\), represents an observed genuine tie on all components. Since both individuals have been observed alive without any nonfatal event through \(L\), the RCW scheme is already applicable. The regression estimator \(\widehat{\bm\beta}_L^{\mathrm{CTW}}\) is defined as the solution to $\bm\Psi_n^{\mathrm{CTW}}(\bm\beta_L)=\frac{2}{n(n-1)}\sum_{(i,j)\in\mathcal I_n}\bm U_{ij}^{\mathrm{CTW}}(\bm\beta_L) = 0$.

To evaluate the conditional tie probabilities, we specify a joint survival model for death and the nonfatal components. Denote the joint survival function by \(\mathcal S(t_0,t_1,\ldots,t_Q\mid\bm x)=P\{T_{0i}>t_0,\ldots,T_{Qi}>t_Q\mid\bm X_i=\bm x\}\), which will be estimated under a copula specification. That is, we posit \(\mathcal S(t_0,\ldots,t_Q\mid\bm x)=\mathcal C_\theta\left\{e^{-\Lambda_0(t_0\mid\bm x)},e^{-\Lambda_1(t_1\mid\bm x)},\ldots,e^{-\Lambda_Q(t_Q\mid\bm x)}\right\}\), for \(0\le t_r\le t_0\), \(r=1,\ldots,Q\), where \(\mathcal C_\theta\) is a survival copula with parameter \(\theta\) controlling dependence among the components and \(\Lambda_0,\ldots,\Lambda_Q\) are cumulative hazard functions for its margins. The semi-competing risks structure allows death to preclude any subsequent nonfatal event, while a nonfatal event does not preclude death.
Under the typical proportional hazards specifications, \(\Lambda_k(t\mid\bm x)=\Lambda_{0k}(t)\exp(\bm\gamma_k^\top\bm x)\), for \(k=0,\ldots,Q\). The baseline cumulative hazards may have parametric forms, such as Weibull, or remain unspecified. For \(Q=1\), these choices include established parametric and semiparametric copula models for semi-competing risks \citep[e.g.][]{Chen2012Semicompeting}. The conditional tie probabilities are then calculated from \(\mathcal S(\bullet)\) conditioning on \(\mathcal O_i(L)\). Write \(\dot{\mathcal S}_{r}=\partial\mathcal S/\partial t_r\), and in the simplest case of \(Q=1\) and when individual \(i\) is censored at \(c<L\), we can derive
\begin{equation*}
\omega_i^{(1)}(L)=
\begin{cases}
\dfrac{\mathcal S(L,c\mid\bm X_i)}{\mathcal S(c,c\mid\bm X_i)},
&\text{if no nonfatal event is observed by }c,\\[6pt]
\dfrac{-\dot{\mathcal S}_{1}(L,v\mid\bm X_i)}{-\dot{\mathcal S}_{1}(c,v\mid\bm X_i)},
&\text{if the nonfatal event is observed at }v<c.
\end{cases}
\end{equation*}
As a more concrete example, we consider that the Gumbel copula is used to parameterize the joint survival function with dependence parameter \(\theta\ge1\). When \(\theta=1\), we have \(\omega_i^{(1)}(L)=\exp\{-\Lambda_0(L\mid\bm X_i)+\Lambda_0(c\mid\bm X_i)\}\), so the observed nonfatal event time \(v\) provides no additional information about survival beyond censoring at \(c\).
For \(\theta>1\), dependence between the components allows \(v\) to inform \(\omega_i^{(1)}(L)\), even though survival after \(c\) is unobserved. CTW uses this information to weight a nonfatal comparison when censoring leaves the fatal comparison indeterminate. For \(Q>1\), suppose individual \(i\) is censored but alive at \(c<L\), and let \(\mathcal J_i=\{r\in\{1,\ldots,Q\}:T_{ri}<c\}\) denote the set of observed nonfatal events. Write \(\dot{\mathcal S}_{\mathcal J_i}={\partial^{|\mathcal J_i|}\mathcal S}/{\prod_{r\in\mathcal J_i}\partial t_r}\), with \(\dot{\mathcal S}_{\varnothing}=\mathcal S\). For \(q=1,\ldots,Q\), if no higher-priority nonfatal event has been observed, so that \(\mathcal J_i\cap\{1,\ldots,q-1\}=\varnothing\), then a general formula is given by
\begin{equation*}
\omega_i^{(q)}(L)=\frac{\dot{\mathcal S}_{\mathcal J_i}\bigl(L,\overbrace{L,\ldots,L}^{q-1},T_{qi}\wedge c,\ldots,T_{Qi}\wedge c\mid\bm X_i\bigr)}{\dot{\mathcal S}_{\mathcal J_i}\bigl(c,T_{1i}\wedge c,\ldots,T_{Qi}\wedge c\mid\bm X_i\bigr)}.
\end{equation*}
A full derivation is provided in Appendix \ref{supp:joint-model-choices}, where the closed-form conditional tie probabilities are also characterized under three typical Archimedean copulas (Gumbel, Clayton, and Frank). Under correctly specified censoring and joint event models, iterated expectation and inverse censoring weighting give
\begin{equation}
E\{\bm U_{ij}^{\mathrm{CTW}}(\bm\beta_L) \mid \bm X_i,\bm X_j\}=\bm K_{ij}(\bm\beta_L)\{E(H_{ij}(L)\mid\bm X_i,\bm X_j)-\mu_{ij}(\bm\beta_L)\}.
\label{eq:ctw-mean}
\end{equation}

Therefore, Equation \eqref{eq:ctw-mean} implies that CTW targets the same regression parameter \(\bm\beta_L^0\) as CHW and RCW. The proof for any fixed \(Q\) is given in Appendix \ref{supp:ctw-identification}. A schematic comparison of the pairs utilized by CHW, RCW and CTW is given in Figure \ref{fig:weight-information-progression}.

\section{Asymptotic theory and efficiency implications}
\label{sec:theory}

For \(\star\in\mathcal A=\{\mathrm{CHW},\mathrm{RCW},\mathrm{CTW}\}\), let \(\bm U_{ij}^\star(\bm\beta_L)\) and \(\widehat{\bm U}_{ij}^\star(\bm\beta_L)\) denote the pair estimating functions based on \(S_C(\bullet)\) and its estimator \(\widehat S_C(\bullet)\), respectively. For \(\star=\mathrm{CTW}\), they additionally use \(\omega_i^{(q)}(L)\omega_j^{(q)}(L)\) and \(\widehat \omega_i^{(q)}(L)\widehat \omega_j^{(q)}(L)\), respectively, for \(q=1,\ldots,Q\). Define the estimating function and its solution by \(\widehat{\bm\Psi}_n^\star(\bm\beta_L)=\frac{2}{n(n-1)}\sum_{(i,j)\in\mathcal I_n}\widehat{\bm U}_{ij}^\star(\bm\beta_L)\) and \(\widehat{\bm\Psi}_n^\star(\widehat{\bm\beta}_L^\star)=\bm0\).  Let \(m_n=\lfloor n/2\rfloor\) be the maximum number of disjoint pairs. Under regularity conditions C.1--C.7 in Appendix \ref{supp:theory} for CHW and RCW and regularity conditions G.1--G.4 in Appendix \ref{supp:ctw-conditions} for CTW, the effect of estimating the weights satisfies
\[
\widehat{\bm\Psi}_n^\star(\bm\beta_L^0)=\bm\Psi_n^\star(\bm\beta_L^0)+\frac1n\sum_{i=1}^n\bm b_L^\star(\mathcal O_i)+o_p(m_n^{-1/2}),
\]
where \(E\{\bm b_L^\star(\mathcal O_i)\}=\bm0\). For CHW and RCW, \(\bm b_L^\star\) accounts for estimation of \(S_C\), with the expansion established in Appendix Lemma \ref{supp:lem:nuisance}. CTW additionally requires estimation of the conditional tie probabilities, giving \(\bm b_L^{\mathrm{CTW}}(\mathcal O_i)=\bm b_{C,L}^{\mathrm{CTW}}(\mathcal O_i)+\bm b_{J,L}^{\mathrm{CTW}}(\mathcal O_i)\), where the two terms account for estimation of the censoring distribution and the joint survival model, respectively, with \(E\{\bm b_{C,L}^{\mathrm{CTW}}(\mathcal O_i)\}=E\{\bm b_{J,L}^{\mathrm{CTW}}(\mathcal O_i)\}=\bm0\).

Regularity conditions G.1--G.4 require consistent estimation of the conditional tie probabilities and ensure that the joint model estimation error contributes through \(\bm b_{J,L}^{\mathrm{CTW}}\) in the expansion above. They accommodate finite-dimensional models and models with unspecified baseline functions. For a copula model, this contribution includes estimation of the dependence parameters, regression coefficients, and baseline functions. The detailed parametric and semiparametric constructions are given in Appendices \ref{supp:ctw-theory} and \ref{supp:cox-theory}. The joint model may use follow-up through a fixed time \(\tau_E\ge L\), allowing later observations to inform estimation and the same fit to serve several restriction times. In this case, \(\bm b_{J,L}^{\mathrm{CTW}}(\mathcal O_i)\) depends on \(\mathcal O_i(\tau_E)\), even though the conditional tie probabilities continue to condition on \(\mathcal O_i(L)\).

For each method, define the adjusted pair contribution \(\bm\zeta_{ij}^\star=\bm U_{ij}^\star(\bm\beta_L^0)+\frac12\{\bm b_L^\star(\mathcal O_i)+\bm b_L^\star(\mathcal O_j)\}\).  Each individual appears in \(n-1\) pairs, so \(\widehat{\bm\Psi}_n^\star(\bm\beta_L^0)=\frac{2}{n(n-1)}\sum_{(i,j)\in\mathcal I_n}\bm\zeta_{ij}^\star+o_p(m_n^{-1/2})\), which incorporates the effects of estimating the weights into the pair contributions.
Each \(\bm\zeta_{ij}^\star\) depends only on \(\mathcal O_i\) and \(\mathcal O_j\), so contributions from disjoint pairs are independent, whereas those sharing an individual may be correlated. A given pair \((i,j)\) overlaps with \(n-2\) other pairs containing \(i\) and \(n-2\) containing \(j\). Including itself gives \(M_n=2n-3\) overlapping pairs, with \(\{2M_n\}/\{n(n-1)\}\longrightarrow0\) and \(\{2M_nm_n\}/\{n(n-1)\}=O(1)\). These relations naturally satisfy the sparse correlation structure of \citet{Thas2012} without weighting.

The three methods share the derivative matrix \(\bm A_L=-\left.\left[\partial E\{\bm U_{ij}^\star(\bm\beta_L)\}\right]/\partial\bm\beta_L^\top\right|_{\bm\beta_L=\bm\beta_L^0}\). Let \(\varphi_{ijkl}=I\{\{i,j\}\cap\{k,l\}\ne\varnothing\}\). Since \(E(\bm\zeta_{ij}^\star)=\bm0\), define
\begin{equation}
\bm B_L^\star=\lim_{n\to\infty}\frac{4m_n}{n^2(n-1)^2}\sum_{(i,j)\in\mathcal I_n}\sum_{(k,l)\in\mathcal I_n}\varphi_{ijkl}E\{\bm\zeta_{ij}^\star(\bm\zeta_{kl}^\star)^\top\}.
\label{eq:B-sparse}
\end{equation}
Terms with \((i,j)=(k,l)\) give individual pair variances, and the remaining nonzero terms give covariances between overlapping pairs. We then obtain the following result.
\begin{theorem}
\label{thm:asymptotics}
Let \(Q\) be fixed and \(\star\in\mathcal A\). Suppose that the censoring model is correctly specified. For \(\star\in\{\mathrm{CHW},\mathrm{RCW}\}\), assume regularity conditions C.1--C.7 in Appendix \ref{supp:theory}. For \(\star=\mathrm{CTW}\), suppose that the joint survival model is also correctly specified and regularity conditions G.1--G.4 in Appendix \ref{supp:ctw-conditions} hold. Then a local solution \(\widehat{\bm\beta}_L^\star\) exists with probability tending to one. As \(n\to\infty\), \(\widehat{\bm\beta}_L^\star\overset{p}{\longrightarrow}\bm\beta_L^0\), and
\begin{equation}
\sqrt{m_n}(\widehat{\bm\beta}_L^\star-\bm\beta_L^0)\overset{d}{\longrightarrow}N\left(\bm0,\bm A_L^{-1}\bm B_L^\star\bm A_L^{-\top}\right).
\label{eq:asymptotic-normal}
\end{equation}
\end{theorem}
The proofs are given in Appendix \ref{supp:theory}. If the PIM is misspecified, \(\bm\beta_L^0\) is interpreted as the root of the full-data population estimating equations. If the censoring survival function and, for CTW, the conditional tie probabilities are known, then \(\bm b_L^\star(\mathcal O_i)=\bm0\) and \(\bm\zeta_{ij}^\star=\bm U_{ij}^\star(\bm\beta_L^0)\). Under heavy censoring, small estimated censoring survival probabilities can produce large inverse weights, allowing a few observed pairs to dominate the estimating equations and increasing sampling variability. Truncation caps these weights at a prespecified value to limit their influence and improve finite-sample stability. Theorem \ref{thm:asymptotics} applies to untruncated weights.

For variance estimation, let \(\widehat{\bm b}_{L,i}^\star\) estimate \(\bm b_L^\star(\mathcal O_i)\), and set \(\widehat{\bm\zeta}_{ij}^\star=\widehat{\bm U}_{ij}^\star(\widehat{\bm\beta}_L^\star)+\frac12(\widehat{\bm b}_{L,i}^\star+\widehat{\bm b}_{L,j}^\star)\). The expressions for \(\widehat{\bm b}_{L,i}^\star\) are given in Appendix Equation \eqref{supp:eq:b-hat} for CHW and RCW, and in Appendix Equation \eqref{supp:eq:ctw-censor-if-hat} for CTW. Replacing the expected products in \eqref{eq:B-sparse} by their empirical counterparts gives
\[
\widehat{\bm B}_L^\star=\frac{4m_n}{n^2(n-1)^2}\sum_{(i,j)\in\mathcal I_n}\sum_{(k,l)\in\mathcal I_n}\varphi_{ijkl} \widehat{\bm\zeta}_{ij}^\star(\widehat{\bm\zeta}_{kl}^\star)^\top.
\]
Holding the fitted weights fixed, and define \(\widehat{\bm A}_L^\star=-\left.\frac{\partial\widehat{\bm\Psi}_n^\star(\bm\beta_L)}{\partial\bm\beta_L^\top}\right|_{\bm\beta_L=\widehat{\bm\beta}_L^\star}\).
The sandwich covariance estimator is \(\widehat{\operatorname{Var}}(\widehat{\bm\beta}_L^\star)=\frac1{m_n}(\widehat{\bm A}_L^\star)^{-1}\widehat{\bm B}_L^\star(\widehat{\bm A}_L^\star)^{-\top}\).  Under Theorem \ref{thm:asymptotics} and the corresponding conditions on the estimated influence contributions, we show in Appendix \ref{supp:theory} and Appendix \ref{supp:variance} that the variance estimator is consistent, such that \(m_n\widehat{\operatorname{Var}}(\widehat{\bm\beta}_L^\star)\xrightarrow{p}\bm A_L^{-1}\bm B_L^\star\bm A_L^{-\top}\). Once the estimated pair contributions are available, \(\widehat{\bm B}_L^\star\) can be computed in \(O(n^2)\) operations for a fixed number of regression coefficients. The calculation is given in Appendix \ref{supp:pair-covariance-computation}.
We compare the conditional variances of the pair estimating functions evaluated using the true censoring survival function and, for CTW, the true joint survival function.
For \(\star\in\mathcal A\), define \(\bm\Gamma^\star=\operatorname{Var}\{\bm U_{ij}^\star(\bm\beta_L^0)\mid\bm X_i,\bm X_j\}\).
\begin{theorem}
\label{thm:efficiency-comparison}
Under regularity conditions in Appendix \ref{supp:efficiency}, suppose that the true censoring survival function and conditional tie probabilities are used and the pair estimating functions have finite second moments. Then
\begin{align}
\bm\Gamma^{\mathrm{CHW}}-\bm\Gamma^{\mathrm{RCW}}
=\bm K_{ij}(\bm\beta_L^0)^{\otimes 2} E\Bigg[\{H_{ij}(L)-\mu_{ij}(\bm\beta_L^0)\}^2\left\{\frac1{\pi_{ij}^{\mathrm{CHW}}(L)}-\frac1{\pi_{ij}^{\mathrm{RCW}}(L)}\right\}\Biggm|\bm X_i,\bm X_j\Bigg]\succeq\bm0,
\label{eq:chw-rcw-pair-variance}
\end{align}
\begin{align}
\bm\Gamma^{\mathrm{RCW}}-\bm\Gamma^{\mathrm{CTW}}={}&\bm K_{ij}(\bm\beta_L^0)^{\otimes 2} E\Bigg[\sum_{q=1}^{Q}R_{ij}^{\mathrm{CTW},(q)}(L)\{\widetilde H_{ij}^{(q)}(L)-\mu_{ij}(\bm\beta_L^0)\}^2\nonumber\\
&\quad\times\frac{\omega_i^{(q)}(L)\omega_j^{(q)}(L)}{\pi_{ij}^{\mathrm{CTW},(q)}(L)}\left\{\frac1{S_C(L\mid\bm X_i^C)S_C(L\mid\bm X_j^C)}-\frac{\omega_i^{(q)}(L)\omega_j^{(q)}(L)}{\pi_{ij}^{\mathrm{CTW},(q)}(L)}\right\}\Biggm|\bm X_i,\bm X_j\Bigg]\succeq\bm0.
\label{eq:rcw-ctw-pair-variance}
\end{align}
The variance reduction in \eqref{eq:rcw-ctw-pair-variance} reflects the use of observed nonfatal comparisons, including partially observed lower-priority-resolved pairs excluded by RCW. Thus,
\begin{equation}
\bm\Gamma^{\mathrm{CHW}}\succeq\bm\Gamma^{\mathrm{RCW}}\succeq
\bm\Gamma^{\mathrm{CTW}}.
\label{eq:pair-variance-comparison}
\end{equation}
\end{theorem}
The first inequality follows from \(\pi_{ij}^{\mathrm{CHW}}(L)\le\pi_{ij}^{\mathrm{RCW}}(L)\). For the second, \(0\le \omega_i^{(q)}(L)\omega_j^{(q)}(L)\le1\) and, whenever \(R_{ij}^{\mathrm{CTW},(q)}(L)=1\), \(\pi_{ij}^{\mathrm{CTW},(q)}(L)\ge S_C(L\mid\bm X_i^C)S_C(L\mid\bm X_j^C)\).  Thus, both differences are positive semi-definite.
RCW reduces pairwise variation by resolving fatal comparisons before both restricted histories are complete. The further reduction with CTW consists of two nonnegative terms. The first comes from evaluating censoring survival functions at the earlier nonfatal event time. The second comes from replacing the indeterminate higher-priority genuine-tie indicator by its conditional probability. For nonfatal comparisons already resolved by RCW, this probability is one, so only the first contribution remains. Partially observed comparisons can also benefit from the second when their tie probability lies strictly between zero and one.
Appendix \ref{supp:efficiency} derives the decomposition results and detailed proof.%
\textbf{blue}{Since the three methods share \(\bm A_L\), a reduction in \(\bm B_L^\star\) yields a corresponding reduction in the asymptotic covariance \(\bm A_L^{-1}\bm B_L^\star\bm A_L^{-\top}\). Whether the pairwise gains carry over to coefficient efficiency also depends on covariances between overlapping pairs and estimation of the censoring and joint survival functions, which enter \(\bm B_L^\star\).}

\section{Simulation studies}
\label{sec:simulation}

We conducted a simulation study to evaluate the finite-sample performance of CHW, RCW, and CTW for a hierarchical endpoint consisting of death and one nonfatal event, with death given priority. We considered Gumbel dependence parameters $\alpha\in\{1.25,2,4\}$, sample sizes $n\in\{200,500,1000\}$, and restriction times $L\in\{2.0,2.8,4.5\}$. For each combination of $\alpha$ and $n$, we generated 2,000 datasets and applied all three methods to each dataset at the three restriction times. We also included an unweighted indeterminate as tie (IAT) benchmark. It assigns response \(1/2\) to genuine ties and to comparisons that remain indeterminate under the prespecified hierarchy, and fits the same PIM to all pairs with unit weights. We present results for the logit and identity links with $n=500$. Results for the other sample sizes and the probit link, together with additional variance comparisons, are provided in Appendix \ref{supp:simulation-results}. To generate the covariates, we drew independent standard normal variables $V_{i1}$ and $E_i$ and set $V_{i2}=-0.7V_{i1}+\sqrt{0.51}E_i$, $X_{i1}=2\Phi(V_{i1})-1$, $B_i=I\{\Phi(V_{i2})>0.8\}$, and $X_{i2}=B_i-0.5$, resulting in a uniform $(-1,1)$ distribution for $X_{i1}$ and a Bernoulli distribution with probability 0.2 for $B_i$. The censoring covariate $W_i\sim\operatorname{Bernoulli}(0.5)-0.5$ was generated independently, and censoring times followed the cumulative hazard $$\Lambda_C(c\mid W_i)=0.560(c-1)_+\exp(0.05W_i).$$
Event histories were generated under the Gumbel model for semi-competing risks described in Appendix \ref{supp:joint-model-choices}, with
\begin{align*}
H_D(d\mid\bm X_i)&=\{\lambda_{D1}\min(d,1)+\lambda_{D2}(d-1)_+\}\exp(-0.5X_{i1}-0.8X_{i2}),\\
H_T(t\mid\bm X_i)&=k_\alpha t^{1.2}\exp(-1.6X_{i1}-3.5X_{i2}),
\end{align*}
where $(\lambda_{D1},\lambda_{D2})=(0.132,0.020)$ and $k_\alpha=1.020$, $1.098$, and $1.121$ for $\alpha=1.25$, 2, and 4, respectively. These parameters were calibrated to give $P(D>2.842)=0.80$ and $P(T_1\le1\mid D>2.842)=0.80$ after averaging over the covariates.
Mean censoring proportions were approximately 35.6\%, 53.0\%, and 69.8\% at the three restriction times \(L\).
We used $\bm Z_{ij}=(X_{i1}-X_{j1},X_{i2}-X_{j2})^\top$ and working variance $\mu_{ij}(1-\mu_{ij})$ for the logit and probit links,  $Z_{ij}=B_i-B_j$ and a constant working variance for the identity link.
For each link and each $(\alpha,L)$ combination, $\bm\beta_L^0$ is obtained by solving the full-data estimating equation based on 2,000,000 independent individuals. In each replication, we fitted a Cox censoring model with covariate $W_i$ and estimated the baseline cumulative hazard by Breslow's method. The primary CTW analysis additionally fitted the Gumbel copula model with sandwich standard errors obtained from Theorem \ref{thm:asymptotics}, accounting for estimation of the censoring distribution and the joint event model.

We report absolute bias (Abias), Monte Carlo standard deviation (MCSD), average estimated standard error (ASE), and empirical coverage probability (ECP) of nominal 95\% Wald intervals. The tables use \(c_L\) for the mean realized censoring proportion.
The VR (\%) column gives \(100\{1-(\mathrm{MCSD}_{\mathrm{CTW}}/\mathrm{MCSD}_{\star})^2\}\).

\subsection{Logit and probit links}
\label{sec:simulation-links}

Tables \ref{tab:sim-logit-n500} and \ref{tab:sim-logit-n500-binary} show the simulation results under the logit link with $n=500$. The maximum absolute bias over both coefficients was 0.037 for CHW, 0.034 for RCW, and 0.045 for CTW.
\begin{table}[!htb]
\caption{Finite-sample performance of complete history weighting (CHW), resolved comparison weighting (RCW), conditional tie weighting (CTW), and indeterminate as tie (IAT) for the continuous covariate coefficient \(\beta_{L1}\) under the logit link with \(n=500\). Death is prioritized over one nonfatal event. Results are based on 2,000 replications for each Gumbel dependence parameter \(\alpha\) and restriction time \(L\).}
\label{tab:sim-logit-n500}
\centering
\begingroup
\renewcommand{\baselinestretch}{1}
\fontsize{9.5}{10.5}\selectfont
\renewcommand{\arraystretch}{1}
\setlength{\tabcolsep}{5pt}
\begin{tabular*}{\textwidth}{@{\extracolsep{\fill}}ccclrrrrrr@{}}
\toprule
$\alpha$ & $L$ & $c_L$ & Method & Truth & Abias & MCSD & ASE & ECP & VR (\%)\\
\midrule
\multirow{4}{*}{1.25} & \multirow{4}{*}{2.0} & \multirow{4}{*}{0.357} & CHW & \multirow{4}{*}{1.097} & 0.013 & 0.142 & 0.139 & 0.947 & 22.8\\
 &  &  & RCW &  & 0.012 & 0.140 & 0.136 & 0.948 & 20.0\\
 &  &  & CTW &  & 0.011 & 0.125 & 0.123 & 0.948 & \textemdash\\
 &  &  & IAT &  & 0.690 & 0.077 & 0.077 & 0.000 & \textemdash\\
\addlinespace[2pt]
\multirow{4}{*}{1.25} & \multirow{4}{*}{2.8} & \multirow{4}{*}{0.530} & CHW & \multirow{4}{*}{1.065} & 0.012 & 0.163 & 0.160 & 0.949 & 36.5\\
 &  &  & RCW &  & 0.008 & 0.158 & 0.152 & 0.945 & 32.5\\
 &  &  & CTW &  & 0.010 & 0.130 & 0.125 & 0.944 & \textemdash\\
 &  &  & IAT &  & 0.811 & 0.071 & 0.071 & 0.000 & \textemdash\\
\addlinespace[2pt]
\multirow{4}{*}{1.25} & \multirow{4}{*}{4.5} & \multirow{4}{*}{0.699} & CHW & \multirow{4}{*}{1.003} & 0.015 & 0.229 & 0.223 & 0.947 & 60.4\\
 &  &  & RCW &  & 0.007 & 0.206 & 0.199 & 0.937 & 51.4\\
 &  &  & CTW &  & 0.010 & 0.144 & 0.137 & 0.934 & \textemdash\\
 &  &  & IAT &  & 0.823 & 0.070 & 0.069 & 0.000 & \textemdash\\
\addlinespace[4pt]
\multirow{4}{*}{2} & \multirow{4}{*}{2.0} & \multirow{4}{*}{0.356} & CHW & \multirow{4}{*}{1.176} & 0.006 & 0.141 & 0.142 & 0.943 & 18.6\\
 &  &  & RCW &  & 0.004 & 0.138 & 0.139 & 0.948 & 15.0\\
 &  &  & CTW &  & 0.007 & 0.128 & 0.128 & 0.939 & \textemdash\\
 &  &  & IAT &  & 0.756 & 0.077 & 0.077 & 0.000 & \textemdash\\
\addlinespace[2pt]
\multirow{4}{*}{2} & \multirow{4}{*}{2.8} & \multirow{4}{*}{0.529} & CHW & \multirow{4}{*}{1.138} & 0.009 & 0.160 & 0.161 & 0.947 & 35.1\\
 &  &  & RCW &  & 0.007 & 0.152 & 0.154 & 0.945 & 27.9\\
 &  &  & CTW &  & 0.009 & 0.129 & 0.130 & 0.950 & \textemdash\\
 &  &  & IAT &  & 0.879 & 0.071 & 0.071 & 0.000 & \textemdash\\
\addlinespace[2pt]
\multirow{4}{*}{2} & \multirow{4}{*}{4.5} & \multirow{4}{*}{0.698} & CHW & \multirow{4}{*}{1.067} & 0.010 & 0.227 & 0.221 & 0.946 & 60.2\\
 &  &  & RCW &  & 0.001 & 0.200 & 0.198 & 0.949 & 48.9\\
 &  &  & CTW &  & 0.012 & 0.143 & 0.141 & 0.947 & \textemdash\\
 &  &  & IAT &  & 0.886 & 0.070 & 0.069 & 0.000 & \textemdash\\
\addlinespace[4pt]
\multirow{4}{*}{4} & \multirow{4}{*}{2.0} & \multirow{4}{*}{0.356} & CHW & \multirow{4}{*}{1.238} & 0.005 & 0.147 & 0.144 & 0.940 & 16.2\\
 &  &  & RCW &  & 0.003 & 0.144 & 0.142 & 0.946 & 13.3\\
 &  &  & CTW &  & 0.005 & 0.134 & 0.131 & 0.944 & \textemdash\\
 &  &  & IAT &  & 0.808 & 0.077 & 0.077 & 0.000 & \textemdash\\
\addlinespace[2pt]
\multirow{4}{*}{4} & \multirow{4}{*}{2.8} & \multirow{4}{*}{0.530} & CHW & \multirow{4}{*}{1.197} & 0.013 & 0.166 & 0.162 & 0.948 & 30.0\\
 &  &  & RCW &  & 0.009 & 0.159 & 0.155 & 0.945 & 23.9\\
 &  &  & CTW &  & 0.011 & 0.139 & 0.134 & 0.947 & \textemdash\\
 &  &  & IAT &  & 0.933 & 0.071 & 0.071 & 0.000 & \textemdash\\
\addlinespace[2pt]
\multirow{4}{*}{4} & \multirow{4}{*}{4.5} & \multirow{4}{*}{0.699} & CHW & \multirow{4}{*}{1.125} & 0.011 & 0.229 & 0.221 & 0.940 & 55.3\\
 &  &  & RCW &  & 0.002 & 0.203 & 0.197 & 0.944 & 43.5\\
 &  &  & CTW &  & 0.012 & 0.153 & 0.146 & 0.936 & \textemdash\\
 &  &  & IAT &  & 0.942 & 0.070 & 0.069 & 0.000 & \textemdash\\
\bottomrule
\end{tabular*}
\endgroup
\par\smallskip
\begin{minipage}{\textwidth}
\renewcommand{\baselinestretch}{1}
\fontsize{9}{11}\selectfont\raggedright
\(c_L\) is the mean realized censoring proportion, Abias denotes absolute bias, MCSD the Monte Carlo standard deviation, ASE the average estimated standard error, and ECP the empirical coverage probability of nominal \(95\%\) Wald intervals.
\par\smallskip
For a CHW or RCW row, VR (\%) is \(100\{1-(\mathrm{MCSD}_{\mathrm{CTW}}/\mathrm{MCSD}_{\star})^2\}\), where \(\star\) denotes that row's method. Positive values favor CTW. IAT assigns response \(1/2\) to every unresolved observed pair and gives every pair unit weight.
\end{minipage}
\end{table}

ASE generally agreed with MCSD, and coverage was generally close to the nominal \(95\%\) level. As expected, RCW yielded a smaller MCSD than CHW under every configuration, and CTW further reduced the MCSD for both coefficients.
The VR columns in Tables \ref{tab:sim-logit-n500} and \ref{tab:sim-logit-n500-binary} show the resulting variance reductions.
For example, at $\alpha=1.25$ and $L=4.5$, the MCSDs of the continuous covariate coefficient were 0.229, 0.206, and 0.144 for CHW, RCW, and CTW, respectively. The corresponding variance reductions with CTW were 60.4\% relative to CHW and 51.4\% relative to RCW. Across all settings, the reductions relative to RCW ranged from 6.5\% to 51.4\%, and
the gains increased with the restriction time and were larger for the continuous-covariate coefficient than for the binary-covariate coefficient. The probit results in Appendix \ref{supp:simulation-results} showed the same pattern of MCSD. At $n=500$, the maximum absolute biases were 0.020, 0.018, and 0.025 for CHW, RCW, and CTW, respectively. Their ECP ranges were 0.935--0.949, 0.933--0.948, and 0.929--0.946.
and its variance reductions relative to RCW ranged from 5.3\% to 50.8\%.
Across all sample sizes and configurations, CTW had smaller MCSD than both CHW and RCW.

\begin{table}[!htb]
\caption{Finite-sample performance of complete history weighting (CHW), resolved comparison weighting (RCW), conditional tie weighting (CTW), and indeterminate as tie (IAT) for the binary covariate coefficient \(\beta_{L2}\) under the logit link with \(n=500\). Death is prioritized over one nonfatal event. Results are based on 2,000 replications for each Gumbel dependence parameter \(\alpha\) and restriction time \(L\).}
\label{tab:sim-logit-n500-binary}
\centering
\begingroup
\renewcommand{\baselinestretch}{1}
\fontsize{9.5}{10.5}\selectfont
\renewcommand{\arraystretch}{1}
\setlength{\tabcolsep}{5pt}
\begin{tabular*}{\textwidth}{@{\extracolsep{\fill}}ccclrrrrrr@{}}
\toprule
$\alpha$ & $L$ & $c_L$ & Method & Truth & Abias & MCSD & ASE & ECP & VR (\%)\\
\midrule
\multirow{4}{*}{1.25} & \multirow{4}{*}{2.0} & \multirow{4}{*}{0.357} & CHW & \multirow{4}{*}{2.355} & 0.027 & 0.286 & 0.278 & 0.938 & 18.0\\
 &  &  & RCW &  & 0.025 & 0.278 & 0.272 & 0.940 & 13.6\\
 &  &  & CTW &  & 0.022 & 0.259 & 0.255 & 0.947 & \textemdash\\
 &  &  & IAT &  & 1.564 & 0.120 & 0.120 & 0.000 & \textemdash\\
\addlinespace[2pt]
\multirow{4}{*}{1.25} & \multirow{4}{*}{2.8} & \multirow{4}{*}{0.530} & CHW & \multirow{4}{*}{2.285} & 0.019 & 0.324 & 0.310 & 0.943 & 31.2\\
 &  &  & RCW &  & 0.017 & 0.308 & 0.296 & 0.942 & 23.7\\
 &  &  & CTW &  & 0.019 & 0.269 & 0.261 & 0.944 & \textemdash\\
 &  &  & IAT &  & 1.836 & 0.101 & 0.101 & 0.000 & \textemdash\\
\addlinespace[2pt]
\multirow{4}{*}{1.25} & \multirow{4}{*}{4.5} & \multirow{4}{*}{0.699} & CHW & \multirow{4}{*}{2.140} & 0.011 & 0.433 & 0.413 & 0.942 & 52.7\\
 &  &  & RCW &  & 0.008 & 0.386 & 0.363 & 0.934 & 40.5\\
 &  &  & CTW &  & 0.021 & 0.298 & 0.282 & 0.933 & \textemdash\\
 &  &  & IAT &  & 1.860 & 0.094 & 0.094 & 0.000 & \textemdash\\
\addlinespace[4pt]
\multirow{4}{*}{2} & \multirow{4}{*}{2.0} & \multirow{4}{*}{0.356} & CHW & \multirow{4}{*}{2.539} & 0.022 & 0.306 & 0.292 & 0.938 & 12.8\\
 &  &  & RCW &  & 0.020 & 0.299 & 0.286 & 0.939 & 9.1\\
 &  &  & CTW &  & 0.023 & 0.285 & 0.273 & 0.941 & \textemdash\\
 &  &  & IAT &  & 1.714 & 0.125 & 0.122 & 0.000 & \textemdash\\
\addlinespace[2pt]
\multirow{4}{*}{2} & \multirow{4}{*}{2.8} & \multirow{4}{*}{0.529} & CHW & \multirow{4}{*}{2.451} & 0.025 & 0.338 & 0.321 & 0.939 & 26.2\\
 &  &  & RCW &  & 0.022 & 0.320 & 0.305 & 0.942 & 17.6\\
 &  &  & CTW &  & 0.030 & 0.291 & 0.279 & 0.941 & \textemdash\\
 &  &  & IAT &  & 1.989 & 0.105 & 0.101 & 0.000 & \textemdash\\
\addlinespace[2pt]
\multirow{4}{*}{2} & \multirow{4}{*}{4.5} & \multirow{4}{*}{0.698} & CHW & \multirow{4}{*}{2.283} & 0.019 & 0.439 & 0.419 & 0.949 & 46.2\\
 &  &  & RCW &  & 0.014 & 0.387 & 0.366 & 0.946 & 30.8\\
 &  &  & CTW &  & 0.042 & 0.322 & 0.301 & 0.937 & \textemdash\\
 &  &  & IAT &  & 1.999 & 0.097 & 0.094 & 0.000 & \textemdash\\
\addlinespace[4pt]
\multirow{4}{*}{4} & \multirow{4}{*}{2.0} & \multirow{4}{*}{0.356} & CHW & \multirow{4}{*}{2.664} & 0.037 & 0.310 & 0.304 & 0.945 & 10.2\\
 &  &  & RCW &  & 0.034 & 0.304 & 0.297 & 0.948 & 6.5\\
 &  &  & CTW &  & 0.035 & 0.294 & 0.287 & 0.942 & \textemdash\\
 &  &  & IAT &  & 1.815 & 0.122 & 0.124 & 0.000 & \textemdash\\
\addlinespace[2pt]
\multirow{4}{*}{4} & \multirow{4}{*}{2.8} & \multirow{4}{*}{0.530} & CHW & \multirow{4}{*}{2.576} & 0.035 & 0.341 & 0.332 & 0.945 & 21.0\\
 &  &  & RCW &  & 0.031 & 0.325 & 0.315 & 0.947 & 13.0\\
 &  &  & CTW &  & 0.038 & 0.303 & 0.294 & 0.944 & \textemdash\\
 &  &  & IAT &  & 2.105 & 0.101 & 0.102 & 0.000 & \textemdash\\
\addlinespace[2pt]
\multirow{4}{*}{4} & \multirow{4}{*}{4.5} & \multirow{4}{*}{0.699} & CHW & \multirow{4}{*}{2.414} & 0.017 & 0.450 & 0.427 & 0.945 & 45.3\\
 &  &  & RCW &  & 0.010 & 0.383 & 0.369 & 0.948 & 24.7\\
 &  &  & CTW &  & 0.045 & 0.333 & 0.317 & 0.938 & \textemdash\\
 &  &  & IAT &  & 2.127 & 0.094 & 0.094 & 0.000 & \textemdash\\
\bottomrule
\end{tabular*}
\endgroup
\par\smallskip
\begin{minipage}{\textwidth}
\renewcommand{\baselinestretch}{1}
\fontsize{9}{11}\selectfont\raggedright
\(c_L\) is the mean realized censoring proportion, Abias denotes absolute bias, MCSD the Monte Carlo standard deviation, ASE the average estimated standard error, and ECP the empirical coverage probability of nominal \(95\%\) Wald intervals.
\par\smallskip
For a CHW or RCW row, VR (\%) is \(100\{1-(\mathrm{MCSD}_{\mathrm{CTW}}/\mathrm{MCSD}_{\star})^2\}\), where \(\star\) denotes that row's method. Positive values favor CTW. IAT assigns response \(1/2\) to every unresolved observed pair and gives every pair unit weight.
\end{minipage}
\end{table}

\FloatBarrier
\subsection{Identity link and interval coverage}
\label{sec:simulation-identity}

Table \ref{tab:sim-identity-n500} reports the identity-link results at $n=500$. The maximum absolute bias was 0.013 for CHW, 0.008 for RCW, and 0.003 for CTW. The largest biases for CHW and RCW occurred at later restriction time $L=4.493$, whereas CTW showed small absolute bias across all configurations. Similarly, RCW had smaller MCSD than CHW, and CTW had the smallest MCSD throughout.
The ECP across all settings for CHW, RCW, and CTW ranges from 0.928--0.945, 0.927--0.941, and 0.919--0.944, respectively. The VR column in Table \ref{tab:sim-identity-n500} shows CTW variance reductions ranging from 15.0\% to 59.2\% relative to CHW and from 11.1\% to 47.2\% relative to RCW. The gains increased with the restriction time.%
The additional  \(n=200\) and \(n=1000\) results also gave smaller MCSD for CTW than for CHW or RCW (Appendix \ref{supp:simulation-results}). IAT showed substantial bias toward the null under all three links, with zero coverage of the full data target in every configuration at \(n=500\).

\begin{table}[!htb]
\caption{Finite-sample performance of complete history weighting (CHW), resolved comparison weighting (RCW), conditional tie weighting (CTW), and indeterminate as tie (IAT) for the binary covariate coefficient \(\beta_{L,B}\) under the identity link with \(n=500\). Death is prioritized over one nonfatal event. Results are based on 2,000 replications for each Gumbel dependence parameter \(\alpha\) and restriction time \(L\).}
\label{tab:sim-identity-n500}
\centering
\begingroup
\renewcommand{\baselinestretch}{1}
\fontsize{9.5}{10.5}\selectfont
\renewcommand{\arraystretch}{1}
\setlength{\tabcolsep}{5pt}
\begin{tabular*}{\textwidth}{@{\extracolsep{\fill}}ccclrrrrrr@{}}
\toprule
$\alpha$ & $L$ & $c_L$ & Method & Truth & Abias & MCSD & ASE & ECP & VR (\%)\\
\midrule
\multirow{4}{*}{1.25} & \multirow{4}{*}{2.0} & \multirow{4}{*}{0.357} & CHW & \multirow{4}{*}{0.617} & 0.001 & 0.070 & 0.070 & 0.945 & 19.6\\
 &  &  & RCW &  & 0.001 & 0.068 & 0.068 & 0.941 & 15.7\\
 &  &  & CTW &  & 0.001 & 0.063 & 0.063 & 0.938 & \textemdash\\
 &  &  & IAT &  & 0.374 & 0.047 & 0.048 & 0.000 & \textemdash\\
\addlinespace[2pt]
\multirow{4}{*}{1.25} & \multirow{4}{*}{2.8} & \multirow{4}{*}{0.530} & CHW & \multirow{4}{*}{0.605} & 0.004 & 0.083 & 0.081 & 0.938 & 35.8\\
 &  &  & RCW &  & 0.003 & 0.079 & 0.077 & 0.938 & 28.5\\
 &  &  & CTW &  & 0.002 & 0.067 & 0.066 & 0.944 & \textemdash\\
 &  &  & IAT &  & 0.471 & 0.041 & 0.042 & 0.000 & \textemdash\\
\addlinespace[2pt]
\multirow{4}{*}{1.25} & \multirow{4}{*}{4.5} & \multirow{4}{*}{0.699} & CHW & \multirow{4}{*}{0.579} & 0.013 & 0.122 & 0.118 & 0.935 & 59.2\\
 &  &  & RCW &  & 0.008 & 0.108 & 0.105 & 0.930 & 47.2\\
 &  &  & CTW &  & 0.002 & 0.078 & 0.075 & 0.926 & \textemdash\\
 &  &  & IAT &  & 0.502 & 0.038 & 0.038 & 0.000 & \textemdash\\
\addlinespace[4pt]
\multirow{4}{*}{2} & \multirow{4}{*}{2.0} & \multirow{4}{*}{0.356} & CHW & \multirow{4}{*}{0.647} & 0.001 & 0.069 & 0.067 & 0.935 & 16.7\\
 &  &  & RCW &  & 0.001 & 0.068 & 0.065 & 0.937 & 12.4\\
 &  &  & CTW &  & 0.000 & 0.063 & 0.061 & 0.936 & \textemdash\\
 &  &  & IAT &  & 0.392 & 0.049 & 0.048 & 0.000 & \textemdash\\
\addlinespace[2pt]
\multirow{4}{*}{2} & \multirow{4}{*}{2.8} & \multirow{4}{*}{0.529} & CHW & \multirow{4}{*}{0.633} & 0.003 & 0.082 & 0.077 & 0.931 & 33.6\\
 &  &  & RCW &  & 0.002 & 0.077 & 0.074 & 0.933 & 25.0\\
 &  &  & CTW &  & 0.001 & 0.067 & 0.064 & 0.934 & \textemdash\\
 &  &  & IAT &  & 0.495 & 0.043 & 0.042 & 0.000 & \textemdash\\
\addlinespace[2pt]
\multirow{4}{*}{2} & \multirow{4}{*}{4.5} & \multirow{4}{*}{0.698} & CHW & \multirow{4}{*}{0.604} & 0.010 & 0.118 & 0.112 & 0.928 & 55.5\\
 &  &  & RCW &  & 0.005 & 0.104 & 0.098 & 0.927 & 42.2\\
 &  &  & CTW &  & 0.003 & 0.079 & 0.074 & 0.919 & \textemdash\\
 &  &  & IAT &  & 0.526 & 0.040 & 0.038 & 0.000 & \textemdash\\
\addlinespace[4pt]
\multirow{4}{*}{4} & \multirow{4}{*}{2.0} & \multirow{4}{*}{0.356} & CHW & \multirow{4}{*}{0.663} & 0.003 & 0.066 & 0.065 & 0.933 & 15.0\\
 &  &  & RCW &  & 0.003 & 0.064 & 0.064 & 0.932 & 11.1\\
 &  &  & CTW &  & 0.003 & 0.061 & 0.060 & 0.940 & \textemdash\\
 &  &  & IAT &  & 0.402 & 0.047 & 0.048 & 0.000 & \textemdash\\
\addlinespace[2pt]
\multirow{4}{*}{4} & \multirow{4}{*}{2.8} & \multirow{4}{*}{0.530} & CHW & \multirow{4}{*}{0.652} & 0.002 & 0.077 & 0.075 & 0.934 & 31.8\\
 &  &  & RCW &  & 0.001 & 0.073 & 0.071 & 0.935 & 22.9\\
 &  &  & CTW &  & 0.002 & 0.064 & 0.064 & 0.936 & \textemdash\\
 &  &  & IAT &  & 0.511 & 0.042 & 0.042 & 0.000 & \textemdash\\
\addlinespace[2pt]
\multirow{4}{*}{4} & \multirow{4}{*}{4.5} & \multirow{4}{*}{0.699} & CHW & \multirow{4}{*}{0.626} & 0.011 & 0.116 & 0.108 & 0.928 & 57.8\\
 &  &  & RCW &  & 0.006 & 0.097 & 0.094 & 0.933 & 40.0\\
 &  &  & CTW &  & 0.002 & 0.075 & 0.073 & 0.922 & \textemdash\\
 &  &  & IAT &  & 0.547 & 0.038 & 0.038 & 0.000 & \textemdash\\
\bottomrule
\end{tabular*}
\endgroup
\par\smallskip
\begin{minipage}{\textwidth}
\renewcommand{\baselinestretch}{1}
\fontsize{9}{11}\selectfont\raggedright
The coefficient \(\beta_{L,B}\) corresponds to the binary-covariate contrast \(B_i-B_j\) and is on the marginal net benefit scale. \(c_L\) is the mean realized censoring proportion, Abias denotes absolute bias, MCSD the Monte Carlo standard deviation, ASE the average estimated standard error, and ECP the empirical coverage probability of nominal \(95\%\) Wald intervals.
\par\smallskip
For a CHW or RCW row, VR (\%) is \(100\{1-(\mathrm{MCSD}_{\mathrm{CTW}}/\mathrm{MCSD}_{\star})^2\}\), where \(\star\) denotes that row's method. Positive values favor CTW. IAT assigns response \(1/2\) to every unresolved observed pair and gives every pair unit weight.
\end{minipage}
\end{table}

We also fitted Clayton and Frank copulas for CTW as sensitivity analysis at \(n=200\) and \(n=500\) under the same data generating process. Bias and ECP showed little changes across the fitted copulas, and CTW retained its variance reductions relative to CHW and RCW in all configurations (Appendix \ref{supp:copula-sensitivity}).

\FloatBarrier
\section{Application to the  HF-ACTION study}
\label{sec:hfaction}

HF-ACTION (Heart Failure: A Controlled Trial Investigating Outcomes of Exercise Training) was a multicenter randomized clinical trial of patients with chronic heart failure who were enrolled between April 2003 and February 2007 and assigned to usual care alone or usual care plus aerobic exercise training \citep{flynn2009effects}. We reanalyzed the study under a prioritized semicompeting risks formulation in which death was the fatal endpoint and time to first hospitalization before death was the lower priority nonfatal endpoint. Two participants were compared first according to their restricted death times and then, when their restricted death times were tied, according to their restricted times to first hospitalization.

The analytic dataset contained \(2{,}109\) participants with complete baseline covariates and observed endpoint histories, including \(1{,}051\) assigned to exercise training and \(1{,}058\) assigned to usual care. The median follow-up was \(2.59\) years. There were \(168\) deaths in the exercise-training group and \(182\) in the usual-care group, corresponding to mortality proportions of \(16.0\%\) and \(17.2\%\), respectively. First hospitalization was observed for \(657\) exercise-training participants and \(692\) usual-care participants, corresponding to \(62.5\%\) and \(65.4\%\), respectively. In addition to randomized treatment, the regression model included ischemic etiology, sex, geographic region, history of atrial fibrillation or flutter, history of diabetes, age, and cardiopulmonary exercise (CPX) test duration. Binary covariates were coded as exercise training versus usual care, ischemic versus nonischemic etiology, male versus female, outside the US versus the US, and presence versus absence of the corresponding medical history.

We considered CHW, RCW, and CTW with logit, probit, and identity links, using the same baseline covariates in the PIM and censoring models. The censoring survival function was estimated by Cox regression with a Breslow baseline. The restriction time grid contained values from \(0.95\) to \(3.86\) years, corresponding to the \(25\)th and \(99\)th percentiles of observed death times. Over this range, the censoring proportion increased from \(1.8\%\) to \(65.9\%\). For CTW, the Gumbel-copula model uses a piecewise-constant baseline death hazard with prespecified knots at 1, 2, and 3 years and a Weibull nonfatal component. Appendix \ref{supp:hfaction-model} gives the detailed joint model.

\begin{figure}[!htbp]
\centering
\includegraphics[width=0.98\textwidth,height=0.72\textheight,keepaspectratio]{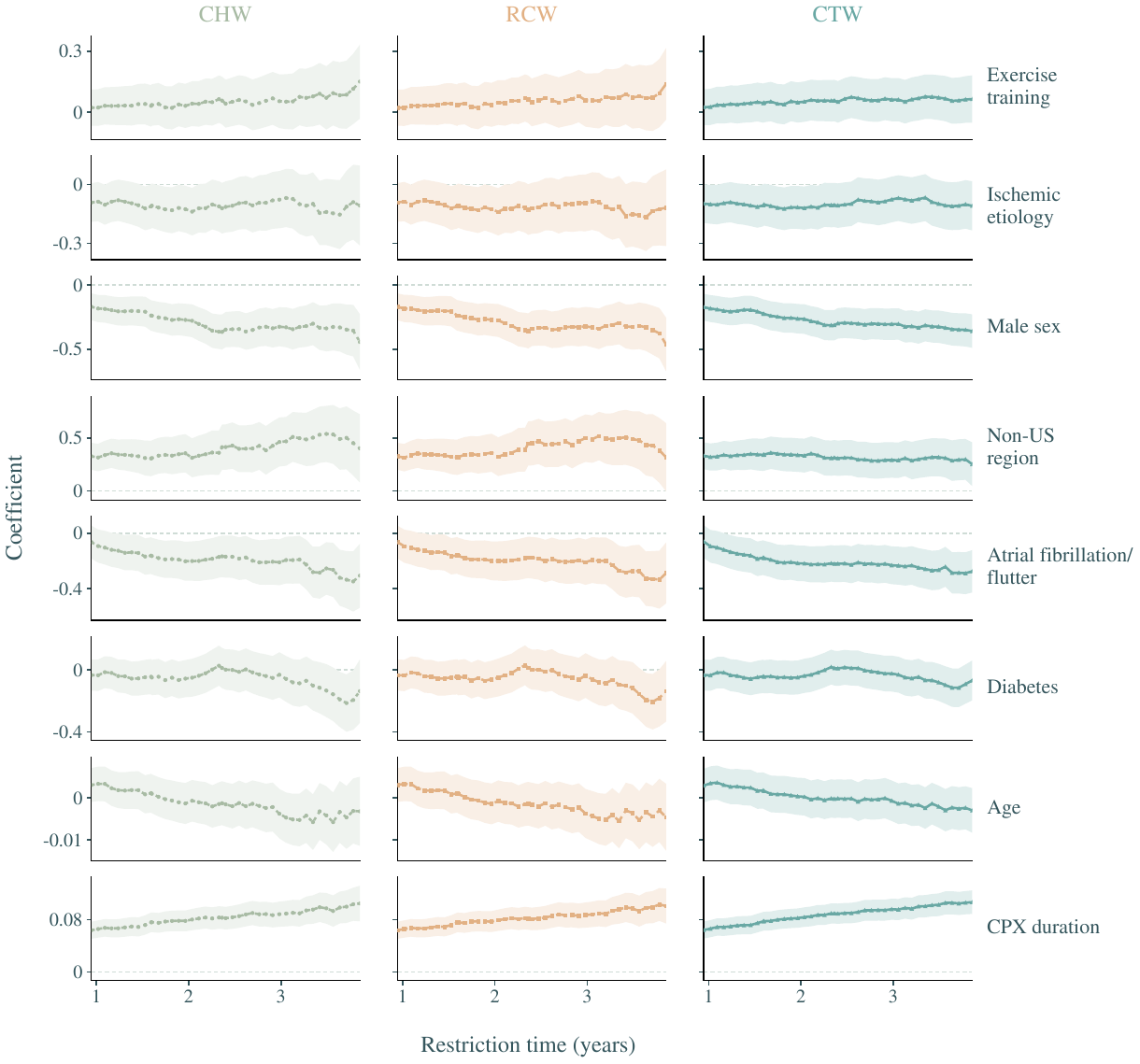}
\caption{HF-ACTION logit coefficient estimates and pointwise \(95\%\) confidence bands over \(L=0.945\)--\(3.855\) years. Columns identify methods, and vertical scales are shared within covariate rows. Coefficients are on the conditional log win-odds scale.}
\label{fig:hfaction-logit-trajectories}
\end{figure}

At \(L=3.86\) years, CHW kept \(258{,}840\) of the \(2{,}222{,}886\) unordered pairs (\(11.6\%\)), and RCW resolved \(600{,}349\) (\(27.0\%\)). CTW assigned a positive contribution to \(1{,}736{,}748\) pairs (\(78.1\%\)), including \(1{,}136{,}399\) additional lower-priority contributions whose full hierarchical outcome remained unresolved under RCW. These counts indicate the observed information used by each equation. Fractional tie weights do not turn those additional contributions into fully observed outcomes. Figure \ref{fig:hfaction-logit-trajectories} presents the primary logit-link analysis. The three methods broadly agree on the direction of the principal covariate associations, with differences in their magnitudes. Their treatment estimates were positive at all \(43\) horizons, but every pointwise treatment interval included zero. At \(L=3.86\) years, the treatment coefficients were \(0.151\), \(0.139\), and \(0.065\) for CHW, RCW, and CTW, respectively, with all three pointwise confidence intervals including zero. The CTW coefficient had standard error \(0.059\) and \(95\%\) confidence interval \((-0.051,0.181)\), corresponding to conditional win odds of \(1.07\) with interval \((0.95,1.20)\). The corresponding RCW win odds were \(1.15\) with interval \((0.96,1.37)\). Thus the smaller CTW standard error accompanied a treatment estimate closer to zero. The probit analysis gave similar covariate associations, and the centered-identity treatment intervals included zero over \(L=0.945\)--\(1.455\) years, see Appendix \ref{supp:hfaction-results} in details.

At the final restriction time, male sex and atrial fibrillation or flutter were unfavorably associated with the prioritized outcome under all three methods, whereas longer CPX duration was favorable. The CTW conditional win odds were \(0.70\) for male versus female participants, with interval \((0.61,0.79)\), and \(0.76\) for atrial fibrillation or flutter, with interval \((0.65,0.89)\). For each additional minute of CPX duration, the CTW win odds were \(1.11\), with interval \((1.09,1.13)\). The non-US region coefficient was positive under all three methods. Its final interval excluded zero under CHW and CTW but narrowly included zero under RCW. The final intervals for ischemic etiology, diabetes, and age contained zero under all three methods. Table \ref{tab:hfaction-efficiency-two-times} reports percentage reductions in estimated variance at two restriction times for each link. For example, the CTW treatment variance estimate under the logit link was about \(2.0\%\) smaller than those of CHW and RCW at \(L=1.018\) years, and the reductions were \(59.0\%\) and \(56.0\%\) at \(L=3.855\) years, respectively. Across the eight coefficients at this later time, CTW estimated variances were \(51.4\%\)--\(61.0\%\) smaller than RCW's. These comparisons depend on the working event model and the recorded-time conventions.

\FloatBarrier
\section{Discussion}
\label{sec:discussion}

We formulated PIM for hierarchical survival outcomes by defining the regression estimand in the restricted full data. This formulation makes explicit that the interpretation of a regression coefficient depends on the clinical hierarchy, the restriction time \(L\), and the allocation of genuine ties. These choices determine which aspects of the clinical histories are summarized and are therefore part of the scientific question. In particular, \(L\) specifies the period over which the outcomes are compared and can affect the relative contributions of fatal and nonfatal events. Extending \(L\), for example, may allow death to determine a comparison that was previously determined by a nonfatal event. Associations at different restriction times describe different clinical comparisons, and their variation should be interpreted in relation to the underlying hierarchy. The allocation of genuine ties also has a substantive role, since it determines how histories with no difference under the comparison rule contribute to the probabilistic index. Together with the choice of link, this gives the regression coefficients their interpretation, including conditional win odds and net benefit under the logit and identity links. Specifying these features independently of censoring is important because the follow-up available in a study can otherwise influence the quantity being estimated \citep{mao2024defining}. Our formulation follows the distinction between an estimand and an estimator emphasized in ICH E9(R1) \citep{ICH2019E9R1}. It defines the association of interest before introducing assumptions about incomplete observation, while allowing that association to vary across restriction times without a proportionality assumption.

To estimate this full data target, we constructed CHW, RCW, and CTW by explicitly distinguishing the information available from different observed pairs. The population identities establish that their weighted estimating equations recover the same full data equation under their respective assumptions. RCW relaxes the complete history requirement of CHW when the hierarchical comparison is already determined, whereas CTW further incorporates \emph{partially observed lower-priority-resolved pairs} through the conditional probabilities of the required higher-priority ties. These probabilities weight the complete regression residual, allowing uncertainty about the higher-priority comparisons to be incorporated while preserving the full data target. Our variance comparison also clarifies that efficiency depends on how comparisons are weighted as well as which pairs contribute. In particular, CTW reduces the inverse censoring weights for nonfatal comparisons already resolved by RCW by evaluating censoring survival at the earlier event time.
The additional joint survival model required by CTW introduces a trade-off between potential variance reduction and bias from misspecified conditional tie probabilities. In our simulation evaluation, bias and coverage were similar across the fitted Gumbel, Clayton, and Frank copulas in our simulations, suggesting limited sensitivity to copula choice in the settings examined. Implementations are available in the \texttt{gpim} package and accompanying R/C++ code, with a worked tutorial in Appendix \ref{supp:gpim-tutorial}. Further work could extend the methods to hierarchies combining survival outcomes with binary, ordinal, or continuous measurements.

\section*{Data Availability Statement}
The HF-ACTION data are available through the NHLBI Biologic Specimen and Data Repository Information Coordinating Center under an approved data use agreement. The CHW, RCW, and CTW estimators are implemented in the R package \texttt{gpim}, available at \url{https://github.com/fancy575/gpim}. A reproducible example is provided in Appendix \ref{supp:gpim-tutorial}. The accompanying R/C++ simulation and HF-ACTION analysis code implements all three methods, including CTW and its copula sensitivity analysis.

\clearpage

\pdfbookmark[1]{References}{main-references}

\clearpage
\appendix
\setcounter{secnumdepth}{3}
\renewcommand{\thesection}{A\arabic{section}}
\renewcommand{\thesubsection}{\thesection.\arabic{subsection}}
\renewcommand{\thesubsubsection}{\thesubsection.\arabic{subsubsection}}
\setcounter{equation}{0}
\setcounter{theorem}{0}
\setcounter{lemma}{0}
\setcounter{proposition}{0}
\setcounter{condition}{0}
\setcounter{remark}{0}
\setcounter{table}{0}
\setcounter{figure}{0}
\renewcommand{\theequation}{A.\arabic{equation}}
\renewcommand{\thefigure}{S\arabic{figure}}
\renewcommand{\thetable}{S\arabic{table}}
\renewcommand{\theHsection}{app.\arabic{section}}
\renewcommand{\theHequation}{app.\arabic{equation}}
\renewcommand{\theHtheorem}{app.\arabic{theorem}}
\renewcommand{\theHlemma}{app.\arabic{lemma}}
\renewcommand{\theHproposition}{app.\arabic{proposition}}
\renewcommand{\theHcondition}{app.\arabic{condition}}
\renewcommand{\theHremark}{app.\arabic{remark}}
\renewcommand{\theHfigure}{app.\arabic{figure}}
\renewcommand{\theHtable}{app.\arabic{table}}
\etocdepthtag.toc{appendix}
\pdfbookmark[0]{Appendix}{appendix-start}
\begin{center}
{\Large\bfseries Appendix\par}
\medskip
{\bfseries Supplementary material for\par}
{\bfseries Probabilistic index models for hierarchical survival outcomes\par}
\medskip
Xi Fang$^{1,\dagger}$, Zhiqiang Cao$^{2,\dagger}$, Lu Mao$^{3}$, and Fan Li$^{4,*}$\par
$^{1}$Medical College of Wisconsin, $^{2}$Yale School of Medicine\par
$^{3}$University of Wisconsin-Madison, $^{4}$Yale School of Public Health\par
$^{*}$\href{mailto:fan.f.li@yale.edu}{\nolinkurl{fan.f.li@yale.edu}}\par
$^{\dagger}$These authors contributed equally to this work
\end{center}

\etocsettagdepth{main}{none}
\etocsettagdepth{appendix}{subsection}
\etocsettocstyle{\section*{Appendix contents}}{}
\tableofcontents
\clearpage

\section*{Overview of the appendix}

Section \ref{supp:connections} describes the relationship with existing regression models. Sections \ref{supp:weights} and \ref{supp:unbiased} give the observation requirements and population identities for CHW, RCW, and CTW, followed by a common symmetry argument and the comparison of the pairwise covariance matrices. Section \ref{supp:cox-if} derives the censoring weight expansions used by all three methods. Section \ref{supp:joint-model-choices} gives the joint likelihood, copula formulas, and contributions from fitting parametric joint models and copula models with Cox marginal functions.

\begin{figure}[H]
\centering
\begin{tikzpicture}[
 x=1cm,y=1cm,
 source/.style={draw=black,text=black,rounded corners=2pt,fill=black!4,align=center,text width=3.45cm,minimum height=1.35cm,inner sep=5pt,line width=0.65pt,font=\scriptsize},
 proc/.style={draw=black,text=black,rounded corners=2pt,align=center,text width=3.80cm,minimum height=1.05cm,inner sep=5pt,line width=0.65pt,font=\scriptsize},
 conclusion/.style={draw=black,text=black,rounded corners=2pt,fill=black!4,align=center,text width=5.50cm,minimum height=1.05cm,inner sep=5pt,line width=0.75pt,font=\scriptsize},
 arrow/.style={draw=black,-{Stealth[length=2.2mm,width=1.45mm]},line width=0.65pt}
]
\node[source] (population) at (-4.45,10.00) {{\bfseries Sections \ref*{supp:weights}--\ref*{supp:unbiased}}\\Observation, population mean, and pair symmetry\\CHW, RCW, CTW};
\node[source] (censoring) at (0,10.00) {{\bfseries Section \ref*{supp:cox-if}}\\Censoring model and weight expansions\\CHW, RCW, CTW};
\node[source] (joint) at (4.45,10.00) {{\bfseries Section \ref*{supp:joint-model-choices}}\\Joint survival models and influence functions\\CTW};

\node[proc] (conditions) at (0,7.35) {{\bfseries Section \ref*{supp:regularity}}\\Conditions C.1--C.8 for CHW and RCW; G.1--G.4 for CTW};
\node[proc] (consistency) at (-2.65,4.95) {{\bfseries Section \ref*{supp:consistency}}\\Consistency of the estimating equation and its local root};
\node[proc] (expansion) at (2.65,4.95) {{\bfseries Section \ref*{supp:weight-expansion}}\\Contribution from fitting the censoring and joint models};
\node[conclusion] (normality) at (0,2.45) {{\bfseries Section \ref*{supp:asymptotic-normality}}\\Common Hoeffding decomposition and asymptotic normality};
\node[conclusion] (covariance) at (0,0) {{\bfseries Section \ref*{supp:variance}}\\Estimation of the influence functions and consistency of the covariance estimator};

\draw[arrow] (population.south) to[out=-90,in=150] (conditions.north west);
\draw[arrow] (censoring.south) -- (conditions.north);
\draw[arrow] (joint.south) to[out=-90,in=30] (conditions.north east);
\draw[arrow] (conditions.south west) to[out=-120,in=90] (consistency.north);
\draw[arrow] (conditions.south east) to[out=-60,in=90] (expansion.north);
\draw[arrow] (consistency.south) to[out=-90,in=145] (normality.north west);
\draw[arrow] (expansion.south) to[out=-90,in=35] (normality.north east);
\draw[arrow] (normality.south) -- (covariance.north);
\end{tikzpicture}
\caption{Organization of the proofs for CHW, RCW, and CTW. The censoring and joint model calculations give the model estimation contributions in the common expansion. The resulting subject influence functions determine the asymptotic covariance and its estimator.}
\label{fig:supp-roadmap}
\end{figure}
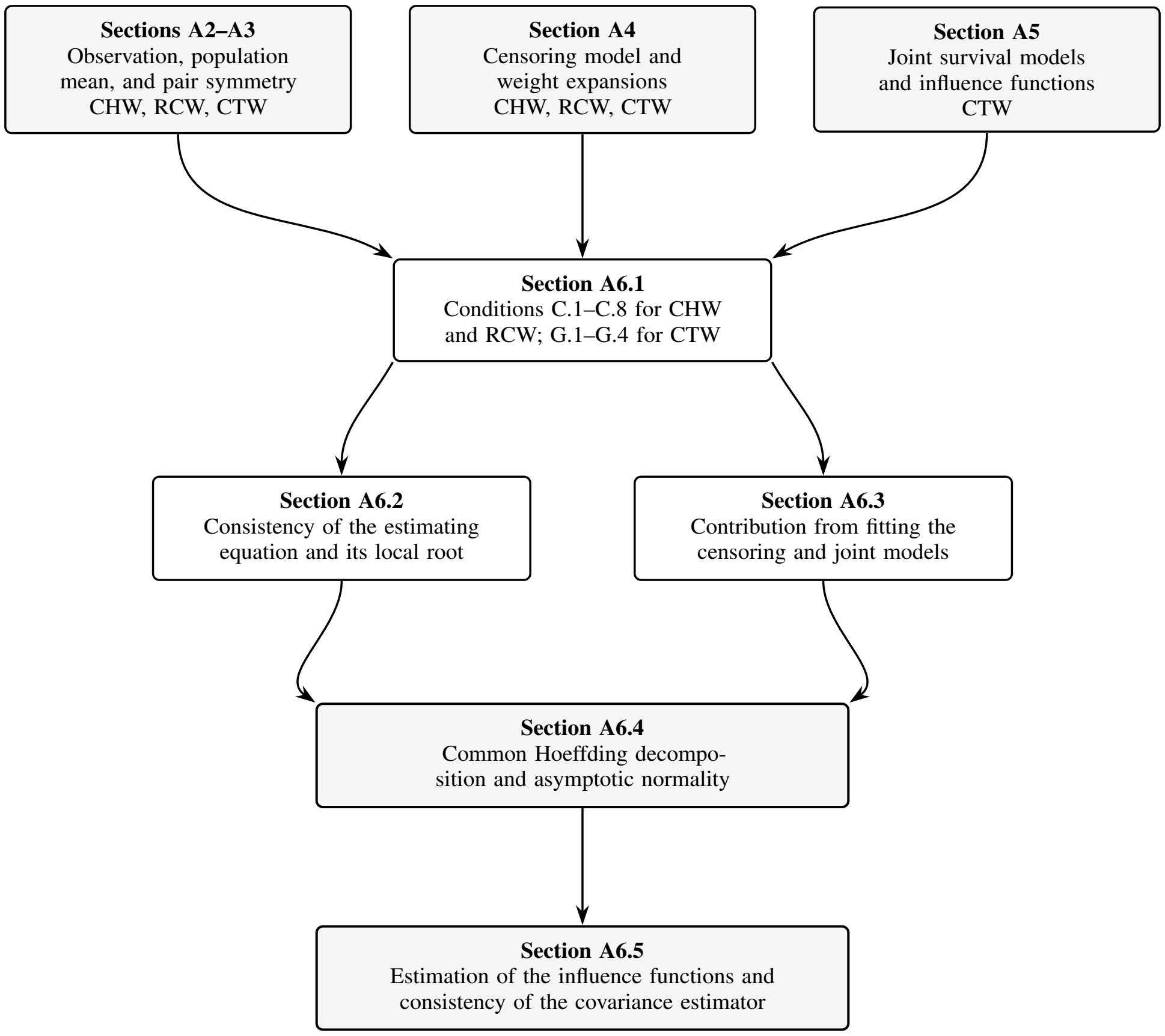

Figure \ref{fig:supp-roadmap} shows how these calculations enter the common asymptotic argument in Section \ref{supp:theory}. That section treats consistency, the effects of model estimation, asymptotic normality, and covariance estimation together for the three methods. It also relates the \(\sqrt n\) and \(\sqrt{m_n}\) normalizations and explains the factor \(1/2\) in the adjusted pair contribution. Sections \ref{supp:simulations}--\ref{supp:gpim-tutorial} contain the additional simulation results, HF-ACTION analyses, and the \texttt{gpim} tutorial.

\clearpage

\section{Relationship with existing regression models}
\label{supp:connections}

\subsection{Link functions and related regression models}
\label{supp:related-models}

Table \ref{tab:regression-comparison} compares related regression models and their assumptions. The win odds models of \citet{Song2023} and \citet{Wang2026general_wo} have the conditional mean specification in \eqref{eq:pim-model} with a logit link when the pair response and covariates are defined in the same way. Estimation also depends on the pairs included and the treatment of censoring. For example, \citet{Wang2026general_wo} use treatment--control pairs, whereas our estimating equations use the unordered pairs in \(\mathcal{I}_n\).

\begin{table}[!htbp]
\centering
\caption{Comparison of regression models for pairwise outcomes.}
\label{tab:regression-comparison}
\begin{tabularx}{\textwidth}{@{} >{\raggedright\arraybackslash}p{0.25\textwidth} >{\raggedright\arraybackslash}p{0.22\textwidth} >{\raggedright\arraybackslash}X @{}}
\toprule
Method & Regression quantity & Specification and censoring \\
\midrule
Probabilistic index model \citep{Thas2012} & Conditional probabilistic index & A link function relates the probabilistic index for a univariate outcome to the covariates of the two individuals. \\
\addlinespace
Win odds regression \citep{Song2023} & Conditional win odds & Logit PIM for a univariate outcome without right censoring, with one half assigned to ties. \\
\addlinespace
Proportional win-fractions regression \citep{Mao2021} & Conditional win ratio & Assumes proportional conditional win and loss fractions over follow-up. Under this assumption and conditionally independent censoring, estimation does not require a fitted censoring distribution. \\
\addlinespace
Stratified proportional win-fractions regression \citep{Wang2022} & Conditional win ratio within strata & Uses within-stratum comparisons, with common regression coefficients across strata. The stratifying factors need not satisfy the proportionality assumption. \\
\addlinespace
Generalized win odds regression \citep{Wang2026general_wo} & Conditional win odds for hierarchical outcomes & The basic model is a logit PIM for treatment--control comparisons. Proposes inverse probability of censoring weighting for censored comparisons. Section \ref{supp:wang-comparison} examines the resulting estimating equation. \\
\addlinespace
Proposed model & Conditional probabilistic index at \(L\) & Allows logit, probit, and centered identity links, with coefficients that may vary with \(L\). CHW and RCW target the same estimating equation for the full data under conditionally independent censoring. CTW targets that equation when the joint model is also correctly specified for a fixed hierarchy with continuous finite event times. \\
\bottomrule
\end{tabularx}
\end{table}
\FloatBarrier

The treatment of genuine ties distinguishes the proposed PIM from proportional win-fractions regression. For the same comparison rule and covariate differences, the latter model assumes that the ratio of the conditional win and loss probabilities equals \(\exp(\bm\theta^\top\bm Z_{ij})\) over follow-up, with \(\bm\theta\) independent of time \(t\). Let \(p_{T,ij}(L)=P\{H_{ij}(L)=1/2\mid\bm X_i^{Y},\bm X_j^{Y}\}\). Because the conditional win and loss probabilities sum to \(1-p_{T,ij}(L)\),
\begin{align*}
E\{H_{ij}(L)\mid\bm X_i^{Y},\bm X_j^{Y}\}
&=E\{\mathcal W(\bm Y_i,\bm Y_j)(L)\mid\bm X_i^{Y},\bm X_j^{Y}\}
  +\frac12p_{T,ij}(L)\\
&=\{1-p_{T,ij}(L)\}\operatorname{expit}(\bm\theta^\top\bm Z_{ij})
  +\frac12p_{T,ij}(L).
\end{align*} This mean generally differs from \(\operatorname{expit}(\bm\theta^\top\bm Z_{ij})\) when genuine ties have positive probability. They agree when ties are absent or when \(\bm\theta^\top\bm Z_{ij}=0\). Thus, a win ratio that is constant over follow-up need not imply a win odds that is constant over follow-up, because the probability of a tie can change. Model \eqref{eq:pim-model} in main text is specified at the chosen restriction time and does not impose proportionality over follow-up.

The reversal identity determines the allocation of one half to a \emph{genuine tie}. Consider the response \(H_{ij}^{(a)}(L)=\mathcal W(\bm Y_i,\bm Y_j)(L)+aI\{\bm Y_i(L)=\bm Y_j(L)\}\). Then \(H_{ij}^{(a)}(L)+H_{ji}^{(a)}(L)=1+(2a-1)I\{\bm Y_i(L)=\bm Y_j(L)\}\), so the sum equals one for every pair only when \(a=1/2\) if genuine ties have positive probability. Under \(\bm Z_{ji}=-\bm Z_{ij}\), taking conditional expectations therefore gives \(g^{-1}(-v)=1-g^{-1}(v)\), or equivalently \(g(1-u)=-g(u)\). The link choices in Section \ref{sec:target} satisfy this restriction. For the centered identity link, the net benefit interpretation follows directly from
\begin{align*}
2E\{H_{ij}(L)\mid\bm X_i^Y,\bm X_j^Y\}-1
&=2P\{\mathcal W(\bm Y_i,\bm Y_j)(L)=1\mid\bm X_i^Y,\bm X_j^Y\}+p_{T,ij}(L)-1\\
&=P\{\mathcal W(\bm Y_i,\bm Y_j)(L)=1\mid\bm X_i^Y,\bm X_j^Y\}\\
&\quad-P\{\mathcal W(\bm Y_j,\bm Y_i)(L)=1\mid\bm X_i^Y,\bm X_j^Y\}.
\end{align*}

The censoring factors also occur in estimators of win statistics. Consider continuous event times, zero equivalence margins, and death as the highest-priority outcome. For a comparison determined by an observed death at time $s<L$, the RCW denominator is $S_C(s\mid\bm X_i^{C})S_C(s\mid\bm X_j^{C})$. With group-specific censoring survival functions and between-group pairs, this is the censoring correction at the earlier event time in equations (9a)--(9b) of \citet{Dong2020IPCW}. For an observed lower-priority comparison, the fatal tie requires both individuals to be observed alive through $L$, so the denominator is $S_C(L\mid\bm X_i^{C})S_C(L\mid\bm X_j^{C})$, corresponding to the censoring correction at the restriction time with zero equivalence margins in Section 2.1 of \citet{Cui2025}. These correspondences concern the censoring factors. The PIM equation additionally weights the fitted mean and includes the genuine-tie residual $1/2-\mu_{ij}$. The regression estimator and the marginal win-statistic estimator can therefore differ in finite samples.

\subsection{Weighted win odds estimating equation}
\label{supp:wang-comparison}

The basic mean model in \citet{Wang2026general_wo} is a logit PIM. With \(\mu_{ij}(\bm\beta_L)=\operatorname{expit}(\bm\beta_L^\top\bm Z_{ij})\), \(\bm Z_{ij}=\bm X_i^{Y}-\bm X_j^{Y}\), and working variance \(\mu_{ij}(1-\mu_{ij})\), the multiplier is \(\bm K_{ij}=\bm Z_{ij}\). We express their weighted contribution in our notation and sum over \(\mathcal{I}_n\), while their original analysis uses between-group comparisons.

Consider one fatal and one nonfatal component, with continuous event times satisfying \(P(0<T_{1i}<D_i<L)=1\). The full-data comparison is determined by death, so \(H_{ij}(L)=I(D_i>D_j)\) and \(A_{ij}(L)=D_i\wedge D_j\). Define \(H_{ij}^{(1)}(L)=I(T_{1i}>T_{1j})\) and \(A_{ij}^{(1)}(L)=T_{1i}\wedge T_{1j}\). To use the censoring weights specified in their paper, assume that censoring is independent of the outcomes and covariates, with common survival function \(S_C\) satisfying \(S_C(L)\ge \epsilon>0\). Under these assumptions, the event time and censoring time for the same subject are equal with probability zero. The nonfatal comparison is observed when \(R_{ij}^{(1)}(L)=I\{C_i\ge A_{ij}^{(1)}(L),C_j\ge A_{ij}^{(1)}(L)\}=1\), with probability \(\pi_{ij}^{(1)}(L)=S_C\{A_{ij}^{(1)}(L)\}^2\). The probability of observing the fatal comparison is \(\pi_{ij}^{\mathrm{RCW}}(L)=S_C\{A_{ij}(L)\}^2\). Since \(A_{ij}^{(1)}(L)<A_{ij}(L)\), \(R_{ij}^{\mathrm{RCW}}(L)\le R_{ij}^{(1)}(L)\) and \(\pi_{ij}^{\mathrm{RCW}}(L)\le\pi_{ij}^{(1)}(L)\).

In equation (3) of \citet{Wang2026general_wo}, a lower-priority component is used when no winner is observed on higher-priority components. Thus, when censoring prevents the death comparison but the nonfatal comparison is observed, the nonfatal result is used. With zero equivalence margins, their weight becomes \(\Delta_{ij}^{\mathrm W}=R_{ij}^{\mathrm{RCW}}/\pi_{ij}^{\mathrm{RCW}}+(R_{ij}^{(1)}-R_{ij}^{\mathrm{RCW}})/\pi_{ij}^{(1)}\), and the weighted residual in their equation (4) is
\begin{align}
\bm U_{ij}^{\mathrm W}(\bm\beta_L)
=\bm Z_{ij}\bigg[&\frac{R_{ij}^{\mathrm{RCW}}}{\pi_{ij}^{\mathrm{RCW}}}
 \{H_{ij}(L)-\mu_{ij}(\bm\beta_L)\}+\frac{R_{ij}^{(1)}-R_{ij}^{\mathrm{RCW}}}{\pi_{ij}^{(1)}}
 \{H_{ij}^{(1)}(L)-\mu_{ij}(\bm\beta_L)\}\bigg].
\label{eq:wang-score}
\end{align}
Each term is evaluated only when its numerator is nonzero, using the corresponding observed comparison and event time. Write \(\widehat{\bm\Psi}_n^{\mathrm W}(\bm\beta_L)=\frac{2}{n(n-1)}\sum_{(i,j)\in \mathcal{I}_n}\widehat{\bm U}_{ij}^{\mathrm W}(\bm\beta_L)\).

For this subsection, suppress the fixed argument \(L\) except in the pair responses, and let \(\mathcal F_{ij}^{F}=\sigma(\bm Y_i,\bm Y_j,\bm X_i,\bm X_j)\). Hats denote substitution of \(\widehat S_C\) for \(S_C\). Conditional on \(\mathcal F_{ij}^{F}\), the event \(R_{ij}^{(1)}-R_{ij}^{\mathrm{RCW}}=1\) has probability \(\pi_{ij}^{(1)}-\pi_{ij}^{\mathrm{RCW}}\). Therefore,
\begin{align*}
E\left\{\left.\frac{R_{ij}^{(1)}-R_{ij}^{\mathrm{RCW}}}{\pi_{ij}^{(1)}}
 \right|\mathcal F_{ij}^{F}\right\}
=\frac{E(R_{ij}^{(1)}\mid\mathcal F_{ij}^{F})
 -E(R_{ij}^{\mathrm{RCW}}\mid\mathcal F_{ij}^{F})}{\pi_{ij}^{(1)}}=1-\frac{\pi_{ij}^{\mathrm{RCW}}}{\pi_{ij}^{(1)}}.
\end{align*}
If the full-data logit PIM is correctly specified, the first term of \eqref{eq:wang-score} has expectation zero at \(\bm\beta_L^0\). Conditioning first on \(\mathcal F_{ij}^{F}\) consequently gives
\begin{align}
E\{\bm U_{ij}^{\mathrm W}(\bm\beta_L^0)\}
&=E\{\bm Z_{ij}(H_{ij}(L)-\mu_{ij}(\bm\beta_L^0))\}+E\left[\bm Z_{ij}\left(1-\frac{\pi_{ij}^{\mathrm{RCW}}}{\pi_{ij}^{(1)}}\right)
 \{H_{ij}^{(1)}(L)-\mu_{ij}(\bm\beta_L^0)\}\right]\nonumber\\
&=E\left[\bm Z_{ij}\left(1-\frac{\pi_{ij}^{\mathrm{RCW}}}{\pi_{ij}^{(1)}}\right)
 \{H_{ij}^{(1)}(L)-\mu_{ij}(\bm\beta_L^0)\}\right].
\label{eq:wang-extra-term}
\end{align}
The proposition below gives sufficient conditions for this expectation to be nonzero.

\begin{proposition}
\label{prop:wang-inconsistency}
Suppose the full-data logit PIM is correctly specified, \(E\|\bm Z_{ij}\|^2<\infty\), and, for a vector \(\bm\beta_L^{(1)}\), we have
\begin{equation}
E\{H_{ij}^{(1)}(L)\mid\bm X_i,\bm X_j,A_{ij},A_{ij}^{(1)}\}
=\mu_{ij}(\bm\beta_L^{(1)}).
\label{eq:wang-nonfatal-mean}
\end{equation}
If
\begin{equation}
P\left\{\pi_{ij}^{\mathrm{RCW}}<\pi_{ij}^{(1)},\
(\bm\beta_L^{(1)}-\bm\beta_L^0)^\top\bm Z_{ij}\ne0\right\}>0,
\label{eq:wang-positive-probability}
\end{equation}
then \(E\{\bm U_{ij}^{\mathrm W}(\bm\beta_L^0)\}\ne\bm0\). Under positivity and uniform consistency of \(\widehat S_C\) on \([0,L]\), a solution of \(\widehat{\bm\Psi}_n^{\mathrm W}(\bm\beta_L)=\bm0\) cannot be consistent for \(\bm\beta_L^0\).
\end{proposition}

\begin{proof}
Write \(\bm d=\bm\beta_L^{(1)}-\bm\beta_L^0\). The ratio \(\pi_{ij}^{\mathrm{RCW}}/\pi_{ij}^{(1)}\) is determined by the two comparison times. Conditioning further in \eqref{eq:wang-extra-term} and using \eqref{eq:wang-nonfatal-mean} yields
\begin{align*}
E\{\bm U_{ij}^{\mathrm W}(\bm\beta_L^0)\}
&=E\left[\bm Z_{ij}\left(1-\frac{\pi_{ij}^{\mathrm{RCW}}}{\pi_{ij}^{(1)}}\right)
 \{\mu_{ij}(\bm\beta_L^{(1)})-\mu_{ij}(\bm\beta_L^0)\}\right].
\end{align*}
The difference of logistic means satisfies
\begin{align*}
\mu_{ij}(\bm\beta_L^{(1)})-\mu_{ij}(\bm\beta_L^0)
&=(\bm d^\top\bm Z_{ij})
 \int_0^1\operatorname{expit}'\{(\bm\beta_L^0+t\bm d)^\top\bm Z_{ij}\}\,dt.
\end{align*}
Since \(\operatorname{expit}'(v)=\operatorname{expit}(v)\{1-\operatorname{expit}(v)\}>0\),
\begin{align}
\bm d^\top E\{\bm U_{ij}^{\mathrm W}(\bm\beta_L^0)\}
&=E\bigg[\left(1-\frac{\pi_{ij}^{\mathrm{RCW}}}{\pi_{ij}^{(1)}}\right)
 (\bm d^\top\bm Z_{ij})^2\int_0^1
 \operatorname{expit}'\{(\bm\beta_L^0+t\bm d)^\top\bm Z_{ij}\}\,dt\bigg]>0.
\label{eq:wang-strict-inequality}
\end{align}
The integrand is nonnegative, and condition \eqref{eq:wang-positive-probability} makes it positive with positive probability. This proves the nonzero expectation.

For the corresponding estimator, let \(\mathcal B\) be a compact neighborhood of \(\bm\beta_L^0\), and define \(\bm\Psi^{\mathrm W}(\bm\beta_L)=E\{\bm U_{12}^{\mathrm W}(\bm\beta_L)\}\). Positivity bounds the weights, while the logistic function and its derivative are bounded. Hence, for a finite constant \(C\), and any two parameter values \(\bm\beta_L,\bm\beta_L'\in\mathcal B\), we have
\begin{align*}
\sup_{\bm\beta_L\in\mathcal B}
 \|\bm U_{ij}^{\mathrm W}(\bm\beta_L)\|
&\le C\|\bm Z_{ij}\|,\\
\|\bm U_{ij}^{\mathrm W}(\bm\beta_L)
 -\bm U_{ij}^{\mathrm W}(\bm\beta_L')\|
&\le C\|\bm Z_{ij}\|^2\|\bm\beta_L-\bm\beta_L'\|.
\end{align*}
Apply convergence of U-statistic averages at the finitely many points of an \(\varepsilon\)-net of \(\mathcal B\). The second bound controls the difference between each parameter and its nearest net point by \(C\varepsilon\{\frac{2}{n(n-1)}\sum_{(i,j)\in \mathcal{I}_n}\|\bm Z_{ij}\|^2+E\|\bm Z_{12}\|^2\}\). This expression is \(O_p(\varepsilon)\). Letting \(n\) go to infinity, then \(\varepsilon\) goes to zero, whice gives \(\sup_{\bm\beta_L\in\mathcal B}\|\bm\Psi_n^{\mathrm W}(\bm\beta_L)-\bm\Psi^{\mathrm W}(\bm\beta_L)\|=o_p(1)\).

With probability tending to one, \(\inf_{0\le t\le L}\widehat S_C(t)\ge \epsilon/2\). On this event the reciprocal weights are uniformly Lipschitz functions of \(S_C\), giving
\begin{align*}
\sup_{\bm\beta_L\in\mathcal B}
 \|\widehat{\bm\Psi}_n^{\mathrm W}(\bm\beta_L)
 -\bm\Psi_n^{\mathrm W}(\bm\beta_L)\|
&\le C\sup_{0\le t\le L}|\widehat S_C(t)-S_C(t)|
 \frac{2}{n(n-1)}\sum_{(i,j)\in \mathcal{I}_n}\|\bm Z_{ij}\|=o_p(1).
\end{align*}
Therefore \(\widehat{\bm\Psi}_n^{\mathrm W}(\cdot)\) converges uniformly to the continuous function \(\bm\Psi^{\mathrm W}(\cdot)\) on \(\mathcal B\). If a sequence of roots \(\widehat{\bm\beta}_L^{\mathrm W}\) converged to \(\bm\beta_L^0\), then
\begin{align*}
\bm0=\widehat{\bm\Psi}_n^{\mathrm W}(\widehat{\bm\beta}_L^{\mathrm W})=\bm\Psi^{\mathrm W}(\widehat{\bm\beta}_L^{\mathrm W})+o_p(1)=\bm\Psi^{\mathrm W}(\bm\beta_L^0)+o_p(1),
\end{align*}
contradicting \eqref{eq:wang-strict-inequality}.
\end{proof}

Condition \eqref{eq:wang-nonfatal-mean} is stronger than a logit model for the nonfatal comparison given only the covariates, because it is also conditioned on both comparison times. Condition \eqref{eq:wang-positive-probability} requires a positive probability that the nonfatal comparison is observed while the fatal comparison is unresolved, for covariate contrasts with different nonfatal and full-data mean specifications. Under these sufficient conditions, the additional expectation in \eqref{eq:wang-extra-term} remains nonzero even when \(S_C\) is known. Thus independent censoring and consistent estimation of its survival function alone do not ensure that this weighted equation targets the full-data regression coefficient. This conclusion applies to distributions satisfying the stated sufficient conditions.

\section{Observation of pairwise comparisons}
\label{supp:weights}

For the observed records \(\mathcal O_i\) and \(\mathcal O_i(L)\) defined in Section \ref{sec:ctw} of main text, write \(U_i=D_i\wedge C_i\wedge L\) and \(\delta_i^D=I(D_i\le C_i\wedge L)\). Pairwise comparisons use \(\mathcal O_i(L)\), whereas model estimation may use a longer observed record. Events at a stopping time are recorded before censoring. Thus \(U_i<L\) and \(\delta_i^D=0\) establish that \(D_i>U_i\), including when another individual dies at that time. Assigning a midpoint to an unresolved comparison gives a \emph{censoring-induced tie}, distinct from a \emph{genuine tie}.

\subsection{Complete histories and resolved comparisons}
\label{supp:chw-rcw-observation}

An individual history is complete after an observed death or follow-up through \(L\). Define \(J_i(L)=I\{U_i=L\text{ or }\delta_i^D=1\}\) and \(R_{ij}^{\mathrm{CHW}}(L)=J_i(L)J_j(L)\). For the RCW observation rule, let \(s_{ij}=U_i\wedge U_j\) and define 
\(R_{ij}^{\mathrm{RCW}}(L) =I\!\left[\{s_{ij}=L\}\cup \bigcup_{k\in\{i,j\}}\{U_k=s_{ij}<L,\ \delta_k^D=1\}\right].\)
If exactly one death is observed at \(s_{ij}<L\), that individual loses on the fatal component. The conclusion also holds when the other individual is censored at the same time, because its death indicator is zero. If both deaths occur at \(s_{ij}<L\), or both individuals are followed through \(L\), apply the complete hierarchy to the observed restricted outcomes. On \(R_{ij}^{\mathrm{RCW}}(L)=1\), let \(\widetilde H_{ij}(L)\) equal \(1\), \(0\), or \(1/2\) for the resulting win, loss, or genuine tie. On \(R_{ij}^{\mathrm{RCW}}(L)=0\), set \(\widetilde H_{ij}(L)=1/2\) to define the response on every observed record. This assigned value indicates that censoring prevents determination of the comparison.

The common determination time \(A_{ij}(L)\) is defined in Section \ref{sec:estimating-equation} of main text. Since every nonfatal component is truncated by death, its equivalent expression for any \(q\in\{0,\ldots,Q\}\) is
\begin{equation}
A_{ij}(L)=Y_{0i}(L)\wedge Y_{0j}(L)=\max_{0\le r\le q}\{Y_{ri}(L)\wedge Y_{rj}(L)\}.
\label{supp:eq:Aq}
\end{equation}

\begin{lemma}[Observation of a resolved comparison]
\label{supp:lem:comparison-observability}
\leavevmode\newline
The observed indicators satisfy \(J_i(L)=I\{C_i\ge Y_{0i}(L)\}\) and \(R_{ij}^{\mathrm{RCW}}(L)=I\{C_i\ge A_{ij}(L),C_j\ge A_{ij}(L)\}\). On \(R_{ij}^{\mathrm{RCW}}(L)=1\), \(s_{ij}=A_{ij}(L)\) and \(\widetilde H_{ij}(L)=H_{ij}(L)\).
Consequently, \(R_{ij}^{\mathrm{RCW}}(L)\{\widetilde H_{ij}(L)-H_{ij}(L)\}=0\) and \(R_{ij}^{\mathrm{CHW}}(L)\{\widetilde H_{ij}(L)-H_{ij}(L)\}=0\).
\end{lemma}

\begin{proof}
The event defining a complete individual history satisfies
\begin{align*}
\{C_i\ge Y_{0i}(L)\}
&=\{D_i\le C_i,D_i\le L\}\cup\{D_i> L,C_i\ge L\}\\
&=\{\delta_i^D=1\}\cup\{U_i=L\}
=\{J_i(L)=1\}.
\end{align*}
Write \(a=D_i\wedge D_j\wedge L\). If \(C_i,C_j\ge a\), then \(s_{ij}=D_i\wedge D_j\wedge C_i\wedge C_j\wedge L=a\). When \(a=L\), the first event in the definition of \(R_{ij}^{\mathrm{RCW}}(L)\) holds. When \(a<L\), at least one individual has \(D_k=a\le C_k\), so \(U_k=s_{ij}=a\) and \(\delta_k^D=1\). Thus \(R_{ij}^{\mathrm{RCW}}(L)=1\).

Conversely, if \(s_{ij}=L\), then \(D_i,D_j,C_i,C_j\ge L\), giving \(a=L\) and \(C_i,C_j\ge a\). If \(U_k=s_{ij}<L\) and \(\delta_k^D=1\) for some \(k\), then \(D_k=s_{ij}\). The other record extends through \(s_{ij}\), so its death and censoring times are at least \(s_{ij}\). Hence \(a=s_{ij}\) and \(C_i,C_j\ge a\). This proves the identity for \(R_{ij}^{\mathrm{RCW}}(L)\).

For the response, exactly one death at \(a<L\) determines a strict fatal comparison. If both deaths occur at \(a<L\), all restricted components are observed because \(Y_{rk}(L)\le D_k=a\) for each \(r\) and \(k=i,j\). If \(a=L\), all components are observed through \(L\). Applying the hierarchy therefore gives \(\widetilde H_{ij}(L)=H_{ij}(L)\). Since \(A_{ij}(L)\le Y_{0k}(L)\) for \(k=i,j\), \(R_{ij}^{\mathrm{CHW}}(L)=1\) implies \(R_{ij}^{\mathrm{RCW}}(L)=1\). Multiplying by either observation indicator proves both assertions.
\end{proof}

Let \(I_{ij}^{(q)}(L)\) and \(I_{ji}^{(q)}(L)\) indicate an observed strict win in the corresponding direction determined by component \(q\), and let \(I_{ij}^{(T)}(L)\) indicate an observed genuine tie. All these indicators are zero when \(R_{ij}^{\mathrm{RCW}}(L)=0\). For the following identities, suppress the fixed argument \(L\) except in the pair responses. Write \(\mathcal W_{ij}=\mathcal W(\bm Y_i,\bm Y_j)(L)\) and \(\Delta_{ij}=\mathcal W_{ij}+\mathcal W_{ji}\). The indicators for the determining components in Section \ref{sec:target} give \(I_{ij}^{(q)}=R_{ij}^{\mathrm{RCW}}\kappa_{ij}^{(q)}\mathcal W_{ij}\). Reversing the individuals gives \(I_{ji}^{(q)}=R_{ij}^{\mathrm{RCW}}\kappa_{ij}^{(q)}\mathcal W_{ji}\), while a genuine tie has indicator \(I_{ij}^{(T)}=R_{ij}^{\mathrm{RCW}}\{1-\Delta_{ij}\}\). Since \(\sum_{q=0}^Q\kappa_{ij}^{(q)}=\Delta_{ij}\) and \(\Delta_{ij}^2=\Delta_{ij}\), 
\begin{equation*}
\sum_{q=0}^Q\{I_{ij}^{(q)}+I_{ji}^{(q)}\} +I_{ij}^{(T)} =R_{ij}^{\mathrm{RCW}}\{\Delta_{ij}^2+1-\Delta_{ij}\} =R_{ij}^{\mathrm{RCW}}.
\end{equation*}
Similarly, \(\Delta_{ij}\mathcal W_{ij}=\mathcal W_{ij}\) and Lemma \ref{supp:lem:comparison-observability} yield 
\begin{equation*}
\sum_{q=0}^QI_{ij}^{(q)}+\tfrac12I_{ij}^{(T)} =R_{ij}^{\mathrm{RCW}}\{\mathcal W_{ij}+\tfrac12(1-\Delta_{ij})\} =R_{ij}^{\mathrm{RCW}}H_{ij}(L) =R_{ij}^{\mathrm{RCW}}\widetilde H_{ij}(L).
\end{equation*}
In particular, \(I_{ij}^{(T)}=R_{ij}^{\mathrm{RCW}}I\{\widetilde H_{ij}(L)=1/2\}\) distinguishes a genuine tie from the assigned midpoint on an unresolved pair.

For \(\star\in\{\mathrm{CHW},\mathrm{RCW}\}\) and a positive candidate censoring survival function \(S\), write \(\pi_{ij}^{\mathrm{CHW}}(L;S)=S\{Y_{0i}(L)\mid\bm X_i^{C}\}S\{Y_{0j}(L)\mid\bm X_j^{C}\}\) and \(\pi_{ij}^{\mathrm{RCW}}(L;S)=S\{A_{ij}(L)\mid\bm X_i^{C}\}S\{A_{ij}(L)\mid\bm X_j^{C}\}\). At \(S=S_C\), these are the conditional observation probabilities and the argument \(S_C\) is suppressed. Define \(W_{ij}^{C,\star}(L;S)\) as the reciprocal of the corresponding two survival factors when \(R_{ij}^{\star}(L)=1\), and as zero otherwise. For CHW, the factors are evaluated at the observed completion times \(U_i\) and \(U_j\). For RCW, both are evaluated at \(s_{ij}\). Thus \(W_{ij}^{C,\star}(L;S)=R_{ij}^{\star}(L)/\pi_{ij}^{\star}(L;S)\) can be calculated from the observed times. Set \(W_{ij}^{C,\star}(L)=W_{ij}^{C,\star}(L;S_C)\) and \(\widehat W_{ij}^{C,\star}(L)=W_{ij}^{C,\star}(L;\widehat S_C)\). With the restriction time fixed, \(L\) is suppressed below except in the pair responses. We write \(\bm U_{ij}^{\star}(\bm\beta_L;S)\) and \(\bm\Psi_n^{\star}(\bm\beta_L;S)\) when the argument specifying the censoring survival function is needed. As in the main text, \(\bm U_{ij}^{\star}(\bm\beta_L)\) and \(\bm\Psi_n^{\star}(\bm\beta_L)\) use the true \(S_C\), and hats denote substitution of \(\widehat S_C\).

\begin{lemma}[Ordering of the observation requirements]
\label{supp:lem:dl-ch-observation-comparison}
For \(k=i,j\), \(A_{ij}(L)\le Y_{0k}(L)\le L\). Consequently, \(R_{ij}^{\mathrm{CHW}}(L)\le R_{ij}^{\mathrm{RCW}}(L)\) and \(\pi_{ij}^{\mathrm{CHW}}(L)\le\pi_{ij}^{\mathrm{RCW}}(L)\).
\end{lemma}

\begin{proof}
Equation \eqref{supp:eq:Aq} gives \(A_{ij}(L)=Y_{0i}(L)\wedge Y_{0j}(L)\), proving the time inequalities. It follows that \(\{C_i\ge Y_{0i}(L),C_j\ge Y_{0j}(L)\} \subseteq\{C_i\ge A_{ij}(L),C_j\ge A_{ij}(L)\}\), which proves the indicator inequality. Since \(S_C(t\mid\bm x)\) is nonincreasing in \(t\),
\begin{align}
\pi_{ij}^{\mathrm{CHW}}(L)
&=S_C\{Y_{0i}(L)\mid\bm X_i^{C}\}S_C\{Y_{0j}(L)\mid\bm X_j^{C}\}\nonumber\\
&\le S_C\{A_{ij}(L)\mid\bm X_i^{C}\}S_C\{A_{ij}(L)\mid\bm X_j^{C}\}
=\pi_{ij}^{\mathrm{RCW}}(L).
\label{supp:eq:orderings}
\end{align}
\end{proof}

Every pair observed under CHW is observed under RCW. RCW also includes pairs whose comparison is determined by an earlier death when the other individual is subsequently censored before its history is complete. Positivity and \eqref{supp:eq:orderings} imply \((\pi_{ij}^{\mathrm{RCW}})^{-1}\le(\pi_{ij}^{\mathrm{CHW}})^{-1}\). This does not give a pointwise ordering of \(R_{ij}^{\mathrm{RCW}}/\pi_{ij}^{\mathrm{RCW}}\) and \(R_{ij}^{\mathrm{CHW}}/\pi_{ij}^{\mathrm{CHW}}\), because the CHW indicator may be zero when the RCW indicator is one.

\subsection{Conditional tie weighting}
\label{supp:ctw-observation}

The following indicators apply under the continuous-time assumptions in Section \ref{supp:ctw-identification}. Let \(\delta_{qi}^N\) indicate an observed nonfatal event of type \(q\) before \(U_i\), and set \(E_{qi}=T_{qi}\) when \(\delta_{qi}^N=1\) and \(E_{qi}=\infty\) otherwise. For \(q\ge1\), the first observed event time for that component is \(t_{ij}^{(q)}=E_{qi}\wedge E_{qj}\). The nonfatal observation indicator \(R_{ij}^{\mathrm{CTW},(q)}(L)\) is one if \(t_{ij}^{(q)}<L\), \(E_{qi}\ne E_{qj}\), and each \(k\in\{i,j\}\) satisfies either \(U_k>t_{ij}^{(q)}\) or \(U_k=t_{ij}^{(q)}\) with \(\delta_k^D=0\). It is zero otherwise. The latter condition allows censoring alive at the comparison time but excludes death at that time. When this indicator is one, \(\widetilde H_{ij}^{(q)}(L)=I(E_{qi}>E_{qj})\). The indicator is zero if neither event is recorded before \(L\), and its associated time and response need not then be evaluated. These observations include the \emph{partially observed lower-priority-resolved pairs} described in Section \ref{sec:ctw}.

Let \(s_{ij}=U_i\wedge U_j\). The fatal indicator \(R_{ij}^{\mathrm{CTW},(0)}(L)\) is one if \(s_{ij}<L\) and exactly one of \(I(U_i=s_{ij},\delta_i^D=1)\) and \(I(U_j=s_{ij},\delta_j^D=1)\) is one. On this event, \(t_{ij}^{(0)}=s_{ij}=A_{ij}(L)\) and \(\widetilde H_{ij}^{(0)}(L)=I(U_j=s_{ij},\delta_j^D=1)\). The genuine tie indicator \(R_{ij}^{\mathrm{CTW},T}(L)\) is one when \(U_i=U_j=L\), \(\delta_i^D=\delta_j^D=0\), and \(\delta_{ri}^N=\delta_{rj}^N=0\) for every \(r\in\{1,\ldots,Q\}\). It is zero otherwise. These indicators in \eqref{eq:ctw-score} of main text are determined by the recorded data.

\section{Properties of the three estimating functions}
\label{supp:unbiased}

The three estimating functions have the same conditional mean given the covariates under their respective model assumptions. The proofs below establish this identity, symmetry under pair reversal, and the covariance comparison in Theorem \ref{thm:efficiency-comparison}.

\subsection{Complete history and resolved comparison weighting}
\label{supp:chw-rcw-mean}

The expectation identities in Section \ref{sec:estimation} of main text follow by conditioning on the full data. Fix \(L\) and a candidate \(\bm\beta_L\), and write \(\mu_{ij}=\mu_{ij}(\bm\beta_L)\) and \(\bm K_{ij}=\bm K_{ij}(\bm\beta_L)\). Both are functions of the covariates. Assume independent sampling, conditionally independent censoring, and positivity as in Section \ref{sec:complete-history}, and finite expectations of the estimating functions.

For fixed \(u,v\in[0,L]\), independence across individuals and conditional independence of censoring give
\begin{align*}
P(C_i\ge u,C_j\ge v\mid\mathcal F_{ij}^{F})
&=P(C_i\ge u\mid\bm Y_i,\bm X_i)P(C_j\ge v\mid\bm Y_j,\bm X_j)\\
&=S_C(u\mid\bm X_i^{C})S_C(v\mid\bm X_j^{C}).
\end{align*}
The times \(Y_{0i}(L)\), \(Y_{0j}(L)\), and \(A_{ij}(L)\) are \(\mathcal F_{ij}^{F}\)-measurable. Evaluating this conditional survival function at the required times and applying Lemma \ref{supp:lem:comparison-observability} therefore yields
\begin{align*}
E(R_{ij}^{\mathrm{CHW}}\mid\mathcal F_{ij}^{F})
&=S_C\{Y_{0i}(L)\mid\bm X_i^{C}\}S_C\{Y_{0j}(L)\mid\bm X_j^{C}\}
=\pi_{ij}^{\mathrm{CHW}},\\
E(R_{ij}^{\mathrm{RCW}}\mid\mathcal F_{ij}^{F})
&=S_C\{A_{ij}(L)\mid\bm X_i^{C}\}S_C\{A_{ij}(L)\mid\bm X_j^{C}\}
=\pi_{ij}^{\mathrm{RCW}}.
\end{align*}
Hence \(E(R_{ij}^{\star}\mid\mathcal F_{ij}^{F})=\pi_{ij}^{\star}\ge \epsilon^2>0\) for every \(\star\in\{\mathrm{CHW},\mathrm{RCW}\}\). The probabilities are \(\mathcal F_{ij}^{F}\)-measurable, and \(0\le W_{ij}^{C,\star}\le \epsilon^{-2}\).

Although the observed response need not be \(\mathcal F_{ij}^{F}\)-measurable, \(R_{ij}^{\star}\widetilde H_{ij}(L)=R_{ij}^{\star}H_{ij}(L)\). Taking conditional expectations gives
\begin{align*}
E(W_{ij}^{C,\star}\mid\mathcal F_{ij}^{F})
&=\frac{E(R_{ij}^{\star}\mid\mathcal F_{ij}^{F})}{\pi_{ij}^{\star}}
=1,\\
E(W_{ij}^{C,\star}\widetilde H_{ij}(L)\mid\mathcal F_{ij}^{F})
&=\frac{E(R_{ij}^{\star}\widetilde H_{ij}(L)\mid\mathcal F_{ij}^{F})}{\pi_{ij}^{\star}}
=\frac{E(R_{ij}^{\star}H_{ij}(L)\mid\mathcal F_{ij}^{F})}{\pi_{ij}^{\star}}\\
&=H_{ij}(L)\frac{E(R_{ij}^{\star}\mid\mathcal F_{ij}^{F})}{\pi_{ij}^{\star}}
=H_{ij}(L).
\end{align*}

For CHW and RCW, the entire residual is multiplied by the weight. Since \(H_{ij}(L)\), \(\mu_{ij}\), and \(\bm K_{ij}\) are fixed under conditioning on \(\mathcal F_{ij}^{F}\), for \(\star\in\{\mathrm{CHW},\mathrm{RCW}\}\),
\begin{align*}
E\{\bm U_{ij}^{\star}(\bm\beta_L)\mid\mathcal F_{ij}^{F}\}
&=\frac{\bm K_{ij}}{\pi_{ij}^{\star}}
 E\{R_{ij}^{\star}(\widetilde H_{ij}(L)-\mu_{ij})\mid\mathcal F_{ij}^{F}\}=\frac{\bm K_{ij}}{\pi_{ij}^{\star}}
 E\{R_{ij}^{\star}(H_{ij}(L)-\mu_{ij})\mid\mathcal F_{ij}^{F}\}\\
&=\bm K_{ij}(H_{ij}(L)-\mu_{ij})
 \frac{E(R_{ij}^{\star}\mid\mathcal F_{ij}^{F})}{\pi_{ij}^{\star}}=\bm K_{ij}(H_{ij}(L)-\mu_{ij}).
\end{align*}
Iterated expectation and measurability of \(\bm K_{ij}\) and \(\mu_{ij}\) with respect to the covariates now imply
\begin{align*}
E\{\bm U_{ij}^{\star}(\bm\beta_L)\mid\bm X_i,\bm X_j\}
&=E\left[E\{\bm U_{ij}^{\star}(\bm\beta_L)\mid\mathcal F_{ij}^{F}\}
 \mid\bm X_i,\bm X_j\right]\\
&=\bm K_{ij}(\bm\beta_L)
 \left[E\{H_{ij}(L)\mid\bm X_i,\bm X_j\}-\mu_{ij}(\bm\beta_L)\right].
\end{align*}
Linearity of expectation gives
\begin{align*}
E\{\bm\Psi_n^{\star}(\bm\beta_L)\}
&=\frac{2}{n(n-1)}\sum_{(i,j)\in \mathcal{I}_n}
 E\{\bm U_{ij}^{\star}(\bm\beta_L)\}=E\left[\bm K_{12}(\bm\beta_L)
 \{H_{12}(L)-\mu_{12}(\bm\beta_L)\}\right].
\end{align*}
Both estimating functions therefore have the same expectation as the estimating function for the full data at every candidate parameter. Under the conditional PIM, \(E\{H_{ij}(L)\mid\bm X_i^{Y},\bm X_j^{Y}\}=\mu_{ij}(\bm\beta_L^0)\). Since \(\bm K_{ij}\) and \(\mu_{ij}\) depend only on these PIM covariates, iterated expectation gives \(E\{\bm U_{ij}^{\star}(\bm\beta_L^0)\}=\bm0\). If the PIM is a working model, its full-data population root instead satisfies \(E[\bm K_{12}(\bm\beta_L^0)\{H_{12}(L)-\mu_{12}(\bm\beta_L^0)\}]=\bm0\). In either case, \(E\{\bm\Psi_n^{\star}(\bm\beta_L^0)\}=\bm0\), and the two equations have the same population root whenever that root is unique. These identities establish unbiasedness of the estimating functions evaluated at the true censoring probabilities. Replacing \(S_C\) by \(\widehat S_C\) gives the asymptotic results in Sections \ref{supp:two-weight-expansions} and \ref{supp:theory}. Finite-sample unbiasedness of the regression estimators is a separate property. Remark \ref{supp:rem:truncation} discusses fixed truncation of the weights.

The RCW equation can also be written as a sum over the determining components. For a positive candidate \(S\), define
\begin{align*}
\bm U_{ij}^{(q)}(\bm\beta_L;S)
&=\frac{\bm K_{ij}}{\pi_{ij}^{\mathrm{RCW}}(L;S)}
 \left[I_{ij}^{(q)}-\mu_{ij}
 \{I_{ij}^{(q)}+I_{ji}^{(q)}\}\right].
\end{align*}
The genuine-tie contribution is \(\bm U_{ij}^{(T)}(\bm\beta_L;S)=\bm K_{ij}I_{ij}^{(T)}(1/2-\mu_{ij})/\pi_{ij}^{\mathrm{RCW}}(L;S)\). The partition in \ref{supp:weights} gives
\begin{align*}
\sum_{q=0}^Q\bm U_{ij}^{(q)}(\bm\beta_L;S)+\bm U_{ij}^{(T)}(\bm\beta_L;S)
&=\frac{\bm K_{ij}}{\pi_{ij}^{\mathrm{RCW}}(L;S)}
 \left[\sum_{q=0}^QI_{ij}^{(q)}+\tfrac12I_{ij}^{(T)}-\mu_{ij}R_{ij}^{\mathrm{RCW}}\right]\\
&=\bm K_{ij}W_{ij}^{C,\mathrm{RCW}}(L;S)
 (\widetilde H_{ij}(L)-\mu_{ij})\\
&=\bm U_{ij}^{\mathrm{RCW}}(\bm\beta_L;S).
\end{align*}
At \(S=S_C\), conditioning on \(\mathcal F_{ij}^{F}\) replaces \(I_{ij}^{(q)}/\pi_{ij}^{\mathrm{RCW}}\) by \(\kappa_{ij}^{(q)}\mathcal W_{ij}\), \(I_{ji}^{(q)}/\pi_{ij}^{\mathrm{RCW}}\) by \(\kappa_{ij}^{(q)}\mathcal W_{ji}\), and \(I_{ij}^{(T)}/\pi_{ij}^{\mathrm{RCW}}\) by \(1-\Delta_{ij}\) in expectation. Summing the resulting residuals gives
\begin{align*}
\sum_{q=0}^Q\kappa_{ij}^{(q)}
 (\mathcal W_{ij}-\mu_{ij}\Delta_{ij})
+(1-\Delta_{ij})(1/2-\mu_{ij})
&=\mathcal W_{ij}+\tfrac12(1-\Delta_{ij})-\mu_{ij}=H_{ij}(L)-\mu_{ij}.
\end{align*}
The sum includes every strict comparison and the residual \(1/2-\mu_{ij}\) for a genuine tie.

\subsection{Identification under conditional tie weighting}
\label{supp:ctw-identification}

For the CTW estimating function in \eqref{eq:ctw-score} of main text, use the observed records and conditional tie probabilities in Section \ref{sec:ctw} of main text. Throughout, \(Q\) is fixed, independent individuals have no equal finite component times with positive probability, and no finite component time equals \(L\) with positive probability. Censoring satisfies Condition C.1, and \(P(T_{qi}=C_i<D_i\wedge L)=0\) for \(q=1,\ldots,Q\), so the strict event indicators in \(\mathcal O_i(L)\) agree almost surely with the convention that an event is recorded before censoring at the same time. The observation indicators are given in Section \ref{supp:ctw-observation}. The joint model is consistent with these histories and gives the true conditional tie probabilities.

For \(q=1,\ldots,Q+1\), put \(J_i^{(q)}(L)=I\{D_i>L,\ T_{ri}>L\text{ for every }r\in\{1,\ldots,q-1\}\}\), where the condition on nonfatal events is empty when \(q=1\). Write \(J_{ij}^{(q)}(L)=J_i^{(q)}(L)J_j^{(q)}(L)\). Continuity implies \(I\{\bigcap_{r=0}^{q-1}\mathcal E_{ij}^{(r)}(L)\}=J_{ij}^{(q)}(L)\) almost surely. In particular, \(J_{ij}^{(Q+1)}(L)\) indicates a \emph{genuine tie}. For \(q=1,\ldots,Q\), independence across individuals gives
\begin{equation}
E\{J_{ij}^{(q)}(L)\mid\mathcal O_i(L),\mathcal O_j(L)\}
=E\{J_i^{(q)}(L)\mid\mathcal O_i(L)\}E\{J_j^{(q)}(L)\mid\mathcal O_j(L)\}
=\omega_i^{(q)}(L)\omega_j^{(q)}(L).
\label{supp:eq:ctw-general-posterior}
\end{equation}
Independence is required across individuals. The components within an individual may be dependent.

\begin{proposition}[CTW population equation]
\label{supp:prop:ctw-mean}
Under the preceding assumptions, for every candidate \(\bm\beta_L\) with finite score expectations, we have
\begin{equation}
E\{\bm U_{ij}^{\mathrm{CTW}}(\bm\beta_L)\mid\bm X_i,\bm X_j\}
=\bm K_{ij}(\bm\beta_L)\left[E\{H_{ij}(L)\mid\bm X_i,\bm X_j\}-\mu_{ij}(\bm\beta_L)\right].
\label{supp:eq:ctw-general-mean}
\end{equation}
\end{proposition}

\begin{proof}
Suppress \(L\) except in the pair responses, and suppress \(\bm\beta_L\) where no ambiguity arises. For \(q\ge1\), let \(a_{ij}^{(q)}=T_{qi}\wedge T_{qj}\), \(N_{ij}^{(q)}=I\{a_{ij}^{(q)}<L,\ T_{qi}\ne T_{qj}\}\), and \(H_{ij}^{(q)}(L)=I(T_{qi}>T_{qj})\) on \(N_{ij}^{(q)}=1\). Also put \(N_{ij}^{(0)}=I\{D_i\wedge D_j<L,\ D_i\ne D_j\}\) and \(H_{ij}^{(0)}(L)=I(D_i>D_j)\) on this event. The determining component indicators satisfy \(\kappa_{ij}^{(0)}=N_{ij}^{(0)}\) and \(\kappa_{ij}^{(q)}=J_{ij}^{(q)}N_{ij}^{(q)}\) for \(q\ge1\). Every product below is defined as zero when its indicator is zero. Consequently,
\begin{equation}
H_{ij}(L)-\mu_{ij}
=N_{ij}^{(0)}(H_{ij}^{(0)}(L)-\mu_{ij})
+\sum_{q=1}^{Q}J_{ij}^{(q)}N_{ij}^{(q)}(H_{ij}^{(q)}(L)-\mu_{ij})
+J_{ij}^{(Q+1)}(1/2-\mu_{ij}).
\label{supp:eq:ctw-residual-partition}
\end{equation}

For each nonfatal component, define the observed quantity
\(\bm F_{ij}^{(q)} =\bm K_{ij}\frac{R_{ij}^{\mathrm{CTW},(q)}(L)}{\pi_{ij}^{\mathrm{CTW},(q)}(L)}(\widetilde H_{ij}^{(q)}(L)-\mu_{ij}),\)
with value zero when the observation indicator is zero. Since \(\bm F_{ij}^{(q)}\) is measurable with respect to \(\mathcal O_i(L)\) and \(\mathcal O_j(L)\), \eqref{supp:eq:ctw-general-posterior} gives
\begin{align*}
E\{\bm F_{ij}^{(q)}\omega_i^{(q)}\omega_j^{(q)}\mid\bm X_i,\bm X_j\}
&=E\left[\bm F_{ij}^{(q)}E\{J_{ij}^{(q)}\mid\mathcal O_i(L),\mathcal O_j(L)\}\mid\bm X_i,\bm X_j\right]\\
&=E\{\bm F_{ij}^{(q)}J_{ij}^{(q)}\mid\bm X_i,\bm X_j\}.
\end{align*}
On \(J_{ij}^{(q)}N_{ij}^{(q)}=1\), both individuals survive beyond \(L\). The component ordering is observed exactly when \(C_i \wedge C_j\ge a_{ij}^{(q)}\), with \(t_{ij}^{(q)}=a_{ij}^{(q)}\) and \(\widetilde H_{ij}^{(q)}(L)=H_{ij}^{(q)}(L)\). Conditional on \(\mathcal F_{ij}^{F}\),
\begin{align*}
E\{\bm F_{ij}^{(q)}J_{ij}^{(q)}\mid\mathcal F_{ij}^{F}\}
&=\bm K_{ij}J_{ij}^{(q)}N_{ij}^{(q)}(H_{ij}^{(q)}(L)-\mu_{ij})
\frac{P\{C_i \wedge C_j\ge a_{ij}^{(q)}\mid\mathcal F_{ij}^{F}\}}{S_C\{a_{ij}^{(q)}\mid\bm X_i^C\}S_C\{a_{ij}^{(q)}\mid\bm X_j^C\}}\\
&=\bm K_{ij}J_{ij}^{(q)}N_{ij}^{(q)}(H_{ij}^{(q)}(L)-\mu_{ij}).
\end{align*}
When \(J_{ij}^{(q)}N_{ij}^{(q)}=0\), both sides are zero. The fatal and genuine tie terms have the same observation requirements as RCW, so the same cancellation at \(A_{ij}(L)\) gives
\begin{align*}
E\{\bm U_{ij}^{\mathrm{CTW}}(\bm\beta_L)\mid\bm X_i,\bm X_j\}
&=\bm K_{ij}E\left[N_{ij}^{(0)}(H_{ij}^{(0)}(L)-\mu_{ij})
+\sum_{q=1}^{Q}J_{ij}^{(q)}N_{ij}^{(q)}(H_{ij}^{(q)}(L)-\mu_{ij})\right.\\
&\hspace{1.7cm}\left.+J_{ij}^{(Q+1)}(1/2-\mu_{ij})\mid\bm X_i,\bm X_j\right]\\
&=\bm K_{ij}\left[E\{H_{ij}(L)\mid\bm X_i,\bm X_j\}-\mu_{ij}\right],
\end{align*}
by \eqref{supp:eq:ctw-residual-partition}.
\end{proof}

At most one weighted component contributes for a given observed pair. If \(R_{ij}^{\mathrm{CTW},(q)}(L)=1\), an event of type \(q\) has been recorded, so \(\omega_i^{(r)}(L)\omega_j^{(r)}(L)=0\) for every \(r>q\). After an observed death, every product of conditional tie probabilities for a nonfatal comparison is zero. An observed genuine tie has no recorded component event. The efficiency comparison uses this property.

The iterated expectation step uses that the covariates are included in the observed records. It need not hold after additionally conditioning on \(\bm Y_i,\bm Y_j\), because the full event histories are not determined by those records. After averaging over the event histories, \eqref{supp:eq:ctw-general-mean} gives the same population estimating equation and derivative \(\bm A_L\) for CHW, RCW, and CTW.

\subsection{Symmetry under pair reversal}
\label{supp:pair-symmetry}

\begin{lemma}[Symmetry under pair reversal]
\label{supp:lem:score-symmetry}
\leavevmode\newline
Suppose \(\bm Z_{ji}=-\bm Z_{ij}\), \(g^{-1}(-v)=1-g^{-1}(v)\), and \(V(u)=V(1-u)\). For every positive candidate censoring survival function \(S\), the RCW component contributions satisfy \(\bm U_{ji}^{(q)}(\bm\beta_L;S)=\bm U_{ij}^{(q)}(\bm\beta_L;S)\) for each \(q\) and \(\bm U_{ji}^{(T)}(\bm\beta_L;S)=\bm U_{ij}^{(T)}(\bm\beta_L;S)\). For each of CHW, RCW, and CTW, \(\bm U_{ji}^{\star}(\bm\beta_L;S)=\bm U_{ij}^{\star}(\bm\beta_L;S)\), with the same fixed joint model used for both individuals in CTW.
\end{lemma}

\begin{proof}
The inverse link identity implies
\begin{align*}
\mu_{ji}(\bm\beta_L)
&=g^{-1}(-\bm\beta_L^\top\bm Z_{ij})
=1-\mu_{ij}(\bm\beta_L),\\
\bm K_{ji}(\bm\beta_L)
&=\frac{-\partial\mu_{ij}(\bm\beta_L)/\partial\bm\beta_L}
 {V\{1-\mu_{ij}(\bm\beta_L)\}}
=-\bm K_{ij}(\bm\beta_L).
\end{align*}
For a strict comparison determined by component \(q\),
\begin{align*}
I_{ji}^{(q)}-\mu_{ji}(I_{ji}^{(q)}+I_{ij}^{(q)})
&=I_{ji}^{(q)}-(1-\mu_{ij})
 (I_{ji}^{(q)}+I_{ij}^{(q)})\\
&=-I_{ij}^{(q)}+\mu_{ij}
 (I_{ij}^{(q)}+I_{ji}^{(q)})\\
&=-\{I_{ij}^{(q)}-\mu_{ij}
 (I_{ij}^{(q)}+I_{ji}^{(q)})\}.
\end{align*}
The observation time and the product of survival probabilities are unchanged when the two individuals are exchanged. Both the vector multiplier and the residual therefore change sign, proving symmetry of \(\bm U_{ij}^{(q)}\). For a genuine tie, \(I_{ji}^{(T)}=I_{ij}^{(T)}\) and \(1/2-\mu_{ji}=-(1/2-\mu_{ij})\), which proves symmetry of \(\bm U_{ij}^{(T)}\).

The definition of the observed response also gives \(\widetilde H_{ji}(L)=1-\widetilde H_{ij}(L)\). This follows from reversal of a resolved comparison and from the assigned value \(1/2\) on an unresolved comparison. The censoring weights are unchanged when \(i\) and \(j\) are exchanged. For \(\star\in\{\mathrm{CHW},\mathrm{RCW}\}\),
\begin{align*}
\bm U_{ji}^{\star}(\bm\beta_L;S)
&=(-\bm K_{ij})W_{ij}^{C,\star}(L;S)
 \{(1-\widetilde H_{ij}(L))-(1-\mu_{ij})\}\\
&=\bm K_{ij}W_{ij}^{C,\star}(L;S)(\widetilde H_{ij}(L)-\mu_{ij})
=\bm U_{ij}^{\star}(\bm\beta_L;S).
\end{align*}

For CTW, reversal leaves every observation indicator, censoring probability, and product of conditional tie probabilities unchanged. It changes \(\widetilde H_{ij}^{(q)}(L)-\mu_{ij}\) to its negative and \(\bm K_{ij}\) to \(-\bm K_{ij}\). The genuine tie residual has the same sign change, proving symmetry.
\end{proof}

With fixed censoring and joint model quantities, the three estimating functions are symmetric U-statistics of order two. Section \ref{supp:weight-expansion} accounts for estimation of these quantities from the same sample.

\subsection{Efficiency comparison of the three estimating functions}
\label{supp:efficiency}

Theorem \ref{thm:efficiency-comparison} of main text concerns the true censoring survival function and conditional tie probabilities. Note that all scores are evaluated at \(\bm\beta_L^0\). As in the main text, \(\bm\Gamma^{\star}=\operatorname{Var}\{\bm U_{ij}^{\star}(\bm\beta_L^0)\mid\bm X_i,\bm X_j\}\). For this proof, write \(\bm K_{ij}=\bm K_{ij}(\bm\beta_L^0)\), \(\mu_{ij}=\mu_{ij}(\bm\beta_L^0)\), and \(p_{ij,L}=S_C(L\mid\bm X_i^C)S_C(L\mid\bm X_j^C)\). The population identities in Sections \ref{supp:unbiased} and \ref{supp:ctw-identification} give the same conditional mean for all three scores. Their conditional covariance differences therefore equal the differences between their conditional second moments.

\begin{proof}
For \(\star\in\{\mathrm{CHW},\mathrm{RCW}\}\), Lemma \ref{supp:lem:comparison-observability}, \((R_{ij}^{\star})^2=R_{ij}^{\star}\), and iterated expectation give
\begin{align*}
E\{\bm U_{ij}^{\star}(\bm U_{ij}^{\star})^\top\mid\bm X_i,\bm X_j\}
&=\bm K_{ij}\bm K_{ij}^\top E\left[\frac{R_{ij}^{\star}(L)}{\{\pi_{ij}^{\star}(L)\}^2}\{H_{ij}(L)-\mu_{ij}\}^2\mid\bm X_i,\bm X_j\right]\\
&=\bm K_{ij}\bm K_{ij}^\top E\left[\frac{\{H_{ij}(L)-\mu_{ij}\}^2}{\pi_{ij}^{\star}(L)}\mid\bm X_i,\bm X_j\right].
\end{align*}
Subtracting the RCW expression from the CHW expression proves \eqref{eq:chw-rcw-pair-variance}. Its nonnegative definiteness follows from \(\pi_{ij}^{\mathrm{CHW}}(L)\le\pi_{ij}^{\mathrm{RCW}}(L)\). The difference is zero for nonfatal comparisons and genuine ties, which require both individuals to survive through \(L\). The reduction arises from fatal comparisons that can be determined before both restricted histories are complete, including \emph{partially observed death-resolved pairs}.

For the CTW comparison, a recorded event on component \(q\) makes \(\omega_i^{(r)}(L)\omega_j^{(r)}(L)=0\) for every \(r>q\). After a recorded death, the product of conditional tie probabilities is zero for every nonfatal component, and an observed genuine tie has no recorded event before \(L\). Hence at most one component of the CTW score is nonzero. The same property holds for RCW because its determining components partition the resolved comparisons. The fatal and genuine-tie contributions agree between CTW and RCW, so only the squared nonfatal contributions remain in the difference between their second moments.

For \(q=1,\ldots,Q\), the iterated expectation argument in Section \ref{supp:ctw-identification} also applies to the squared residual. In particular,
\begin{align*}
&E\left[\frac{R_{ij}^{\mathrm{CTW},(q)}(L)\omega_i^{(q)}(L)\omega_j^{(q)}(L)}{\pi_{ij}^{\mathrm{CTW},(q)}(L)}\{\widetilde H_{ij}^{(q)}(L)-\mu_{ij}\}^2\mid\bm X_i,\bm X_j\right]\\
&\hspace{1em}=E\left[\kappa_{ij}^{(q)}(L)\{H_{ij}(L)-\mu_{ij}\}^2\mid\bm X_i,\bm X_j\right].
\end{align*}
To obtain this equality, first replace the product of conditional tie probabilities inside the expectation by the indicator of the required higher-priority genuine ties, and then condition on the full event histories to cancel the inverse censoring factor. On \(\kappa_{ij}^{(q)}(L)=1\), \(A_{ij}(L)=L\), so the squared RCW contribution has conditional expectation
\begin{align*}
&E\left[\frac{R_{ij}^{\mathrm{RCW}}(L)\kappa_{ij}^{(q)}(L)}{p_{ij,L}^2}\{H_{ij}(L)-\mu_{ij}\}^2\mid\bm X_i,\bm X_j\right]\\
&\hspace{1em}=\frac1{p_{ij,L}}E\left[\kappa_{ij}^{(q)}(L)\{H_{ij}(L)-\mu_{ij}\}^2\mid\bm X_i,\bm X_j\right]\\
&\hspace{1em}=E\left[\frac{R_{ij}^{\mathrm{CTW},(q)}(L)\omega_i^{(q)}(L)\omega_j^{(q)}(L)}{p_{ij,L}\pi_{ij}^{\mathrm{CTW},(q)}(L)}\{\widetilde H_{ij}^{(q)}(L)-\mu_{ij}\}^2\mid\bm X_i,\bm X_j\right].
\end{align*}
The squared CTW contribution instead has the factor 
\begin{equation*}
R_{ij}^{\mathrm{CTW},(q)}(L)\{\omega_i^{(q)}(L)\omega_j^{(q)}(L)\}^2/\{\pi_{ij}^{\mathrm{CTW},(q)}(L)\}^2.
\end{equation*}
Subtracting it, summing over \(q=1,\ldots,Q\), and multiplying by \(\bm K_{ij}\bm K_{ij}^\top\) yields \eqref{eq:rcw-ctw-pair-variance}.

The factor multiplying each squared residual separates into two nonnegative terms,
\begin{align*}
&\frac{\omega_i^{(q)}(L)\omega_j^{(q)}(L)}{\pi_{ij}^{\mathrm{CTW},(q)}(L)}\left\{\frac1{p_{ij,L}}-\frac{\omega_i^{(q)}(L)\omega_j^{(q)}(L)}{\pi_{ij}^{\mathrm{CTW},(q)}(L)}\right\}\\
&\hspace{1em}=\frac{\omega_i^{(q)}(L)\omega_j^{(q)}(L)}{\pi_{ij}^{\mathrm{CTW},(q)}(L)}\left\{\frac1{p_{ij,L}}-\frac1{\pi_{ij}^{\mathrm{CTW},(q)}(L)}\right\}+\frac{\omega_i^{(q)}(L)\omega_j^{(q)}(L)\{1-\omega_i^{(q)}(L)\omega_j^{(q)}(L)\}}{\{\pi_{ij}^{\mathrm{CTW},(q)}(L)\}^2}.
\end{align*}
The first term is nonnegative because the nonfatal event time precedes \(L\), giving \(\pi_{ij}^{\mathrm{CTW},(q)}(L)\ge p_{ij,L}\). It reflects the earlier time at which the censoring survival functions are evaluated. The second is nonnegative because the product of conditional tie probabilities lies in \([0,1]\). It follows from replacing the indicator of the required higher-priority genuine ties by its conditional probability. This proves \(\bm\Gamma^{\mathrm{CHW}}\succeq\bm\Gamma^{\mathrm{RCW}}\succeq\bm\Gamma^{\mathrm{CTW}}\).
\end{proof}

For a nonfatal comparison already resolved by RCW, \(\omega_i^{(q)}(L)\omega_j^{(q)}(L)=1\), so only the first term contributes to the reduction. For \emph{partially observed lower-priority-resolved pairs}, the second term also contributes whenever the required higher-priority ties have conditional probability strictly between zero and one.

The comparison concerns a single pair evaluated at the true censoring survival function and conditional tie probabilities. The covariance of a fitted estimator additionally depends on overlapping pairs and model estimation. Let \(\bm\xi_L^{\star}(o)=E(\bm\zeta_{12}^{\star}\mid\mathcal O_1=o)\) be the first projection of the adjusted pair score defined in Section \ref{supp:theory}, including both model contributions for CTW, and, for \(\star,\star'\in\mathcal A=\{\mathrm{CHW},\mathrm{RCW},\mathrm{CTW}\}\), write \(\bm d_L^{\star,\star'}(o)=\bm\xi_L^{\star}(o)-\bm\xi_L^{\star'}(o)\). The representation \(\bm B_L^{\star}=2E\{\bm\xi_L^{\star}(\mathcal O_1)\bm\xi_L^{\star}(\mathcal O_1)^\top\}\) gives
\begin{align}
\bm B_L^{\star}-\bm B_L^{\star'}
&=2E\left[\{\bm\xi_L^{\star'}+\bm d_L^{\star,\star'}\}\{\bm\xi_L^{\star'}+\bm d_L^{\star,\star'}\}^\top-\bm\xi_L^{\star'}(\bm\xi_L^{\star'})^\top\right]\nonumber\\
&=2E\left[\bm\xi_L^{\star'}(\bm d_L^{\star,\star'})^\top+\bm d_L^{\star,\star'}(\bm\xi_L^{\star'})^\top+\bm d_L^{\star,\star'}(\bm d_L^{\star,\star'})^\top\right].
\label{supp:eq:efficiency-projection}
\end{align}
The cross terms in \eqref{supp:eq:efficiency-projection} need not have a definite sign. Thus the pairwise comparison alone does not establish an ordering of the covariance matrices of the fitted estimators \(\bm A_L^{-1}\bm B_L^{\star}\bm A_L^{-\top}\).

\section{Estimation of the censoring weights}
\label{supp:cox-if}

The same estimated censoring survival function enters all three estimating equations. We derive its influence function, give the expansions of the CHW and RCW weights, and apply these calculations to each CTW component.

\subsection{Cox model and influence function}
\label{supp:censoring-influence}

For the censoring predictors \(\bm X_i^C\) in Section \ref{sec:estimation}, let \(\bm\gamma^0\) denote the true coefficient vector. The superscript distinguishes this coefficient from \(\bm\gamma_0\), the coefficient vector for the marginal death model. The following uniform influence representation is used for the fitted weights in Appendix \ref{supp:two-weight-expansions}.

Let \(\tau_C\ge L\) be the fixed horizon through which the censoring model is fitted. The censoring counting processes are based on \(\mathcal O_i(\tau_C)\), and the pairwise comparisons on \(\mathcal O_i(L)\). Set \(\widetilde D_i^C=D_i\wedge C_i\wedge\tau_C\) and \(\delta_i^C=I(C_i<D_i,C_i\le\tau_C)\). For \(0\le t\le\tau_C\), define the at-risk process \(\mathcal R_i^C(t)=I(\widetilde D_i^C\ge t)\) and the censoring counting process \(N_i^C(t)=I(\widetilde D_i^C\le t,\delta_i^C=1)\). Under the correctly specified Cox model \(\lambda_i^C(t\mid\bm X_i^C)=\lambda_0^C(t)e^{(\bm\gamma^0)^\top\bm X_i^C}\), the martingale increment is \(dM_i^C(t)=dN_i^C(t)-\mathcal R_i^C(t)e^{(\bm\gamma^0)^\top\bm X_i^C}d\Lambda_0^C(t)\), where \(\Lambda_0^C(t)=\int_0^t\lambda_0^C(u)du\). The censoring survival function is \(S_C(t\mid\bm x)=\exp\{-\Lambda_0^C(t)e^{(\bm\gamma^0)^\top\bm x}\}\). If follow-up is available only through \(L\), take \(\tau_C=L\).

For \(k=0,1,2\), write \(r_C^{(k)}(t)=E\{\mathcal R_i^C(t)e^{(\bm\gamma^0)^\top\bm X_i^C}(\bm X_i^C)^{\otimes k}\}\), where \(\bm x^{\otimes0}=1\), \(\bm x^{\otimes1}=\bm x\), and \(\bm x^{\otimes2}=\bm x\bm x^\top\). Define \(\overline{\bm X}_C(t)=r_C^{(1)}(t)/r_C^{(0)}(t)\) and \(\bm a_C(t)=\int_0^t\overline{\bm X}_C(u)d\Lambda_0^C(u)\). The population information for the censoring coefficient is
\begin{align*}
\bm\Omega_C
&=E\left[\int_0^{\tau_C}
\left\{\frac{r_C^{(2)}(u)}{r_C^{(0)}(u)}
-\overline{\bm X}_C(u)\overline{\bm X}_C(u)^\top\right\}dN_i^C(u)\right]\\
&=\int_0^{\tau_C}
\left\{r_C^{(2)}(u)
-\frac{r_C^{(1)}(u)r_C^{(1)}(u)^\top}{r_C^{(0)}(u)}\right\}d\Lambda_0^C(u).
\end{align*}

Assume conditionally independent censoring, a continuous cumulative censoring hazard, bounded censoring covariates, \(\inf_{u\le\tau_C}r_C^{(0)}(u)>0\), \(\Lambda_0^C(\tau_C)<\infty\), and nonsingular \(\bm\Omega_C\). The partial likelihood estimator is assumed to be a consistent local root in a neighborhood of the finite true value \(\bm\gamma^0\). The empirical risk-set quantities converge uniformly on a fixed neighborhood of \(\bm\gamma^0\), where bounded covariates and monotone at-risk paths give integrable envelopes. The outcome PIM may be a working model under these conditions. All population martingale integrands below are predictable, and the fitted empirical integrals are evaluated pathwise.

Write \(\ell_C(t,\bm x)=\log S_C(t\mid\bm x)\) and \(\widehat\ell_C(t,\bm x)=-\widehat\Lambda_0^C(t-)e^{\widehat{\bm\gamma}^\top\bm x}\). The left limit \(\widehat\Lambda_0^C(t-)=\lim_{u\uparrow t}\widehat\Lambda_0^C(u)\) excludes an increment in the estimated censoring hazard at \(t\), under the event and censoring convention in \ref{supp:weights}.

\begin{proposition}[Uniform Cox and Breslow expansion]
\label{supp:prop:cox-breslow}
The Cox partial likelihood estimator and the Breslow estimator satisfy
\begin{align}
\widehat{\bm\gamma}-\bm\gamma^0
&=\frac1n\sum_{i=1}^n\operatorname{IF}_{\gamma,i}+o_p(n^{-1/2}),
\label{supp:eq:gamma-expansion}\\
\widehat\Lambda_0^C(t)-\Lambda_0^C(t)
&=\frac1n\sum_{i=1}^n\operatorname{IF}_{\Lambda,i}(t)+o_p(n^{-1/2}),\nonumber
\end{align}
with the second remainder uniform over \(t\in[0,\tau_C]\). The influence functions are
\begin{align}
\operatorname{IF}_{\gamma,i}
&=\bm\Omega_C^{-1}\int_0^{\tau_C}
\{\bm X_i^C-\overline{\bm X}_C(u)\}dM_i^C(u),\nonumber\\
\operatorname{IF}_{\Lambda,i}(t)
&=\int_0^t\frac{dM_i^C(u)}{r_C^{(0)}(u)}
-\bm a_C(t)^\top\operatorname{IF}_{\gamma,i}.
\label{supp:eq:IF-Lambda}
\end{align}
Consequently, \(\widehat\ell_C(t,\bm x)-\ell_C(t,\bm x) =\frac1n\sum_{i=1}^n\mathcal I(\mathcal O_i;t,\bm x)+o_p(n^{-1/2})\) uniformly over \(t\in[0,L]\) and the bounded covariate support, where
\begin{align}
\mathcal I(\mathcal O_i;t,\bm x)
&=-e^{(\bm\gamma^0)^\top\bm x}\left[
\int_{[0,t)}\frac{dM_i^C(u)}{r_C^{(0)}(u)}
+\{\Lambda_0^C(t)\bm x-\bm a_C(t)\}^\top
\operatorname{IF}_{\gamma,i}\right].
\label{supp:eq:IF-logS}
\end{align}
\end{proposition}

\begin{proof}
For \(k=0,1,2\), let \(S_{C,n}^{(k)}(u;\bm\gamma)=n^{-1}\sum_{i=1}^n\mathcal R_i^C(u)e^{\bm\gamma^\top\bm X_i^C}(\bm X_i^C)^{\otimes k}\) and \(\overline{\bm X}_{C,n}(u;\bm\gamma)=S_{C,n}^{(1)}(u;\bm\gamma)/S_{C,n}^{(0)}(u;\bm\gamma)\). Let \(\bm U_{C,n}(\bm\gamma)\) denote the normalized partial likelihood score. At the true coefficient, the compensator terms cancel because
\begin{align*}
\frac1n\sum_{i=1}^n
\mathcal R_i^C(u)e^{(\bm\gamma^0)^\top\bm X_i^C}
\{\bm X_i^C-\overline{\bm X}_{C,n}(u;\bm\gamma^0)\}
&=S_{C,n}^{(1)}(u;\bm\gamma^0)
-S_{C,n}^{(0)}(u;\bm\gamma^0)
\overline{\bm X}_{C,n}(u;\bm\gamma^0)\\
&=\bm0.
\end{align*}
Hence
\begin{align*}
\bm U_{C,n}(\bm\gamma^0)
&=\frac1n\sum_{i=1}^n\int_0^{\tau_C}
\{\bm X_i^C-\overline{\bm X}_{C,n}(u;\bm\gamma^0)\}dN_i^C(u)\\
&=\frac1n\sum_{i=1}^n\int_0^{\tau_C}
\{\bm X_i^C-\overline{\bm X}_{C,n}(u;\bm\gamma^0)\}dM_i^C(u)\\
&=\frac1n\sum_{i=1}^n\int_0^{\tau_C}
\{\bm X_i^C-\overline{\bm X}_C(u)\}dM_i^C(u)
+o_p(n^{-1/2}).
\end{align*}
To justify the final equality, set \(\bm D_n(u)=\overline{\bm X}_{C,n}(u;\bm\gamma^0)-\overline{\bm X}_C(u)\). Independence across individuals and the continuous censoring hazard give orthogonal censoring martingales. The trace of the predictable quadratic variation of \(n^{-1/2}\sum_{i=1}^n\int_0^{\tau_C}\bm D_n(u)dM_i^C(u)\) is bounded by
\begin{align*}
\int_0^{\tau_C}\|\bm D_n(u)\|^2
S_{C,n}^{(0)}(u;\bm\gamma^0)d\Lambda_0^C(u)
&\le \sup_{u\le\tau_C}\|\bm D_n(u)\|^2
\int_0^{\tau_C}S_{C,n}^{(0)}(u;\bm\gamma^0)d\Lambda_0^C(u)=o_p(1).
\end{align*}
The martingale inequality gives the remainder bound. Bounded covariates, a risk-set denominator bounded away from zero, and finite cumulative hazard also give \(\bm U_{C,n}(\bm\gamma^0)=O_p(n^{-1/2})\).

The negative derivative of the score is the empirical risk-set covariance integrated against \(d\overline N_n^C\), where \(\overline N_n^C=n^{-1}\sum_{i=1}^nN_i^C\). Uniform risk-set convergence and the counting-process decomposition give
\begin{align*}
-\frac{\partial\bm U_{C,n}(\bm\gamma)}{\partial\bm\gamma^\top}
&=\int_0^{\tau_C}\left\{
\frac{S_{C,n}^{(2)}(u;\bm\gamma)}{S_{C,n}^{(0)}(u;\bm\gamma)}
-\overline{\bm X}_{C,n}(u;\bm\gamma)
\overline{\bm X}_{C,n}(u;\bm\gamma)^\top
\right\}d\overline N_n^C(u)=\bm\Omega_C+o_p(1)
\end{align*}
uniformly on any shrinking neighborhood of \(\bm\gamma^0\). Applying the integral form of Taylor's theorem to the consistent root gives
\begin{align*}
\bm0&=\bm U_{C,n}(\widehat{\bm\gamma})=\bm U_{C,n}(\bm\gamma^0)
-\{\bm\Omega_C+o_p(1)\}(\widehat{\bm\gamma}-\bm\gamma^0),\\
\widehat{\bm\gamma}-\bm\gamma^0
&=\bm\Omega_C^{-1}\bm U_{C,n}(\bm\gamma^0)+o_p(n^{-1/2})=\frac1n\sum_{i=1}^n\operatorname{IF}_{\gamma,i}+o_p(n^{-1/2}),
\end{align*}
which proves \eqref{supp:eq:gamma-expansion} and the root-\(n\) rate.

Let \(\overline M_n^C=n^{-1}\sum_{i=1}^nM_i^C\). For fixed \(\bm\gamma\), define \(\widehat\Lambda_0^C(t;\bm\gamma)=\int_0^t\{S_{C,n}^{(0)}(u;\bm\gamma)\}^{-1}d\overline N_n^C(u)\), so that \(\widehat\Lambda_0^C(t)=\widehat\Lambda_0^C(t;\widehat{\bm\gamma})\). With probability tending to one, the empirical denominator is bounded away from zero over the fitting interval. On this event,
\begin{align*}
\widehat\Lambda_0^C(t;\bm\gamma^0)-\Lambda_0^C(t)
&=\int_0^t\frac{d\overline M_n^C(u)
+S_{C,n}^{(0)}(u;\bm\gamma^0)d\Lambda_0^C(u)}
{S_{C,n}^{(0)}(u;\bm\gamma^0)}-\Lambda_0^C(t)=\int_0^t\frac{d\overline M_n^C(u)}
{S_{C,n}^{(0)}(u;\bm\gamma^0)}\\
&=\frac1n\sum_{i=1}^n\int_0^t
\frac{dM_i^C(u)}{r_C^{(0)}(u)}+o_p(n^{-1/2}).
\end{align*}
The final remainder is uniform in \(t\le\tau_C\). After multiplication by \(\sqrt n\), its predictable quadratic variation is bounded by a constant times \(\sup_{u\le\tau_C}|1/S_{C,n}^{(0)}(u;\bm\gamma^0)-1/r_C^{(0)}(u)|^2=o_p(1)\). The same martingale inequality bounds the leading process uniformly by \(O_p(n^{-1/2})\).

Differentiation with respect to the censoring coefficient gives
\begin{align*}
\frac{\partial\widehat\Lambda_0^C(t;\bm\gamma)}{\partial\bm\gamma}
&=-\int_0^t
\frac{\overline{\bm X}_{C,n}(u;\bm\gamma)}
{S_{C,n}^{(0)}(u;\bm\gamma)}d\overline N_n^C(u)=-\int_0^t\overline{\bm X}_C(u)d\Lambda_0^C(u)+o_p(1)=-\bm a_C(t)+o_p(1)
\end{align*}
uniformly in \(t\le\tau_C\) and in a shrinking neighborhood of \(\bm\gamma^0\). Bounded covariates and the lower bound on the denominator imply that the second derivatives are uniformly bounded in probability. A further Taylor expansion therefore yields
\begin{align*}
\widehat\Lambda_0^C(t)-\Lambda_0^C(t)
&=\{\widehat\Lambda_0^C(t;\widehat{\bm\gamma})
-\widehat\Lambda_0^C(t;\bm\gamma^0)\}
+\{\widehat\Lambda_0^C(t;\bm\gamma^0)-\Lambda_0^C(t)\}\\
&=-\bm a_C(t)^\top(\widehat{\bm\gamma}-\bm\gamma^0)
+\frac1n\sum_{i=1}^n\int_0^t
\frac{dM_i^C(u)}{r_C^{(0)}(u)}+o_p(n^{-1/2})\\
&=\frac1n\sum_{i=1}^n\operatorname{IF}_{\Lambda,i}(t)
+o_p(n^{-1/2})
\end{align*}
uniformly in \(t\le\tau_C\). This proves \eqref{supp:eq:IF-Lambda} and the uniform root-\(n\) rate for the baseline estimator.

Taking left limits preserves the uniform remainder. Since \(\Lambda_0^C\) and \(\bm a_C\) are continuous, write \(\Delta_\Lambda(t)=\widehat\Lambda_0^C(t-)-\Lambda_0^C(t)\) and \(\Delta_\gamma=\widehat{\bm\gamma}-\bm\gamma^0\). Bounded covariates imply \(e^{\widehat{\bm\gamma}^\top\bm x}=e^{(\bm\gamma^0)^\top\bm x}(1+\bm x^\top\Delta_\gamma)+O_p(n^{-1})\) uniformly in \(\bm x\). Substitution into the fitted log survival gives
\begin{align*}
\widehat\ell_C(t,\bm x)-\ell_C(t,\bm x)
&=-e^{(\bm\gamma^0)^\top\bm x}
\{\Delta_\Lambda(t)+\Lambda_0^C(t)\bm x^\top\Delta_\gamma\}
+O_p(n^{-1})\\
&=-e^{(\bm\gamma^0)^\top\bm x}\frac1n\sum_{i=1}^n
\left[\int_{[0,t)}\frac{dM_i^C(u)}{r_C^{(0)}(u)}
+\{\Lambda_0^C(t)\bm x-\bm a_C(t)\}^\top\operatorname{IF}_{\gamma,i}\right]
+o_p(n^{-1/2})\\
&=\frac1n\sum_{i=1}^n\mathcal I(\mathcal O_i;t,\bm x)
+o_p(n^{-1/2}).
\end{align*}
The product of the baseline and coefficient errors is \(O_p(n^{-1})\), because \(\sup_{t\le L}|\Delta_\Lambda(t)|=O_p(n^{-1/2})\) and \(\|\Delta_\gamma\|=O_p(n^{-1/2})\). All remainders are uniform over the stated time interval and covariate support. For each fixed \((t,\bm x)\), the martingale integrals have expectation zero, so \(E\{\mathcal I(\mathcal O_i;t,\bm x)\}=0\). Bounded covariates, a risk-set denominator bounded away from zero, finite cumulative hazard, and nonsingular \(\bm\Omega_C\) also bound the influence functions uniformly. The Cox and Breslow decompositions are the standard counting-process expansions of \citet{AndersenGill1982}.
\end{proof}

The Cox coefficient influence integrates through \(\tau_C\), including an observed censoring jump at that endpoint. The baseline term in \eqref{supp:eq:IF-logS} instead integrates over \([0,t)\), as required by the fitted log survival at \(t\). An individual influence function may jump at the observed censoring time, so its left and right limits must be distinguished. With an empirical risk-set denominator bounded away from zero and continuous censoring times, the largest Breslow jump is \(O_p(n^{-1})\). A right-continuous survival estimator would therefore differ uniformly by \(O_p(n^{-1})\). The same \([0,t)\) convention applies to the population and empirical influence functions for the log survival. If \(\tau_C>L\), the coefficient influence and information use \(\tau_C\), and all pairwise survival evaluations are at times no greater than \(L\).

\subsection{Estimation of the censoring influence function}
\label{supp:estimated-censoring-influence}

\begin{lemma}[Uniform consistency of the empirical censoring influence]
\label{supp:lem:empirical-censoring-if}
\leavevmode\newline
Let \(\widehat r_C^{(k)}(u)=n^{-1}\sum_{i=1}^n\mathcal R_i^C(u)e^{\widehat{\bm\gamma}^\top\bm X_i^C}(\bm X_i^C)^{\otimes k}\), \(\widehat{\overline{\bm X}}_C=\widehat r_C^{(1)}/\widehat r_C^{(0)}\), and let \(\widehat{\bm\Omega}_C\) be the observed Cox information divided by \(n\). Define \(d\widehat M_i^C(u)=dN_i^C(u)-\mathcal R_i^C(u)e^{\widehat{\bm\gamma}^\top\bm X_i^C}d\widehat\Lambda_0^C(u)\) and \(\widehat{\bm a}_C(t)=\int_0^t\widehat{\overline{\bm X}}_C(u)d\widehat\Lambda_0^C(u)\). The empirical influence functions are
\begin{align*}
\widehat{\operatorname{IF}}_{\gamma,i}
&=\widehat{\bm\Omega}_C^{-1}\int_0^{\tau_C}
\{\bm X_i^C-\widehat{\overline{\bm X}}_C(u)\}d\widehat M_i^C(u),\\
\widehat{\mathcal I}(\mathcal O_i;t,\bm x)
&=-e^{\widehat{\bm\gamma}^\top\bm x}\left[
\int_{[0,t)}\frac{d\widehat M_i^C(u)}{\widehat r_C^{(0)}(u)}
+\{\widehat\Lambda_0^C(t-)\bm x-\widehat{\bm a}_C(t-)\}^\top
\widehat{\operatorname{IF}}_{\gamma,i}\right].
\end{align*}
Under the preceding Cox conditions, \(\max_{1\le i\le n}\sup_{t\le L,\bm x} |\widehat{\mathcal I}(\mathcal O_i;t,\bm x)-\mathcal I(\mathcal O_i;t,\bm x)| =o_p(1)\). Moreover, \(|\mathcal I(\mathcal O_i;t,\bm x)|\le C\) uniformly in \(i\), \(t\), and \(\bm x\), for a finite deterministic constant \(C\). The empirical influences satisfy \(\max_i\sup_{t,\bm x}|\widehat{\mathcal I}(\mathcal O_i;t,\bm x)|=\allowbreak O_p(1)\).
\end{lemma}

\begin{proof}
Uniform convergence of the empirical risk-set quantities and consistency of \(\widehat{\bm\gamma}\) imply \(\max_{0\le k\le2}\|\widehat r_C^{(k)}-r_C^{(k)}\|_\infty=o_p(1)\). The positive lower bound on \(r_C^{(0)}\) then gives uniform convergence of \(1/\widehat r_C^{(0)}\) and \(\widehat{\overline{\bm X}}_C\) to their population counterparts. The score derivative in the preceding proof gives \(\widehat{\bm\Omega}_C\overset{p}{\longrightarrow}\bm\Omega_C\), and hence \(\widehat{\bm\Omega}_C^{-1}\overset{p}{\longrightarrow}\bm\Omega_C^{-1}\).

Each coordinate of \(r_C^{(k)}\) has bounded variation on \([0,\tau_C]\), because it is the expectation of a bounded covariate multiple of a monotone at-risk indicator. Positivity implies the same property for \(1/r_C^{(0)}\) and \(\overline{\bm X}_C\). Thus both \(\mathcal R_i^C(u)e^{(\bm\gamma^0)^\top\bm X_i^C}\{\bm X_i^C-\overline{\bm X}_C(u)\}\) and \(\mathcal R_i^C(u)e^{(\bm\gamma^0)^\top\bm X_i^C}/r_C^{(0)}(u)\) have uniformly bounded suprema and total variations over \(i\). Multiplication by \(I(u<t)\) preserves a bound uniform in \(t\). Let \(f_i(u)\) be either \(\bm X_i^C-\overline{\bm X}_C(u)\) or \(1/r_C^{(0)}(u)\). Integration by parts gives
\begin{align*}
\max_i\sup_{t\le\tau_C}
\left|\int_{[0,t)}\mathcal R_i^C(u)e^{(\bm\gamma^0)^\top\bm X_i^C}
f_i(u)d\{\widehat\Lambda_0^C(u)-\Lambda_0^C(u)\}\right|
&\le C\|\widehat\Lambda_0^C-\Lambda_0^C\|_\infty=o_p(1),
\end{align*}
where a vector integral is bounded coordinatewise.

Let \(\widehat f_i\) be the fitted version of \(f_i\). Expanding the empirical and population martingales gives the exact difference
\begin{align*}
\int_{[0,t)}\widehat f_i(u)d\widehat M_i^C(u)
-\int_{[0,t)}f_i(u)dM_i^C(u)
&=\int_{[0,t)}\{\widehat f_i(u)-f_i(u)\}dN_i^C(u)\\
&\quad-\int_{[0,t)}\mathcal R_i^C(u)
\{\widehat f_i(u)e^{\widehat{\bm\gamma}^\top\bm X_i^C}
-f_i(u)e^{(\bm\gamma^0)^\top\bm X_i^C}\}d\widehat\Lambda_0^C(u)\\
&\quad-\int_{[0,t)}\mathcal R_i^C(u)e^{(\bm\gamma^0)^\top\bm X_i^C}
f_i(u)d\{\widehat\Lambda_0^C(u)-\Lambda_0^C(u)\}.
\end{align*}
Since \(N_i^C\) has at most one jump, the first term is bounded uniformly by \(\max_i\|\widehat f_i-f_i\|_\infty=o_p(1)\). The second is bounded by
\begin{align*}
\widehat\Lambda_0^C(\tau_C)
\max_i\sup_{u\le\tau_C}
\|\widehat f_i(u)e^{\widehat{\bm\gamma}^\top\bm X_i^C}
-f_i(u)e^{(\bm\gamma^0)^\top\bm X_i^C}\|
&=O_p(1)o_p(1)=o_p(1).
\end{align*}
The third is uniformly \(o_p(1)\) by the integration-by-parts bound. The same bounds apply to the closed integral through \(\tau_C\) used for the coefficient influence. Consequently,
\begin{align*}
\max_i\|\widehat{\operatorname{IF}}_{\gamma,i}
-\operatorname{IF}_{\gamma,i}\|
&=o_p(1),\\
\max_i\sup_{t\le L}\left|
\int_{[0,t)}\frac{d\widehat M_i^C(u)}{\widehat r_C^{(0)}(u)}
-\int_{[0,t)}\frac{dM_i^C(u)}{r_C^{(0)}(u)}\right|
&=o_p(1).
\end{align*}
Likewise, \(\sup_{t\le L}\|\widehat{\bm a}_C(t)-\bm a_C(t)\|=o_p(1)\) follows by separating the difference in \(\overline{\bm X}_C\) from the difference in the cumulative hazard and applying the same two bounds. Substitution into the definition of \(\widehat{\mathcal I}\) proves the asserted uniform convergence.

Finally, for any bounded integrand \(f_i\),
\begin{align*}
\left\|\int_0^{\tau_C}f_i(u)dM_i^C(u)\right\|
&\le \|f_i\|_\infty
\left\{N_i^C(\tau_C)
+\int_0^{\tau_C}\mathcal R_i^C(u)e^{(\bm\gamma^0)^\top\bm X_i^C}d\Lambda_0^C(u)\right\}\\
&\le \|f_i\|_\infty\{1+C\Lambda_0^C(\tau_C)\}.
\end{align*}
The bounds on the integrands and \(\bm\Omega_C^{-1}\) give the deterministic bound for \(\mathcal I\). Uniform convergence then gives the empirical bound.
\end{proof}

\subsection{Evaluation times and derivatives of the weights}
\label{supp:censoring-evaluation}

For \(\star\in\{\mathrm{CHW},\mathrm{RCW}\}\), define the evaluation times as follows. When \(R_{ij}^{\mathrm{CHW}}(L)=1\), let \(t_{ij,i}^{\mathrm{CHW}}(L)\) and \(t_{ij,j}^{\mathrm{CHW}}(L)\) be the observed completion times, equal to \(Y_{0i}(L)\) and \(Y_{0j}(L)\). For RCW, when \(R_{ij}^{\star}(L)=1\), both times equal the observed common determination time \(A_{ij}(L)\). Set both times to zero when \(R_{ij}^{\star}(L)=0\). Every evaluation time is then a measurable function of \((\mathcal O_i,\mathcal O_j)\), and its assigned value is immaterial when the observation indicator is zero. The weights in \ref{supp:weights} can therefore be written, for each \(\star\in\{\mathrm{CHW},\mathrm{RCW}\}\), as
\begin{equation}
W_{ij}^{C,\star}(L;S)
=\frac{R_{ij}^{\star}(L)}
{S\{t_{ij,i}^{\star}(L)\mid\bm X_i^{C}\}
 S\{t_{ij,j}^{\star}(L)\mid\bm X_j^{C}\}}.
\label{supp:eq:common-pair-weight}
\end{equation}
On an observed pair, the denominator is the appropriate observation probability when \(S=S_C\).

To differentiate the scores with respect to the censoring weights, hold the observed pair and \(\bm\beta_L\) fixed and write \(\ell_{ij}^{\star}(S)=-\sum_{k\in\{i,j\}}\log S\{t_{ij,k}^{\star}(L)\mid\bm X_k^{C}\}\). Then \(W_{ij}^{C,\star}(L;S)=R_{ij}^{\star}(L)e^{\ell_{ij}^{\star}(S)}\) and \(\partial W_{ij}^{C,\star}/\partial\ell_{ij}^{\star}=W_{ij}^{C,\star}\). For CHW and RCW, suppressing the other arguments gives
\begin{align*}
\frac{\partial\bm U_{ij}^{\star}}{\partial\ell_{ij}^{\star}}
&=\bm K_{ij}\{\widetilde H_{ij}(L)-\mu_{ij}\}
\frac{\partial W_{ij}^{C,\star}}{\partial\ell_{ij}^{\star}}=\bm K_{ij}W_{ij}^{C,\star}
\{\widetilde H_{ij}(L)-\mu_{ij}\}=\bm U_{ij}^{\star}.
\end{align*}
For \(\star\in\{\mathrm{CHW},\mathrm{RCW}\}\), denote these derivatives by
\begin{equation}
\bm V_{ij}^{\star}(\bm\beta_L;S)
=\bm U_{ij}^{\star}(\bm\beta_L;S).
\label{supp:eq:common-nuisance-multiplier}
\end{equation}
Set \(\bm V_{ij}^{\star}=\bm V_{ij}^{\star}(\bm\beta_L^0)\) and \(\widehat{\bm V}_{ij}^{\star}=\widehat{\bm V}_{ij}^{\star}(\widehat{\bm\beta}_L^{\star})\). These derivatives are functions of the observed records and vanish when the corresponding observation indicator is zero. The equality of observed and full-data responses on an observed pair is used in the expectation calculation in \ref{supp:unbiased}.

The empirical subject contribution in \eqref{supp:eq:b-hat} can be computed by first summing over pairs. Let \(\widehat e_i=e^{\widehat{\bm\gamma}^\top\bm X_i^C}\). Substitution of the empirical log-survival influence into \eqref{supp:eq:b-hat}, followed by interchange of the finite sum and integral, gives
\begin{align}
\widehat{\bm J}_L^{\star}(u)
&=\frac{2}{n(n-1)\widehat r_C^{(0)}(u)}
\sum_{(j,k)\in\mathcal I_n}\widehat{\bm V}_{jk}^{\star}
\left[\widehat e_j I\{u<t_{jk,j}^{\star}(L)\}
+\widehat e_k I\{u<t_{jk,k}^{\star}(L)\}\right],
\label{supp:eq:J-aggregation}\\
\widehat{\bm D}_{\gamma,L}^{\star}
&=\frac{2}{n(n-1)}\sum_{(j,k)\in\mathcal I_n}\widehat{\bm V}_{jk}^{\star}
\left[\widehat e_j
\{\widehat\Lambda_0^C\{t_{jk,j}^{\star}(L)-\}\bm X_j^C
-\widehat{\bm a}_C\{t_{jk,j}^{\star}(L)-\}\}^\top\right.\nonumber\\
&\left.\hspace{2cm}+\widehat e_k
\{\widehat\Lambda_0^C\{t_{jk,k}^{\star}(L)-\}\bm X_k^C
-\widehat{\bm a}_C\{t_{jk,k}^{\star}(L)-\}\}^\top\right].
\label{supp:eq:Dgamma-aggregation}
\end{align}
Consequently,
\begin{align*}
\widehat{\bm b}_{L,i}^{\star}
&=\int_0^L\widehat{\bm J}_L^{\star}(u)d\widehat M_i^C(u)
+\widehat{\bm D}_{\gamma,L}^{\star}\widehat{\operatorname{IF}}_{\gamma,i}\\
&=\int_0^L\widehat{\bm J}_L^{\star}(u)dN_i^C(u)
-\widehat e_i\int_0^L\widehat{\bm J}_L^{\star}(u)
\mathcal R_i^C(u)d\widehat\Lambda_0^C(u)
+\widehat{\bm D}_{\gamma,L}^{\star}\widehat{\operatorname{IF}}_{\gamma,i}.
\end{align*}
The pair sums require order \(n^2\) operations. For \(\widehat{\bm J}_L^{\star}\), contributions are grouped by observed evaluation time and accumulated over the ordered times. The martingale integrals are then evaluated for each subject.

\subsection{Consistency and expansion of the fitted weights}
\label{supp:two-weight-expansions}

The censoring survival expansion gives an expansion of the weights in \eqref{supp:eq:common-pair-weight}. The evaluation times in \ref{supp:cox-if} are observed for pairs with nonzero weights and set to zero otherwise. Thus all quantities below are functions of the observed records. RCW evaluates both censoring survival functions at the common determination time, and CHW uses the two individual completion times.

\begin{proposition}[Uniform consistency of the fitted censoring weights]
\label{supp:prop:weight-consistency}
\leavevmode\newline
Suppose positivity holds and
\begin{equation}
\|\widehat\ell_C-\ell_C\|_\infty
=\sup_{t\in[0,L],\bm x}
|\log\widehat S_C(t\mid\bm x)-\log S_C(t\mid\bm x)|
=o_p(1).
\label{supp:eq:log-survival-consistency}
\end{equation}
Then \(\max_{\star\in\{\mathrm{CHW},\mathrm{RCW}\}}\max_{(i,j)\in\mathcal I_n}|\widehat W_{ij}^{C,\star}(L)-W_{ij}^{C,\star}(L)|=o_p(1)\). If the left side of \eqref{supp:eq:log-survival-consistency} is \(O_p(n^{-1/2})\), the maximum weight difference is also \(O_p(n^{-1/2})\).
\end{proposition}

\begin{proof}
Let \(\delta_n(t,\bm x)=\widehat\ell_C(t,\bm x)-\ell_C(t,\bm x)\) and \(\varepsilon_n=\|\delta_n\|_\infty\). For each pair, define
\begin{equation}
d_{ij,n}^{\star}
=\delta_n\{t_{ij,i}^{\star}(L),\bm X_i^{C}\}
+\delta_n\{t_{ij,j}^{\star}(L),\bm X_j^{C}\}.
\label{supp:eq:common-log-weight-error}
\end{equation}
The two times lie in \([0,L]\), so \(|d_{ij,n}^{\star}|\le2\varepsilon_n\). When the observation indicator is zero, both times equal zero and \(S_C(0\mid\bm x)=\widehat S_C(0\mid\bm x)=1\), giving \(d_{ij,n}^{\star}=0\). Equation \eqref{supp:eq:common-pair-weight} yields the exact identity
\begin{equation}
\widehat W_{ij}^{C,\star}(L)
=W_{ij}^{C,\star}(L)e^{-d_{ij,n}^{\star}}.
\label{supp:eq:exact-weight-ratio}
\end{equation}
Using positivity and \(|e^{-x}-1|\le e^{|x|}|x|\), we obtain
\begin{align*}
|\widehat W_{ij}^{C,\star}(L)-W_{ij}^{C,\star}(L)|
=W_{ij}^{C,\star}(L)|e^{-d_{ij,n}^{\star}}-1|
\le \epsilon^{-2}e^{|d_{ij,n}^{\star}|}|d_{ij,n}^{\star}|\le2\epsilon^{-2}e^{2\varepsilon_n}\varepsilon_n.
\end{align*}
The final bound holds uniformly over pairs and methods. Since \(e^{2\varepsilon_n}=1+o_p(1)\), it is \(o_p(1)\) when \(\varepsilon_n=o_p(1)\) and \(O_p(n^{-1/2})\) when \(\varepsilon_n=O_p(n^{-1/2})\). This proves both assertions.
\end{proof}

For an observed record \(o\), define \(\Gamma_{ij}^{\star}(o)=W_{ij}^{C,\star}(L)[\mathcal I\{o;t_{ij,i}^{\star}(L),\bm X_i^{C}\}+\mathcal I\{o;t_{ij,j}^{\star}(L),\bm X_j^{C}\}]\). This quantity enters the score expansion in Lemma \ref{supp:lem:nuisance}.

\begin{proposition}[Asymptotic expansion of the fitted censoring weights]
\label{supp:prop:weight-first-order}
\leavevmode\newline
Under positivity and the uniform influence representation in Proposition \ref{supp:prop:cox-breslow}, for each \(\star\in\{\mathrm{CHW},\mathrm{RCW}\}\),
\begin{equation}
\widehat W_{ij}^{C,\star}(L)-W_{ij}^{C,\star}(L)
=-\frac1n\sum_{r=1}^n\Gamma_{ij}^{\star}(\mathcal O_r)
+\mathcal R_{ij,n}^{\star},
\label{supp:eq:common-weight-if}
\end{equation}
where \(\max_{\star\in\{\mathrm{CHW},\mathrm{RCW}\}}\max_{(i,j)\in\mathcal I_n}|\mathcal R_{ij,n}^{\star}|=o_p(n^{-1/2})\).
\end{proposition}

\begin{proof}
Write the uniform log-survival expansion as \(\delta_n(t,\bm x)=n^{-1}\sum_{r=1}^n\mathcal I(\mathcal O_r;t,\bm x)+r_n(t,\bm x)\). Substitution into \eqref{supp:eq:common-log-weight-error} gives
\begin{align*}
d_{ij,n}^{\star}
&=\frac1n\sum_{r=1}^n\left[
\mathcal I\{\mathcal O_r;t_{ij,i}^{\star}(L),\bm X_i^{C}\}
+\mathcal I\{\mathcal O_r;t_{ij,j}^{\star}(L),\bm X_j^{C}\}\right]
+r_{ij,n}^{\star},
\end{align*}
where \(r_{ij,n}^{\star}=\sum_{s\in\{i,j\}}r_n\{t_{ij,s}^{\star}(L),\bm X_s^{C}\}\). The uniform bound on these remainders follows from \(\|r_n\|_\infty=o_p(n^{-1/2})\), since \(\max_{\star\in\{\mathrm{CHW},\mathrm{RCW}\}}\max_{(i,j)\in\mathcal I_n}|r_{ij,n}^{\star}|\le2\|r_n\|_\infty\). Uniformity of the influence expansion and its remainder bound permits evaluation at the random pair times.

Taylor's theorem applied to \eqref{supp:eq:exact-weight-ratio} gives, for some \(\theta_{ij,n}^{\star}\in(0,1)\),
\begin{align*}
\widehat W_{ij}^{C,\star}(L)-W_{ij}^{C,\star}(L)
&=-W_{ij}^{C,\star}(L)d_{ij,n}^{\star}
+\frac12W_{ij}^{C,\star}(L)
e^{-\theta_{ij,n}^{\star}d_{ij,n}^{\star}}(d_{ij,n}^{\star})^2\\
&=-\frac1n\sum_{r=1}^n\Gamma_{ij}^{\star}(\mathcal O_r)
-W_{ij}^{C,\star}(L)r_{ij,n}^{\star}
+\frac12W_{ij}^{C,\star}(L)
e^{-\theta_{ij,n}^{\star}d_{ij,n}^{\star}}(d_{ij,n}^{\star})^2.
\end{align*}
If \(d_{ij,n}^{\star}=0\), any value of \(\theta_{ij,n}^{\star}\) in \((0,1)\) gives the same identity. The quadratic term satisfies
\begin{align*}
0\le\frac12W_{ij}^{C,\star}(L)
e^{-\theta_{ij,n}^{\star}d_{ij,n}^{\star}}(d_{ij,n}^{\star})^2\le\frac12\epsilon^{-2}e^{|d_{ij,n}^{\star}|}|d_{ij,n}^{\star}|^2\le2\epsilon^{-2}e^{2\varepsilon_n}\varepsilon_n^2.
\end{align*}
Thus the remainder in \eqref{supp:eq:common-weight-if} is \(\mathcal R_{ij,n}^{\star} =-W_{ij}^{C,\star}(L)r_{ij,n}^{\star} +\frac12W_{ij}^{C,\star}(L) e^{-\theta_{ij,n}^{\star}d_{ij,n}^{\star}}(d_{ij,n}^{\star})^2\), with uniform bound
\begin{align*}
\max_{\star\in\{\mathrm{CHW},\mathrm{RCW}\}}\max_{(i,j)\in\mathcal I_n}|\mathcal R_{ij,n}^{\star}|
&\le2\epsilon^{-2}\|r_n\|_\infty
+2\epsilon^{-2}e^{2\varepsilon_n}\varepsilon_n^2=o_p(n^{-1/2})+O_p(n^{-1})=o_p(n^{-1/2}),
\end{align*}
because \(\varepsilon_n=O_p(n^{-1/2})\) and \(\|r_n\|_\infty=o_p(n^{-1/2})\). The bound holds for all pairs and both methods.
\end{proof}

Both weights have the same uniform convergence rate and form of asymptotic expansion. Their evaluation times determine the censoring contribution to the score expansion.

\subsection{Application to conditional tie weighting}
\label{supp:ctw-censoring-weights}

The preceding calculations apply separately to the fatal, nonfatal, and genuine-tie contributions in \eqref{eq:ctw-score}. For component \(q\in\{0,\ldots,Q\}\), use \(R_{ij}^{\mathrm{CTW},(q)}(L)\) as the observation indicator and evaluate both censoring survival functions at \(t_{ij}^{(q)}\). For the genuine-tie contribution, use \(R_{ij}^{\mathrm{CTW},T}(L)\) and evaluate both functions at \(L\). When an indicator is zero, set its two evaluation times to zero and its contribution to zero, as above.

The bound in Proposition \ref{supp:prop:weight-consistency} and the expansion in Proposition \ref{supp:prop:weight-first-order} hold uniformly over these components. For a nonfatal component, multiplication by the fixed product \(\omega_i^{(q)}(L)\omega_j^{(q)}(L)\in[0,1]\) preserves the bounds. Since \(Q\) is fixed, summing over components gives the CTW contribution from estimating the censoring distribution in Section \ref{supp:weight-expansion}. The additional contribution from estimating the conditional tie probabilities is derived from the joint models in Section \ref{supp:joint-model-choices} and combined with the censoring contribution in the same expansion.

\section{Joint survival models and estimation}
\label{supp:joint-model-choices}

\subsection{Likelihood and conditional tie probabilities}
\label{supp:joint-likelihood}

The likelihood for the joint survival function in Section \ref{sec:ctw} of main text is obtained from the observed event histories. First consider \(Q=1\). As defined in the main text, \(T_1=\infty\) when the nonfatal event does not occur before death.

\subsubsection{Event histories and likelihood}

For \(0<t<d\), the density of a nonfatal event at \(t\) followed by death at \(d\) is \(f(d,t\mid\bm x)=\dot{\mathcal S}_{01}(d,t\mid\bm x)\), where \(\dot{\mathcal S}_{01}=\partial^2\mathcal S/\partial t_0\partial t_1\). The separate density of death without a preceding nonfatal event is
\begin{equation}
f_0(d\mid\bm x)=-\dot{\mathcal S}_0(d,d\mid\bm x).
\label{supp:eq:ctw-event-distribution}
\end{equation}
Both derivatives are evaluated on the region specified in Section \ref{sec:ctw}. Assume that the densities are nonnegative, \(\mathcal S(0,0\mid\bm x)=1\), and \(\mathcal S(d,t\mid\bm x)\to0\) as \(d\to\infty\) for fixed \(t\). Suppressing the covariates, their total density at death time \(d\) is \(f_0(d)+\int_0^d f(d,t)\,dt=-\dot{\mathcal S}_0(d,0)\), which integrates to one. Moreover,
\begin{align*}
P(D>d,T_1>t\mid\bm x)
&=\int_d^\infty\left\{-\dot{\mathcal S}_0(z,z)+\int_t^z\dot{\mathcal S}_{01}(z,v)\,dv\right\}dz\\
&=\int_d^\infty-\dot{\mathcal S}_0(z,t)\,dz
=\mathcal S(d,t).
\end{align*}
These densities therefore define the joint distribution of the event histories. Its nonfatal cumulative incidence is \(P(T_1\le t\mid\bm x)=\int_0^t-\dot{\mathcal S}_1(s,s\mid\bm x)\,ds\).

For a record \(\mathcal O_i(a)\) restricted at a fixed time \(a\), write \(U_i=D_i\wedge C_i\wedge a\), \(V_i=T_{1i}\wedge U_i\), \(\delta_i^D=I(D_i\le C_i,D_i\le a)\), and \(\delta_i^N=I(T_{1i}\le C_i,T_{1i}<D_i\wedge a)\). Under conditional independence of censoring and event histories, the likelihood factors into contributions from censoring and the event histories. The latter contribution is
\begin{equation}
\mathcal L\{\mathcal O_i(a)\}=
\begin{cases}
\dot{\mathcal S}_{01}(U_i,V_i\mid\bm X_i),
 &\delta_i^D=1,\ \delta_i^N=1,\\
-\dot{\mathcal S}_1(U_i,V_i\mid\bm X_i),
 &\delta_i^D=0,\ \delta_i^N=1,\\
-\dot{\mathcal S}_0(U_i,U_i\mid\bm X_i),
 &\delta_i^D=1,\ \delta_i^N=0,\\
\mathcal S(U_i,U_i\mid\bm X_i),
 &\delta_i^D=0,\ \delta_i^N=0.
\end{cases}
\label{supp:eq:ctw-four-pattern}
\end{equation}
These contributions include individuals who die without a preceding nonfatal event \citep{Chen2012Semicompeting,Wei2023Copula,Sun2024Copula}. In particular, the third line uses the partial derivative in \eqref{supp:eq:ctw-event-distribution}. Differentiating \(\mathcal S(d,d\mid\bm x)\) along the diagonal would also include first nonfatal events.

The two ratios for \(\omega_i^{(1)}(L)\) in Section \ref{sec:ctw} follow by conditioning on \(D_i>c,T_{1i}>c\) when no nonfatal event is observed and on \(D_i>c,T_{1i}=v\) when that event is observed at \(v<c<L\). The likelihood may use a longer fixed horizon \(\tau_E\ge L\). The conditional tie probability continues to use \(\mathcal O_i(L)\), as required by the population identity in Section \ref{supp:ctw-identification}.

\subsubsection{Copula derivatives and margin functions}

For the bivariate copula in Section \ref{sec:ctw}, put \(u=\exp\{-\Lambda_0(d\mid\bm x)\}\) and \(v=\exp\{-\Lambda_1(t\mid\bm x)\}\). Let \(\partial_0\mathcal C_\theta\) and \(\partial_1\mathcal C_\theta\) denote derivatives in the death and nonfatal arguments, respectively, and let \(\partial_{01}\mathcal C_\theta\) be their mixed derivative. The following derivatives are evaluated at interior survival probabilities and at times where the hazards exist, with endpoint values defined by limits. With \(\lambda_k(t\mid\bm x)=\partial\Lambda_k(t\mid\bm x)/\partial t\), the chain rule gives
\begin{align}
-\dot{\mathcal S}_0(d,t\mid\bm x)
&=\lambda_0(d\mid\bm x)u\,\partial_0\mathcal C_\theta(u,v),\nonumber\\
-\dot{\mathcal S}_1(d,t\mid\bm x)
&=\lambda_1(t\mid\bm x)v\,\partial_1\mathcal C_\theta(u,v),\nonumber\\
\dot{\mathcal S}_{01}(d,t\mid\bm x)
&=\lambda_0(d\mid\bm x)\lambda_1(t\mid\bm x)uv\,\partial_{01}\mathcal C_\theta(u,v).
\label{supp:eq:ctw-copula-derivatives}
\end{align}
Substitution gives the four likelihood contributions in \eqref{supp:eq:ctw-four-pattern}. Write \(u_{ks}=\exp\{-\Lambda_k(s\mid\bm X_i)\}\), for \(k=0,1\), when calculating the conditional tie probability for individual \(i\). For a subject censored alive at \(c<L\), the conditional probability is
\begin{equation}
\omega_i^{(1)}(L)=
\begin{cases}
\dfrac{\mathcal C_\theta(u_{0L},u_{1c})}{\mathcal C_\theta(u_{0c},u_{1c})},
 &\delta_i^N=0,\\[6pt]
\dfrac{\partial_1\mathcal C_\theta(u_{0L},u_{1v})}{\partial_1\mathcal C_\theta(u_{0c},u_{1v})},
 &\delta_i^N=1,\ V_i=v<c.
\end{cases}
\label{supp:eq:ctw-copula-q}
\end{equation}
In the second line, the common factor \(\lambda_1(v\mid\bm X_i)u_{1v}\) cancels. The conditional tie probabilities therefore depend on the copula and its first derivative.

The first copula margin is the marginal death survival function because \(\mathcal C_\theta(u,1)=u\). The second margin function parameterizes the joint survival function on \(t\le d\). With the convention \(T_1=\infty\) after death without a nonfatal event, it generally differs from \(P(T_1>t\mid\bm x)\). The nonfatal cumulative incidence is given by the integral above. Under this specification, every finite nonfatal event time precedes death.

The proportional hazards model is \(\Lambda_k(t\mid\bm x)=\Lambda_{0k}(t)\exp(\bm\gamma_k^\top\bm x)\). A Weibull baseline has \(\Lambda_{0k}(t)=\kappa_k t^{\nu_k}\). With unspecified \(\Lambda_{00}\) and \(\Lambda_{01}\), the two margin functions follow Cox proportional hazards models \citep{Chen2012Semicompeting,Sun2024Copula,Arachchige2025Copula}. The likelihood in \eqref{supp:eq:ctw-four-pattern} estimates the nonfatal component together with the dependence parameter, accounting for death ending nonfatal follow-up. Alternatively, the marginal death Cox model can be fitted first, followed by maximization of the joint likelihood over the remaining parameters. Section \ref{supp:cox-theory} gives the influence contributions for joint estimation and this two-stage procedure, including both baseline cumulative hazard functions.

\subsubsection{Gumbel copula}

For \(\theta\ge1\), the Gumbel copula is \(\mathcal C_\theta(u,v)=\exp[-\{(-\log u)^\theta+(-\log v)^\theta\}^{1/\theta}]\). Let \(a=-\log u\), \(b=-\log v\), and \(r=(a^\theta+b^\theta)^{1/\theta}\). Direct differentiation gives
\begin{align*}
\partial_0\mathcal C_\theta(u,v)
&=\frac{e^{-r}}u a^{\theta-1}r^{1-\theta},\\
\partial_1\mathcal C_\theta(u,v)
&=\frac{e^{-r}}v b^{\theta-1}r^{1-\theta},\\
\partial_{01}\mathcal C_\theta(u,v)
&=\frac{e^{-r}}{uv}a^{\theta-1}b^{\theta-1}r^{1-2\theta}(r+\theta-1).
\end{align*}
Substitution into \eqref{supp:eq:ctw-copula-derivatives} yields the likelihood derivatives. The resulting joint survival function is
\begin{equation}
\mathcal S(d,t\mid\bm x)
=\exp\!\left[-\{\Lambda_0(d\mid\bm x)^\theta+\Lambda_1(t\mid\bm x)^\theta\}^{1/\theta}\right].
\label{supp:eq:ctw-gumbel}
\end{equation}
For \(r_i(s,t)=\{\Lambda_0(s\mid\bm X_i)^\theta+\Lambda_1(t\mid\bm X_i)^\theta\}^{1/\theta}\), equation \eqref{supp:eq:ctw-copula-q} gives
\begin{equation*}
\omega_i^{(1)}(L)=
\begin{cases}
\exp\{r_i(c,c)-r_i(L,c)\},
 &\delta_i^N=0,\\[4pt]
\exp\{r_i(c,v)-r_i(L,v)\}
\left\{\dfrac{r_i(L,v)}{r_i(c,v)}\right\}^{1-\theta},
 &\delta_i^N=1,\ V_i=v<c.
\end{cases}
\end{equation*}
At \(\theta=1\), both expressions reduce to \(\exp\{-\Lambda_0(L\mid\bm X_i)+\Lambda_0(c\mid\bm X_i)\}\). This is the conditional death survival probability when the copula is the independence copula.

\subsubsection{Clayton copula}

For \(\theta>0\), put \(w=u^{-\theta}+v^{-\theta}-1\). The Clayton copula is \(\mathcal C_\theta(u,v)=w^{-1/\theta}\), with derivatives
\begin{align*}
\partial_0\mathcal C_\theta(u,v)
&=u^{-\theta-1}w^{-1/\theta-1},\\
\partial_1\mathcal C_\theta(u,v)
&=v^{-\theta-1}w^{-1/\theta-1},\\
\partial_{01}\mathcal C_\theta(u,v)
&=(1+\theta)(uv)^{-\theta-1}w^{-1/\theta-2}.
\end{align*}
Define \(w_i(s,t)=\exp\{\theta\Lambda_0(s\mid\bm X_i)\}+\exp\{\theta\Lambda_1(t\mid\bm X_i)\}-1\). The conditional tie probabilities are
\begin{equation*}
\omega_i^{(1)}(L)=
\begin{cases}
\left\{\dfrac{w_i(L,c)}{w_i(c,c)}\right\}^{-1/\theta},
 &\delta_i^N=0,\\[6pt]
\left\{\dfrac{w_i(L,v)}{w_i(c,v)}\right\}^{-1/\theta-1},
 &\delta_i^N=1,\ V_i=v<c.
\end{cases}
\end{equation*}
The limit as \(\theta\to0\) is the independence copula and gives the same conditional death survival ratio as above. For a positive dependence parameter, the copula is valid for any fixed number of components.

\subsubsection{Frank copula}

For \(\theta>0\), write \(p_\theta=1-e^{-\theta}\), \(a_\theta(u)=1-e^{-\theta u}\), and \(w_\theta(u,v)=p_\theta-a_\theta(u)a_\theta(v)\). The Frank copula is \(\mathcal C_\theta(u,v)=-\theta^{-1}\log\{w_\theta(u,v)/p_\theta\}\). Its derivatives are
\begin{align*}
\partial_0\mathcal C_\theta(u,v)
&=\frac{e^{-\theta u}a_\theta(v)}{w_\theta(u,v)},\\
\partial_1\mathcal C_\theta(u,v)
&=\frac{e^{-\theta v}a_\theta(u)}{w_\theta(u,v)},\\
\partial_{01}\mathcal C_\theta(u,v)
&=\frac{\theta p_\theta e^{-\theta(u+v)}}{w_\theta(u,v)^2}.
\end{align*}
Consequently,
\begin{equation*}
\omega_i^{(1)}(L)=
\begin{cases}
\dfrac{\log\{w_\theta(u_{0L},u_{1c})/p_\theta\}}{\log\{w_\theta(u_{0c},u_{1c})/p_\theta\}},
 &\delta_i^N=0,\\[6pt]
\dfrac{a_\theta(u_{0L})w_\theta(u_{0c},u_{1v})}{a_\theta(u_{0c})w_\theta(u_{0L},u_{1v})},
 &\delta_i^N=1,\ V_i=v<c.
\end{cases}
\end{equation*}
The limit \(\theta\to0\) again gives the independence copula. For \(Q=1\), the Frank formulas also hold for \(\theta<0\). We use \(\theta>0\) for the multivariate copula below. These three copula families and their generators are standard choices \citep{Nelsen2006Copulas}.

\subsubsection{Several nonfatal components}

The formula in the main text for \(\omega_i^{(q)}(L)\) uses the mixed derivative associated with the set \(\mathcal J_i\) of observed nonfatal events. For an Archimedean survival copula, the mixed derivative can be expressed in terms of its generator. Write \(\mathcal C_\theta(u_0,\ldots,u_Q)=\psi_\theta\{\sum_{k=0}^Q\varphi_\theta(u_k)\}\), where \(u_k=\exp\{-\Lambda_k(t_k\mid\bm X_i)\}\) and \(\varphi_\theta=\psi_\theta^{-1}\), and set \(z=\sum_{k=0}^Q\varphi_\theta(u_k)\). The generators for the three families are \(\psi_\theta(z)=e^{-z^{1/\theta}}\) for Gumbel, \(\psi_\theta(z)=(1+z)^{-1/\theta}\) for Clayton, and \(\psi_\theta(z)=-\theta^{-1}\log(1-p_\theta e^{-z})\) for Frank. Their inverses are \(\varphi_\theta(u)=(-\log u)^\theta\), \(\varphi_\theta(u)=u^{-\theta}-1\), and \(\varphi_\theta(u)=-\log\{a_\theta(u)/p_\theta\}\), respectively.

Let \(m=|\mathcal J_i|\). Since each coordinate is differentiated at most once,
\begin{equation*}
\dot{\mathcal S}_{\mathcal J_i}(t_0,\ldots,t_Q\mid\bm X_i)
=\psi_\theta^{(m)}(z)
\prod_{r\in\mathcal J_i}
\left\{\varphi_\theta'(u_r)\frac{\partial u_r}{\partial t_r}\right\}.
\end{equation*}
Suppose the individual is censored alive at \(c<L\) and \(\mathcal J_i\cap\{1,\ldots,q-1\}=\varnothing\). Denote the sums in the generator evaluated at the denominator and numerator arguments of that formula by \(z_{i,c}\) and \(z_{i,L}^{(q)}\), respectively. The observed event times are the same in the numerator and denominator, so their derivative factors cancel and
\begin{equation}
\omega_i^{(q)}(L)
=\frac{\psi_\theta^{(m)}\{z_{i,L}^{(q)}\}}{\psi_\theta^{(m)}(z_{i,c})}.
\label{supp:eq:ctw-archimedean-q}
\end{equation}
For \(m=0\), use \(\psi_\theta^{(0)}=\psi_\theta\). If a higher-priority event is observed, the conditional tie probability is zero, as in the main text.

For Clayton, \((-1)^m\psi_\theta^{(m)}(z)=\{\prod_{r=0}^{m-1}(\theta^{-1}+r)\}(1+z)^{-1/\theta-m}\), with an empty product equal to one. Therefore \(\omega_i^{(q)}(L)=\{(1+z_{i,L}^{(q)})/(1+z_{i,c})\}^{-1/\theta-m}\). For Gumbel, let \(b=1/\theta\) and define \(P_0(x)=1\) and \(P_{m+1}(x)=(m+bx)P_m(x)-bxP_m'(x)\). Repeated differentiation gives \((-1)^m\psi_\theta^{(m)}(z)=e^{-z^b}z^{-m}P_m(z^b)\). Thus \eqref{supp:eq:ctw-archimedean-q} is an exponential times a ratio of powers and polynomials. For example, \(P_1(x)=bx\) and \(P_2(x)=b(1-b)x+b^2x^2\), yielding the formulas above when at most one nonfatal event is observed.

For Frank, put \(x=p_\theta e^{-z}\). For \(m\ge1\), \((-1)^m\psi_\theta^{(m)}(z)=\theta^{-1}xA_{m-1}(x)/(1-x)^m\), where \(A_0(x)=1\) and \(A_{r+1}(x)=(1+rx)A_r(x)+x(1-x)A_r'(x)\) are the Eulerian polynomials. Substitution into \eqref{supp:eq:ctw-archimedean-q} gives a ratio of rational functions. The first two derivative factors are \(\theta^{-1}x/(1-x)\) and \(\theta^{-1}x/(1-x)^2\). These derivatives are available for any fixed \(Q\) and allow dependence among all components within an individual.

\subsubsection{Multistate model}

A multistate model specifies the transition intensities \(\lambda^{01}(t\mid\bm x)\), \(\lambda^{02}(t\mid\bm x)\), and \(\lambda^{12}(d\mid t,\bm x)\). States 0, 1, and 2 denote alive before the nonfatal event, alive after that event, and dead. The last intensity may depend on the observed nonfatal time \(t\) \citep{Xu2010IllnessDeath}. Put \(S_0(t\mid\bm x)=\exp[-\int_0^t\{\lambda^{01}(s\mid\bm x)+\lambda^{02}(s\mid\bm x)\}\,ds]\) and \(S_1(d\mid t,\bm x)=\exp\{-\int_t^d\lambda^{12}(s\mid t,\bm x)\,ds\}\). The corresponding densities are \(f(d,t\mid\bm x)=S_0(t\mid\bm x)\lambda^{01}(t\mid\bm x)S_1(d\mid t,\bm x)\lambda^{12}(d\mid t,\bm x)\) and \(f_0(d\mid\bm x)=S_0(d\mid\bm x)\lambda^{02}(d\mid\bm x)\).

For a subject censored alive at \(c<L\) after a nonfatal event at \(v\), the conditional tie probability is \(\omega_i^{(1)}(L)=S_1(L\mid v,\bm X_i)/S_1(c\mid v,\bm X_i)\). If no nonfatal event has been observed, then
\begin{equation*}
\omega_i^{(1)}(L)
=\frac{S_0(L\mid\bm X_i)+\int_c^L S_0(s\mid\bm X_i)\lambda^{01}(s\mid\bm X_i)S_1(L\mid s,\bm X_i)\,ds}{S_0(c\mid\bm X_i)}.
\end{equation*}
The numerator includes survival without a nonfatal event and survival after a nonfatal event between \(c\) and \(L\). The conditional tie probabilities can therefore be estimated using parametric or semiparametric models for the transition intensities.

\subsubsection{The parametric model used in the analyses}

The simulations use the Gumbel specification in \eqref{supp:eq:ctw-gumbel}, with its dependence parameter denoted by \(\alpha\) in the simulation design, so \(\theta=\alpha\). The two event covariates are \(x_1,x_2\). The death baseline has a fixed knot at time 1, with \(\Lambda_{00}(d)=\lambda_{D1}\min(d,1)+\lambda_{D2}(d-1)_+\), and the nonfatal baseline is \(\Lambda_{01}(t)=\kappa_Tt^{\nu_T}\). Their proportional hazards factors are \(\exp(\gamma_{01}x_1+\gamma_{02}x_2)\) and \(\exp(\gamma_{11}x_1+\gamma_{12}x_2)\), respectively. The vector of nine parameters is
\begin{equation}
\bm\vartheta=(\log\lambda_{D1},\log\lambda_{D2},\gamma_{01},\gamma_{02},\log\kappa_T,\log\nu_T,\gamma_{11},\gamma_{12},\theta)^\top.
\label{supp:eq:ctw-nuisance-parameter}
\end{equation}
The parameters are estimated from the observed records using \eqref{supp:eq:ctw-four-pattern}. The HF-ACTION parameterization is given in Section \ref{supp:hfaction}. Identification and asymptotic inference depend on the joint model and the estimation procedure. The moment and information conditions in Sections \ref{supp:ctw-theory} and \ref{supp:cox-theory} include the behavior of derivatives near time zero and at parameter boundaries.

\subsection{Parametric joint models}
\label{supp:ctw-theory}

For a parametric joint model, denote the vector of dependence, regression, and baseline parameters by \(\bm\vartheta\). The copula dependence parameter \(\theta\) and the margin parameters in the main text are components of this vector. The model with nine parameters in \eqref{supp:eq:ctw-nuisance-parameter} is one example. The likelihood influence function determines the joint model contribution \(\bm b_{J,L}^{\mathrm{CTW}}\) in the common theory of Section \ref{supp:theory}.

\paragraph{Regularity conditions for the joint model}
The sampling, regression, and censoring conditions are stated in Section \ref{supp:regularity}. The following conditions concern a joint model with a fixed number of parameters and a likelihood for the observed histories. It may be a copula model, a parametric multistate model, or another identifiable joint model. The number of event types, hazard intervals, and covariates in the joint model is fixed.

Fix \(\tau_E\ge L\) and let \(\ell_{\bm\vartheta}(\mathcal O_i)=\log\mathcal L_{\bm\vartheta}\{\mathcal O_i(\tau_E)\}\). For \(Q=1\), use the likelihood in \eqref{supp:eq:ctw-four-pattern}. The likelihood score and its influence use \(\mathcal O_i(\tau_E)\), whereas the pair contribution and the derivatives of its conditional tie probabilities use \(\mathcal O_i(L)\). The same joint model estimates can be used at a fixed prespecified set of restriction times.
\setcounter{condition}{0}
\renewcommand{\thecondition}{J.\arabic{condition}}
\providecommand{\theHcondition}{\arabic{condition}}
\renewcommand{\theHcondition}{J.\arabic{condition}}

\begin{condition}[Joint likelihood]
\label{supp:cond:joint-likelihood}
The joint model is correctly specified. Its true parameter \(\bm\vartheta_0\) lies in the interior of a fixed-dimensional parameter space, with a compact neighborhood \(\mathcal H\) contained in that interior. The expected log likelihood is uniquely maximized at \(\bm\vartheta_0\), and the likelihood estimator used in the analysis \(\widehat{\bm\vartheta}\) is consistent and interior with probability tending to one. On a compact identifiable parameter set, uniform convergence of the empirical log likelihood and separation of the expected maximum imply consistency. For the Gumbel model, an interior parameter has \(\theta>1\), positive death rates, and positive nonfatal scale and shape.

The log likelihood has three continuous parameter derivatives on \(\mathcal H\), almost surely, and differentiation may be interchanged with expectation. The score \(\bm s_{\vartheta}(\mathcal O)=\partial\ell_{\bm\vartheta_0}(\mathcal O)/\partial\bm\vartheta\) has mean zero. Its information matrix,
\begin{equation*}
\bm I_{\vartheta}=-E\left\{\frac{\partial^2\ell_{\bm\vartheta_0}(\mathcal O_1)}{\partial\bm\vartheta\partial\bm\vartheta^\top}\right\},
\end{equation*}
is positive definite. Envelopes for the absolute log likelihood, score, Hessian, and third derivative, uniformly on \(\mathcal H\), have finite first, \((2+\delta)\)th, second, and first moments, respectively, for some \(\delta>0\). The baseline hazard may have fixed knots under the continuous event-time model.
\end{condition}

\begin{condition}[CTW derivative moments and nondegeneracy]
\label{supp:cond:joint-weight}
The CTW pair contribution and its components, evaluated at \(S_C\), are twice continuously differentiable in \(\bm\beta_L\) and \(\bm\vartheta\) on the neighborhoods under consideration. Their values, derivatives through order two, and mixed first derivatives have a symmetric envelope \(M_*(\mathcal O_1,\mathcal O_2)\), uniformly over \(\mathcal B\times\mathcal H\), with \(E(M_*^{2+\delta})<\infty\). Differentiation may be interchanged with expectation. The covariance of the CTW subject contribution in \eqref{supp:eq:subject-influence} is positive definite.
\end{condition}

These conditions accommodate likelihood derivatives such as \(\log V_i\) from a Weibull component. The CTW pair contribution itself is bounded under bounded regression predictors, censoring positivity, and \(0\le \omega_i^{(q)}(L)\le1\).

\begin{lemma}[Influence function for the likelihood estimator]
\label{supp:lem:joint-likelihood-if}
Under J.1, define \(\bm\phi_{\vartheta}(\mathcal O)=\bm I_{\vartheta}^{-1}\bm s_{\vartheta}(\mathcal O)\). Then
\begin{equation}
\widehat{\bm\vartheta}-\bm\vartheta_0
=\frac1n\sum_{i=1}^n\bm\phi_{\vartheta}(\mathcal O_i)+o_p(n^{-1/2}).
\label{supp:eq:joint-likelihood-if}
\end{equation}
Let \(\widehat{\bm I}_{\vartheta}\) be the negative fitted likelihood Hessian divided by \(n\), and define \(\widehat{\bm\phi}_{\vartheta,i}=\widehat{\bm I}_{\vartheta}^{-1}\partial\ell_{\widehat{\bm\vartheta}}(\mathcal O_i)/\partial\bm\vartheta\). Then \(\widehat{\bm I}_{\vartheta}\overset{p}{\longrightarrow}\bm I_{\vartheta}\) and \(n^{-1}\sum_{i=1}^n\|\widehat{\bm\phi}_{\vartheta,i}-\bm\phi_{\vartheta}(\mathcal O_i)\|^2=o_p(1)\).
\end{lemma}

\begin{proof}
The mean score at \(\bm\vartheta_0\) is \(O_p(n^{-1/2})\). Consistency and uniform Hessian convergence give
\(\bm0=\frac1n\sum_{i=1}^n\bm s_{\vartheta}(\mathcal O_i) -\{\bm I_{\vartheta}+o_p(1)\}(\widehat{\bm\vartheta}-\bm\vartheta_0),\)
so nonsingularity gives the \(n^{-1/2}\) rate and \eqref{supp:eq:joint-likelihood-if}. If \(L_2\) is the square-integrable Hessian envelope, the mean value inequality gives
\begin{equation*}
\frac1n\sum_{i=1}^n\left\|\frac{\partial\ell_{\widehat{\bm\vartheta}}(\mathcal O_i)}{\partial\bm\vartheta}-\bm s_{\vartheta}(\mathcal O_i)\right\|^2
\le\|\widehat{\bm\vartheta}-\bm\vartheta_0\|^2\frac1n\sum_{i=1}^nL_2(\mathcal O_i)^2=o_p(1).
\end{equation*}
This bound, together with \(\widehat{\bm I}_{\vartheta}^{-1}\to_p\bm I_{\vartheta}^{-1}\) and the finite second moment of the score, proves convergence of the estimated influence functions.
\end{proof}

\paragraph{Influence of joint model estimation}
For the finite parameter vector, let \(\dot{\bm\omega}_i^{(q)}\) denote the derivative of \(\omega_i^{(q)}(L;\bm\vartheta)\) with respect to \(\bm\vartheta\), evaluated at \(\bm\vartheta_0\). The matrix corresponding to the directional derivative in \eqref{supp:eq:ctw-general-event-derivative} is
\begin{align}
\bm D_{\vartheta,L}
&=E\left\{\left.\frac{\partial\bm U_{12}^{\mathrm{CTW}}(\bm\beta_L^0;\bm\vartheta)}{\partial\bm\vartheta^\top}\right|_{\bm\vartheta=\bm\vartheta_0}\right\}\nonumber\\
&=E\left[\sum_{q=1}^{Q}\frac{\bm K_{12}(\bm\beta_L^0)R_{12}^{\mathrm{CTW},(q)}(L)\{\widetilde H_{12}^{(q)}(L)-\mu_{12}(\bm\beta_L^0)\}}{\pi_{12}^{\mathrm{CTW},(q)}(L)}
\{\omega_2^{(q)}(L)\dot{\bm\omega}_1^{(q)\top}+\omega_1^{(q)}(L)\dot{\bm\omega}_2^{(q)\top}\}\right].
\label{supp:eq:ctw-joint-derivative-explicit}
\end{align}
The derivative is taken with respect to the joint model parameters, holding \(S_C\) fixed. The derivatives of the conditional probabilities are zero when the observed record establishes a death, a higher-priority nonfatal event, or complete follow-up through \(L\). Lemma \ref{supp:lem:joint-likelihood-if} gives
\begin{equation}
\bm b_{J,L}^{\mathrm{CTW}}(o)
=\bm D_{\vartheta,L}\bm\phi_{\vartheta}(o)
=\bm D_{\vartheta,L}\bm I_{\vartheta}^{-1}\bm s_{\vartheta}(o).
\label{supp:eq:parametric-joint-influence}
\end{equation}
This contribution has mean zero and a finite second moment by J.1--J.2. It is added to the censoring contribution in \eqref{supp:eq:ctw-general-censor-influence}.

\begin{lemma}[Contribution from parametric model estimation]
\label{supp:lem:ctw-substitution}
Let \(\bm\Psi_{n,J}^{\mathrm{CTW}}(\bm\beta_L^0)\) be the pair average calculated with \(\widehat{\bm\vartheta}\) and the true \(S_C\). Under J.1--J.2,
\begin{align}
\bm\Psi_{n,J}^{\mathrm{CTW}}(\bm\beta_L^0)-\bm\Psi_n^{\mathrm{CTW}}(\bm\beta_L^0)
&=\bm D_{\vartheta,L}(\widehat{\bm\vartheta}-\bm\vartheta_0)+o_p(n^{-1/2})\nonumber\\
&=\frac1n\sum_{i=1}^n\bm b_{J,L}^{\mathrm{CTW}}(\mathcal O_i)+o_p(n^{-1/2}).
\label{supp:eq:ctw-substitution}
\end{align}
The same joint model contribution applies when \(S_C\) is replaced by \(\widehat S_C\), provided \(\|\widehat\ell_C-\ell_C\|_\infty=O_p(n^{-1/2})\).
\end{lemma}

\begin{proof}
Condition J.2 and the law of large numbers for U-statistics give
\begin{equation*}
\frac{2}{n(n-1)}\sum_{(i,j)\in\mathcal I_n}\frac{\partial\bm U_{ij}^{\mathrm{CTW}}(\bm\beta_L^0;\bm\vartheta_0)}{\partial\bm\vartheta^\top}\overset{p}{\longrightarrow}\bm D_{\vartheta,L}.
\end{equation*}
The pair average of the second derivative envelope is \(O_p(1)\). Taylor expansion and Lemma \ref{supp:lem:joint-likelihood-if} therefore give
\begin{align*}
\bm\Psi_{n,J}^{\mathrm{CTW}}(\bm\beta_L^0)-\bm\Psi_n^{\mathrm{CTW}}(\bm\beta_L^0)
&=\left\{\bm D_{\vartheta,L}+o_p(1)\right\}(\widehat{\bm\vartheta}-\bm\vartheta_0)+O_p(\|\widehat{\bm\vartheta}-\bm\vartheta_0\|^2)\\
&=\bm D_{\vartheta,L}(\widehat{\bm\vartheta}-\bm\vartheta_0)+o_p(n^{-1/2})\\
&=\frac1n\sum_{i=1}^n\bm b_{J,L}^{\mathrm{CTW}}(\mathcal O_i)+o_p(n^{-1/2}).
\end{align*}
The population Taylor expansion gives the same leading term. Subtracting it from the empirical expansion proves the stochastic equicontinuity requirement in G.4, and \eqref{supp:eq:ctw-substitution} verifies the joint model expansion in G.3.

Positivity and uniform log-survival consistency permit the same derivative convergence with fitted censoring weights. Let \(\varepsilon_n=\|\widehat\ell_C-\ell_C\|_\infty\). The quadratic and mixed remainders are bounded in probability by a constant times
\begin{equation*}
\{\|\widehat{\bm\vartheta}-\bm\vartheta_0\|^2+\varepsilon_n^2+\varepsilon_n\|\widehat{\bm\vartheta}-\bm\vartheta_0\|\}
\frac{2}{n(n-1)}\sum_{(i,j)\in\mathcal I_n}M_*(\mathcal O_i,\mathcal O_j)=O_p(n^{-1}).
\end{equation*}
Thus simultaneous estimation of the censoring and joint models does not change \eqref{supp:eq:parametric-joint-influence}, including when both fits use the same sample.
\end{proof}

On the compact neighborhoods in J.1--J.2, the derivative envelopes, consistency of \(\widehat{\bm\vartheta}\), and uniform convergence of the censoring weights also give uniform convergence of the fitted pair equation and its regression derivative. Under the sampling and censoring assumptions in G.1--G.2, the preceding lemma verifies the remaining joint model conditions in Section \ref{supp:ctw-conditions}. Consistency and asymptotic normality then follow from Theorem \ref{supp:thm:asymptotic-normality}.

\paragraph{Estimation of the joint model contribution}
Estimate \(\bm D_{\vartheta,L}\) by the fitted pair average of the derivative in \eqref{supp:eq:ctw-joint-derivative-explicit}, evaluated at \(\widehat{\bm\beta}_L^{\mathrm{CTW}}\), \(\widehat S_C\), and \(\widehat{\bm\vartheta}\). With \(\widehat{\bm\phi}_{\vartheta,i}\) from Lemma \ref{supp:lem:joint-likelihood-if}, the fitted contribution is
\begin{equation*}
\widehat{\bm b}_{J,L,i}^{\mathrm{CTW}}
=\widehat{\bm D}_{\vartheta,L}\widehat{\bm\phi}_{\vartheta,i}
=\widehat{\bm D}_{\vartheta,L}\widehat{\bm I}_{\vartheta}^{-1}
\frac{\partial\ell_{\widehat{\bm\vartheta}}(\mathcal O_i)}{\partial\bm\vartheta}.
\end{equation*}
Condition J.2 gives \(\widehat{\bm D}_{\vartheta,L}\to_p\bm D_{\vartheta,L}\). Together with the empirical mean square convergence in Lemma \ref{supp:lem:joint-likelihood-if} and \(n^{-1}\sum_{i=1}^n\|\bm\phi_{\vartheta}(\mathcal O_i)\|^2=O_p(1)\),
\begin{align*}
\frac1n\sum_{i=1}^n\|\widehat{\bm b}_{J,L,i}^{\mathrm{CTW}}-\bm b_{J,L}^{\mathrm{CTW}}(\mathcal O_i)\|^2
&\le 2\|\widehat{\bm D}_{\vartheta,L}\|^2\frac1n\sum_{i=1}^n\|\widehat{\bm\phi}_{\vartheta,i}-\bm\phi_{\vartheta}(\mathcal O_i)\|^2\\
&\quad+2\|\widehat{\bm D}_{\vartheta,L}-\bm D_{\vartheta,L}\|^2\frac1n\sum_{i=1}^n\|\bm\phi_{\vartheta}(\mathcal O_i)\|^2
=o_p(1).
\end{align*}
For the pair contributions, the mean value inequality bounds their fitted differences by an envelope times \(\|\widehat{\bm\beta}_L^{\mathrm{CTW}}-\bm\beta_L^0\|+\|\widehat{\bm\vartheta}-\bm\vartheta_0\|+\|\widehat\ell_C-\ell_C\|_\infty\). The pair average of the squared envelope is \(O_p(1)\), so
\begin{equation*}
\frac{2}{n(n-1)}\sum_{(i,j)\in\mathcal I_n}\|\widehat{\bm U}_{ij}^{\mathrm{CTW}}(\widehat{\bm\beta}_L^{\mathrm{CTW}})-\bm U_{ij}^{\mathrm{CTW}}(\bm\beta_L^0)\|^2=o_p(1),
\end{equation*}
for both the total pair contribution and each component. These bounds verify the joint model and pair contribution requirements of Theorem \ref{supp:thm:covariance}. Combining the fitted joint model contribution with \eqref{supp:eq:ctw-censor-if-hat} gives a consistent covariance estimator under the corresponding censoring conditions. The empirical subject covariance includes the covariances among the pair projection, censoring estimation, and joint model estimation.

\subsection{Copula models with Cox marginal functions}
\label{supp:cox-theory}

For one fatal and one nonfatal component, unspecified baseline cumulative hazards give a semiparametric joint model for the conditional tie probabilities. Likelihood estimation for copulas with semiparametric survival regressions was studied by \citet{Chen2012Semicompeting}, and \citet{Arachchige2025Copula} describe estimation in two stages. The CTW influence function must then account for estimation of both baseline functions.

\paragraph{Model and likelihood}
The joint model is parameterized by \((\theta,\bm\gamma_0,\bm\gamma_1,\Lambda_{00},\Lambda_{01})\), with \(\Lambda_k(t\mid\bm x)=\Lambda_{0k}(t)\exp(\bm\gamma_k^\top\bm x)\), for \(k=0,1\), as in the main text. Here \(\Lambda_{00}\) and \(\Lambda_{01}\) are unspecified continuous baseline cumulative hazards. Write \(u(d)=\exp\{-\Lambda_0(d\mid\bm x)\}\) and \(v(t)=\exp\{-\Lambda_1(t\mid\bm x)\}\). On the region \(0\le t\le d\),
\(\mathcal S(d,t\mid\bm x)=\mathcal C_\theta\{u(d),v(t)\}.\)
The two margin functions and the four likelihood contributions are described in Section \ref{supp:joint-model-choices}, particularly \eqref{supp:eq:ctw-four-pattern}. The finite parameters and both baselines are estimated jointly from this likelihood, which includes deaths without a preceding nonfatal event.

\paragraph{Counting processes and martingales}
Let states 0, 1, and 2 represent being alive without the nonfatal event, being alive after that event, and death, respectively. State 2 is absorbing. Using the copula derivatives from \eqref{supp:eq:ctw-copula-derivatives}, define \(F_{01}(u,v)=v\partial_1\mathcal C_\theta(u,v)/\mathcal C_\theta(u,v)\), \(F_{02}(u,v)=u\partial_0\mathcal C_\theta(u,v)/\mathcal C_\theta(u,v)\), and \(F_{12}(u,v)=u\partial_{01}\mathcal C_\theta(u,v)/\partial_1\mathcal C_\theta(u,v)\), on positive support. If \(\lambda_k(s\mid\bm x)=\partial\Lambda_k(s\mid\bm x)/\partial s\), the three transition intensities are
\begin{align}
\lambda^{01}(s)&=\lambda_1(s)F_{01}\{u(s),v(s)\},\nonumber\\
\lambda^{02}(s)&=\lambda_0(s)F_{02}\{u(s),v(s)\},\nonumber\\
\lambda^{12}(s\mid V)&=\lambda_0(s)F_{12}\{u(s),v(V)\}.
\label{supp:cox-transition-intensities}
\end{align}
The third intensity is evaluated for \(s>V\), where \(V\) is the recorded nonfatal event time. Dependence on \(V\) allows the death intensity to depend on the time of the preceding nonfatal event.

To verify these expressions, the probability of remaining in state 0 is \(\mathcal S(s,s)\). Its nonfatal and fatal exit intensities are \(-\dot{\mathcal S}_1(s,s)/\mathcal S(s,s)\) and \(-\dot{\mathcal S}_0(s,s)/\mathcal S(s,s)\), respectively. After a nonfatal event at \(V\),
\(P(D>s\mid T_1=V,D>V,\bm X=\bm x) =\frac{-\dot{\mathcal S}_1(s,V\mid\bm x)}{-\dot{\mathcal S}_1(V,V\mid\bm x)},\)
and the death intensity is \(\dot{\mathcal S}_{01}(s,V)/\{-\dot{\mathcal S}_1(s,V)\}\). The copula derivatives give \eqref{supp:cox-transition-intensities}.

Let \(N_i^{01}\), \(N_i^{02}\), and \(N_i^{12}\) count the three observed transitions through \(\tau_E\), and let \(U_i=D_i\wedge C_i\wedge\tau_E\). Define \(Y_i^0(s)=I(U_i\ge s)I\{N_i^{01}(s-)=0\}\) and \(Y_i^1(s)=I(U_i\ge s)I\{N_i^{01}(s-)=1\}\), using predictable left limits. Put \(Y_i^{01}=Y_i^{02}=Y_i^0\) and \(Y_i^{12}=Y_i^1\). Up to a factor independent of the joint model parameters, the likelihood in counting process form is
\begin{equation}
\mathcal L(\mathcal O_i)=\exp\left[\sum_{ab\in\{01,02,12\}}
\left\{\int_0^{\tau_E}\log\lambda_i^{ab}(s)\,dN_i^{ab}(s)
-\int_0^{\tau_E}Y_i^{ab}(s)\lambda_i^{ab}(s)\,ds\right\}\right].
\label{supp:cox-likelihood}
\end{equation}
Multiplying the probability of remaining in state 0 by the transition intensity and, when appropriate, survival and the death intensity after the nonfatal event gives the four contributions in \eqref{supp:eq:ctw-four-pattern}. Under the correctly specified joint model and conditionally independent censoring,
\(M_i^{ab}(t)=N_i^{ab}(t)-\int_0^tY_i^{ab}(s)\lambda_i^{ab}(s)\,ds\)
is a martingale for every \(ab\in\{01,02,12\}\). The predictable covariation processes are zero because transitions cannot occur simultaneously under the continuous-time model. Each process is stopped at death or censoring.

\paragraph{Score directions and information}
Consider a path with directions \(a_\theta,\bm a_0,\bm a_1\) for the finite parameters and \(h_0^*,h_1^*\) for the baselines. Along this path, \(\theta_\epsilon=\theta+\epsilon a_\theta\), \(\bm\gamma_{k,\epsilon}=\bm\gamma_k+\epsilon\bm a_k\), and \(d\Lambda_{0k,\epsilon}(s)=\exp\{\epsilon h_k^*(s)\}\,d\Lambda_{0k}(s)\), for \(k=0,1\). Denote the full direction by \(h\). The derivative of the cumulative hazard is
\(A_{k,h}(t)=e^{\bm\gamma_k^\top\bm x}\left\{\Lambda_{0k}(t)\bm x^\top\bm a_k+\int_0^th_k^*(s)\,d\Lambda_{0k}(s)\right\}.\)
Thus the derivatives of \(\log u(t)\) and \(\log v(t)\) are \(-A_{0,h}(t)\) and \(-A_{1,h}(t)\). The baseline perturbations are defined through their increments, so these derivatives are expressed in terms of cumulative hazards.

For each transition, define \(G_{ab,0}=u\partial\log F_{ab}/\partial u\), \(G_{ab,1}=v\partial\log F_{ab}/\partial v\), and \(G_{ab,\theta}=\partial\log F_{ab}/\partial\theta\). The derivatives of the log intensities are
\begin{align*}
Z_{01}[h](s)&=\bm x^\top\bm a_1+h_1^*(s)+G_{01,\theta}a_\theta-G_{01,0}A_{0,h}(s)-G_{01,1}A_{1,h}(s),\\
Z_{02}[h](s)&=\bm x^\top\bm a_0+h_0^*(s)+G_{02,\theta}a_\theta-G_{02,0}A_{0,h}(s)-G_{02,1}A_{1,h}(s),\\
Z_{12}[h](s\mid V)&=\bm x^\top\bm a_0+h_0^*(s)+G_{12,\theta}a_\theta-G_{12,0}A_{0,h}(s)-G_{12,1}A_{1,h}(V).
\end{align*}
The first two rows use \(\{u(s),v(s)\}\), and the third uses \(\{u(s),v(V)\}\). Differentiation of \eqref{supp:cox-likelihood} gives the subject score
\begin{equation}
\mathbb S[h](\mathcal O)=\sum_{ab\in\{01,02,12\}}\int_0^{\tau_E}Z_{ab}[h](s)\,dM^{ab}(s).
\label{supp:cox-score-operator}
\end{equation}
The derivative of each event term is \(\int Z\,dN\), and that of its compensator is \(\int Y\lambda Z\,ds\), giving the martingale integral. It includes the score directions for the finite parameters and both baseline functions.

At the true joint model, martingale isometry gives
\begin{equation*}
\mathbb I(h,k)=E\{\mathbb S[h](\mathcal O_1)\mathbb S[k](\mathcal O_1)\}
=E\left\{\sum_{ab\in\{01,02,12\}}\int_0^{\tau_E}Y^{ab}(s)\lambda^{ab}(s)Z_{ab}[h](s)Z_{ab}[k](s)\,ds\right\}.
\end{equation*}
This information form includes the cross information between the finite parameters and the baselines. Let \(\mathbb H_I\) be the completion of admissible directions for the finite parameters and bounded directions for the baselines in the norm \(\|h\|_I=\{\mathbb I(h,h)\}^{1/2}\), after identifying directions with the same score almost surely.

\paragraph{Derivative of the conditional tie probability}
For an individual censored while alive at \(c<L\), put \(w=c\) if no nonfatal event was observed and \(w=V\) otherwise. Set \(G_0=\mathcal C_\theta\) and \(G_1=\partial_1\mathcal C_\theta\). The conditional probabilities in the main text become
\begin{equation}
\omega_i^{(1)}(L)=\frac{G_{\delta_i^N}\{u(L),v(w)\}}{G_{\delta_i^N}\{u(c),v(w)\}}.
\label{supp:cox-q-ratio}
\end{equation}
For an observed nonfatal event, the factor \(\lambda_1(V)v(V)\) cancels from the ratio. Its derivative therefore depends on cumulative baseline hazards.

For \(m=0,1\), define \(A_m(u,v)=u\partial\log G_m(u,v)/\partial u\), \(B_m(u,v)=v\partial\log G_m(u,v)/\partial v\), and \(K_m(u,v)=\partial\log G_m(u,v)/\partial\theta\). Superscripts \((L,w)\) and \((c,w)\) indicate evaluation at the numerator and denominator arguments of \eqref{supp:cox-q-ratio}. For \(m=\delta_i^N\), the derivative of \(\log \omega_i^{(1)}(L)\) is
\begin{align*}
\dot\ell_{\omega,i}[h]
={}&\{K_m^{(L,w)}-K_m^{(c,w)}\}a_\theta
-A_m^{(L,w)}A_{0,h}(L)+A_m^{(c,w)}A_{0,h}(c)-\{B_m^{(L,w)}-B_m^{(c,w)}\}A_{1,h}(w).
\end{align*}
Hence \(\dot \omega_i^{(1)}[h]=\omega_i^{(1)}(L)\dot\ell_{\omega,i}[h]\). On an observed death, the probability and its derivative are zero. On observed survival through \(L\), the probability is one and its derivative is zero. All these quantities use \(\mathcal O_i(L)\).

For \(Q=1\), the derivative of the population CTW equation with respect to the joint model is
\begin{equation*}
\bm d_L[h]=E\left[\frac{\bm K_{12}(\bm\beta_L^0)R_{12}^{\mathrm{CTW},(1)}(L)\{\widetilde H_{12}^{(1)}(L)-\mu_{12}(\bm\beta_L^0)\}}{\pi_{12}^{\mathrm{CTW},(1)}(L)}
\{\omega_2^{(1)}(L)\dot \omega_1^{(1)}[h]+\omega_1^{(1)}(L)\dot \omega_2^{(1)}[h]\}\right].
\end{equation*}
This is \(\bm D_{J,L}[h]\) from \eqref{supp:eq:ctw-general-event-derivative}, evaluated along the paths of the model with Cox margins. Let \(d_{Lr}\) be its \(r\)th coordinate.

\paragraph{Regularity conditions for semiparametric inference}
Write \(\widehat\Delta=(\widehat\theta-\theta,\widehat{\bm\gamma}_0-\bm\gamma_0,\widehat{\bm\gamma}_1-\bm\gamma_1,\widehat\Lambda_{00}-\Lambda_{00},\widehat\Lambda_{01}-\Lambda_{01})\) for the difference between the estimated and true model parameters. Let \(\widehat{\mathbb S}[h]\) be the score evaluated at the fitted model, and \(\mathbb S[h]\) its value at the true model. Assume the following conditions on the likelihood and the CTW estimating function.

\emph{(SP.1) Identification and moments.}
The observations are independent, the joint model is correctly specified and identifiable on the fitted parameter set, and censoring is conditionally independent of the event history. The follow-up period used for estimation is fixed, finite, and at least \(L\). Covariates are bounded and omit intercepts confounded with baseline scales. The true finite parameters are interior, and the true baselines are absolutely continuous. The at-risk probabilities satisfy \(\inf_{s\le\tau_E}E\{Y^0(s)\}>0\) and \(\inf_{s\le\tau_E}E\{Y^0(s)+Y^1(s)\}>0\). The likelihood scores used below have finite second moments, and the score and CTW derivatives have integrable envelopes. All ratios are evaluated where their denominators are positive.

\emph{(SP.2) Consistency and quadratic remainders.}
The likelihood estimator used in the analysis is consistent. Its error \(r_n\), in a metric for the finite parameters and both cumulative baseline functions, satisfies \(r_n=o_p(n^{-1/4})\). The mean likelihood score and the population CTW functional have linear expansions with remainders \(O_p(r_n^2)\). Alternatively, their remainders may be required directly to be \(o_p(n^{-1/2})\) under a sharper expansion for the chosen model. Uniform convergence of the empirical likelihood and separation of the expected maximum imply consistency on a suitable compact parameter set. The rate and remainder conditions must also hold.

\emph{(SP.3) Information and CTW directions.}
Each \(d_{Lr}\) is a continuous linear functional on \(\mathbb H_I\). Its Riesz representer \(h_{Lr}\), defined up to directions with the same score, satisfies, for every \(h\in\mathbb H_I\),
\begin{equation}
\mathbb I(h_{Lr},h)=d_{Lr}[h].
\label{supp:cox-riesz}
\end{equation}
The Riesz representer therefore has finite information norm.

\emph{(SP.4) Score equations and approximation.}
Let \(P_n\) be the empirical measure of the independent subjects and \(P_0\) expectation at the true model. There are deterministic admissible directions \(h_{Lr,n}\) in the fitted score spaces such that \(\|h_{Lr,n}-h_{Lr}\|_I\to0\) and \(P_n\widehat{\mathbb S}[h_{Lr,n}]=o_p(n^{-1/2})\). Let \(\mathcal J(h,\Delta)\) be minus the derivative of the expected score along the model increment \(\Delta\), comprising changes in the finite parameters and cumulative baseline hazards. On an increment generated by a regular direction \(k\), it equals \(\mathbb I(h,k)\). Both \(\mathcal J(h,\cdot)\) and the CTW derivative extend continuously to increments of the estimated cumulative baseline hazards. Denoting the latter extension by \(D_{Lr}\), require
\begin{equation}
|D_{Lr}[\widehat\Delta]-\mathcal J(h_{Lr,n},\widehat\Delta)|=o_p(n^{-1/2}).
\label{supp:cox-direction-approximation}
\end{equation}
An approximation error for the derivative, measured in the dual norm and denoted by \(\epsilon_n\), is sufficient when \(\epsilon_n r_n=o_p(n^{-1/2})\). The quadratic score remainder in SP.2 holds uniformly for these directions. For a sieve fit, approximation bias in the CTW functional is \(o(n^{-1/2})\).

\emph{(SP.5) Empirical process remainders.}
For every coordinate,
\begin{equation*}
\sqrt n(P_n-P_0)\{\widehat{\mathbb S}[h_{Lr,n}]-\mathbb S[h_{Lr,n}]\}=o_p(1).
\end{equation*}

The corresponding empirical process remainder for the fitted CTW U-statistic of order two is \(o_p(n^{-1/2})\). Sufficient conditions include a Donsker score class, square-integrable envelopes, score convergence in \(L_2(P_0)\), and local equicontinuity for the pair contribution. The mixed derivative with respect to the joint model and log censoring survival, using the model norm and sup norm respectively, has an integrable envelope over pairs. Its empirical pair average is \(O_p(1)\), so the interaction of the two model fits is \(O_p(n^{-1/2}r_n)=o_p(n^{-1/2})\).

The conditions depend on the copula and estimation procedure. For Gumbel dependence, higher copula derivatives can be singular near \((u,v)=(1,1)\), so SP.1 requires the corresponding weighted integrability near time zero.

\begin{theorem}[Influence of semiparametric model estimation]
\label{supp:cox-directional-theorem}
\leavevmode\newline
Under G.1--G.2 in Section \ref{supp:ctw-conditions} and SP.1--SP.5, the contribution from estimating the joint model satisfies
\begin{equation}
\bm D_L[\widehat\Delta]=\frac1n\sum_{i=1}^n\bm b_{J,L}^{\mathrm{CTW}}(\mathcal O_i)+o_p(n^{-1/2}),
\label{supp:cox-event-influence}
\end{equation}
where \(b_{J,L,r}^{\mathrm{CTW}}(\mathcal O)=\mathbb S[h_{Lr}](\mathcal O)\). Its coordinates have mean zero and finite second moments. Theorem \ref{supp:thm:asymptotic-normality} therefore applies with this contribution from estimating the joint model and the censoring contribution in \eqref{supp:eq:ctw-general-censor-influence}.
\end{theorem}

\begin{proof}
For regular directions \(h,k\), let \(\mathcal S_{\epsilon,k}\) be the submodel generated by \(k\), and \(\mathbb S_{\epsilon,k}[h]\) its directional score. Differentiating the identity \(E_{\epsilon,k}\{\mathbb S_{\epsilon,k}[h]\}=0\) gives
\begin{equation*}
P_0\left\{\left.\frac{\partial\mathbb S_{\epsilon,k}[h]}{\partial\epsilon}\right|_{\epsilon=0}\right\}
=-E\{\mathbb S[h]\mathbb S[k]\}=-\mathbb I(h,k).
\end{equation*}
This is the information identity in SP.4. Its continuous extension applies to increments of the estimated cumulative baseline hazards. For each \(r\), the fitted score equation, SP.2, and SP.5 give
\begin{align*}
o_p(n^{-1/2})
&=P_n\widehat{\mathbb S}[h_{Lr,n}]\\
&=(P_n-P_0)\mathbb S[h_{Lr,n}]+P_0\widehat{\mathbb S}[h_{Lr,n}]+o_p(n^{-1/2})\\
&=P_n\mathbb S[h_{Lr,n}]-\mathcal J(h_{Lr,n},\widehat\Delta)+o_p(n^{-1/2}).
\end{align*}
Equation \eqref{supp:cox-direction-approximation} therefore gives
\(D_{Lr}[\widehat\Delta] =P_n\mathbb S[h_{Lr,n}]+o_p(n^{-1/2}) =P_n\mathbb S[h_{Lr}]+o_p(n^{-1/2}).\)
For the second equality, the score difference associated with the deterministic direction approximation has mean zero, and the variance of its sample average is \(n^{-1}\|h_{Lr,n}-h_{Lr}\|_I^2=o(n^{-1})\). The finite number of regression coefficients gives \eqref{supp:cox-event-influence}. Mean zero and finite variance follow from \eqref{supp:cox-score-operator} and \(\|h_{Lr}\|_I<\infty\).

The population CTW expansion and its quadratic remainder establish G.3. The empirical process condition for the pair contributions establishes G.4, and the remainder involving both model estimates is \(o_p(n^{-1/2})\). The common result in Theorem \ref{supp:thm:asymptotic-normality} therefore applies with the joint model contribution in \eqref{supp:cox-event-influence}.
\end{proof}

\paragraph{Parametric models as a special case}
For fixed parametric baselines, \(\mathbb S[h]=h^\top\bm s_{\vartheta}\), \(\mathbb I(h,k)=h^\top\bm I_{\vartheta}k\), and \(d_{Lr}[h]=\bm D_{\vartheta,L,r}h\). Thus \eqref{supp:cox-riesz} gives \(h_{Lr}=\bm I_{\vartheta}^{-1}\bm D_{\vartheta,L,r}^\top\), and hence \(\bm b_{J,L}^{\mathrm{CTW}}=\bm D_{\vartheta,L}\bm I_{\vartheta}^{-1}\bm s_{\vartheta}\), recovering \eqref{supp:eq:parametric-joint-influence}. With unspecified baselines, their directions enter the same information equation.

\paragraph{Estimation of the influence function}
The information equations can be solved in the score space used to fit the model. Let its coordinates be \(\bm\vartheta_n\), including the copula and regression parameters and both sets of baseline coefficients or increments. Define \(\widehat{\bm s}_i=\partial\ell_i/\partial\bm\vartheta_n\), \(\widehat{\bm I}_n=-n^{-1}\sum_{i=1}^n\partial^2\ell_i/\partial\bm\vartheta_n\partial\bm\vartheta_n^\top\), and let \(\widehat{\bm D}_n\) be the average over pairs of the fitted CTW derivative with respect to \(\bm\vartheta_n\). Solving \(\widehat{\bm I}_n\widehat h_{Lr}=\widehat{\bm D}_{n,r}^\top\) gives
\(\widehat{\bm b}_{J,L,i}^{\mathrm{CTW}}=\widehat{\bm D}_n\widehat{\bm I}_n^{-1}\widehat{\bm s}_i.\)
For baselines estimated by jumps, the scores and Hessian come from the likelihood expressed in terms of baseline increments, with SP.4--SP.5 assumed for those scores. The full matrix includes both baseline functions. For a sieve fit, its dimension and approximation error must satisfy the likelihood and Riesz conditions, including \(o(n^{-1/2})\) bias for the CTW functional and convergence of the fitted score directions in information norm.

For covariance consistency, require \(n^{-1}\sum_{i=1}^n\|\widehat{\bm b}_{J,L,i}^{\mathrm{CTW}}-\bm b_{J,L}^{\mathrm{CTW}}(\mathcal O_i)\|^2=o_p(1)\), together with the conditions on the pair contributions and censoring influence function in Theorem \ref{supp:thm:covariance}. Sufficient conditions include consistency of the empirical information operator in the relevant directions, convergence of fitted Riesz directions, and empirical mean square convergence of their scores. Adding all three subject contributions gives \(n^{-1}\sum_{i=1}^n\|\widehat{\bm\psi}_{L,i}-\bm\psi_L^{\mathrm{CTW}}(\mathcal O_i)\|^2=o_p(1)\). Cauchy--Schwarz and consistency of \(\widehat{\bm A}_L^{\mathrm{CTW}}\) then give covariance consistency by Theorem \ref{supp:thm:covariance}. This covariance estimator includes the covariances between the two model estimates and the pair comparisons.

\paragraph{Estimation in two stages}
First fit the marginal Cox model for death, then estimate the nonfatal component and copula jointly, as in \citet{Arachchige2025Copula}. The first-stage parameters are \((\bm\gamma_0,\Lambda_{00})\), and the second-stage parameters are \((\bm\gamma_1,\theta,\Lambda_{01})\).

Let \(N_i^D=N_i^{02}+N_i^{12}\), \(Y_i^D(s)=I(U_i\ge s)\), and \(d\overline M_i^D(s)=dN_i^D(s)-Y_i^D(s)e^{\bm\gamma_0^\top\bm X_i}d\Lambda_{00}(s)\). This is a martingale in the filtration generated by death, independent censoring, and baseline covariates. This filtration is used for the first-stage Cox analysis. For \(k=0,1,2\), define \(r_D^{(k)}(s)=E\{Y^D(s)e^{\bm\gamma_0^\top\bm X}\bm X^{\otimes k}\}\) and \(\overline{\bm x}_D(s)=r_D^{(1)}(s)/r_D^{(0)}(s)\). With \(\bm\Omega_D=\int_0^{\tau_E}\{r_D^{(2)}(s)/r_D^{(0)}(s)-\overline{\bm x}_D(s)^{\otimes2}\}r_D^{(0)}(s)\,d\Lambda_{00}(s)\), the coefficient influence is \(\bm\phi_{\gamma_0}(\mathcal O)=\bm\Omega_D^{-1}\int_0^{\tau_E}\{\bm X-\overline{\bm x}_D(s)\}\,d\overline M^D(s)\). Put \(\bm a_D(t)=\int_0^t\overline{\bm x}_D(s)\,d\Lambda_{00}(s)\). The baseline influence is \(\phi_{\Lambda_{00}}(\mathcal O,t)=\int_0^t\{r_D^{(0)}(s)\}^{-1}\,d\overline M^D(s)-\bm a_D(t)^\top\bm\phi_{\gamma_0}(\mathcal O)\). These expressions follow from the Cox--Breslow expansion for the death model, by the same argument as in Section \ref{supp:cox-if}.

In the second stage, maximize the joint likelihood over \((\bm\gamma_1,\theta,\Lambda_{01})\) with \((\widehat{\bm\gamma}_0,\widehat\Lambda_{00})\) fixed. Denote its directional score by \(\mathbb S_2[h]\), and put \(\mathbb I_{22}(h,k)=E\{\mathbb S_2[h]\mathbb S_2[k]\}\). Let \(\mathcal J_{22}(h,\Delta_2)\) and \(\mathbb I_{21}(h,\Delta_D)\) be minus the derivatives of \(P_0\mathbb S_2[h]\) in the second-stage and first-stage increments, respectively. Let \(d_{2,Lr}\) be the CTW derivative in a second-stage direction. Its Riesz representer satisfies \(\mathbb I_{22}(h_{2,Lr},h)=d_{2,Lr}[h]\) for every admissible \(h\). Under the corresponding SP.1--SP.5 conditions, the contribution from estimating the joint model is
\begin{equation}
b_{J,L,r}^{\mathrm{CTW}}(\mathcal O)
=\mathbb S_2[h_{2,Lr}](\mathcal O)
+\{D_{D,Lr}-\mathbb I_{21}(h_{2,Lr},\cdot)\}[\bm\phi_D(\mathcal O)],
\label{supp:cox-two-stage-influence}
\end{equation}
where \(\bm\phi_D\) contains the influence functions for the first-stage regression coefficients and cumulative baseline hazard and \(D_{D,Lr}\) is the direct CTW derivative in the first-stage increment.

To verify \eqref{supp:cox-two-stage-influence}, write the fitted increments as \(\widehat\Delta_D\) and \(\widehat\Delta_2\). The second-stage score expansion gives \(0=P_n\mathbb S_2[h]-\mathcal J_{22}(h,\widehat\Delta_2)-\mathbb I_{21}(h,\widehat\Delta_D)+o_p(n^{-1/2})\). Evaluating it at \(h_{2,Lr}\), applying the Riesz identity, and adding the direct derivative with respect to the first-stage parameters gives
\begin{align*}
D_{D,Lr}[\widehat\Delta_D]+D_{2,Lr}[\widehat\Delta_2]
&=P_n\mathbb S_2[h_{2,Lr}]+\{D_{D,Lr}-\mathbb I_{21}(h_{2,Lr},\cdot)\}[\widehat\Delta_D]+o_p(n^{-1/2})\\
&=\frac1n\sum_{i=1}^nb_{J,L,r}^{\mathrm{CTW}}(\mathcal O_i)+o_p(n^{-1/2}).
\end{align*}
The first-stage estimates affect CTW directly through the conditional tie probability and indirectly through the second-stage estimates. Substituting \eqref{supp:cox-two-stage-influence} into the total subject score gives the covariance estimator for CTW fitted in two stages.

\section{Consistency, asymptotic normality, and covariance estimation}
\label{supp:theory}

The three estimating equations have the same population limit. Their asymptotic distributions differ through the pair contributions and the effects of fitting the censoring and joint models. The proofs below use uniform convergence of U-statistics, the local root argument of \citet{Foutz1977}, and the Hoeffding decomposition described by \citet{kowalski2008modern}. Uniform convergence can also be obtained from the U-process results of \citet{NolanPollard1987}.

Use \(\mathcal A=\{\mathrm{CHW},\mathrm{RCW},\mathrm{CTW}\}\) and \(\mathcal I_n=\{(i,j):1\le i<j\le n\}\), as in the main text. For \(\star\in\mathcal A\), write \(\bm U_{ij}^{\star}(\bm\beta_L)\) for the contribution evaluated at the true models and \(\widehat{\bm U}_{ij}^{\star}(\bm\beta_L)\) for its fitted counterpart. For \(\star\in\{\mathrm{CHW},\mathrm{RCW}\}\) and a positive candidate censoring survival function \(S\), the score function is
\begin{equation}
\bm U_{ij}^{\star}(\bm\beta_L;S)
=\bm K_{ij}(\bm\beta_L)W_{ij}^{C,\star}(L;S)
\{\widetilde H_{ij}(L)-\mu_{ij}(\bm\beta_L)\}.
\label{supp:eq:compact-oracle-score}
\end{equation}
For CTW, the component contributions are those in \eqref{eq:ctw-score} of main text. The fatal and \emph{genuine tie} contributions use only the censoring model, whereas the nonfatal contributions additionally use the conditional tie probabilities. Each contribution is zero when its observation requirement is not satisfied, without evaluating an unobserved event time or a probability denominator.

Let \(\mathcal O_i\) contain the observed record used by all fitted models. If the joint model uses follow-up through a fixed time \(\tau_E\ge L\), its influence function may depend on \(\mathcal O_i(\tau_E)\). The conditional tie probabilities remain conditional on \(\mathcal O_i(L)\). With the true models, for every \(\star\in\mathcal A\), the population identities in Sections \ref{supp:unbiased} and \ref{supp:ctw-identification} give,
\begin{equation}
E\{\bm U_{12}^{\star}(\bm\beta_L)\}
=E\left[\bm K_{12}(\bm\beta_L)
\{H_{12}(L)-\mu_{12}(\bm\beta_L)\}\right]
\equiv\bm\Psi(\bm\beta_L).
\label{supp:eq:common-population-score}
\end{equation}
Thus the target \(\bm\beta_L^0\) and derivative \(\bm A_L=-\dot{\bm\Psi}(\bm\beta_L^0)\) are common to all three methods, including when the PIM is a working model. A dot denotes differentiation with respect to the regression parameter transpose. Denote the local root of \(\bm\Psi_n^{\star}(\bm\beta_L)=\frac{2}{n(n-1)}\sum_{(i,j)\in\mathcal I_n}\bm U_{ij}^{\star}(\bm\beta_L)\)={\bf 0} by \(\bm\beta_L^{\star,\mathrm{or}}\) and the local root of the fitted equation by \(\widehat{\bm\beta}_L^{\star}\). Before deriving the consistency and asymptotic normality of \(\widehat{\bm\beta}_L^{\star}\), we first provide regularity conditions. 

\subsection{Regularity conditions}
\label{supp:regularity}

Conditions C.1--C.7 are sufficient for consistency and asymptotic normality of CHW and RCW. Condition C.8 gives a sufficient condition for estimating their influence functions. Conditions G.1--G.4 are the corresponding requirements for CTW, including the contribution from fitting the joint model.

\subsubsection{Conditions for CHW and RCW}
\setcounter{condition}{0}
\renewcommand{\thecondition}{C.\arabic{condition}}
\renewcommand{\theHcondition}{C.\arabic{condition}}

\begin{condition}[Sampling and positivity]
The observations are independent and identically distributed. For \(t\in[0,L]\), assume \(P(C_i\ge t\mid\bm Y_i,\bm X_i)=S_C(t\mid\bm X_i^{C})\) and \(S_C(t\mid\bm X_i^{C})\ge \epsilon>0\) almost surely. If a nuisance model is fitted through a longer fixed horizon, the same censoring assumption holds throughout that interval. The event and censoring tie convention is fixed in advance and is used consistently in the observed histories and the censoring counting process.
\end{condition}

\begin{condition}[Parameter space and identification]
The parameter space \(\mathcal B\subset\mathbb R^p\) is compact and convex, and \(\bm\beta_L^0\) is an interior point. The population function \(\bm\Psi\) is continuous and has the unique zero \(\bm\beta_L^0\). For every \(\varepsilon>0\), \(\inf_{\bm\beta_L\in\mathcal B:\,\|\bm\beta_L-\bm\beta_L^0\|\ge\varepsilon}\|\bm\Psi(\bm\beta_L)\|>0\). For the identity link, the target and a fixed neighborhood of the target are strictly separated from the boundary of the admissible predictor range.
\end{condition}

\begin{condition}[Smoothness, boundedness, and moments]
\leavevmode\newline
The dimension \(p\) and the number of nonfatal components \(Q\) are fixed, and \(\bm Z_{ij}\) is bounded. The inverse link \(g^{-1}\) has three continuous derivatives on an open set containing the predictor range generated by \(\mathcal B\). The working variance \(V\) has three continuous derivatives on an open set containing the corresponding mean range and is bounded away from zero there. There is a nonnegative symmetric envelope \(M(\mathcal O_i,\mathcal O_j)\) such that, almost surely,
\begin{equation*}
\max_{\star\in\{\mathrm{CHW},\mathrm{RCW}\}}\max_{0\le k\le2}
\sup_{\bm\beta_L\in\mathcal B}
\left\|\nabla_{\bm\beta_L}^{k}
\bm U_{ij}^{\star}(\bm\beta_L)\right\|
\le M(\mathcal O_i,\mathcal O_j),
\end{equation*}
with \(E\{M(\mathcal O_1,\mathcal O_2)^{2+\delta}\}<\infty\) for some \(\delta>0\), where \(\nabla_{\bm\beta_L}^{k}\) denotes the \(k\)th parameter derivative. The same envelope, enlarged if necessary, bounds the value and first two parameter derivatives of \(\bm V_{ij}^{\star}(\bm\beta_L)\) in \eqref{supp:eq:common-nuisance-multiplier}, uniformly over \(\star\in\{\mathrm{CHW},\mathrm{RCW}\}\) and \(\bm\beta_L\in\mathcal B\). Differentiation may be interchanged with expectation near \(\bm\beta_L^0\).
\end{condition}

\begin{condition}[Dependence among pair contributions]
For known \(S_C\), each contribution indexed by \(\mathcal{I}_n\) is a symmetric measurable function of the two observations in that pair. Contributions from pairs with disjoint subject sets are therefore independent. For \((i,j)\in \mathcal{I}_n\), define \(\mathcal N_{ij,n}=\{(k,l)\in \mathcal{I}_n:\{i,j\}\cap\{k,l\}\ne\varnothing\}\).
\end{condition}

\begin{condition}[Derivative matrix]
The Jacobian \(\dot{\bm\Psi}\) is continuous around \(\bm\beta_L^0\), and \(\bm A_L=-\dot{\bm\Psi}(\bm\beta_L^0)\) is nonsingular.
\end{condition}

\begin{condition}[Nondegenerate limiting covariance]
For each of CHW and RCW, the covariance limit for the adjusted pair contributions in \eqref{supp:eq:adjusted-pair-score}, computed by summing over overlapping pairs, exists, which is finite and positive definite. The adjusted pair contributions have finite \((2+\delta)\)th moments for the \(\delta\) in Condition C.3.
\end{condition}

\begin{condition}[Uniform asymptotic linearity of the censoring estimator]
\label{supp:cond:censoring-if}
\leavevmode\newline
Let \(\ell_C(t,\bm x)=\log S_C(t\mid\bm x)\). There is a measurable influence function \(\mathcal I(\mathcal O;t,\bm x)\), with mean zero for every \((t,\bm x)\), such that \(\widehat\ell_C(t,\bm x)-\ell_C(t,\bm x) =\frac1n\sum_{r=1}^n\mathcal I(\mathcal O_r;t,\bm x)+r_n(t,\bm x)\) uniformly over \(t\in[0,L]\) and the support of \(\bm X_i^{C}\). The remainder satisfies \(\|r_n\|_\infty=o_p(n^{-1/2})\), and \(\|\widehat\ell_C-\ell_C\|_\infty=O_p(n^{-1/2})\). Also, \(E\{\sup_{t,\bm x}|\mathcal I(\mathcal O;t,\bm x)|^{2+\delta}\}<\infty\). Proposition \ref{supp:prop:cox-breslow} verifies these conditions for the correctly specified Cox and Breslow estimator.
\end{condition}

\begin{condition}[Consistency of the fitted censoring influence function]
\leavevmode\newline
Let \(\mathcal X_C\) be the support of \(\bm X_i^{C}\). Assume \(|\mathcal I(\mathcal O;t,\bm x)|\le C_{\mathcal I}\) almost surely, uniformly over \(t\in[0,L]\) and \(\bm x\in\mathcal X_C\), for a fixed finite constant \(C_{\mathcal I}\). The empirical influence function satisfies 
\begin{equation*}
\Delta_{\mathcal I,n} =\max_{1\le i\le n}\sup_{t\in[0,L],\,\bm x\in\mathcal X_C} |\widehat{\mathcal I}(\mathcal O_i;t,\bm x)-\mathcal I(\mathcal O_i;t,\bm x)| =o_p(1).
\end{equation*}
Both influence functions correspond to the log survival evaluated just before \(t\), following the convention in \ref{supp:cox-if}. The Cox coefficient influence includes its endpoint increment, and all fitted quantities use the horizon and endpoint convention of the Cox fit. Lemma \ref{supp:lem:empirical-censoring-if} verifies these requirements under the stated Cox and Breslow assumptions. Consistency of the empirical subject averages is proved below under these conditions.
\end{condition}

Conditions C.1--C.5 concern the oracle equations. Condition C.6 gives moment and nondegeneracy requirements for the adjusted pair contributions. Condition C.7 gives the first-order contribution from estimating the censoring distribution with the same sample. Condition C.8 is additionally used to establish consistency of the sandwich estimator.

\subsubsection{Conditions for CTW}
\label{supp:ctw-conditions}

The derivatives \(\dot{\bm U}_{C,ij}^{\mathrm{CTW}}\) and \(\bm D_{J,L}\) below are defined in Section \ref{supp:weight-expansion}. They describe perturbations of log censoring survival and of the joint survival function, respectively.

\paragraph{Condition G.1}
The independent and identically distributed sampling and censoring requirements in C.1 hold. The assumptions of Proposition \ref{supp:prop:ctw-mean} hold, with fixed finite \(Q\), and the joint model correctly specifies the conditional tie probabilities. The full data population equation \(\bm\Psi(\bm\beta_L)=E[\bm K_{12}(\bm\beta_L)\{H_{12}(L)-\mu_{12}(\bm\beta_L)\}]\) has an interior root \(\bm\beta_L^0\), with nonsingular derivative \(\bm A_L=-\partial\bm\Psi(\bm\beta_L^0)/\partial\bm\beta_L^\top\). On a fixed convex neighborhood of the root, the PIM functions satisfy the boundedness and differentiability requirements in C.2--C.3, and the oracle pair contribution is symmetric and measurable as in C.4. The neighborhood is contained in the interior of the parameter space. The fitted CTW estimating function and its regression derivative converge uniformly there to their population counterparts. The pair contribution evaluated at the true models has a finite second moment.

\paragraph{Condition G.2}
The censoring estimator satisfies \(\widehat\ell_C-\ell_C=n^{-1}\sum_{i=1}^n\mathcal I(\mathcal O_i;\,\cdot,\cdot)+o_p(n^{-1/2})\) uniformly, with \(\|\widehat\ell_C-\ell_C\|_\infty=O_p(n^{-1/2})\), with a mean zero influence function as in Condition \ref{supp:cond:censoring-if}. For three independent records, \(\dot{\bm U}_{C,12}^{\mathrm{CTW}}[\mathcal I(\mathcal O_3;\,\cdot,\cdot)]\) has a finite second moment. The same expression has a finite first moment when \(\mathcal O_3=\mathcal O_1\) or \(\mathcal O_3=\mathcal O_2\). The fitted conditional tie probabilities satisfy
\begin{equation*}
\frac{2}{n(n-1)}\sum_{(i,j)\in\mathcal I_n}\sum_{q=1}^{Q}|\widehat \omega_i^{(q)}(L)\widehat \omega_j^{(q)}(L)-\omega_i^{(q)}(L)\omega_j^{(q)}(L)|^2=o_p(1).
\end{equation*}

\paragraph{Condition G.3}
For a candidate joint survival function \(\mathcal S^*\), let \(\bm M_L(\mathcal S^*)\) be the expectation of the CTW pair contribution at \(\bm\beta_L^0\), using \(S_C\) and the conditional tie probabilities calculated from \(\mathcal S^*\). Its derivative at \(\mathcal S\) is the continuous linear functional \(\bm D_{J,L}\) in \eqref{supp:eq:ctw-general-event-derivative}. For the fitted joint model, we have
\begin{align*}
\bm M_L(\widehat{\mathcal S})-\bm M_L(\mathcal S)
=\bm D_{J,L}[\widehat{\mathcal S}-\mathcal S]+o_p(n^{-1/2})=\frac1n\sum_{i=1}^n\bm b_{J,L}^{\mathrm{CTW}}(\mathcal O_i)+o_p(n^{-1/2}),
\end{align*}
where \(E\{\bm b_{J,L}^{\mathrm{CTW}}(\mathcal O_1)\}=\bm0\) and its second moment is finite. When the joint model includes cumulative baseline hazards, the increment is expressed in its local parameterization. A quadratic population remainder is negligible if the model error is \(o_p(n^{-1/4})\). The linear function must also satisfy the influence function expansion in the second equality.

\paragraph{Condition G.4}
Let \(\bm M_{n,L}(\mathcal S^*)\) be the empirical pair average corresponding to \(\bm M_L(\mathcal S^*)\), with normalization \(2/\{n(n-1)\}\) and sum over \((i,j)\in\mathcal I_n\). Along the fitted joint model,
\(\{\bm M_{n,L}(\widehat{\mathcal S})-\bm M_L(\widehat{\mathcal S})\} -\{\bm M_{n,L}(\mathcal S)-\bm M_L(\mathcal S)\}=o_p(n^{-1/2}).\)
This condition is a stochastic equicontinuity requirement for the pair contributions evaluated at the fitted conditional tie probabilities. Section \ref{supp:joint-model-contribution} gives sufficient conditions based on an asymptotically linear estimator of the joint model. Parametric and Cox model calculations are given in Sections \ref{supp:ctw-theory} and \ref{supp:cox-theory}, respectively. The derivative \(\bm D_{J,L}\) is generally nonzero, so estimation of the joint model contributes to the first-order variance.

\subsection{Consistency}
\label{supp:consistency}

\begin{proposition}[Consistency of the local roots]
\label{supp:lem:consistency}
For \(\star\in\{\mathrm{CHW},\mathrm{RCW}\}\), assume Conditions C.1--C.7. For \(\star=\mathrm{CTW}\), assume Condition G.1. The oracle and fitted estimating equations each have a unique root in a fixed sufficiently small neighborhood of \(\bm\beta_L^0\), with probability tending to one. These roots satisfy \(\bm\beta_L^{\star,\mathrm{or}}\overset{p}{\longrightarrow}\bm\beta_L^0\) and \(\widehat{\bm\beta}_L^{\star}\overset{p}{\longrightarrow}\bm\beta_L^0\).
\end{proposition}

\begin{proof}
First consider CHW and RCW. For fixed \(\delta>0\), compactness gives a finite set \(\{\bm\beta_1,\ldots,\bm\beta_{K_\delta}\}\subset\mathcal B\) such that every point in \(\mathcal B\) is within distance \(\delta\) of this set. The law of large numbers for integrable U-statistics of order two gives \(\max_{1\le k\le K_\delta}\|\bm\Psi_n^{\star}(\bm\beta_k)-\bm\Psi(\bm\beta_k)\|=o_p(1)\). Let \(\overline M_n=\frac{2}{n(n-1)}\sum_{(i,j)\in\mathcal I_n}M(\mathcal O_i,\mathcal O_j)\). The derivative envelope and the mean value inequality give
\begin{align*}
\sup_{\bm\beta_L\in\mathcal B}
\|\bm\Psi_n^{\star}(\bm\beta_L)-\bm\Psi(\bm\beta_L)\|
\le\max_{1\le k\le K_\delta}
\|\bm\Psi_n^{\star}(\bm\beta_k)-\bm\Psi(\bm\beta_k)\|\quad+\delta\{\overline M_n+E M(\mathcal O_1,\mathcal O_2)\}.
\end{align*}
Since \(\overline M_n\overset{p}{\longrightarrow}E M(\mathcal O_1,\mathcal O_2)<\infty\), first letting \(n\to\infty\) and then \(\delta\downarrow0\) proves uniform convergence of the estimating function. Applying the same argument to its first derivative, using the second derivative envelope, proves uniform convergence of the Jacobian near the target.

Proposition \ref{supp:prop:weight-consistency} gives \(\Delta_{W,n}=\max_{\star\in\{\mathrm{CHW},\mathrm{RCW}\}}\max_{(i,j)\in\mathcal I_n}|\widehat W_{ij}^{C,\star}(L)-W_{ij}^{C,\star}(L)|=o_p(1)\). Bounded predictors, compactness, and smoothness bound \(\bm K_{ij}\), \(\mu_{ij}\), and their required derivatives. The observed response is bounded, so a finite constant \(C^{\star}\) satisfies
\begin{align*}
\sup_{\bm\beta_L\in\mathcal B}
\|\widehat{\bm\Psi}_n^{\star}(\bm\beta_L)-\bm\Psi_n^{\star}(\bm\beta_L)\|
&\le C^{\star}\Delta_{W,n}=o_p(1),\\
\sup_{\bm\beta_L\in\mathcal B}
\|\dot{\widehat{\bm\Psi}}_n^{\star}(\bm\beta_L)-\dot{\bm\Psi}_n^{\star}(\bm\beta_L)\|
&\le C^{\star}\Delta_{W,n}=o_p(1).
\end{align*}
Unobserved pairs have zero contributions to these differences. For CTW, G.1 gives uniform convergence of the fitted estimating function and its derivative. The bounded PIM functions, fixed \(Q\), and positivity give an integrable derivative envelope for the oracle CTW contribution on a compact neighborhood of the root. The same finite set argument proves oracle convergence there.

For any \(\star\in\mathcal A=\{\mathrm{CHW},\mathrm{RCW},\mathrm{CTW}\}\), write \(\bm A=\bm A_L\). Continuity of \(\dot{\bm\Psi}\) and \(\dot{\bm\Psi}(\bm\beta_L^0)=-\bm A\) give a radius \(r>0\) such that \(\mathcal C_r=\{\bm\beta_L:\|\bm\beta_L-\bm\beta_L^0\|\le r\}\) lies in the stated neighborhood and
\begin{align*}
\sup_{\bm\beta_L\in\mathcal C_r}
\|\bm I_p+\bm A^{-1}\dot{\bm\Psi}(\bm\beta_L)\|
&=\sup_{\bm\beta_L\in\mathcal C_r}
\|\bm A^{-1}\{\dot{\bm\Psi}(\bm\beta_L)-\dot{\bm\Psi}(\bm\beta_L^0)\}\|\\
&\le\|\bm A^{-1}\|
\sup_{\bm\beta_L\in\mathcal C_r}
\|\dot{\bm\Psi}(\bm\beta_L)-\dot{\bm\Psi}(\bm\beta_L^0)\|
\le\tfrac14.
\end{align*}
Uniform convergence implies, with probability tending to one, \(\sup_{\bm\beta_L\in\mathcal C_r}\|\bm I_p+\bm A^{-1}\dot{\widehat{\bm\Psi}}_n^{\star}(\bm\beta_L)\|\le1/2\) and \(\|\bm A^{-1}\widehat{\bm\Psi}_n^{\star}(\bm\beta_L^0)\|\le r/2\). On this event, the map \(\widehat T_n^{\star}(\bm\beta_L)=\bm\beta_L+\bm A^{-1}\widehat{\bm\Psi}_n^{\star}(\bm\beta_L)\) has Lipschitz constant at most \(1/2\), and
\begin{align*}
\|\widehat T_n^{\star}(\bm\beta_L)-\bm\beta_L^0\|
\le\|\widehat T_n^{\star}(\bm\beta_L)-\widehat T_n^{\star}(\bm\beta_L^0)\|
+\|\widehat T_n^{\star}(\bm\beta_L^0)-\bm\beta_L^0\|\le\tfrac12\|\bm\beta_L-\bm\beta_L^0\|+\tfrac r2
\le r.
\end{align*}
Thus \(\widehat T_n^{\star}\) maps \(\mathcal C_r\) into itself. Its unique fixed point solves the fitted estimating equation and satisfies
\begin{align*}
\|\widehat{\bm\beta}_L^{\star}-\bm\beta_L^0\|
&\le\tfrac12\|\widehat{\bm\beta}_L^{\star}-\bm\beta_L^0\|
+\|\bm A^{-1}\widehat{\bm\Psi}_n^{\star}(\bm\beta_L^0)\|,\\
\|\widehat{\bm\beta}_L^{\star}-\bm\beta_L^0\|
&\le2\|\bm A^{-1}\widehat{\bm\Psi}_n^{\star}(\bm\beta_L^0)\|
=o_p(1).
\end{align*}
Replacing \(\widehat{\bm\Psi}_n^{\star}\) by \(\bm\Psi_n^{\star}\) gives the same contraction and error bound for the oracle root. Since \(\mathcal A\) is finite, the conclusions hold simultaneously whenever the conditions for all three methods hold.
\end{proof}

\subsection{Effects of fitting the models}
\label{supp:weight-expansion}

The fitted pair contributions share estimates of the censoring distribution and, for CTW, the joint survival model. Their dependence is captured by the influence functions of these estimates. All derivatives in this subsection are evaluated at \(\bm\beta_L^0\) and the true models.

\subsubsection{Derivatives and influence contributions}

For \(\star\in\{\mathrm{CHW},\mathrm{RCW}\}\), use the multiplier \(\bm V_{ij}^{\star}\) in \eqref{supp:eq:common-nuisance-multiplier} and the observed evaluation times \(t_{ij,i}^{\star}(L)\) and \(t_{ij,j}^{\star}(L)\) from Section \ref{supp:cox-if}. For a perturbation \(h_C\) of \(\ell_C=\log S_C\), the derivative of the pair contribution is \(\dot{\bm U}_{C,ij}^{\star}[h_C]=-\bm V_{ij}^{\star}\sum_{s\in\{i,j\}}h_C\{t_{ij,s}^{\star}(L),\bm X_s^C\}\).

For CTW, let \(\bm U_{ij}^{\mathrm{CTW},(q)}\) and \(\bm U_{ij}^{\mathrm{CTW},T}\) denote the component contributions in \eqref{eq:ctw-score}. Their censoring survival functions are evaluated at \(t_{ij}^{(q)}\), for \(q=0,\ldots,Q\), and \(L\), respectively. On an observed fatal comparison, \(t_{ij}^{(0)}=A_{ij}(L)\). On an observed \emph{genuine tie}, \(A_{ij}(L)=L\). Differentiating the inverse censoring products gives
\begin{align}
\dot{\bm U}_{C,ij}^{\mathrm{CTW}}[h_C]
={}&-\sum_{q=0}^{Q}\bm U_{ij}^{\mathrm{CTW},(q)}
\{h_C(t_{ij}^{(q)},\bm X_i^C)+h_C(t_{ij}^{(q)},\bm X_j^C)\}\nonumber\\
&-\bm U_{ij}^{\mathrm{CTW},T}
\{h_C(L,\bm X_i^C)+h_C(L,\bm X_j^C)\}.
\label{supp:eq:ctw-censoring-direction}
\end{align}
A derivative contribution is zero whenever the corresponding observation requirement is not satisfied.

For \(\star\in\mathcal A=\{\mathrm{CHW},\mathrm{RCW},\mathrm{CTW}\}\), define \(\bm q_L^{\star}(o_1,o_2,o_3)=\dot{\bm U}_{C,12}^{\star}[\mathcal I(o_3;\,\cdot,\cdot)]\), where the pair quantities are evaluated at \((o_1,o_2)\), and
\begin{equation}
\bm b_{C,L}^{\star}(o)
=E\{\bm q_L^{\star}(\mathcal O_1,\mathcal O_2,o)\}
=E\{\dot{\bm U}_{C,12}^{\star}[\mathcal I(o;\,\cdot,\cdot)]\}.
\label{supp:eq:ctw-general-censor-influence}
\end{equation}
The expectation is over an independent pair, with \(o\) fixed. For an independent third record, the censoring influence has conditional mean zero given the pair. Therefore
\begin{align*}
E\{\bm q_L^{\star}(\mathcal O_1,\mathcal O_2,\mathcal O_3)\mid\mathcal O_1,\mathcal O_2\}
&=\dot{\bm U}_{C,12}^{\star}[E\{\mathcal I(\mathcal O_3;\,\cdot,\cdot)\mid\mathcal O_1,\mathcal O_2\}]
=\bm0,\\
E\{\bm b_{C,L}^{\star}(\mathcal O_3)\}
&=E\{\bm q_L^{\star}(\mathcal O_1,\mathcal O_2,\mathcal O_3)\}
=\bm0.
\end{align*}

For a smooth path \(\mathcal S_\epsilon\) through the joint survival function \(\mathcal S\), with direction \(h\), write \(\dot \omega_i^{(q)}[h]=\left.\partial \omega_i^{(q)}(L;\mathcal S_\epsilon)/\partial\epsilon\right|_{\epsilon=0}\). Differentiation of the CTW nonfatal contributions gives
\begin{equation}
\bm D_{J,L}[h]=E\left[\bm K_{12}(\bm\beta_L^0)\sum_{q=1}^{Q}
\frac{R_{12}^{\mathrm{CTW},(q)}(L)\{\widetilde H_{12}^{(q)}(L)-\mu_{12}(\bm\beta_L^0)\}}
{\pi_{12}^{\mathrm{CTW},(q)}(L)}
\{\omega_2^{(q)}(L)\dot \omega_1^{(q)}[h]+\omega_1^{(q)}(L)\dot \omega_2^{(q)}[h]\}\right].
\label{supp:eq:ctw-general-event-derivative}
\end{equation}
The distribution of the observed records is held fixed when this derivative is taken. For every \(\star\in\mathcal A\), write \(\bm b_L^{\star}=\bm b_{C,L}^{\star}+\bm b_{J,L}^{\star}\), where \(\bm b_{J,L}^{\mathrm{CHW}}=\bm b_{J,L}^{\mathrm{RCW}}=\bm0\), and  \(\bm b_{J,L}^{\mathrm{CTW}}\) is the mean zero contribution in G.3, 

\subsubsection{Projection of the terms from model estimation}

Substituting an asymptotically linear model estimate into a pair average produces a sum over three observations. The following calculation applies to the censoring estimate for every method and to a joint model estimate with an influence representation.

\begin{lemma}[Projection of an average over three observations]
\label{supp:lem:three-observation-projection}
Let \(\bm h(o_1,o_2;o_3)\) be symmetric in its first two arguments, with \(E\{\bm h(\mathcal O_1,\mathcal O_2;\mathcal O_3)\mid\mathcal O_1,\mathcal O_2\}=\bm0\) for independent records. Suppose its second moment is finite for three independent records and its first moment is finite when its third argument equals either of the first two. For \(\bm b_h(o)=E\{\bm h(\mathcal O_1,\mathcal O_2;o)\}\),
\begin{equation}
\frac{2}{n^2(n-1)}\sum_{(i,j)\in\mathcal I_n}\sum_{r=1}^n
\bm h(\mathcal O_i,\mathcal O_j;\mathcal O_r)
=\frac1n\sum_{i=1}^n\bm b_h(\mathcal O_i)+O_p(n^{-1}).
\label{supp:eq:three-observation-projection}
\end{equation}
\end{lemma}

\begin{proof}
There are \(n(n-1)\) terms with \(r=i\) or \(r=j\). Their finite first moments give
\begin{align*}
&E\left\|\frac{2}{n^2(n-1)}\sum_{(i,j)\in\mathcal I_n}
\{\bm h(\mathcal O_i,\mathcal O_j;\mathcal O_i)+\bm h(\mathcal O_i,\mathcal O_j;\mathcal O_j)\}\right\|\\
&\le\frac1n\left\{E\|\bm h(\mathcal O_1,\mathcal O_2;\mathcal O_1)\|
+E\|\bm h(\mathcal O_1,\mathcal O_2;\mathcal O_2)\|\right\}
=O(n^{-1}).
\end{align*}
Markov's inequality makes this part \(O_p(n^{-1})\).

For three distinct records, define \(\mathcal I_{n,3}=\{(i,j,k):1\le i<j<k\le n\}\) and the symmetric kernel
\begin{align*}
\overline{\bm h}(o_1,o_2,o_3)
=\tfrac13\bigl\{&\bm h(o_1,o_2;o_3)+\bm h(o_1,o_3;o_2)
+\bm h(o_2,o_3;o_1)\bigr\}.
\end{align*}
Every unordered triple contributes these three terms, and \(6\binom n3/\{n^2(n-1)\}=(n-2)/n\). Hence
\begin{equation}
\frac{2}{n^2(n-1)}\sum_{r=1}^n
\sum_{\substack{(i,j)\in\mathcal I_n\\r\notin\{i,j\}}}
\bm h(\mathcal O_i,\mathcal O_j;\mathcal O_r)
=\frac{n-2}{n}\binom n3^{-1}
\sum_{(i,j,k)\in\mathcal I_{n,3}}\overline{\bm h}(\mathcal O_i,\mathcal O_j,\mathcal O_k).
\label{supp:eq:distinct-triple-U}
\end{equation}
The first projection of this square-integrable kernel is
\begin{align*}
E\{\overline{\bm h}(o,\mathcal O_2,\mathcal O_3)\}
&=\tfrac13\bigl[E\{\bm h(o,\mathcal O_2;\mathcal O_3)\}
+E\{\bm h(o,\mathcal O_3;\mathcal O_2)\}
+E\{\bm h(\mathcal O_2,\mathcal O_3;o)\}\bigr]\\
&=\tfrac13\{\bm0+\bm0+\bm b_h(o)\}
=\tfrac13\bm b_h(o).
\end{align*}
The first two expectations vanish upon conditioning on the first two arguments. The Hoeffding decomposition then gives
\begin{align*}
\binom n3^{-1}\sum_{(i,j,k)\in\mathcal I_{n,3}}
\overline{\bm h}(\mathcal O_i,\mathcal O_j,\mathcal O_k)
&=\frac3n\sum_{i=1}^n\frac{\bm b_h(\mathcal O_i)}3+O_p(n^{-1})\\
&=\frac1n\sum_{i=1}^n\bm b_h(\mathcal O_i)+O_p(n^{-1}).
\end{align*}
The canonical terms of orders two and three have variances \(O(n^{-2})\) and \(O(n^{-3})\), respectively, by square integrability. Their orders are \(O_p(n^{-1})\) and \(O_p(n^{-3/2})\). Jensen's inequality gives \(E\|\bm b_h(\mathcal O_1)\|^2<\infty\), and iterated expectation gives \(E\bm b_h(\mathcal O_1)=\bm0\). Thus
\begin{align*}
\left(\frac{n-2}{n}-1\right)\frac1n\sum_{i=1}^n\bm b_h(\mathcal O_i)
&=-\frac2n\left\{\frac1n\sum_{i=1}^n\bm b_h(\mathcal O_i)\right\}
=O_p(n^{-3/2}).
\end{align*}
Combining this bound with \eqref{supp:eq:distinct-triple-U} and the repeated-record terms proves \eqref{supp:eq:three-observation-projection}.
\end{proof}

\subsubsection{Expansion of the fitted estimating functions}

\begin{lemma}[Effect of fitting the censoring and joint models]
\label{supp:lem:nuisance}
For CHW and RCW, assume Conditions C.1--C.7. For CTW, assume Conditions G.1--G.4. Then, for each \(\star\in\mathcal A\),
\begin{equation}
\widehat{\bm\Psi}_n^{\star}(\bm\beta_L^0)
=\bm\Psi_n^{\star}(\bm\beta_L^0)
+\frac1n\sum_{i=1}^n\bm b_L^{\star}(\mathcal O_i)
+o_p(n^{-1/2}).
\label{supp:eq:score-expansion}
\end{equation}
\end{lemma}

\begin{proof}
We first fit the censoring distribution while fixing the conditional tie probabilities at their true values. Denote this pair average by \(\bm\Psi_{n,C}^{\star}\), which  equals the fully fitted estimating function for CHW and RCW. Let \(\delta_n=\widehat\ell_C-\ell_C\). For \(\star\in\{\mathrm{CHW},\mathrm{RCW}\}\), the exact weight ratio in \eqref{supp:eq:exact-weight-ratio} gives
\begin{equation}
\widehat{\bm U}_{ij}^{\star}(\bm\beta_L^0)
=\bm U_{ij}^{\star}(\bm\beta_L^0)-\bm V_{ij}^{\star} d_{ij,n}^{\star}
+\bm r_{ij,n}^{\star,W},
\label{supp:eq:pair-score-weight-expansion}
\end{equation}
where \(d_{ij,n}^{\star}\) is defined in \eqref{supp:eq:common-log-weight-error} and \(\bm r_{ij,n}^{\star,W}=\bm V_{ij}^{\star} e^{-\theta_{ij,n}^{\star} d_{ij,n}^{\star}}(d_{ij,n}^{\star})^2/2\) for some \(\theta_{ij,n}^{\star}\in(0,1)\). This identity also holds for an unobserved pair, whose weighted contribution is zero. The evaluation times lie in \([0,L]\), so
\begin{align*}
\max_{\star\in\{\mathrm{CHW},\mathrm{RCW}\}}\max_{(i,j)\in\mathcal I_n}|d_{ij,n}^{\star}|
&\le2\|\delta_n\|_\infty=O_p(n^{-1/2}),\\
\|\bm r_{ij,n}^{\star,W}\|
&\le\tfrac12\|\bm V_{ij}^{\star}\|e^{|d_{ij,n}^{\star}|}|d_{ij,n}^{\star}|^2\le2e^{2\|\delta_n\|_\infty}\|\delta_n\|_\infty^2M(\mathcal O_i,\mathcal O_j).
\end{align*}
For CTW, the same exponential expansion is applied to each censoring product in \eqref{eq:ctw-score} of main text. The linear terms add to \(\dot{\bm U}_{C,ij}^{\mathrm{CTW}}[\delta_n]\). Since \(Q\) is fixed, the sum of the component norms is bounded by an integrable envelope, also denoted by \(M\). For every method, the pair average of the quadratic remainder is consequently bounded by
\begin{align*}
\frac{2}{n(n-1)}\sum_{(i,j)\in\mathcal I_n}\|\bm r_{ij,n}^{\star,W}\|
&\le2e^{2\|\delta_n\|_\infty}\|\delta_n\|_\infty^2\overline M_n=O_p(n^{-1})=o_p(n^{-1/2}).
\end{align*}

Average \eqref{supp:eq:pair-score-weight-expansion} and the corresponding CTW component expansions, and substitute the uniform censoring influence representation into their linear terms. The uniform remainder contributes at most \(2\|r_n\|_\infty\overline M_n=o_p(n^{-1/2})\), after enlarging the envelope for the finite CTW sum. Thus
\begin{equation}
\bm\Psi_{n,C}^{\star}-\bm\Psi_n^{\star}(\bm\beta_L^0)
=\frac{2}{n^2(n-1)}\sum_{(i,j)\in\mathcal I_n}\sum_{r=1}^n
\bm q_L^{\star}(\mathcal O_i,\mathcal O_j,\mathcal O_r)+o_p(n^{-1/2}).
\label{supp:eq:score-triple-sum}
\end{equation}
For CHW and RCW, let \(\mathcal I_*(o)=\sup_{t\in[0,L],\,\bm x\in\mathcal X_C}|\mathcal I(o;t,\bm x)|\). Conditions C.3 and C.7 give
\begin{align*}
E\|\bm q_L^{\star}(\mathcal O_1,\mathcal O_2,\mathcal O_3)\|^2
&\le4E\{M(\mathcal O_1,\mathcal O_2)^2\mathcal I_*(\mathcal O_3)^2\}\\
&=4E\{M(\mathcal O_1,\mathcal O_2)^2\}E\{\mathcal I_*(\mathcal O_3)^2\}<\infty,\\
E\|\bm q_L^{\star}(\mathcal O_1,\mathcal O_2,\mathcal O_1)\|
&\le2\{E M(\mathcal O_1,\mathcal O_2)^2\}^{1/2}
\{E\mathcal I_*(\mathcal O_1)^2\}^{1/2}<\infty.
\end{align*}
The same first-moment bound holds when the third argument is \(\mathcal O_2\). For CTW, these moments are assumed in G.2. Pair symmetry and the conditional mean zero identity above verify the other assumptions of Lemma \ref{supp:lem:three-observation-projection}. For every \(\star\in\mathcal A\), applying Lemma \ref{supp:lem:three-observation-projection} to \eqref{supp:eq:score-triple-sum} gives
\begin{equation}
\bm\Psi_{n,C}^{\star}-\bm\Psi_n^{\star}(\bm\beta_L^0)
=\frac1n\sum_{i=1}^n\bm b_{C,L}^{\star}(\mathcal O_i)+o_p(n^{-1/2}).
\label{supp:eq:censoring-contribution-expansion}
\end{equation}

For CTW, keeping \(S_C\) fixed, fit the joint model and write the resulting average as \(\bm\Psi_{n,J}^{\mathrm{CTW}}=\bm M_{n,L}(\widehat{\mathcal S})\). Conditions G.3--G.4 give
\begin{align}
\bm\Psi_{n,J}^{\mathrm{CTW}}-\bm\Psi_n^{\mathrm{CTW}}(\bm\beta_L^0)
&=\bm M_L(\widehat{\mathcal S})-\bm M_L(\mathcal S)+o_p(n^{-1/2})\nonumber\\
&=\frac1n\sum_{i=1}^n\bm b_{J,L}^{\mathrm{CTW}}(\mathcal O_i)+o_p(n^{-1/2}).
\label{supp:eq:ctw-general-joint-expansion}
\end{align}
The difference between simultaneous substitution and the sum of the two separate substitutions satisfies
\begin{align*}
&\|\widehat{\bm\Psi}_n^{\mathrm{CTW}}(\bm\beta_L^0)
-\bm\Psi_{n,C}^{\mathrm{CTW}}-\bm\Psi_{n,J}^{\mathrm{CTW}}
+\bm\Psi_n^{\mathrm{CTW}}(\bm\beta_L^0)\|\\
&\le C e^{2\|\delta_n\|_\infty}\|\delta_n\|_\infty
\left\{\frac{2}{n(n-1)}\sum_{(i,j)\in\mathcal I_n}\sum_{q=1}^{Q}
|\widehat \omega_i^{(q)}(L)\widehat \omega_j^{(q)}(L)-\omega_i^{(q)}(L)\omega_j^{(q)}(L)|^2\right\}^{1/2}\\
&=o_p(n^{-1/2}),
\end{align*}
where the last equality follows from Condition G.2, boundedness of the PIM functions, and positivity. This bound allows both models to be fitted to the same individuals. Combining \eqref{supp:eq:censoring-contribution-expansion} and \eqref{supp:eq:ctw-general-joint-expansion} proves \eqref{supp:eq:score-expansion} for CTW. For CHW and RCW, \(\bm b_{J,L}^{\star}=\bm0\), so \eqref{supp:eq:censoring-contribution-expansion} already gives the result.
\end{proof}

\subsubsection{Asymptotically linear estimates of the joint model}
\label{supp:joint-model-contribution}

A sufficient condition for G.3--G.4 is an influence representation for the fitted joint model. Suppose \(\widehat{\mathcal S}-\mathcal S=n^{-1}\sum_{i=1}^n\phi_J(\mathcal O_i)+r_{J,n}\) in a normed space of local perturbations, with \(\|r_{J,n}\|=o_p(n^{-1/2})\), \(\|\widehat{\mathcal S}-\mathcal S\|=O_p(n^{-1/2})\), and \(E\|\phi_J(\mathcal O_1)\|^2<\infty\). Assume the influence has mean zero in every derivative direction. Suppose also that the pair contribution has a quadratic remainder bounded by an integrable envelope times the squared model error, and its first derivative has an integrable bound on its operator norm. Define \(\bm h_J(o_1,o_2;o_3)=\dot{\bm U}_{J,12}[\phi_J(o_3)]\). Assume a finite second moment for three independent records and finite first moments when the third argument equals either member of the pair. Then
\begin{equation}
\bm\Psi_{n,J}^{\mathrm{CTW}}-\bm\Psi_n^{\mathrm{CTW}}(\bm\beta_L^0)
=\frac1n\sum_{i=1}^n\bm D_{J,L}[\phi_J(\mathcal O_i)]+o_p(n^{-1/2}),
\label{supp:eq:ctw-general-full-nuisance-al}
\end{equation}
so \(\bm b_{J,L}^{\mathrm{CTW}}=\bm D_{J,L}[\phi_J]\).

To verify this expansion, the quadratic pair remainder is \(O_p(n^{-1})\), and the contribution of \(r_{J,n}\) is \(o_p(n^{-1/2})\). The remaining term is the average on the left side of \eqref{supp:eq:three-observation-projection}, with \(\bm h=\bm h_J\) and \(\bm b_h(o)=\bm D_{J,L}[\phi_J(o)]\). Lemma \ref{supp:lem:three-observation-projection} gives \eqref{supp:eq:ctw-general-full-nuisance-al}. Applying the population derivative to the model expansion gives the same leading term, and hence verifies G.3--G.4. The directional expansion in G.3 is sufficient when the full joint model does not have such a representation. The Cox model result in Section \ref{supp:cox-theory} uses this directional form.

\subsection{Hoeffding projection and asymptotic normality}
\label{supp:asymptotic-normality}

For every \(\star\in\mathcal A\), define the adjusted pair contribution
\begin{equation}
\bm\zeta_{ij}^{\star}
=\bm U_{ij}^{\star}(\bm\beta_L^0)
+\tfrac12\{\bm b_L^{\star}(\mathcal O_i)+\bm b_L^{\star}(\mathcal O_j)\}.
\label{supp:eq:adjusted-pair-score}
\end{equation}
Each individual occurs in \(n-1\) unordered pairs. The factor one half makes the added terms average to the contribution from model estimation,
\begin{align*}
\frac{2}{n(n-1)}\sum_{(i,j)\in\mathcal I_n}\bm\zeta_{ij}^{\star}
&=\bm\Psi_n^{\star}(\bm\beta_L^0)
+\frac{1}{n(n-1)}\sum_{(i,j)\in\mathcal I_n}
\{\bm b_L^{\star}(\mathcal O_i)+\bm b_L^{\star}(\mathcal O_j)\}\\
&=\bm\Psi_n^{\star}(\bm\beta_L^0)
+\frac{1}{n(n-1)}\sum_{i=1}^n(n-1)\bm b_L^{\star}(\mathcal O_i)\\
&=\bm\Psi_n^{\star}(\bm\beta_L^0)+\frac1n\sum_{i=1}^n\bm b_L^{\star}(\mathcal O_i).
\end{align*}
Lemma \ref{supp:lem:nuisance} therefore gives
\begin{equation}
\widehat{\bm\Psi}_n^{\star}(\bm\beta_L^0)
=\frac{2}{n(n-1)}\sum_{(i,j)\in\mathcal I_n}\bm\zeta_{ij}^{\star}+o_p(n^{-1/2}).
\label{supp:eq:adjusted-score-expansion}
\end{equation}

\begin{lemma}[Dependence among oracle and adjusted pair contributions]
\label{supp:lem:sparse-adjusted}
\leavevmode\newline
Under the assumptions of Lemma \ref{supp:lem:nuisance}, both the oracle and adjusted pair contributions are independent for pairs with disjoint subject sets. For each \((i,j)\in\mathcal I_n\), let \(\mathcal N_{ij,n}=\{(k,l)\in\mathcal I_n:\{i,j\}\cap\{k,l\}\ne\varnothing\}\). The largest neighborhood has size \(M_n=2n-3\), the largest collection of disjoint pairs has size \(m_n=\lfloor n/2\rfloor\), and \(M_nm_n=O(n^2)\).
\end{lemma}

\begin{proof}
With the true models, \(\bm U_{ij}^{\star}(\bm\beta_L^0)\) is measurable with respect to \(\sigma(\mathcal O_i,\mathcal O_j)\). Each of \(\bm b_L^{\star}(\mathcal O_i)\) and \(\bm b_L^{\star}(\mathcal O_j)\) is a function of the indicated record, so \(\bm\zeta_{ij}^{\star}\) is measurable with respect to the same sigma field. Independent sampling gives independence for disjoint subject sets.

The neighborhood of \((i,j)\) contains the pair itself, \(n-2\) other pairs containing \(i\), and \(n-2\) other pairs containing \(j\). These collections are disjoint, giving \(M_n=1+(n-2)+(n-2)=2n-3\). A collection of disjoint pairs has size at most \(\lfloor n/2\rfloor\), attained by \((1,2),(3,4),\ldots\). Consequently,
\begin{align*}
\frac{2M_n}{n(n-1)}&=\frac{2(2n-3)}{n(n-1)}\longrightarrow0,\\
\frac{2M_nm_n}{n(n-1)}
&=\frac{(2n-3)\lfloor n/2\rfloor}{n(n-1)/2}
\le\frac{2n-3}{n-1}\le2.
\end{align*}
These are the sparse correlation bounds of \citet{Thas2012}.
\end{proof}

Equation \eqref{supp:eq:common-population-score} and the mean zero model influences imply \(E\bm U_{12}^{\star}(\bm\beta_L^0)=E\bm\zeta_{12}^{\star}=\bm0\). Define \(\bm r_L^{\star}(o)=E\{\bm U_{12}^{\star}(\bm\beta_L^0)\mid\mathcal O_1=o\}\). By \eqref{supp:eq:adjusted-pair-score}, the first projection of the adjusted pair contribution is \(\bm\xi_L^{\star}(o)=E(\bm\zeta_{12}^{\star}\mid\mathcal O_1=o)=\bm r_L^{\star}(o)+\bm b_L^{\star}(o)/2\). The subject influence and its covariance are
\begin{align}
\bm\psi_L^{\star}(o)&=2\bm r_L^{\star}(o)+\bm b_L^{\star}(o)=2\bm\xi_L^{\star}(o),\nonumber\\
\bm\Sigma_{\psi,L}^{\star}&=E\{\bm\psi_L^{\star}(\mathcal O_1)\bm\psi_L^{\star}(\mathcal O_1)^\top\}.
\label{supp:eq:subject-influence}
\end{align}
Both projections and \(\bm\psi_L^{\star}\) have mean zero and finite second moments under the stated conditions.

The covariance \(\bm B_L^{\star}\) in \eqref{eq:B-sparse} can be calculated from these subject contributions. For two distinct overlapping pairs, conditional independence gives
\begin{align*}
E(\bm\zeta_{12}^{\star}\bm\zeta_{13}^{\star\top})
&=E\left\{E(\bm\zeta_{12}^{\star}\mid\mathcal O_1)
E(\bm\zeta_{13}^{\star}\mid\mathcal O_1)^\top\right\}\\
&=E\{\bm\xi_L^{\star}(\mathcal O_1)\bm\xi_L^{\star}(\mathcal O_1)^\top\}.
\end{align*}
There are \(n(n-1)/2\) identical pair terms and \(6\binom n3=n(n-1)(n-2)\) ordered distinct overlapping terms. Exchangeability and pair symmetry yield
\begin{align*}
&\frac{4m_n}{n^2(n-1)^2}\sum_{(i,j)\in\mathcal I_n}\sum_{(k,l)\in\mathcal I_n}\varphi_{ijkl}
E\{\bm\zeta_{ij}^{\star}\bm\zeta_{kl}^{\star\top}\}\\
&=\frac{2m_n}{n(n-1)}E(\bm\zeta_{12}^{\star}\bm\zeta_{12}^{\star\top})
+\frac{4m_n(n-2)}{n(n-1)}E\{\bm\xi_L^{\star}(\mathcal O_1)\bm\xi_L^{\star}(\mathcal O_1)^\top\}.
\end{align*}
The first coefficient tends to zero and the second tends to two, because \(m_n/n\to1/2\). Thus
\begin{equation}
\bm B_L^{\star}
=2E\{\bm\xi_L^{\star}(\mathcal O_1)\bm\xi_L^{\star}(\mathcal O_1)^\top\}
=\tfrac12\bm\Sigma_{\psi,L}^{\star}.
\label{supp:eq:B-first-projection}
\end{equation}
Condition C.6 requires this covariance to be positive definite for CHW and RCW. Conditions G.1--G.4 give a finite CTW covariance, which may be singular.

\begin{theorem}[Asymptotic normality of the three estimators]
\label{supp:thm:asymptotic-normality}
For \(\star\in\{\mathrm{CHW},\mathrm{RCW}\}\), assume Conditions C.1--C.7. For \(\star=\mathrm{CTW}\), assume Conditions G.1--G.4. The consistent local root in Proposition \ref{supp:lem:consistency} satisfies
\begin{equation}
\sqrt n(\widehat{\bm\beta}_L^{\star}-\bm\beta_L^0)
=\bm A_L^{-1}\frac1{\sqrt n}\sum_{i=1}^n\bm\psi_L^{\star}(\mathcal O_i)+o_p(1)
\xrightarrow{d}N\{\bm0,\bm A_L^{-1}\bm\Sigma_{\psi,L}^{\star}\bm A_L^{-\top}\}.
\label{supp:eq:common-normal}
\end{equation}
Equivalently, \(\sqrt{m_n}(\widehat{\bm\beta}_L^{\star}-\bm\beta_L^0)\xrightarrow{d}N\{\bm0,\bm A_L^{-1}\bm B_L^{\star}\bm A_L^{-\top}\}\), with \(\bm B_L^{\star}\) given by \eqref{supp:eq:B-first-projection}.
\end{theorem}

\begin{proof}
The canonical pair contribution in the Hoeffding decomposition of \eqref{supp:eq:adjusted-score-expansion} is
\begin{align*}
\bm d_{ij}^{\star}
&=\bm U_{ij}^{\star}(\bm\beta_L^0)-\bm r_L^{\star}(\mathcal O_i)-\bm r_L^{\star}(\mathcal O_j)\\
&=\bm\zeta_{ij}^{\star}-\bm\xi_L^{\star}(\mathcal O_i)-\bm\xi_L^{\star}(\mathcal O_j).
\end{align*}
Its conditional expectation is zero given either record. Counting the \(n-1\) occurrences of each subject gives the Hoeffding decomposition
\begin{align}
\bm\Psi_n^{\star}(\bm\beta_L^0)
&=\frac{2}{n(n-1)}\sum_{(i,j)\in\mathcal I_n}
\{\bm r_L^{\star}(\mathcal O_i)+\bm r_L^{\star}(\mathcal O_j)+\bm d_{ij}^{\star}\}\nonumber\\
&=\frac2n\sum_{i=1}^n\bm r_L^{\star}(\mathcal O_i)+\bm R_{n,L}^{\star},
\label{supp:eq:common-hoeffding}
\end{align}
where \(\bm R_{n,L}^{\star}=\frac{2}{n(n-1)}\sum_{(i,j)\in\mathcal I_n}\bm d_{ij}^{\star}\). Products of distinct canonical pair contributions have expectation zero, including pairs that share an observation. Consequently,
\begin{align*}
E\|\bm R_{n,L}^{\star}\|^2
=\frac{4}{n^2(n-1)^2}\sum_{(i,j)\in\mathcal I_n}E\|\bm d_{ij}^{\star}\|^2=\frac{2}{n(n-1)}E\|\bm d_{12}^{\star}\|^2
=O(n^{-2}).
\end{align*}
The finite second moments follow from C.3 and C.6 for CHW and RCW and from G.1--G.3 for CTW. Thus \(\sqrt n\bm R_{n,L}^{\star}=o_p(1)\). Combining \eqref{supp:eq:common-hoeffding} and \eqref{supp:eq:score-expansion} gives
\begin{align*}
\sqrt n\,\widehat{\bm\Psi}_n^{\star}(\bm\beta_L^0)
&=\frac1{\sqrt n}\sum_{i=1}^n
\{2\bm r_L^{\star}(\mathcal O_i)+\bm b_L^{\star}(\mathcal O_i)\}+o_p(1)\\
&=\frac1{\sqrt n}\sum_{i=1}^n\bm\psi_L^{\star}(\mathcal O_i)+o_p(1)
\xrightarrow{d}N(\bm0,\bm\Sigma_{\psi,L}^{\star}).
\end{align*}
The last step is the multivariate central limit theorem for independent subject contributions.

Set \(\bm A_{n,L}^{\star}=-\int_0^1\dot{\widehat{\bm\Psi}}_n^{\star}\{\bm\beta_L^0+t(\widehat{\bm\beta}_L^{\star}-\bm\beta_L^0)\}\,dt\). The integral form of Taylor's formula gives
\begin{align*}
\bm0=\widehat{\bm\Psi}_n^{\star}(\widehat{\bm\beta}_L^{\star})=\widehat{\bm\Psi}_n^{\star}(\bm\beta_L^0)-\bm A_{n,L}^{\star}(\widehat{\bm\beta}_L^{\star}-\bm\beta_L^0).
\end{align*}
Consistency of the root, uniform convergence of the fitted Jacobian, and continuity of the population derivative imply \(\bm A_{n,L}^{\star}\overset{p}{\longrightarrow}\bm A_L\). Nonsingularity makes \(\bm A_{n,L}^{\star}\) invertible with probability tending to one. Therefore
\begin{align*}
\sqrt n(\widehat{\bm\beta}_L^{\star}-\bm\beta_L^0)
=(\bm A_{n,L}^{\star})^{-1}\sqrt n\,\widehat{\bm\Psi}_n^{\star}(\bm\beta_L^0)=\bm A_L^{-1}\frac1{\sqrt n}\sum_{i=1}^n\bm\psi_L^{\star}(\mathcal O_i)+o_p(1).
\end{align*}
This proves \eqref{supp:eq:common-normal}. Multiplying by \(\sqrt{m_n/n}\), and using \(m_n/n\to1/2\) and \eqref{supp:eq:B-first-projection}, proves the result under \(\sqrt{m_n}\) scaling.
\end{proof}

The factor \(2\) in \(\bm\psi_L^{\star}\) is the coefficient of the first Hoeffding projection for a U-statistic of order two. Although there are of order \(n^2\) pairs, the leading term is an average of \(n\) independent subject contributions. This gives the \(\sqrt n\) rate. The factor one half in \(\bm\zeta_{ij}^{\star}\) follows from allocating the contribution of model estimation across pairs, and \(\bm B_L^{\star}=\bm\Sigma_{\psi,L}^{\star}/2\) follows from \(m_n/n\to1/2\). Both normalizations give the same first-order covariance of the estimator, \(n^{-1}\bm A_L^{-1}\bm\Sigma_{\psi,L}^{\star}\bm A_L^{-\top}\). The one-half response assigned to a \emph{genuine tie} in \eqref{eq:pim-response} defines the pairwise outcome.

For methods fitted to the same individuals, stack the three vectors \(\bm\psi_L^{\star}(\mathcal O_i)\) and apply the multivariate central limit theorem once. For \(\star,\star'\in\mathcal A\), the covariance block under \(\sqrt{m_n}\) scaling is \(\bm A_L^{-1}\bm B_L^{\star,\star'}\bm A_L^{-\top}\), where \(\bm B_L^{\star,\star'}=2E\{\bm\xi_L^{\star}(\mathcal O_1)\bm\xi_L^{\star'}(\mathcal O_1)^\top\}=\tfrac12E\{\bm\psi_L^{\star}(\mathcal O_1)\bm\psi_L^{\star'}(\mathcal O_1)^\top\}\). This allows dependence between methods and a singular stacked covariance, as occurs when estimators coincide. It also gives the covariance terms in \eqref{supp:eq:efficiency-projection}.

\subsection{Covariance estimation}
\label{supp:variance}

All fitted pair contributions and regression derivatives in this subsection are evaluated at \(\widehat{\bm\beta}_L^{\star}\). Write \(\widehat{\bm U}_{ij}^{\star}=\widehat{\bm U}_{ij}^{\star}(\widehat{\bm\beta}_L^{\star})\), and extend the pair contributions symmetrically to \(j<i\).

\subsubsection{Estimation of the model contributions}

For \(\star\in\{\mathrm{CHW},\mathrm{RCW}\}\), we can estimate the censoring contribution by
\begin{equation}
\widehat{\bm b}_{L,i}^{\star}
=\widehat{\bm b}_{C,L,i}^{\star}
=-\frac{2}{n(n-1)}\sum_{(j,k)\in\mathcal I_n}\widehat{\bm V}_{jk}^{\star}
\sum_{s\in\{j,k\}}
\widehat{\mathcal I}\{\mathcal O_i;t_{jk,s}^{\star}(L),\bm X_s^{C}\}.
\label{supp:eq:b-hat}
\end{equation}
The aggregation formulas \eqref{supp:eq:J-aggregation} and \eqref{supp:eq:Dgamma-aggregation} evaluate this expression. For CTW, the corresponding estimate is
\begin{align}
\widehat{\bm b}_{C,L,i}^{\mathrm{CTW}}
={}&-\frac{2}{n(n-1)}\sum_{(j,k)\in\mathcal I_n}\sum_{q=0}^{Q}\widehat{\bm U}_{jk}^{\mathrm{CTW},(q)}
\{\widehat{\mathcal I}(\mathcal O_i;t_{jk}^{(q)},\bm X_j^C)+\widehat{\mathcal I}(\mathcal O_i;t_{jk}^{(q)},\bm X_k^C)\}\nonumber\\
&-\frac{2}{n(n-1)}\sum_{(j,k)\in\mathcal I_n}\widehat{\bm U}_{jk}^{\mathrm{CTW},T}
\{\widehat{\mathcal I}(\mathcal O_i;L,\bm X_j^C)+\widehat{\mathcal I}(\mathcal O_i;L,\bm X_k^C)\}.
\label{supp:eq:ctw-censor-if-hat}
\end{align}
Every component includes its complete regression residual. The contributions from fitting the joint model are given in Sections \ref{supp:ctw-theory} and \ref{supp:cox-theory}. For each \(\star\in\mathcal A\), we set \(\widehat{\bm b}_{J,L,i}^{\mathrm{CHW}}=\widehat{\bm b}_{J,L,i}^{\mathrm{RCW}}=\bm0\), and use \(\widehat{\bm b}_{L,i}^{\star}=\widehat{\bm b}_{C,L,i}^{\star}+\widehat{\bm b}_{J,L,i}^{\star}\).

\begin{lemma}[Consistency of the fitted censoring contributions]
\label{supp:lem:empirical-nuisance}
Assume the conditions of Lemma \ref{supp:lem:nuisance} for the method under consideration and Condition C.8. Then the fitted pair contributions are consistent in empirical mean square, and \(n^{-1}\sum_{i=1}^n\|\widehat{\bm b}_{C,L,i}^{\star}-\bm b_{C,L}^{\star}(\mathcal O_i)\|^2=o_p(1)\). For CTW, the pair consistency also holds for each component contribution.
\end{lemma}

\begin{proof}
For CHW and RCW, positivity, bounded predictors, compactness, and smoothness give a constant \(C_V<\infty\) bounding the multiplier and its first regression derivative. Uniform weight convergence and parameter consistency imply
\begin{align*}
\Delta_{V,n}^{\star}
=\max_{(j,k)\in\mathcal I_n}\|\widehat{\bm V}_{jk}^{\star}-\bm V_{jk}^{\star}\|\le C\{\|\widehat{\bm\beta}_L^{\star}-\bm\beta_L^0\|+\Delta_{W,n}\}
=o_p(1).
\end{align*}
The same bound applies to the difference between fitted and true pair contributions. For CTW, the mean value inequality and positivity bound each component difference by a constant times the regression parameter error, \(\|\widehat\ell_C-\ell_C\|_\infty\), and, for a nonfatal component, \(|\widehat \omega_j^{(q)}(L)\widehat \omega_k^{(q)}(L)-\omega_j^{(q)}(L)\omega_k^{(q)}(L)|\). Consistency and G.2 therefore give
\begin{equation*}
\frac{2}{n(n-1)}\sum_{(j,k)\in\mathcal I_n}
\left\{\sum_{q=0}^{Q}\|\widehat{\bm U}_{jk}^{\mathrm{CTW},(q)}-\bm U_{jk}^{\mathrm{CTW},(q)}\|^2
+\|\widehat{\bm U}_{jk}^{\mathrm{CTW},T}-\bm U_{jk}^{\mathrm{CTW},T}\|^2\right\}=o_p(1).
\end{equation*}
Since \(Q\) is fixed, the total fitted CTW contribution has the same empirical mean square consistency. Each term is zero when its observation requirement is not satisfied.

Let \(\bm q_{jk,i}^{\star}=\bm q_L^{\star}(\mathcal O_j,\mathcal O_k,\mathcal O_i)\), and let \(\widehat{\bm q}_{jk,i}^{\star}\) be the corresponding summand in \eqref{supp:eq:b-hat} or \eqref{supp:eq:ctw-censor-if-hat}. Condition C.8 gives \(\max_i\sup_{t,\bm x}|\widehat{\mathcal I}(\mathcal O_i;t,\bm x)|\le C_{\mathcal I}+\Delta_{\mathcal I,n}\). For CHW and RCW, adding and subtracting the product of the true multiplier and fitted censoring influence yields
\begin{align*}
\widehat{\bm q}_{jk,i}^{\star}-\bm q_{jk,i}^{\star}
&=-(\widehat{\bm V}_{jk}^{\star}-\bm V_{jk}^{\star})
\sum_{s\in\{j,k\}}\widehat{\mathcal I}\{\mathcal O_i;t_{jk,s}^{\star}(L),\bm X_s^C\}\\
&\quad-\bm V_{jk}^{\star}\sum_{s\in\{j,k\}}
[\widehat{\mathcal I}\{\mathcal O_i;t_{jk,s}^{\star}(L),\bm X_s^C\}
-\mathcal I\{\mathcal O_i;t_{jk,s}^{\star}(L),\bm X_s^C\}],\\
\max_{1\le i\le n}\max_{(j,k)\in\mathcal I_n}
\|\widehat{\bm q}_{jk,i}^{\star}-\bm q_{jk,i}^{\star}\|
&\le2\Delta_{V,n}^{\star}(C_{\mathcal I}+\Delta_{\mathcal I,n})
+2C_V\Delta_{\mathcal I,n}
=o_p(1).
\end{align*}
For CTW, apply this product decomposition component by component to each term, and average over \((j,k)\in\mathcal I_n\). The same bound holds for the average norm of the kernel difference, with \(\Delta_{V,n}^{\star}\) replaced by the pair average of the sum of component errors. Cauchy--Schwarz and the preceding empirical mean square bound make that average \(o_p(1)\). Positivity and bounded PIM functions bound the sum of the true component norms by a fixed constant, so the bound is uniform in \(i\), including pairs containing \(i\).

For \(\overline{\bm b}_{C,n,i}^{\star}=\frac{2}{n(n-1)}\sum_{(j,k)\in\mathcal I_n}\bm q_L^{\star}(\mathcal O_j,\mathcal O_k,\mathcal O_i)\), it follows for all three methods that \(\max_i\|\widehat{\bm b}_{C,L,i}^{\star}-\overline{\bm b}_{C,n,i}^{\star}\|=o_p(1)\). The true kernel is bounded by a constant \(C_q\), also at coincident arguments. For \(n\ge3\), define
\begin{equation*}
\bm B_{C,n,-i}^{\star}=\binom{n-1}{2}^{-1}
\sum_{\substack{(j,k)\in\mathcal I_n\\i\notin\{j,k\}}}
\bm q_L^{\star}(\mathcal O_j,\mathcal O_k,\mathcal O_i).
\end{equation*}
Conditional on \(\mathcal O_i\), this is a U-statistic of order two based on the remaining \(n-1\) independent observations, with mean \(\bm b_{C,L}^{\star}(\mathcal O_i)\). Only overlapping pairs have nonzero conditional covariance. Counting those pairs gives
\begin{align*}
E\left[\left.\|\bm B_{C,n,-i}^{\star}-\bm b_{C,L}^{\star}(\mathcal O_i)\|^2\right|\mathcal O_i\right]
&\le\frac{4C_q^2(2n-5)}{\binom{n-1}{2}}\le\frac Cn.
\end{align*}
The pair average decomposes exactly as
\begin{align*}
\overline{\bm b}_{C,n,i}^{\star}
&=\frac{n-2}{n}\bm B_{C,n,-i}^{\star}
+\frac{2}{n(n-1)}\sum_{\substack{j=1\\j\ne i}}^n\bm q_L^{\star}(\mathcal O_i,\mathcal O_j,\mathcal O_i),\\
\|\overline{\bm b}_{C,n,i}^{\star}-\bm b_{C,L}^{\star}(\mathcal O_i)\|
&\le\|\bm B_{C,n,-i}^{\star}-\bm b_{C,L}^{\star}(\mathcal O_i)\|
+\frac{2C_q}{n}+\frac{2C_q}{n}.
\end{align*}
The last two terms bound the normalization difference and the repeated-record terms. Squaring and averaging gives
\begin{align*}
E\left\{\frac1n\sum_{i=1}^n
\|\overline{\bm b}_{C,n,i}^{\star}-\bm b_{C,L}^{\star}(\mathcal O_i)\|^2\right\}
&\le\frac2n\sum_{i=1}^nE\|\bm B_{C,n,-i}^{\star}-\bm b_{C,L}^{\star}(\mathcal O_i)\|^2
+\frac{32C_q^2}{n^2}\\
&=O(n^{-1}).
\end{align*}
Markov's inequality and the preceding substitution bound now yield
\begin{align*}
\frac1n\sum_{i=1}^n\|\widehat{\bm b}_{C,L,i}^{\star}-\bm b_{C,L}^{\star}(\mathcal O_i)\|^2
\le2\max_i\|\widehat{\bm b}_{C,L,i}^{\star}-\overline{\bm b}_{C,n,i}^{\star}\|^2+\frac2n\sum_{i=1}^n\|\overline{\bm b}_{C,n,i}^{\star}-\bm b_{C,L}^{\star}(\mathcal O_i)\|^2
=o_p(1).
\end{align*}
This argument allows dependence among the fitted subject contributions.
\end{proof}

When a full joint model influence function is available, its contribution can be estimated by \(\widehat{\bm b}_{J,L,i}^{\mathrm{CTW}}=\frac{2}{n(n-1)}\sum_{(j,k)\in\mathcal I_n}\dot{\widehat{\bm U}}_{J,jk}[\widehat\phi_{J,i}]\). A sufficient condition for empirical mean square consistency is that the fitted influences converge in empirical mean square, the pair average of the squared operator norm of the derivative estimation error tends to zero, and the derivative kernels have square-integrable envelopes for distinct and repeated records. To verify this, replace the fitted influences and derivatives by their true counterparts using Cauchy--Schwarz. Conditional on \(\mathcal O_i\), the remaining average over pairs excluding \(i\) has variance \(O(n^{-1})\). The normalization difference and the pairs containing \(i\) have \(L^2\) norm \(O(n^{-1})\). These bounds establish the required consistency. Sections \ref{supp:ctw-theory} and \ref{supp:cox-theory} give the corresponding verification for parametric and Cox models.

\subsubsection{Sandwich estimators and their consistency}

For each method, calculate \(\widehat{\bm A}_L^{\star}=-\dot{\widehat{\bm\Psi}}_n^{\star}(\widehat{\bm\beta}_L^{\star})\), holding the fitted models fixed. To make the regression derivative explicit, write \(\bm U_{ij}^{\star}=\bm K_{ij}s_{ij}^{\star}\). For CHW and RCW, \(s_{ij}^{\star}=W_{ij}^{C,\star}(L;S_C)(\widetilde H_{ij}(L)-\mu_{ij})\) and \(v_{ij}^{\star}=W_{ij}^{C,\star}(L;S_C)\). For CTW, \(s_{ij}^{\mathrm{CTW}}\) is the expression in brackets in \eqref{eq:ctw-score}, and
\begin{equation*}
v_{ij}^{\mathrm{CTW}}
=\frac{R_{ij}^{\mathrm{CTW},(0)}(L)+R_{ij}^{\mathrm{CTW},T}(L)}{\pi_{ij}^{\mathrm{RCW}}(L)}
+\sum_{q=1}^{Q}\frac{R_{ij}^{\mathrm{CTW},(q)}(L)\omega_i^{(q)}(L)\omega_j^{(q)}(L)}{\pi_{ij}^{\mathrm{CTW},(q)}(L)}.
\end{equation*}
With \(\dot{\bm K}_{ij}=\partial\bm K_{ij}/\partial\bm\beta_L^\top\) and \(\dot{\bm\mu}_{ij}=\partial\mu_{ij}/\partial\bm\beta_L\), all three derivatives have the form \(\partial\bm U_{ij}^{\star}/\partial\bm\beta_L^\top=s_{ij}^{\star}\dot{\bm K}_{ij}-v_{ij}^{\star}\bm K_{ij}\dot{\bm\mu}_{ij}^\top\). In particular, each CTW weight multiplies its full regression residual.

Set \(\widehat{\bm\zeta}_{ij}^{\star}=\widehat{\bm U}_{ij}^{\star}+(\widehat{\bm b}_{L,i}^{\star}+\widehat{\bm b}_{L,j}^{\star})/2\). The covariance based on overlapping pairs is
\begin{align}
\widehat{\bm B}_L^{\star}
&=\frac{4m_n}{n^2(n-1)^2}\sum_{(i,j)\in\mathcal I_n}\sum_{(k,l)\in\mathcal I_n}
\varphi_{ijkl}\widehat{\bm\zeta}_{ij}^{\star}\widehat{\bm\zeta}_{kl}^{\star\top},
\label{supp:eq:B-hat}\\
\widehat{\operatorname{Var}}(\widehat{\bm\beta}_L^{\star})
&=\frac1{m_n}(\widehat{\bm A}_L^{\star})^{-1}\widehat{\bm B}_L^{\star}(\widehat{\bm A}_L^{\star})^{-\top}.
\label{supp:eq:variance-hat}
\end{align}
An equivalent estimate to first order uses the empirical subject influences. Define \(\widehat{\bm\Psi}_n^{\star}=\frac{2}{n(n-1)}\sum_{(i,j)\in\mathcal I_n}\widehat{\bm U}_{ij}^{\star}\) and
\begin{equation*}
\widehat{\bm\psi}_{L,i}^{\star}
=2\left\{\frac1{n-1}\sum_{\substack{j=1\\j\ne i}}^n\widehat{\bm U}_{ij}^{\star}-\widehat{\bm\Psi}_n^{\star}\right\}
+\widehat{\bm b}_{C,L,i}^{\star}+\widehat{\bm b}_{J,L,i}^{\star}.
\end{equation*}
For \(\widehat{\bm v}_{L,i}^{\star}=(\widehat{\bm A}_L^{\star})^{-1}\widehat{\bm\psi}_{L,i}^{\star}\) and \(\overline{\bm v}_L^{\star}=n^{-1}\sum_{i=1}^n\widehat{\bm v}_{L,i}^{\star}\), the projection covariance estimate is
\begin{equation}
\widehat{\operatorname{Var}}_{\mathrm{proj}}(\widehat{\bm\beta}_L^{\star})
=\frac1{n^2}\sum_{i=1}^n(\widehat{\bm v}_{L,i}^{\star}-\overline{\bm v}_L^{\star})(\widehat{\bm v}_{L,i}^{\star}-\overline{\bm v}_L^{\star})^\top.
\label{supp:eq:projection-variance}
\end{equation}
Adding all contributions before taking their outer products includes the covariances between the pair comparisons and the fitted models, including models fitted to the same individuals.

\begin{theorem}[Consistency of the covariance estimators]
\label{supp:thm:covariance}
Assume Theorem \ref{supp:thm:asymptotic-normality} holds for the method under consideration. Suppose
\begin{align}
\frac{2}{n(n-1)}\sum_{(i,j)\in\mathcal I_n}
\|\widehat{\bm U}_{ij}^{\star}-\bm U_{ij}^{\star}(\bm\beta_L^0)\|^2&=o_p(1),\nonumber\\
\frac1n\sum_{i=1}^n\left\{
\|\widehat{\bm b}_{C,L,i}^{\star}-\bm b_{C,L}^{\star}(\mathcal O_i)\|^2
+\|\widehat{\bm b}_{J,L,i}^{\star}-\bm b_{J,L}^{\star}(\mathcal O_i)\|^2\right\}&=o_p(1).
\label{supp:eq:covariance-conditions}
\end{align}
Then \(\widehat{\bm A}_L^{\star}\overset{p}{\longrightarrow}\bm A_L\), \(\widehat{\bm B}_L^{\star}\overset{p}{\longrightarrow}\bm B_L^{\star}\), and \(m_n\widehat{\operatorname{Var}}(\widehat{\bm\beta}_L^{\star})\overset{p}{\longrightarrow}\bm A_L^{-1}\bm B_L^{\star}\bm A_L^{-\top}\). Also,
\begin{equation}
n\widehat{\operatorname{Var}}_{\mathrm{proj}}(\widehat{\bm\beta}_L^{\star})
\overset{p}{\longrightarrow}\bm A_L^{-1}\bm\Sigma_{\psi,L}^{\star}\bm A_L^{-\top},
\label{supp:eq:projection-covariance-limit}
\end{equation}
and the two estimators are equivalent to first order,
\begin{equation}
n\{\widehat{\operatorname{Var}}_{\mathrm{proj}}(\widehat{\bm\beta}_L^{\star})-\widehat{\operatorname{Var}}(\widehat{\bm\beta}_L^{\star})\}\overset{p}{\longrightarrow}\bm0.
\label{supp:eq:covariance-equivalence}
\end{equation}
For CHW and RCW, C.8 verifies \eqref{supp:eq:covariance-conditions}. For CTW, Lemma \ref{supp:lem:empirical-nuisance} verifies the pair and censoring parts under C.8, and the joint model part is verified under the model conditions in Sections \ref{supp:ctw-theory} and \ref{supp:cox-theory}.
\end{theorem}

\begin{proof}
Consistency of the root and uniform derivative convergence give
\begin{align*}
\|\widehat{\bm A}_L^{\star}-\bm A_L\|
\le\sup_{\bm\beta_L\in\mathcal C_r}
\|\dot{\widehat{\bm\Psi}}_n^{\star}(\bm\beta_L)-\dot{\bm\Psi}(\bm\beta_L)\|+\|\dot{\bm\Psi}(\widehat{\bm\beta}_L^{\star})-\dot{\bm\Psi}(\bm\beta_L^0)\|
=o_p(1),
\end{align*}
on the event that the fitted root belongs to \(\mathcal C_r\), whose probability tends to one.

Let \(\bm B_{n,L}^{\star}\) be \eqref{supp:eq:B-hat} with \(\widehat{\bm\zeta}_{ij}^{\star}\) replaced by \(\bm\zeta_{ij}^{\star}\). For an unordered triple, define
\begin{align*}
\overline{\bm G}_{ijk}^{\star}=\tfrac16\bigl\{&\bm\zeta_{ij}^{\star}\bm\zeta_{ik}^{\star\top}
+\bm\zeta_{ik}^{\star}\bm\zeta_{ij}^{\star\top}+\bm\zeta_{ij}^{\star}\bm\zeta_{jk}^{\star\top}
+\bm\zeta_{jk}^{\star}\bm\zeta_{ij}^{\star\top}+\bm\zeta_{ik}^{\star}\bm\zeta_{jk}^{\star\top}
+\bm\zeta_{jk}^{\star}\bm\zeta_{ik}^{\star\top}\bigr\}.
\end{align*}
Separating identical pairs from distinct overlapping pairs gives
\begin{align*}
\bm B_{n,L}^{\star}
&=\frac{2m_n}{n(n-1)}
\left\{\frac{2}{n(n-1)}\sum_{(i,j)\in\mathcal I_n}\bm\zeta_{ij}^{\star}\bm\zeta_{ij}^{\star\top}\right\}+\frac{4m_n(n-2)}{n(n-1)}
\left\{\binom n3^{-1}\sum_{(i,j,k)\in\mathcal I_{n,3}}\overline{\bm G}_{ijk}^{\star}\right\}.
\end{align*}
The first term tends to zero because its coefficient tends to zero and the U-statistic in braces converges to a finite matrix. Cauchy--Schwarz gives integrability of the kernel of order three. Conditioning each of its six products on the shared record gives \(E\overline{\bm G}_{123}^{\star}=E\{\bm\xi_L^{\star}(\mathcal O_1)\bm\xi_L^{\star}(\mathcal O_1)^\top\}\). The law of large numbers for U-statistics and \(4m_n(n-2)/\{n(n-1)\}\to2\) therefore give \(\bm B_{n,L}^{\star}\overset{p}{\longrightarrow}2E\{\bm\xi_L^{\star}(\mathcal O_1)\bm\xi_L^{\star}(\mathcal O_1)^\top\}=\bm B_L^{\star}\).

To bound the effect of estimation, let \(\bm e_{n,i}^{\star}=\widehat{\bm b}_{L,i}^{\star}-\bm b_L^{\star}(\mathcal O_i)\) and \(\bm D_{ij}^{\star}=\widehat{\bm\zeta}_{ij}^{\star}-\bm\zeta_{ij}^{\star}\). Then \(\bm D_{ij}^{\star}=\widehat{\bm U}_{ij}^{\star}-\bm U_{ij}^{\star}(\bm\beta_L^0)+(\bm e_{n,i}^{\star}+\bm e_{n,j}^{\star})/2\). Write \(d_{n,\star}\) and \(z_{n,\star}\) for the square roots of the pair averages of \(\|\bm D_{ij}^{\star}\|^2\) and \(\|\bm\zeta_{ij}^{\star}\|^2\), respectively. Counting the \(n-1\) pairs containing each subject gives
\begin{align*}
d_{n,\star}^2
&=\frac{2}{n(n-1)}\sum_{(i,j)\in\mathcal I_n}\|\bm D_{ij}^{\star}\|^2\\
&\le\frac{4}{n(n-1)}\sum_{(i,j)\in\mathcal I_n}
\|\widehat{\bm U}_{ij}^{\star}-\bm U_{ij}^{\star}(\bm\beta_L^0)\|^2
+\frac{2}{n(n-1)}\sum_{(i,j)\in\mathcal I_n}
\{\|\bm e_{n,i}^{\star}\|^2+\|\bm e_{n,j}^{\star}\|^2\}\\
&=\frac{4}{n(n-1)}\sum_{(i,j)\in\mathcal I_n}
\|\widehat{\bm U}_{ij}^{\star}-\bm U_{ij}^{\star}(\bm\beta_L^0)\|^2
+\frac2n\sum_{i=1}^n\|\bm e_{n,i}^{\star}\|^2
=o_p(1).
\end{align*}
The second-moment conditions and the U-statistic law of large numbers give \(z_{n,\star}^2=O_p(1)\).

For each ordered pair of pairs,
\begin{align*}
\widehat{\bm\zeta}_{ij}^{\star}\widehat{\bm\zeta}_{kl}^{\star\top}-\bm\zeta_{ij}^{\star}\bm\zeta_{kl}^{\star\top}
&=\bm D_{ij}^{\star}\bm\zeta_{kl}^{\star\top}+\bm\zeta_{ij}^{\star}\bm D_{kl}^{\star\top}+\bm D_{ij}^{\star}\bm D_{kl}^{\star\top}.
\end{align*}
Each pair has \(M_n\) neighbors. Cauchy--Schwarz over the set of ordered overlapping pairs yields
\begin{align*}
\sum_{(i,j)\in\mathcal I_n}\sum_{(k,l)\in\mathcal N_{ij,n}}
\|\bm D_{ij}^{\star}\|\,\|\bm\zeta_{kl}^{\star}\|
&\le\left\{M_n\sum_{(i,j)\in\mathcal I_n}\|\bm D_{ij}^{\star}\|^2\right\}^{1/2}
\left\{M_n\sum_{(k,l)\in\mathcal I_n}\|\bm\zeta_{kl}^{\star}\|^2\right\}^{1/2}\\
&=\frac{M_n n(n-1)}{2}d_{n,\star}z_{n,\star}.
\end{align*}
The same inequality for the other two products gives
\begin{align*}
\|\widehat{\bm B}_L^{\star}-\bm B_{n,L}^{\star}\|
&\le\frac{2m_nM_n}{n(n-1)}\{2d_{n,\star}z_{n,\star}+d_{n,\star}^2\}
=o_p(1),
\end{align*}
because \(2m_nM_n/\{n(n-1)\}\) is bounded. Hence \(\widehat{\bm B}_L^{\star}\overset{p}{\longrightarrow}\bm B_L^{\star}\). Consistency of \(\widehat{\bm A}_L^{\star}\) and continuity of matrix inversion prove the limit for \eqref{supp:eq:variance-hat}.

For the projection estimate, condition on \(\mathcal O_i\). The oracle pair contributions involving this individual are independent with common conditional mean \(\bm r_L^{\star}(\mathcal O_i)\). Therefore
\begin{equation*}
E\left[\frac1n\sum_{i=1}^n\left\|\frac1{n-1}\sum_{\substack{j=1\\j\ne i}}^n\bm U_{ij}^{\star}(\bm\beta_L^0)-\bm r_L^{\star}(\mathcal O_i)\right\|^2\right]
\le\frac{E\|\bm U_{12}^{\star}(\bm\beta_L^0)\|^2}{n-1}=O(n^{-1}).
\end{equation*}
Jensen's inequality bounds the average squared error from substituting fitted pair contributions in these conditional averages by the first quantity in \eqref{supp:eq:covariance-conditions}. Also \(\widehat{\bm\Psi}_n^{\star}=o_p(1)\), and it is zero at an exact root. Together with the second condition in \eqref{supp:eq:covariance-conditions}, this yields \(n^{-1}\sum_{i=1}^n\|\widehat{\bm\psi}_{L,i}^{\star}-\bm\psi_L^{\star}(\mathcal O_i)\|^2=o_p(1)\). Consistency of \(\widehat{\bm A}_L^{\star}\) gives the same conclusion for the coefficient influence functions. Cauchy--Schwarz then makes the difference between fitted and true empirical second moments \(o_p(1)\). Their empirical means converge to zero. Centering and the law of large numbers prove \eqref{supp:eq:projection-covariance-limit}. Finally, \(\bm B_L^{\star}=\bm\Sigma_{\psi,L}^{\star}/2\) and \(n/m_n\to2\) give the same \(n\)-scaled limit for both covariance estimators, proving \eqref{supp:eq:covariance-equivalence}.
\end{proof}

The simulation and HF-ACTION analyses use \eqref{supp:eq:projection-variance}, with both model contributions included for CTW.

\subsubsection{Computation of the covariance sum}
\label{supp:pair-covariance-computation}

The neighborhood sum can be evaluated with order \(n^2\) operations. Define \(\widehat{\bm S}_i^{\star}=\sum_{j=1}^{i-1}\widehat{\bm\zeta}_{ji}^{\star}+\sum_{j=i+1}^n\widehat{\bm\zeta}_{ij}^{\star}\). Then
\begin{equation}
\sum_{(i,j)\in\mathcal I_n}\sum_{(k,l)\in\mathcal I_n}\varphi_{ijkl}
\widehat{\bm\zeta}_{ij}^{\star}\widehat{\bm\zeta}_{kl}^{\star\top}
=\sum_{i=1}^n\widehat{\bm S}_i^{\star}\widehat{\bm S}_i^{\star\top}
-\sum_{(i,j)\in\mathcal I_n}\widehat{\bm\zeta}_{ij}^{\star}\widehat{\bm\zeta}_{ij}^{\star\top}.
\label{supp:eq:subject-sum-identity}
\end{equation}
Expanding the subject sums gives
\begin{align*}
\sum_{r=1}^n\widehat{\bm S}_r^{\star}\widehat{\bm S}_r^{\star\top}
&=\sum_{r=1}^n
\left(\sum_{\substack{(i,j)\in\mathcal I_n\\r\in\{i,j\}}}\widehat{\bm\zeta}_{ij}^{\star}\right)
\left(\sum_{\substack{(k,l)\in\mathcal I_n\\r\in\{k,l\}}}\widehat{\bm\zeta}_{kl}^{\star}\right)^\top\\
&=\sum_{(i,j)\in\mathcal I_n}\sum_{(k,l)\in\mathcal I_n}
|\{i,j\}\cap\{k,l\}|
\widehat{\bm\zeta}_{ij}^{\star}\widehat{\bm\zeta}_{kl}^{\star\top}\\
&=\sum_{(i,j)\in\mathcal I_n}\sum_{(k,l)\in\mathcal I_n}\varphi_{ijkl}
\widehat{\bm\zeta}_{ij}^{\star}\widehat{\bm\zeta}_{kl}^{\star\top}
+\sum_{(i,j)\in\mathcal I_n}\widehat{\bm\zeta}_{ij}^{\star}\widehat{\bm\zeta}_{ij}^{\star\top}.
\end{align*}
Two distinct overlapping pairs share exactly one subject and contribute once. An identical pair shares two subjects and contributes twice, whereas \(\varphi_{ijij}=1\). Disjoint pairs contribute zero. Subtracting the final sum proves \eqref{supp:eq:subject-sum-identity}.

\subsection{Truncation}
\label{supp:truncation}

The preceding results concern untruncated weights under positivity. A fixed truncation threshold gives the same asymptotic distribution when it is inactive with probability tending to one.

\begin{remark}[Fixed truncation of estimated pair observation probabilities]
\label{supp:rem:truncation}
\leavevmode\newline
For \(\star\in\{\mathrm{CHW},\mathrm{RCW}\}\), let \(\pi_{ij}^{\star}\) and \(\widehat\pi_{ij}^{\star}\) be the true and fitted pair observation probabilities. A fixed threshold \(c_{\mathrm{tr}}\in(0,1)\) replaces a smaller fitted probability by \(c_{\mathrm{tr}}\), so the resulting weight is at most \(1/c_{\mathrm{tr}}\). With \(x\vee y=\max(x,y)\), define \(w_{ij}^{\star,c_{\mathrm{tr}}}=R_{ij}^{\star}/(\pi_{ij}^{\star}\vee c_{\mathrm{tr}})\) and \(\widehat w_{ij}^{\star,c_{\mathrm{tr}}}=R_{ij}^{\star}/(\widehat\pi_{ij}^{\star}\vee c_{\mathrm{tr}})\). Both weights are zero when \(R_{ij}^{\star}=0\), without evaluating a probability denominator. The full-data probability is used to calculate expectations.

Define \(\bm U_{ij}^{\star,c_{\mathrm{tr}}}(\bm\beta_L)\) by replacing \(W_{ij}^{C,\star}\) with \(w_{ij}^{\star,c_{\mathrm{tr}}}\) in \eqref{supp:eq:compact-oracle-score}, and define its fitted counterpart using \(\widehat w_{ij}^{\star,c_{\mathrm{tr}}}\). Both pair contributions use the observed response. Their weighted residuals are zero when the pair is not observed.

If \(c_{\mathrm{tr}}<\epsilon^2\), positivity gives \(\pi_{ij}^{\star}\ge \epsilon^2\). Uniform consistency of the fitted probabilities then implies
\begin{align*}
P\left\{\min_{\star\in\{\mathrm{CHW},\mathrm{RCW}\}}\min_{\substack{(i,j)\in\mathcal I_n\\R_{ij}^{\star}=1}}
\widehat\pi_{ij}^{\star}>c_{\mathrm{tr}}\right\}
&\ge P\left\{\max_{\star\in\{\mathrm{CHW},\mathrm{RCW}\}}
\max_{\substack{(i,j)\in\mathcal I_n\\R_{ij}^{\star}=1}}
|\widehat\pi_{ij}^{\star}-\pi_{ij}^{\star}|
<\epsilon^2-c_{\mathrm{tr}}\right\}\\
&\longrightarrow1.
\end{align*}
For CTW, every censoring product in the contributions for fatal, nonfatal, and \emph{genuine tie} is also at least \(\epsilon^2\). Uniform log-survival consistency gives uniform convergence of these products at their observed evaluation times. Applying the same bound to each of the fixed \(Q+2\) components makes truncation inactive simultaneously with probability tending to one. Thus the untruncated target, influence function, and covariance estimator apply to all three methods whenever \(c_{\mathrm{tr}}<\epsilon^2\).

For CHW and RCW, if truncation is active with positive probability, put \(a_{ij}^{\star}=\pi_{ij}^{\star}/(\pi_{ij}^{\star}\vee c_{\mathrm{tr}})\). The observation identities in \ref{supp:unbiased} yield
\begin{align*}
E\{w_{ij}^{\star,c_{\mathrm{tr}}}\mid\mathcal F_{ij}^{F}\}
&=\frac{E(R_{ij}^{\star}\mid\mathcal F_{ij}^{F})}
{\pi_{ij}^{\star}\vee c_{\mathrm{tr}}}
=a_{ij}^{\star},\\
E\{w_{ij}^{\star,c_{\mathrm{tr}}}\widetilde H_{ij}(L)
\mid\mathcal F_{ij}^{F}\}
&=\frac{H_{ij}(L)E(R_{ij}^{\star}\mid\mathcal F_{ij}^{F})}
{\pi_{ij}^{\star}\vee c_{\mathrm{tr}}}
=a_{ij}^{\star} H_{ij}(L).
\end{align*}
Taking expectations of the observed pair contributions therefore gives, for \(\star\in\{\mathrm{CHW},\mathrm{RCW}\}\),
\begin{equation}
\bm\Psi^{\star,c_{\mathrm{tr}}}(\bm\beta_L)
=E\left[\bm K_{12}(\bm\beta_L)a_{12}^{\star}
\{H_{12}(L)-\mu_{12}(\bm\beta_L)\}\right].
\label{supp:eq:truncated-population-score}
\end{equation}
The full-data response appears in this population expression, while the sample contribution uses \(\widetilde H_{ij}(L)\). The factor \(a_{ij}^{\star}\) equals one above the threshold and \(\pi_{ij}^{\star}/c_{\mathrm{tr}}\) below it. If \eqref{supp:eq:truncated-population-score} has a unique interior zero \(\bm\beta_L^{\star,c_{\mathrm{tr}}}\), that zero can differ from \(\bm\beta_L^0\). The CHW and RCW truncated targets can also differ because their observation probabilities differ.

To expand the truncated weights, write \(c=c_{\mathrm{tr}}\). On an observed pair, set \(d_{ij,n}^{\star}=\log\widehat\pi_{ij}^{\star}-\log\pi_{ij}^{\star}\) and \(\delta_{\pi,ij,n}^{\star}=\widehat\pi_{ij}^{\star}-\pi_{ij}^{\star}\). Set both errors to zero when \(R_{ij}^{\star}=0\). The evaluation times are needed only for observed pairs. Assume \(P(\pi_{12}^{\star}=c)=0\). A sufficient stronger condition is \(P\{|\pi_{12}^{\star}-c|\le\varepsilon\}\le K\varepsilon\) for sufficiently small \(\varepsilon\).

When both probabilities exceed \(c\), the weight ratio has the same exponential form as in \ref{supp:two-weight-expansions}, so its linear term is \(-w_{ij}^{\star,c}d_{ij,n}^{\star}\). When both probabilities are below \(c\), the weight difference is zero. If a probability crosses the threshold, then \(|\pi_{ij}^{\star}-c|\le|\delta_{\pi,ij,n}^{\star}|\). Moreover, the truncated inverse function satisfies \(|(x\vee c)^{-1}-(y\vee c)^{-1}|\le c^{-2}|x-y|\) for positive \(x,y\). Positivity and uniform consistency make every fitted probability on an observed pair at least \(\epsilon^2/2\) with probability tending to one. The mean value theorem then gives \(|d_{ij,n}^{\star}|\le2\epsilon^{-2}|\delta_{\pi,ij,n}^{\star}|\) uniformly over pairs, so the weight difference and its linear term are both bounded by a constant times the probability error.

Let \(\varepsilon_n^{\star}=\max_{(i,j)\in\mathcal I_n}|\delta_{\pi,ij,n}^{\star}|=O_p(n^{-1/2})\). For fixed \(K_0<\infty\), on \(\{\varepsilon_n^{\star}\le K_0n^{-1/2}\}\), the pair average of the estimating function remainder from threshold crossings is bounded by 
\begin{equation*}
C\varepsilon_n^{\star}\frac{2}{n(n-1)}\sum_{(i,j)\in\mathcal I_n} R_{ij}^{\star} M(\mathcal O_i,\mathcal O_j) I\{|\pi_{ij}^{\star}-c|\le K_0n^{-1/2}\}.
\end{equation*}

The summand is zero on an unobserved pair. Its expectation is bounded by the corresponding expectation with the full-data probability and without \(R_{ij}^{\star}\). Integrability of \(M\) and the absence of an atom at the threshold imply
\begin{align*}
&E\left\{\frac{2}{n(n-1)}\sum_{(i,j)\in\mathcal I_n}
R_{ij}^{\star}M(\mathcal O_i,\mathcal O_j)I\{|\pi_{ij}^{\star}-c|\le K_0n^{-1/2}\}\right\}\\
&=E\left[R_{12}^{\star}M(\mathcal O_1,\mathcal O_2)I\{|\pi_{12}^{\star}-c|\le K_0n^{-1/2}\}\right]\\
&\le E\left[M(\mathcal O_1,\mathcal O_2)I\{|\pi_{12}^{\star}-c|\le K_0n^{-1/2}\}\right]\overset{p}{\longrightarrow}0.
\end{align*}
The last limit follows by dominated convergence, and Markov's inequality implies that the pair average is \(o_p(1)\). The crossing contribution is consequently \(o_p(n^{-1/2})\) on the stated event. Since \(\varepsilon_n^{\star}=O_p(n^{-1/2})\), letting \(K_0\) increase proves the unconditional bound. The smooth remainder away from crossings is \(O_p(n^{-1})\) after averaging against the envelope, as in the proof of Lemma \ref{supp:lem:nuisance}. Thus
\begin{equation}
\widehat w_{ij}^{\star,c}-w_{ij}^{\star,c}
=-I(\pi_{ij}^{\star}>c)w_{ij}^{\star,c}d_{ij,n}^{\star}
+\mathcal R_{ij,n}^{\star,c},
\label{supp:eq:truncated-weight-expansion}
\end{equation}
where the pair average of the remainder multiplied by the regression factor is \(o_p(n^{-1/2})\). Thus the averaged expansion is valid despite nondifferentiability at the threshold.

At a regular interior truncated target, define \(\bm V_{ij}^{\star,c,\mathrm{or}}=\bm U_{ij}^{\star,c}(\bm\beta_L^{\star,c})\) for CHW and RCW. Substitution of Condition C.7 into \eqref{supp:eq:truncated-weight-expansion} gives
\begin{equation*}
 \bm q_L^{\star,c}(o_1,o_2,o_3)=-I(\pi_{12}^{\star}>c)\bm V_{12}^{\star,c,\mathrm{or}}\sum_{s=1}^2\mathcal I\{o_3;t_{12,s}^{\star}(L),\bm X_s^{C}\},   
\end{equation*}
with a zero contribution on an unobserved pair. Define \(\bm b_L^{\star,c}(o)=E\{\bm q_L^{\star,c}(\mathcal O_1,\mathcal O_2,o)\}\). The censoring correction for an actively truncated estimator therefore includes the derivative indicator \(I(\pi_{12}^{\star}>c)\) together with the observed multiplier. Omitting this indicator can change the first order expansion. The asymptotic analysis also requires identification, an interior root, and a nonsingular derivative at the truncated target.

For active CTW truncation, \eqref{supp:eq:truncated-weight-expansion} applies separately to each inverse censoring product. The censoring derivative in \eqref{supp:eq:ctw-censoring-direction} then uses the truncated component contribution and the corresponding indicator that its censoring product exceeds \(c\). The joint model derivative in \eqref{supp:eq:ctw-general-event-derivative} uses the same truncated denominators. The resulting population equation and model contributions can differ from their untruncated counterparts. Applying the common asymptotic argument requires identification and a nonsingular regression derivative at the truncated target, together with the directional expansion for this modified population function.
\end{remark}

\section{Additional simulation results}
\label{supp:simulations}

\subsection{Results by sample size and link}
\label{supp:simulation-results}

The resulsts below report the additional simulation results for the design in Section \ref{sec:simulation}, using the performance measures and IAT benchmark defined in the main paper. On the CHW and RCW rows, VR (\%) gives the percentage reduction in empirical variance with CTW relative to that row's method. The simulations used restriction times \(2.0,2.8,4.5\), respectively.

\subsubsection{Logit link}

Tables \ref{tab:sim-logit-n200}--\ref{tab:sim-logit-n1000-binary} give the logit results for \(n=200\) and \(n=1000\), in addition to the results in Tables \ref{tab:sim-logit-n500} and \ref{tab:sim-logit-n500-binary}. The maximum absolute bias of CTW decreased from 0.093 to 0.025 as the sample size increased. The positive VR entries show that CTW had lower empirical variance than CHW and RCW at both sample sizes.

\begin{table}[!htbp]
\caption{Finite-sample performance of complete history weighting (CHW), resolved comparison weighting (RCW), conditional tie weighting (CTW), and indeterminate as tie (IAT) for the continuous covariate coefficient \(\beta_{L1}\) under the logit link with \(n=200\). Death is prioritized over one nonfatal event. For each Gumbel dependence parameter \(\alpha\) and restriction time \(L\), 2,000 datasets were generated.}
\label{tab:sim-logit-n200}
\centering
\begingroup
\renewcommand{\baselinestretch}{1}
\fontsize{10}{12}\selectfont
\renewcommand{\arraystretch}{1}
\setlength{\tabcolsep}{5pt}
\begin{tabular*}{\textwidth}{@{\extracolsep{\fill}}ccclrrrrrr@{}}
\toprule
$\alpha$ & $L$ & $c_L$ & Method & Truth & Abias & MCSD & ASE & ECP & VR (\%)\\
\midrule
\multirow{4}{*}{1.25} & \multirow{4}{*}{2.0} & \multirow{4}{*}{0.355} & CHW & \multirow{4}{*}{1.097} & 0.023 & 0.233 & 0.223 & 0.940 & 23.9\\
 &  &  & RCW &  & 0.020 & 0.228 & 0.218 & 0.936 & 20.8\\
 &  &  & CTW &  & 0.016 & 0.203 & 0.195 & 0.939 & \textemdash\\
 &  &  & IAT &  & 0.687 & 0.122 & 0.121 & 0.000 & \textemdash\\
\addlinespace[3pt]
\multirow{4}{*}{1.25} & \multirow{4}{*}{2.8} & \multirow{4}{*}{0.530} & CHW & \multirow{4}{*}{1.065} & 0.024 & 0.262 & 0.258 & 0.946 & 37.6\\
 &  &  & RCW &  & 0.021 & 0.252 & 0.245 & 0.942 & 32.5\\
 &  &  & CTW &  & 0.015 & 0.207 & 0.198 & 0.935 & \textemdash\\
 &  &  & IAT &  & 0.809 & 0.111 & 0.112 & 0.000 & \textemdash\\
\addlinespace[3pt]
\multirow{4}{*}{1.25} & \multirow{4}{*}{4.5} & \multirow{4}{*}{0.699} & CHW & \multirow{4}{*}{1.003} & 0.036 & 0.389 & 0.369 & 0.942 & 67.2\\
 &  &  & RCW &  & 0.010 & 0.333 & 0.324 & 0.942 & 55.2\\
 &  &  & CTW &  & 0.017 & 0.223 & 0.214 & 0.940 & \textemdash\\
 &  &  & IAT &  & 0.822 & 0.109 & 0.109 & 0.000 & \textemdash\\
\addlinespace[6pt]
\multirow{4}{*}{2} & \multirow{4}{*}{2.0} & \multirow{4}{*}{0.356} & CHW & \multirow{4}{*}{1.176} & 0.019 & 0.227 & 0.227 & 0.950 & 22.3\\
 &  &  & RCW &  & 0.017 & 0.221 & 0.221 & 0.956 & 18.1\\
 &  &  & CTW &  & 0.014 & 0.200 & 0.201 & 0.950 & \textemdash\\
 &  &  & IAT &  & 0.753 & 0.121 & 0.122 & 0.000 & \textemdash\\
\addlinespace[3pt]
\multirow{4}{*}{2} & \multirow{4}{*}{2.8} & \multirow{4}{*}{0.530} & CHW & \multirow{4}{*}{1.138} & 0.026 & 0.259 & 0.259 & 0.955 & 37.7\\
 &  &  & RCW &  & 0.021 & 0.245 & 0.246 & 0.955 & 30.4\\
 &  &  & CTW &  & 0.021 & 0.205 & 0.204 & 0.950 & \textemdash\\
 &  &  & IAT &  & 0.877 & 0.112 & 0.112 & 0.000 & \textemdash\\
\addlinespace[3pt]
\multirow{4}{*}{2} & \multirow{4}{*}{4.5} & \multirow{4}{*}{0.698} & CHW & \multirow{4}{*}{1.067} & 0.045 & 0.382 & 0.365 & 0.943 & 67.5\\
 &  &  & RCW &  & 0.013 & 0.330 & 0.321 & 0.947 & 56.4\\
 &  &  & CTW &  & 0.026 & 0.218 & 0.219 & 0.949 & \textemdash\\
 &  &  & IAT &  & 0.884 & 0.110 & 0.109 & 0.000 & \textemdash\\
\addlinespace[6pt]
\multirow{4}{*}{4} & \multirow{4}{*}{2.0} & \multirow{4}{*}{0.355} & CHW & \multirow{4}{*}{1.238} & 0.009 & 0.231 & 0.230 & 0.948 & 17.4\\
 &  &  & RCW &  & 0.008 & 0.228 & 0.225 & 0.945 & 15.2\\
 &  &  & CTW &  & 0.012 & 0.210 & 0.207 & 0.945 & \textemdash\\
 &  &  & IAT &  & 0.805 & 0.126 & 0.122 & 0.000 & \textemdash\\
\addlinespace[3pt]
\multirow{4}{*}{4} & \multirow{4}{*}{2.8} & \multirow{4}{*}{0.529} & CHW & \multirow{4}{*}{1.197} & 0.014 & 0.262 & 0.261 & 0.948 & 33.3\\
 &  &  & RCW &  & 0.010 & 0.251 & 0.249 & 0.948 & 27.6\\
 &  &  & CTW &  & 0.017 & 0.214 & 0.211 & 0.943 & \textemdash\\
 &  &  & IAT &  & 0.932 & 0.114 & 0.112 & 0.000 & \textemdash\\
\addlinespace[3pt]
\multirow{4}{*}{4} & \multirow{4}{*}{4.5} & \multirow{4}{*}{0.698} & CHW & \multirow{4}{*}{1.125} & 0.025 & 0.381 & 0.362 & 0.948 & 63.1\\
 &  &  & RCW &  & 0.002 & 0.320 & 0.318 & 0.949 & 47.8\\
 &  &  & CTW &  & 0.025 & 0.232 & 0.226 & 0.940 & \textemdash\\
 &  &  & IAT &  & 0.942 & 0.110 & 0.110 & 0.000 & \textemdash\\
\bottomrule
\end{tabular*}
\endgroup
\par\smallskip
\begin{minipage}{\textwidth}
\renewcommand{\baselinestretch}{1}
\fontsize{9}{11}\selectfont\raggedright
\(c_L\) is the mean realized censoring proportion, Abias denotes absolute bias, MCSD the Monte Carlo standard deviation, ASE the average estimated standard error, and ECP the empirical coverage probability of nominal \(95\%\) Wald intervals.
\par\smallskip
For a CHW or RCW row, VR (\%) is \(100\{1-(\mathrm{MCSD}_{\mathrm{CTW}}/\mathrm{MCSD}_{\star})^2\}\), where \(\star\) denotes that row's method. Positive values favor CTW. IAT assigns response \(1/2\) to every unresolved observed pair and gives every pair unit weight. 
\end{minipage}
\end{table}

\begin{table}[!htbp]
\caption{Finite-sample performance of complete history weighting (CHW), resolved comparison weighting (RCW), conditional tie weighting (CTW), and indeterminate as tie (IAT) for the binary covariate coefficient \(\beta_{L2}\) under the logit link with \(n=200\). Death is prioritized over one nonfatal event. For each Gumbel dependence parameter \(\alpha\) and restriction time \(L\), 2,000 datasets were generated.}
\label{tab:sim-logit-n200-binary}
\centering
\begingroup
\renewcommand{\baselinestretch}{1}
\fontsize{10}{12}\selectfont
\renewcommand{\arraystretch}{1}
\setlength{\tabcolsep}{5pt}
\begin{tabular*}{\textwidth}{@{\extracolsep{\fill}}ccclrrrrrr@{}}
\toprule
$\alpha$ & $L$ & $c_L$ & Method & Truth & Abias & MCSD & ASE & ECP & VR (\%)\\
\midrule
\multirow{4}{*}{1.25} & \multirow{4}{*}{2.0} & \multirow{4}{*}{0.355} & CHW & \multirow{4}{*}{2.355} & 0.051 & 0.461 & 0.446 & 0.937 & 19.5\\
 &  &  & RCW &  & 0.051 & 0.453 & 0.436 & 0.938 & 16.4\\
 &  &  & CTW &  & 0.051 & 0.414 & 0.405 & 0.937 & \textemdash\\
 &  &  & IAT &  & 1.560 & 0.187 & 0.191 & 0.000 & \textemdash\\
\addlinespace[3pt]
\multirow{4}{*}{1.25} & \multirow{4}{*}{2.8} & \multirow{4}{*}{0.530} & CHW & \multirow{4}{*}{2.285} & 0.045 & 0.527 & 0.502 & 0.939 & 32.6\\
 &  &  & RCW &  & 0.046 & 0.506 & 0.476 & 0.933 & 26.8\\
 &  &  & CTW &  & 0.049 & 0.433 & 0.413 & 0.930 & \textemdash\\
 &  &  & IAT &  & 1.834 & 0.160 & 0.159 & 0.000 & \textemdash\\
\addlinespace[3pt]
\multirow{4}{*}{1.25} & \multirow{4}{*}{4.5} & \multirow{4}{*}{0.699} & CHW & \multirow{4}{*}{2.140} & 0.019 & 0.818 & 0.679 & 0.924 & 67.1\\
 &  &  & RCW &  & 0.006 & 0.648 & 0.579 & 0.927 & 47.5\\
 &  &  & CTW &  & 0.060 & 0.469 & 0.434 & 0.920 & \textemdash\\
 &  &  & IAT &  & 1.859 & 0.150 & 0.148 & 0.000 & \textemdash\\
\addlinespace[6pt]
\multirow{4}{*}{2} & \multirow{4}{*}{2.0} & \multirow{4}{*}{0.356} & CHW & \multirow{4}{*}{2.539} & 0.043 & 0.470 & 0.468 & 0.948 & 17.7\\
 &  &  & RCW &  & 0.043 & 0.458 & 0.457 & 0.944 & 13.6\\
 &  &  & CTW &  & 0.045 & 0.426 & 0.432 & 0.946 & \textemdash\\
 &  &  & IAT &  & 1.717 & 0.191 & 0.193 & 0.000 & \textemdash\\
\addlinespace[3pt]
\multirow{4}{*}{2} & \multirow{4}{*}{2.8} & \multirow{4}{*}{0.530} & CHW & \multirow{4}{*}{2.451} & 0.054 & 0.529 & 0.522 & 0.950 & 31.3\\
 &  &  & RCW &  & 0.053 & 0.496 & 0.493 & 0.947 & 22.0\\
 &  &  & CTW &  & 0.062 & 0.438 & 0.440 & 0.943 & \textemdash\\
 &  &  & IAT &  & 1.993 & 0.158 & 0.161 & 0.000 & \textemdash\\
\addlinespace[3pt]
\multirow{4}{*}{2} & \multirow{4}{*}{4.5} & \multirow{4}{*}{0.698} & CHW & \multirow{4}{*}{2.283} & 0.031 & 0.798 & 0.692 & 0.939 & 64.8\\
 &  &  & RCW &  & 0.031 & 0.630 & 0.588 & 0.937 & 43.5\\
 &  &  & CTW &  & 0.089 & 0.474 & 0.461 & 0.927 & \textemdash\\
 &  &  & IAT &  & 2.002 & 0.146 & 0.148 & 0.000 & \textemdash\\
\addlinespace[6pt]
\multirow{4}{*}{4} & \multirow{4}{*}{2.0} & \multirow{4}{*}{0.355} & CHW & \multirow{4}{*}{2.664} & 0.059 & 0.490 & 0.483 & 0.948 & 10.8\\
 &  &  & RCW &  & 0.054 & 0.481 & 0.472 & 0.946 & 7.4\\
 &  &  & CTW &  & 0.063 & 0.463 & 0.453 & 0.938 & \textemdash\\
 &  &  & IAT &  & 1.811 & 0.199 & 0.196 & 0.000 & \textemdash\\
\addlinespace[3pt]
\multirow{4}{*}{4} & \multirow{4}{*}{2.8} & \multirow{4}{*}{0.529} & CHW & \multirow{4}{*}{2.576} & 0.055 & 0.545 & 0.532 & 0.949 & 23.6\\
 &  &  & RCW &  & 0.050 & 0.516 & 0.504 & 0.947 & 14.5\\
 &  &  & CTW &  & 0.073 & 0.477 & 0.462 & 0.932 & \textemdash\\
 &  &  & IAT &  & 2.104 & 0.162 & 0.161 & 0.000 & \textemdash\\
\addlinespace[3pt]
\multirow{4}{*}{4} & \multirow{4}{*}{4.5} & \multirow{4}{*}{0.698} & CHW & \multirow{4}{*}{2.414} & 0.000 & 0.826 & 0.694 & 0.937 & 61.1\\
 &  &  & RCW &  & 0.006 & 0.640 & 0.592 & 0.942 & 35.3\\
 &  &  & CTW &  & 0.093 & 0.515 & 0.485 & 0.927 & \textemdash\\
 &  &  & IAT &  & 2.128 & 0.147 & 0.149 & 0.000 & \textemdash\\
\bottomrule
\end{tabular*}
\endgroup
\par\smallskip
\begin{minipage}{\textwidth}
\renewcommand{\baselinestretch}{1}
\fontsize{9}{11}\selectfont\raggedright
\(c_L\) is the mean realized censoring proportion, Abias denotes absolute bias, MCSD the Monte Carlo standard deviation, ASE the average estimated standard error, and ECP the empirical coverage probability of nominal \(95\%\) Wald intervals.
\par\smallskip
For a CHW or RCW row, VR (\%) is \(100\{1-(\mathrm{MCSD}_{\mathrm{CTW}}/\mathrm{MCSD}_{\star})^2\}\), where \(\star\) denotes that row's method. Positive values favor CTW. IAT assigns response \(1/2\) to every unresolved observed pair and gives every pair unit weight. 
\end{minipage}
\end{table}
\begin{table}[!htbp]
\caption{Finite-sample performance of complete history weighting (CHW), resolved comparison weighting (RCW), conditional tie weighting (CTW), and indeterminate as tie (IAT) for the continuous covariate coefficient \(\beta_{L1}\) under the logit link with \(n=1000\). Death is prioritized over one nonfatal event. Results are based on 2,000 replications for each Gumbel dependence parameter \(\alpha\) and restriction time \(L\).}
\label{tab:sim-logit-n1000}
\centering
\begingroup
\renewcommand{\baselinestretch}{1}
\fontsize{10}{12}\selectfont
\renewcommand{\arraystretch}{1}
\setlength{\tabcolsep}{5pt}
\begin{tabular*}{\textwidth}{@{\extracolsep{\fill}}ccclrrrrrr@{}}
\toprule
$\alpha$ & $L$ & $c_L$ & Method & Truth & Abias & MCSD & ASE & ECP & VR (\%)\\
\midrule
\multirow{4}{*}{1.25} & \multirow{4}{*}{2.0} & \multirow{4}{*}{0.356} & CHW & \multirow{4}{*}{1.097} & 0.002 & 0.099 & 0.098 & 0.949 & 20.2\\
 &  &  & RCW &  & 0.001 & 0.098 & 0.096 & 0.950 & 17.3\\
 &  &  & CTW &  & 0.003 & 0.089 & 0.087 & 0.947 & \textemdash\\
 &  &  & IAT &  & 0.692 & 0.055 & 0.054 & 0.000 & \textemdash\\
\addlinespace[3pt]
\multirow{4}{*}{1.25} & \multirow{4}{*}{2.8} & \multirow{4}{*}{0.530} & CHW & \multirow{4}{*}{1.065} & 0.001 & 0.112 & 0.112 & 0.954 & 35.6\\
 &  &  & RCW &  & 0.003 & 0.107 & 0.107 & 0.947 & 30.5\\
 &  &  & CTW &  & 0.001 & 0.090 & 0.089 & 0.950 & \textemdash\\
 &  &  & IAT &  & 0.813 & 0.052 & 0.050 & 0.000 & \textemdash\\
\addlinespace[3pt]
\multirow{4}{*}{1.25} & \multirow{4}{*}{4.5} & \multirow{4}{*}{0.699} & CHW & \multirow{4}{*}{1.003} & 0.001 & 0.158 & 0.155 & 0.950 & 61.4\\
 &  &  & RCW &  & 0.003 & 0.143 & 0.140 & 0.946 & 52.8\\
 &  &  & CTW &  & 0.002 & 0.098 & 0.097 & 0.947 & \textemdash\\
 &  &  & IAT &  & 0.824 & 0.050 & 0.049 & 0.000 & \textemdash\\
\addlinespace[6pt]
\multirow{4}{*}{2} & \multirow{4}{*}{2.0} & \multirow{4}{*}{0.356} & CHW & \multirow{4}{*}{1.176} & 0.000 & 0.101 & 0.100 & 0.949 & 17.4\\
 &  &  & RCW &  & 0.000 & 0.099 & 0.098 & 0.940 & 14.6\\
 &  &  & CTW &  & 0.001 & 0.092 & 0.090 & 0.946 & \textemdash\\
 &  &  & IAT &  & 0.757 & 0.056 & 0.054 & 0.000 & \textemdash\\
\addlinespace[3pt]
\multirow{4}{*}{2} & \multirow{4}{*}{2.8} & \multirow{4}{*}{0.530} & CHW & \multirow{4}{*}{1.138} & 0.004 & 0.114 & 0.113 & 0.958 & 30.9\\
 &  &  & RCW &  & 0.003 & 0.108 & 0.108 & 0.954 & 23.1\\
 &  &  & CTW &  & 0.004 & 0.095 & 0.092 & 0.943 & \textemdash\\
 &  &  & IAT &  & 0.879 & 0.051 & 0.050 & 0.000 & \textemdash\\
\addlinespace[3pt]
\multirow{4}{*}{2} & \multirow{4}{*}{4.5} & \multirow{4}{*}{0.699} & CHW & \multirow{4}{*}{1.067} & 0.008 & 0.153 & 0.155 & 0.951 & 55.7\\
 &  &  & RCW &  & 0.004 & 0.139 & 0.139 & 0.951 & 46.3\\
 &  &  & CTW &  & 0.006 & 0.102 & 0.101 & 0.945 & \textemdash\\
 &  &  & IAT &  & 0.886 & 0.050 & 0.049 & 0.000 & \textemdash\\
\addlinespace[6pt]
\multirow{4}{*}{4} & \multirow{4}{*}{2.0} & \multirow{4}{*}{0.356} & CHW & \multirow{4}{*}{1.238} & 0.002 & 0.102 & 0.102 & 0.944 & 20.1\\
 &  &  & RCW &  & 0.002 & 0.101 & 0.100 & 0.944 & 17.4\\
 &  &  & CTW &  & 0.003 & 0.091 & 0.093 & 0.952 & \textemdash\\
 &  &  & IAT &  & 0.805 & 0.054 & 0.055 & 0.000 & \textemdash\\
\addlinespace[3pt]
\multirow{4}{*}{4} & \multirow{4}{*}{2.8} & \multirow{4}{*}{0.530} & CHW & \multirow{4}{*}{1.197} & 0.007 & 0.114 & 0.114 & 0.947 & 34.1\\
 &  &  & RCW &  & 0.006 & 0.110 & 0.109 & 0.943 & 29.2\\
 &  &  & CTW &  & 0.009 & 0.092 & 0.095 & 0.951 & \textemdash\\
 &  &  & IAT &  & 0.931 & 0.049 & 0.050 & 0.000 & \textemdash\\
\addlinespace[3pt]
\multirow{4}{*}{4} & \multirow{4}{*}{4.5} & \multirow{4}{*}{0.698} & CHW & \multirow{4}{*}{1.125} & 0.010 & 0.159 & 0.153 & 0.942 & 56.5\\
 &  &  & RCW &  & 0.005 & 0.140 & 0.137 & 0.945 & 44.1\\
 &  &  & CTW &  & 0.012 & 0.105 & 0.105 & 0.944 & \textemdash\\
 &  &  & IAT &  & 0.940 & 0.048 & 0.049 & 0.000 & \textemdash\\
\bottomrule
\end{tabular*}
\endgroup
\par\smallskip
\begin{minipage}{\textwidth}
\renewcommand{\baselinestretch}{1}
\fontsize{9}{11}\selectfont\raggedright
\(c_L\) is the mean realized censoring proportion, Abias denotes absolute bias, MCSD the Monte Carlo standard deviation, ASE the average estimated standard error, and ECP the empirical coverage probability of nominal \(95\%\) Wald intervals.
\par\smallskip
For a CHW or RCW row, VR (\%) is \(100\{1-(\mathrm{MCSD}_{\mathrm{CTW}}/\mathrm{MCSD}_{\star})^2\}\), where \(\star\) denotes that row's method. Positive values favor CTW. IAT assigns response \(1/2\) to every unresolved observed pair and gives every pair unit weight.
\end{minipage}
\end{table}

\begin{table}[!htbp]
\caption{Finite-sample performance of complete history weighting (CHW), resolved comparison weighting (RCW), conditional tie weighting (CTW), and indeterminate as tie (IAT) for the binary covariate coefficient \(\beta_{L2}\) under the logit link with \(n=1000\). Death is prioritized over one nonfatal event. Results are based on 2,000 replications for each Gumbel dependence parameter \(\alpha\) and restriction time \(L\).}
\label{tab:sim-logit-n1000-binary}
\centering
\begingroup
\renewcommand{\baselinestretch}{1}
\fontsize{10}{12}\selectfont
\renewcommand{\arraystretch}{1}
\setlength{\tabcolsep}{5pt}
\begin{tabular*}{\textwidth}{@{\extracolsep{\fill}}ccclrrrrrr@{}}
\toprule
$\alpha$ & $L$ & $c_L$ & Method & Truth & Abias & MCSD & ASE & ECP & VR (\%)\\
\midrule
\multirow{4}{*}{1.25} & \multirow{4}{*}{2.0} & \multirow{4}{*}{0.356} & CHW & \multirow{4}{*}{2.355} & 0.010 & 0.194 & 0.196 & 0.950 & 14.8\\
 &  &  & RCW &  & 0.010 & 0.191 & 0.192 & 0.948 & 12.5\\
 &  &  & CTW &  & 0.011 & 0.179 & 0.181 & 0.954 & \textemdash\\
 &  &  & IAT &  & 1.566 & 0.084 & 0.085 & 0.000 & \textemdash\\
\addlinespace[3pt]
\multirow{4}{*}{1.25} & \multirow{4}{*}{2.8} & \multirow{4}{*}{0.530} & CHW & \multirow{4}{*}{2.285} & 0.002 & 0.214 & 0.218 & 0.957 & 26.1\\
 &  &  & RCW &  & 0.001 & 0.206 & 0.208 & 0.951 & 20.6\\
 &  &  & CTW &  & 0.005 & 0.184 & 0.185 & 0.954 & \textemdash\\
 &  &  & IAT &  & 1.839 & 0.071 & 0.071 & 0.000 & \textemdash\\
\addlinespace[3pt]
\multirow{4}{*}{1.25} & \multirow{4}{*}{4.5} & \multirow{4}{*}{0.699} & CHW & \multirow{4}{*}{2.140} & 0.008 & 0.291 & 0.288 & 0.946 & 50.9\\
 &  &  & RCW &  & 0.011 & 0.261 & 0.256 & 0.944 & 38.7\\
 &  &  & CTW &  & 0.000 & 0.204 & 0.202 & 0.952 & \textemdash\\
 &  &  & IAT &  & 1.862 & 0.066 & 0.066 & 0.000 & \textemdash\\
\addlinespace[6pt]
\multirow{4}{*}{2} & \multirow{4}{*}{2.0} & \multirow{4}{*}{0.356} & CHW & \multirow{4}{*}{2.539} & 0.009 & 0.206 & 0.206 & 0.944 & 10.1\\
 &  &  & RCW &  & 0.009 & 0.201 & 0.202 & 0.949 & 6.0\\
 &  &  & CTW &  & 0.009 & 0.195 & 0.193 & 0.942 & \textemdash\\
 &  &  & IAT &  & 1.716 & 0.087 & 0.087 & 0.000 & \textemdash\\
\addlinespace[3pt]
\multirow{4}{*}{2} & \multirow{4}{*}{2.8} & \multirow{4}{*}{0.530} & CHW & \multirow{4}{*}{2.451} & 0.011 & 0.226 & 0.226 & 0.951 & 22.4\\
 &  &  & RCW &  & 0.011 & 0.214 & 0.215 & 0.950 & 13.3\\
 &  &  & CTW &  & 0.016 & 0.199 & 0.198 & 0.946 & \textemdash\\
 &  &  & IAT &  & 1.992 & 0.073 & 0.072 & 0.000 & \textemdash\\
\addlinespace[3pt]
\multirow{4}{*}{2} & \multirow{4}{*}{4.5} & \multirow{4}{*}{0.699} & CHW & \multirow{4}{*}{2.283} & 0.020 & 0.297 & 0.294 & 0.950 & 45.3\\
 &  &  & RCW &  & 0.016 & 0.265 & 0.258 & 0.948 & 31.0\\
 &  &  & CTW &  & 0.025 & 0.220 & 0.216 & 0.938 & \textemdash\\
 &  &  & IAT &  & 2.001 & 0.067 & 0.066 & 0.000 & \textemdash\\
\addlinespace[6pt]
\multirow{4}{*}{4} & \multirow{4}{*}{2.0} & \multirow{4}{*}{0.356} & CHW & \multirow{4}{*}{2.664} & 0.014 & 0.218 & 0.214 & 0.949 & 9.6\\
 &  &  & RCW &  & 0.014 & 0.214 & 0.209 & 0.947 & 6.5\\
 &  &  & CTW &  & 0.015 & 0.207 & 0.203 & 0.944 & \textemdash\\
 &  &  & IAT &  & 1.818 & 0.088 & 0.087 & 0.000 & \textemdash\\
\addlinespace[3pt]
\multirow{4}{*}{4} & \multirow{4}{*}{2.8} & \multirow{4}{*}{0.530} & CHW & \multirow{4}{*}{2.576} & 0.015 & 0.238 & 0.234 & 0.947 & 20.1\\
 &  &  & RCW &  & 0.014 & 0.228 & 0.222 & 0.944 & 13.0\\
 &  &  & CTW &  & 0.018 & 0.213 & 0.208 & 0.941 & \textemdash\\
 &  &  & IAT &  & 2.107 & 0.073 & 0.072 & 0.000 & \textemdash\\
\addlinespace[3pt]
\multirow{4}{*}{4} & \multirow{4}{*}{4.5} & \multirow{4}{*}{0.698} & CHW & \multirow{4}{*}{2.414} & 0.003 & 0.317 & 0.300 & 0.941 & 42.8\\
 &  &  & RCW &  & 0.003 & 0.277 & 0.261 & 0.943 & 25.2\\
 &  &  & CTW &  & 0.023 & 0.240 & 0.228 & 0.936 & \textemdash\\
 &  &  & IAT &  & 2.128 & 0.067 & 0.067 & 0.000 & \textemdash\\
\bottomrule
\end{tabular*}
\endgroup
\par\smallskip
\begin{minipage}{\textwidth}
\renewcommand{\baselinestretch}{1}
\fontsize{9}{11}\selectfont\raggedright
\(c_L\) is the mean realized censoring proportion, Abias denotes absolute bias, MCSD the Monte Carlo standard deviation, ASE the average estimated standard error, and ECP the empirical coverage probability of nominal \(95\%\) Wald intervals.
\par\smallskip
For a CHW or RCW row, VR (\%) is \(100\{1-(\mathrm{MCSD}_{\mathrm{CTW}}/\mathrm{MCSD}_{\star})^2\}\), where \(\star\) denotes that row's method. Positive values favor CTW. IAT assigns response \(1/2\) to every unresolved observed pair and gives every pair unit weight.
\end{minipage}
\end{table}
\FloatBarrier

\subsubsection{Probit link}

Tables \ref{tab:sim-probit-n200}--\ref{tab:sim-probit-n1000-binary} report the probit results for all three sample sizes. The pattern was similar to that under the logit link, with smaller absolute bias at larger sample sizes and lower MCSD for CTW than for CHW or RCW. The VR columns give the reductions for each coefficient and configuration.

\begin{table}[!htbp]
\caption{Finite-sample performance of complete history weighting (CHW), resolved comparison weighting (RCW), conditional tie weighting (CTW), and indeterminate as tie (IAT) for the continuous covariate coefficient \(\beta_{L1}\) under the probit link with \(n=200\). Death is prioritized over one nonfatal event. For each Gumbel dependence parameter \(\alpha\) and restriction time \(L\), 2,000 datasets were generated.}
\label{tab:sim-probit-n200}
\centering
\begingroup
\renewcommand{\baselinestretch}{1}
\fontsize{10}{10.8}\selectfont
\renewcommand{\arraystretch}{1}
\setlength{\tabcolsep}{5pt}
\begin{tabular*}{\textwidth}{@{\extracolsep{\fill}}ccclrrrrrr@{}}
\toprule
$\alpha$ & $L$ & $c_L$ & Method & Truth & Abias & MCSD & ASE & ECP & VR (\%)\\
\midrule
\multirow{4}{*}{1.25} & \multirow{4}{*}{2.0} & \multirow{4}{*}{0.355} & CHW & \multirow{4}{*}{0.651} & 0.014 & 0.135 & 0.128 & 0.935 & 23.5\\
 &  &  & RCW &  & 0.013 & 0.132 & 0.125 & 0.933 & 20.4\\
 &  &  & CTW &  & 0.010 & 0.118 & 0.112 & 0.934 & \textemdash\\
 &  &  & IAT &  & 0.397 & 0.075 & 0.075 & 0.000 & \textemdash\\
\addlinespace[3pt]
\multirow{4}{*}{1.25} & \multirow{4}{*}{2.8} & \multirow{4}{*}{0.530} & CHW & \multirow{4}{*}{0.634} & 0.014 & 0.152 & 0.148 & 0.942 & 37.0\\
 &  &  & RCW &  & 0.013 & 0.146 & 0.141 & 0.941 & 31.8\\
 &  &  & CTW &  & 0.010 & 0.120 & 0.115 & 0.935 & \textemdash\\
 &  &  & IAT &  & 0.474 & 0.069 & 0.070 & 0.000 & \textemdash\\
\addlinespace[3pt]
\multirow{4}{*}{1.25} & \multirow{4}{*}{4.5} & \multirow{4}{*}{0.699} & CHW & \multirow{4}{*}{0.599} & 0.019 & 0.225 & 0.212 & 0.936 & 66.5\\
 &  &  & RCW &  & 0.005 & 0.193 & 0.187 & 0.942 & 54.5\\
 &  &  & CTW &  & 0.011 & 0.130 & 0.124 & 0.934 & \textemdash\\
 &  &  & IAT &  & 0.486 & 0.068 & 0.068 & 0.000 & \textemdash\\
\addlinespace[6pt]
\multirow{4}{*}{2} & \multirow{4}{*}{2.0} & \multirow{4}{*}{0.356} & CHW & \multirow{4}{*}{0.694} & 0.011 & 0.130 & 0.129 & 0.948 & 21.5\\
 &  &  & RCW &  & 0.011 & 0.127 & 0.126 & 0.950 & 17.2\\
 &  &  & CTW &  & 0.009 & 0.115 & 0.115 & 0.948 & \textemdash\\
 &  &  & IAT &  & 0.432 & 0.074 & 0.075 & 0.000 & \textemdash\\
\addlinespace[3pt]
\multirow{4}{*}{2} & \multirow{4}{*}{2.8} & \multirow{4}{*}{0.530} & CHW & \multirow{4}{*}{0.672} & 0.015 & 0.149 & 0.148 & 0.956 & 36.9\\
 &  &  & RCW &  & 0.012 & 0.141 & 0.141 & 0.952 & 29.6\\
 &  &  & CTW &  & 0.013 & 0.118 & 0.117 & 0.946 & \textemdash\\
 &  &  & IAT &  & 0.509 & 0.069 & 0.069 & 0.000 & \textemdash\\
\addlinespace[3pt]
\multirow{4}{*}{2} & \multirow{4}{*}{4.5} & \multirow{4}{*}{0.698} & CHW & \multirow{4}{*}{0.633} & 0.024 & 0.219 & 0.208 & 0.944 & 67.0\\
 &  &  & RCW &  & 0.006 & 0.190 & 0.184 & 0.946 & 56.0\\
 &  &  & CTW &  & 0.015 & 0.126 & 0.126 & 0.946 & \textemdash\\
 &  &  & IAT &  & 0.519 & 0.068 & 0.068 & 0.000 & \textemdash\\
\addlinespace[6pt]
\multirow{4}{*}{4} & \multirow{4}{*}{2.0} & \multirow{4}{*}{0.355} & CHW & \multirow{4}{*}{0.727} & 0.006 & 0.133 & 0.131 & 0.944 & 16.9\\
 &  &  & RCW &  & 0.005 & 0.131 & 0.128 & 0.940 & 14.7\\
 &  &  & CTW &  & 0.008 & 0.121 & 0.118 & 0.940 & \textemdash\\
 &  &  & IAT &  & 0.459 & 0.077 & 0.075 & 0.001 & \textemdash\\
\addlinespace[3pt]
\multirow{4}{*}{4} & \multirow{4}{*}{2.8} & \multirow{4}{*}{0.529} & CHW & \multirow{4}{*}{0.703} & 0.009 & 0.150 & 0.148 & 0.947 & 32.7\\
 &  &  & RCW &  & 0.007 & 0.145 & 0.141 & 0.943 & 27.0\\
 &  &  & CTW &  & 0.012 & 0.123 & 0.120 & 0.943 & \textemdash\\
 &  &  & IAT &  & 0.538 & 0.071 & 0.070 & 0.000 & \textemdash\\
\addlinespace[3pt]
\multirow{4}{*}{4} & \multirow{4}{*}{4.5} & \multirow{4}{*}{0.698} & CHW & \multirow{4}{*}{0.664} & 0.013 & 0.218 & 0.205 & 0.944 & 62.4\\
 &  &  & RCW &  & 0.001 & 0.184 & 0.181 & 0.947 & 47.1\\
 &  &  & CTW &  & 0.016 & 0.134 & 0.129 & 0.937 & \textemdash\\
 &  &  & IAT &  & 0.549 & 0.068 & 0.068 & 0.000 & \textemdash\\
\bottomrule
\end{tabular*}
\endgroup
\par\smallskip
\begin{minipage}{\textwidth}
\renewcommand{\baselinestretch}{1}
\fontsize{9}{11}\selectfont\raggedright
\(c_L\) is the mean realized censoring proportion, Abias denotes absolute bias, MCSD the Monte Carlo standard deviation, ASE the average estimated standard error, and ECP the empirical coverage probability of nominal \(95\%\) Wald intervals.
\par\smallskip
For a CHW or RCW row, VR (\%) is \(100\{1-(\mathrm{MCSD}_{\mathrm{CTW}}/\mathrm{MCSD}_{\star})^2\}\), where \(\star\) denotes that row's method. Positive values favor CTW. IAT assigns response \(1/2\) to every unresolved observed pair and gives every pair unit weight. 
\end{minipage}
\end{table}

\begin{table}[!htbp]
\caption{Finite-sample performance of complete history weighting (CHW), resolved comparison weighting (RCW), conditional tie weighting (CTW), and indeterminate as tie (IAT) for the binary covariate coefficient \(\beta_{L2}\) under the probit link with \(n=200\). Death is prioritized over one nonfatal event. For each Gumbel dependence parameter \(\alpha\) and restriction time \(L\), 2,000 datasets were generated.}
\label{tab:sim-probit-n200-binary}
\centering
\begingroup
\renewcommand{\baselinestretch}{1}
\fontsize{10}{12}\selectfont
\renewcommand{\arraystretch}{1}
\setlength{\tabcolsep}{5pt}
\begin{tabular*}{\textwidth}{@{\extracolsep{\fill}}ccclrrrrrr@{}}
\toprule
$\alpha$ & $L$ & $c_L$ & Method & Truth & Abias & MCSD & ASE & ECP & VR (\%)\\
\midrule
\multirow{4}{*}{1.25} & \multirow{4}{*}{2.0} & \multirow{4}{*}{0.355} & CHW & \multirow{4}{*}{1.394} & 0.028 & 0.260 & 0.250 & 0.935 & 18.7\\
 &  &  & RCW &  & 0.028 & 0.255 & 0.244 & 0.933 & 15.3\\
 &  &  & CTW &  & 0.029 & 0.235 & 0.228 & 0.932 & \textemdash\\
 &  &  & IAT &  & 0.901 & 0.114 & 0.116 & 0.000 & \textemdash\\
\addlinespace[3pt]
\multirow{4}{*}{1.25} & \multirow{4}{*}{2.8} & \multirow{4}{*}{0.530} & CHW & \multirow{4}{*}{1.354} & 0.023 & 0.296 & 0.280 & 0.936 & 31.3\\
 &  &  & RCW &  & 0.024 & 0.284 & 0.265 & 0.930 & 25.1\\
 &  &  & CTW &  & 0.028 & 0.246 & 0.232 & 0.927 & \textemdash\\
 &  &  & IAT &  & 1.074 & 0.099 & 0.099 & 0.000 & \textemdash\\
\addlinespace[3pt]
\multirow{4}{*}{1.25} & \multirow{4}{*}{4.5} & \multirow{4}{*}{0.699} & CHW & \multirow{4}{*}{1.274} & 0.021 & 0.462 & 0.378 & 0.919 & 66.5\\
 &  &  & RCW &  & 0.005 & 0.363 & 0.322 & 0.928 & 45.7\\
 &  &  & CTW &  & 0.033 & 0.267 & 0.245 & 0.919 & \textemdash\\
 &  &  & IAT &  & 1.098 & 0.093 & 0.092 & 0.000 & \textemdash\\
\addlinespace[6pt]
\multirow{4}{*}{2} & \multirow{4}{*}{2.0} & \multirow{4}{*}{0.356} & CHW & \multirow{4}{*}{1.491} & 0.023 & 0.262 & 0.260 & 0.947 & 16.2\\
 &  &  & RCW &  & 0.023 & 0.256 & 0.253 & 0.944 & 11.8\\
 &  &  & CTW &  & 0.025 & 0.240 & 0.241 & 0.941 & \textemdash\\
 &  &  & IAT &  & 0.982 & 0.116 & 0.118 & 0.000 & \textemdash\\
\addlinespace[3pt]
\multirow{4}{*}{2} & \multirow{4}{*}{2.8} & \multirow{4}{*}{0.530} & CHW & \multirow{4}{*}{1.442} & 0.027 & 0.294 & 0.288 & 0.949 & 29.4\\
 &  &  & RCW &  & 0.027 & 0.275 & 0.272 & 0.944 & 19.6\\
 &  &  & CTW &  & 0.034 & 0.247 & 0.245 & 0.941 & \textemdash\\
 &  &  & IAT &  & 1.157 & 0.098 & 0.100 & 0.000 & \textemdash\\
\addlinespace[3pt]
\multirow{4}{*}{2} & \multirow{4}{*}{4.5} & \multirow{4}{*}{0.698} & CHW & \multirow{4}{*}{1.350} & 0.007 & 0.443 & 0.381 & 0.938 & 63.5\\
 &  &  & RCW &  & 0.009 & 0.347 & 0.323 & 0.932 & 40.4\\
 &  &  & CTW &  & 0.049 & 0.268 & 0.258 & 0.927 & \textemdash\\
 &  &  & IAT &  & 1.174 & 0.091 & 0.093 & 0.000 & \textemdash\\
\addlinespace[6pt]
\multirow{4}{*}{4} & \multirow{4}{*}{2.0} & \multirow{4}{*}{0.355} & CHW & \multirow{4}{*}{1.555} & 0.032 & 0.274 & 0.267 & 0.945 & 9.5\\
 &  &  & RCW &  & 0.030 & 0.268 & 0.261 & 0.943 & 5.9\\
 &  &  & CTW &  & 0.035 & 0.260 & 0.252 & 0.935 & \textemdash\\
 &  &  & IAT &  & 1.028 & 0.121 & 0.119 & 0.000 & \textemdash\\
\addlinespace[3pt]
\multirow{4}{*}{4} & \multirow{4}{*}{2.8} & \multirow{4}{*}{0.529} & CHW & \multirow{4}{*}{1.506} & 0.029 & 0.303 & 0.292 & 0.945 & 22.0\\
 &  &  & RCW &  & 0.027 & 0.286 & 0.277 & 0.943 & 12.5\\
 &  &  & CTW &  & 0.042 & 0.268 & 0.257 & 0.930 & \textemdash\\
 &  &  & IAT &  & 1.211 & 0.100 & 0.100 & 0.000 & \textemdash\\
\addlinespace[3pt]
\multirow{4}{*}{4} & \multirow{4}{*}{4.5} & \multirow{4}{*}{0.698} & CHW & \multirow{4}{*}{1.417} & 0.008 & 0.459 & 0.380 & 0.931 & 60.4\\
 &  &  & RCW &  & 0.003 & 0.352 & 0.323 & 0.938 & 32.6\\
 &  &  & CTW &  & 0.052 & 0.289 & 0.270 & 0.922 & \textemdash\\
 &  &  & IAT &  & 1.238 & 0.092 & 0.093 & 0.000 & \textemdash\\
\bottomrule
\end{tabular*}
\endgroup
\par\smallskip
\begin{minipage}{\textwidth}
\renewcommand{\baselinestretch}{1}
\fontsize{9}{11}\selectfont\raggedright
\(c_L\) is the mean realized censoring proportion, Abias denotes absolute bias, MCSD the Monte Carlo standard deviation, ASE the average estimated standard error, and ECP the empirical coverage probability of nominal \(95\%\) Wald intervals.
\par\smallskip
For a CHW or RCW row, VR (\%) is \(100\{1-(\mathrm{MCSD}_{\mathrm{CTW}}/\mathrm{MCSD}_{\star})^2\}\), where \(\star\) denotes that row's method. Positive values favor CTW. IAT assigns response \(1/2\) to every unresolved observed pair and gives every pair unit weight. 
\end{minipage}
\end{table}
\begin{table}[!htbp]
\caption{Finite-sample performance of complete history weighting (CHW), resolved comparison weighting (RCW), conditional tie weighting (CTW), and indeterminate as tie (IAT) for the continuous covariate coefficient \(\beta_{L1}\) under the probit link with \(n=500\). Death is prioritized over one nonfatal event. Results are based on 2,000 replications for each Gumbel dependence parameter \(\alpha\) and restriction time \(L\).}
\label{tab:sim-probit-n500}
\centering
\begingroup
\renewcommand{\baselinestretch}{1}
\fontsize{10}{12}\selectfont
\renewcommand{\arraystretch}{1}
\setlength{\tabcolsep}{5pt}
\begin{tabular*}{\textwidth}{@{\extracolsep{\fill}}ccclrrrrrr@{}}
\toprule
$\alpha$ & $L$ & $c_L$ & Method & Truth & Abias & MCSD & ASE & ECP & VR (\%)\\
\midrule
\multirow{4}{*}{1.25} & \multirow{4}{*}{2.0} & \multirow{4}{*}{0.357} & CHW & \multirow{4}{*}{0.651} & 0.008 & 0.083 & 0.080 & 0.945 & 22.3\\
 &  &  & RCW &  & 0.007 & 0.081 & 0.079 & 0.944 & 19.8\\
 &  &  & CTW &  & 0.007 & 0.073 & 0.071 & 0.943 & \textemdash\\
 &  &  & IAT &  & 0.399 & 0.047 & 0.047 & 0.000 & \textemdash\\
\addlinespace[3pt]
\multirow{4}{*}{1.25} & \multirow{4}{*}{2.8} & \multirow{4}{*}{0.530} & CHW & \multirow{4}{*}{0.634} & 0.007 & 0.094 & 0.092 & 0.949 & 35.7\\
 &  &  & RCW &  & 0.005 & 0.092 & 0.088 & 0.943 & 31.8\\
 &  &  & CTW &  & 0.006 & 0.076 & 0.073 & 0.941 & \textemdash\\
 &  &  & IAT &  & 0.475 & 0.044 & 0.044 & 0.000 & \textemdash\\
\addlinespace[3pt]
\multirow{4}{*}{1.25} & \multirow{4}{*}{4.5} & \multirow{4}{*}{0.699} & CHW & \multirow{4}{*}{0.599} & 0.008 & 0.133 & 0.129 & 0.945 & 59.8\\
 &  &  & RCW &  & 0.004 & 0.120 & 0.116 & 0.937 & 50.8\\
 &  &  & CTW &  & 0.006 & 0.084 & 0.080 & 0.933 & \textemdash\\
 &  &  & IAT &  & 0.486 & 0.044 & 0.043 & 0.000 & \textemdash\\
\addlinespace[6pt]
\multirow{4}{*}{2} & \multirow{4}{*}{2.0} & \multirow{4}{*}{0.356} & CHW & \multirow{4}{*}{0.694} & 0.003 & 0.081 & 0.081 & 0.943 & 17.9\\
 &  &  & RCW &  & 0.003 & 0.080 & 0.080 & 0.946 & 14.3\\
 &  &  & CTW &  & 0.004 & 0.074 & 0.073 & 0.938 & \textemdash\\
 &  &  & IAT &  & 0.434 & 0.048 & 0.047 & 0.000 & \textemdash\\
\addlinespace[3pt]
\multirow{4}{*}{2} & \multirow{4}{*}{2.8} & \multirow{4}{*}{0.529} & CHW & \multirow{4}{*}{0.672} & 0.006 & 0.092 & 0.092 & 0.948 & 34.5\\
 &  &  & RCW &  & 0.004 & 0.088 & 0.088 & 0.947 & 27.4\\
 &  &  & CTW &  & 0.006 & 0.075 & 0.075 & 0.943 & \textemdash\\
 &  &  & IAT &  & 0.511 & 0.044 & 0.044 & 0.000 & \textemdash\\
\addlinespace[3pt]
\multirow{4}{*}{2} & \multirow{4}{*}{4.5} & \multirow{4}{*}{0.698} & CHW & \multirow{4}{*}{0.633} & 0.005 & 0.131 & 0.127 & 0.943 & 59.8\\
 &  &  & RCW &  & 0.000 & 0.115 & 0.114 & 0.945 & 48.5\\
 &  &  & CTW &  & 0.007 & 0.083 & 0.082 & 0.941 & \textemdash\\
 &  &  & IAT &  & 0.520 & 0.044 & 0.043 & 0.000 & \textemdash\\
\addlinespace[6pt]
\multirow{4}{*}{4} & \multirow{4}{*}{2.0} & \multirow{4}{*}{0.356} & CHW & \multirow{4}{*}{0.727} & 0.003 & 0.084 & 0.082 & 0.941 & 15.5\\
 &  &  & RCW &  & 0.002 & 0.083 & 0.081 & 0.946 & 12.7\\
 &  &  & CTW &  & 0.003 & 0.077 & 0.075 & 0.946 & \textemdash\\
 &  &  & IAT &  & 0.460 & 0.047 & 0.047 & 0.000 & \textemdash\\
\addlinespace[3pt]
\multirow{4}{*}{4} & \multirow{4}{*}{2.8} & \multirow{4}{*}{0.530} & CHW & \multirow{4}{*}{0.703} & 0.008 & 0.095 & 0.093 & 0.944 & 29.0\\
 &  &  & RCW &  & 0.005 & 0.091 & 0.089 & 0.947 & 23.2\\
 &  &  & CTW &  & 0.007 & 0.080 & 0.077 & 0.944 & \textemdash\\
 &  &  & IAT &  & 0.538 & 0.044 & 0.044 & 0.000 & \textemdash\\
\addlinespace[3pt]
\multirow{4}{*}{4} & \multirow{4}{*}{4.5} & \multirow{4}{*}{0.699} & CHW & \multirow{4}{*}{0.664} & 0.006 & 0.131 & 0.126 & 0.935 & 54.5\\
 &  &  & RCW &  & 0.001 & 0.117 & 0.112 & 0.944 & 43.0\\
 &  &  & CTW &  & 0.007 & 0.088 & 0.084 & 0.934 & \textemdash\\
 &  &  & IAT &  & 0.549 & 0.044 & 0.043 & 0.000 & \textemdash\\
\bottomrule
\end{tabular*}
\endgroup
\par\smallskip
\begin{minipage}{\textwidth}
\renewcommand{\baselinestretch}{1}
\fontsize{9}{11}\selectfont\raggedright
\(c_L\) is the mean realized censoring proportion, Abias denotes absolute bias, MCSD the Monte Carlo standard deviation, ASE the average estimated standard error, and ECP the empirical coverage probability of nominal \(95\%\) Wald intervals.
\par\smallskip
For a CHW or RCW row, VR (\%) is \(100\{1-(\mathrm{MCSD}_{\mathrm{CTW}}/\mathrm{MCSD}_{\star})^2\}\), where \(\star\) denotes that row's method. Positive values favor CTW. IAT assigns response \(1/2\) to every unresolved observed pair and gives every pair unit weight.
\end{minipage}
\end{table}

\begin{table}[!htbp]
\caption{Finite-sample performance of complete history weighting (CHW), resolved comparison weighting (RCW), conditional tie weighting (CTW), and indeterminate as tie (IAT) for the binary covariate coefficient \(\beta_{L2}\) under the probit link with \(n=500\). Death is prioritized over one nonfatal event. Results are based on 2,000 replications for each Gumbel dependence parameter \(\alpha\) and restriction time \(L\).}
\label{tab:sim-probit-n500-binary}
\centering
\begingroup
\renewcommand{\baselinestretch}{1}
\fontsize{10}{12}\selectfont
\renewcommand{\arraystretch}{1}
\setlength{\tabcolsep}{5pt}
\begin{tabular*}{\textwidth}{@{\extracolsep{\fill}}ccclrrrrrr@{}}
\toprule
$\alpha$ & $L$ & $c_L$ & Method & Truth & Abias & MCSD & ASE & ECP & VR (\%)\\
\midrule
\multirow{4}{*}{1.25} & \multirow{4}{*}{2.0} & \multirow{4}{*}{0.357} & CHW & \multirow{4}{*}{1.394} & 0.015 & 0.161 & 0.156 & 0.939 & 16.9\\
 &  &  & RCW &  & 0.015 & 0.157 & 0.153 & 0.937 & 12.6\\
 &  &  & CTW &  & 0.013 & 0.147 & 0.144 & 0.943 & \textemdash\\
 &  &  & IAT &  & 0.903 & 0.073 & 0.073 & 0.000 & \textemdash\\
\addlinespace[3pt]
\multirow{4}{*}{1.25} & \multirow{4}{*}{2.8} & \multirow{4}{*}{0.530} & CHW & \multirow{4}{*}{1.354} & 0.010 & 0.182 & 0.174 & 0.943 & 29.7\\
 &  &  & RCW &  & 0.009 & 0.173 & 0.166 & 0.939 & 22.0\\
 &  &  & CTW &  & 0.011 & 0.153 & 0.148 & 0.942 & \textemdash\\
 &  &  & IAT &  & 1.074 & 0.062 & 0.063 & 0.000 & \textemdash\\
\addlinespace[3pt]
\multirow{4}{*}{1.25} & \multirow{4}{*}{4.5} & \multirow{4}{*}{0.699} & CHW & \multirow{4}{*}{1.274} & 0.003 & 0.243 & 0.231 & 0.943 & 51.1\\
 &  &  & RCW &  & 0.002 & 0.216 & 0.203 & 0.933 & 38.1\\
 &  &  & CTW &  & 0.012 & 0.170 & 0.160 & 0.929 & \textemdash\\
 &  &  & IAT &  & 1.099 & 0.058 & 0.058 & 0.000 & \textemdash\\
\addlinespace[6pt]
\multirow{4}{*}{2} & \multirow{4}{*}{2.0} & \multirow{4}{*}{0.356} & CHW & \multirow{4}{*}{1.491} & 0.012 & 0.171 & 0.163 & 0.936 & 11.6\\
 &  &  & RCW &  & 0.011 & 0.167 & 0.160 & 0.938 & 7.9\\
 &  &  & CTW &  & 0.013 & 0.161 & 0.153 & 0.939 & \textemdash\\
 &  &  & IAT &  & 0.980 & 0.076 & 0.074 & 0.000 & \textemdash\\
\addlinespace[3pt]
\multirow{4}{*}{2} & \multirow{4}{*}{2.8} & \multirow{4}{*}{0.529} & CHW & \multirow{4}{*}{1.442} & 0.013 & 0.189 & 0.179 & 0.938 & 24.8\\
 &  &  & RCW &  & 0.011 & 0.178 & 0.170 & 0.941 & 15.9\\
 &  &  & CTW &  & 0.017 & 0.164 & 0.157 & 0.943 & \textemdash\\
 &  &  & IAT &  & 1.154 & 0.065 & 0.063 & 0.000 & \textemdash\\
\addlinespace[3pt]
\multirow{4}{*}{2} & \multirow{4}{*}{4.5} & \multirow{4}{*}{0.698} & CHW & \multirow{4}{*}{1.350} & 0.008 & 0.244 & 0.232 & 0.946 & 44.8\\
 &  &  & RCW &  & 0.005 & 0.215 & 0.202 & 0.945 & 28.5\\
 &  &  & CTW &  & 0.023 & 0.182 & 0.169 & 0.934 & \textemdash\\
 &  &  & IAT &  & 1.172 & 0.060 & 0.059 & 0.000 & \textemdash\\
\addlinespace[6pt]
\multirow{4}{*}{4} & \multirow{4}{*}{2.0} & \multirow{4}{*}{0.356} & CHW & \multirow{4}{*}{1.555} & 0.020 & 0.173 & 0.169 & 0.944 & 9.1\\
 &  &  & RCW &  & 0.018 & 0.170 & 0.165 & 0.944 & 5.3\\
 &  &  & CTW &  & 0.020 & 0.165 & 0.160 & 0.940 & \textemdash\\
 &  &  & IAT &  & 1.030 & 0.074 & 0.075 & 0.000 & \textemdash\\
\addlinespace[3pt]
\multirow{4}{*}{4} & \multirow{4}{*}{2.8} & \multirow{4}{*}{0.530} & CHW & \multirow{4}{*}{1.506} & 0.019 & 0.189 & 0.184 & 0.942 & 19.3\\
 &  &  & RCW &  & 0.017 & 0.181 & 0.174 & 0.945 & 11.1\\
 &  &  & CTW &  & 0.022 & 0.170 & 0.164 & 0.942 & \textemdash\\
 &  &  & IAT &  & 1.212 & 0.063 & 0.063 & 0.000 & \textemdash\\
\addlinespace[3pt]
\multirow{4}{*}{4} & \multirow{4}{*}{4.5} & \multirow{4}{*}{0.699} & CHW & \multirow{4}{*}{1.417} & 0.007 & 0.249 & 0.235 & 0.944 & 43.7\\
 &  &  & RCW &  & 0.004 & 0.212 & 0.203 & 0.948 & 22.4\\
 &  &  & CTW &  & 0.025 & 0.187 & 0.177 & 0.935 & \textemdash\\
 &  &  & IAT &  & 1.238 & 0.058 & 0.059 & 0.000 & \textemdash\\
\bottomrule
\end{tabular*}
\endgroup
\par\smallskip
\begin{minipage}{\textwidth}
\renewcommand{\baselinestretch}{1}
\fontsize{9}{11}\selectfont\raggedright
\(c_L\) is the mean realized censoring proportion, Abias denotes absolute bias, MCSD the Monte Carlo standard deviation, ASE the average estimated standard error, and ECP the empirical coverage probability of nominal \(95\%\) Wald intervals.
\par\smallskip
For a CHW or RCW row, VR (\%) is \(100\{1-(\mathrm{MCSD}_{\mathrm{CTW}}/\mathrm{MCSD}_{\star})^2\}\), where \(\star\) denotes that row's method. Positive values favor CTW. IAT assigns response \(1/2\) to every unresolved observed pair and gives every pair unit weight.
\end{minipage}
\end{table}
\begin{table}[!htbp]
\caption{Finite-sample performance of complete history weighting (CHW), resolved comparison weighting (RCW), conditional tie weighting (CTW), and indeterminate as tie (IAT) for the continuous covariate coefficient \(\beta_{L1}\) under the probit link with \(n=1000\). Death is prioritized over one nonfatal event. Results are based on 2,000 replications for each Gumbel dependence parameter \(\alpha\) and restriction time \(L\).}
\label{tab:sim-probit-n1000}
\centering
\begingroup
\renewcommand{\baselinestretch}{1}
\fontsize{10}{12}\selectfont
\renewcommand{\arraystretch}{1}
\setlength{\tabcolsep}{5pt}
\begin{tabular*}{\textwidth}{@{\extracolsep{\fill}}ccclrrrrrr@{}}
\toprule
$\alpha$ & $L$ & $c_L$ & Method & Truth & Abias & MCSD & ASE & ECP & VR (\%)\\
\midrule
\multirow{4}{*}{1.25} & \multirow{4}{*}{2.0} & \multirow{4}{*}{0.356} & CHW & \multirow{4}{*}{0.651} & 0.001 & 0.058 & 0.057 & 0.949 & 19.8\\
 &  &  & RCW &  & 0.001 & 0.057 & 0.056 & 0.947 & 16.7\\
 &  &  & CTW &  & 0.002 & 0.052 & 0.051 & 0.948 & \textemdash\\
 &  &  & IAT &  & 0.401 & 0.034 & 0.033 & 0.000 & \textemdash\\
\addlinespace[3pt]
\multirow{4}{*}{1.25} & \multirow{4}{*}{2.8} & \multirow{4}{*}{0.530} & CHW & \multirow{4}{*}{0.634} & 0.000 & 0.065 & 0.065 & 0.951 & 35.3\\
 &  &  & RCW &  & 0.001 & 0.062 & 0.062 & 0.947 & 30.1\\
 &  &  & CTW &  & 0.001 & 0.052 & 0.052 & 0.951 & \textemdash\\
 &  &  & IAT &  & 0.476 & 0.032 & 0.031 & 0.000 & \textemdash\\
\addlinespace[3pt]
\multirow{4}{*}{1.25} & \multirow{4}{*}{4.5} & \multirow{4}{*}{0.699} & CHW & \multirow{4}{*}{0.599} & 0.001 & 0.092 & 0.090 & 0.951 & 61.0\\
 &  &  & RCW &  & 0.002 & 0.083 & 0.081 & 0.946 & 52.4\\
 &  &  & CTW &  & 0.001 & 0.058 & 0.057 & 0.946 & \textemdash\\
 &  &  & IAT &  & 0.487 & 0.031 & 0.031 & 0.000 & \textemdash\\
\addlinespace[6pt]
\multirow{4}{*}{2} & \multirow{4}{*}{2.0} & \multirow{4}{*}{0.356} & CHW & \multirow{4}{*}{0.694} & 0.000 & 0.058 & 0.058 & 0.949 & 16.9\\
 &  &  & RCW &  & 0.000 & 0.057 & 0.056 & 0.938 & 14.4\\
 &  &  & CTW &  & 0.001 & 0.053 & 0.052 & 0.946 & \textemdash\\
 &  &  & IAT &  & 0.434 & 0.034 & 0.034 & 0.000 & \textemdash\\
\addlinespace[3pt]
\multirow{4}{*}{2} & \multirow{4}{*}{2.8} & \multirow{4}{*}{0.530} & CHW & \multirow{4}{*}{0.672} & 0.002 & 0.066 & 0.065 & 0.953 & 30.1\\
 &  &  & RCW &  & 0.002 & 0.062 & 0.062 & 0.954 & 22.7\\
 &  &  & CTW &  & 0.003 & 0.055 & 0.053 & 0.942 & \textemdash\\
 &  &  & IAT &  & 0.511 & 0.032 & 0.031 & 0.000 & \textemdash\\
\addlinespace[3pt]
\multirow{4}{*}{2} & \multirow{4}{*}{4.5} & \multirow{4}{*}{0.699} & CHW & \multirow{4}{*}{0.633} & 0.004 & 0.089 & 0.089 & 0.950 & 55.1\\
 &  &  & RCW &  & 0.002 & 0.081 & 0.080 & 0.951 & 45.9\\
 &  &  & CTW &  & 0.004 & 0.059 & 0.058 & 0.946 & \textemdash\\
 &  &  & IAT &  & 0.520 & 0.031 & 0.031 & 0.000 & \textemdash\\
\addlinespace[6pt]
\multirow{4}{*}{4} & \multirow{4}{*}{2.0} & \multirow{4}{*}{0.356} & CHW & \multirow{4}{*}{0.727} & 0.001 & 0.059 & 0.058 & 0.944 & 19.5\\
 &  &  & RCW &  & 0.001 & 0.058 & 0.057 & 0.944 & 16.8\\
 &  &  & CTW &  & 0.002 & 0.053 & 0.053 & 0.949 & \textemdash\\
 &  &  & IAT &  & 0.459 & 0.033 & 0.034 & 0.000 & \textemdash\\
\addlinespace[3pt]
\multirow{4}{*}{4} & \multirow{4}{*}{2.8} & \multirow{4}{*}{0.530} & CHW & \multirow{4}{*}{0.703} & 0.005 & 0.065 & 0.065 & 0.945 & 33.2\\
 &  &  & RCW &  & 0.004 & 0.063 & 0.062 & 0.945 & 28.6\\
 &  &  & CTW &  & 0.005 & 0.053 & 0.054 & 0.948 & \textemdash\\
 &  &  & IAT &  & 0.537 & 0.031 & 0.031 & 0.000 & \textemdash\\
\addlinespace[3pt]
\multirow{4}{*}{4} & \multirow{4}{*}{4.5} & \multirow{4}{*}{0.698} & CHW & \multirow{4}{*}{0.664} & 0.006 & 0.091 & 0.088 & 0.943 & 55.9\\
 &  &  & RCW &  & 0.003 & 0.081 & 0.079 & 0.945 & 43.8\\
 &  &  & CTW &  & 0.007 & 0.060 & 0.060 & 0.940 & \textemdash\\
 &  &  & IAT &  & 0.548 & 0.030 & 0.031 & 0.000 & \textemdash\\
\bottomrule
\end{tabular*}
\endgroup
\par\smallskip
\begin{minipage}{\textwidth}
\renewcommand{\baselinestretch}{1}
\fontsize{9}{11}\selectfont\raggedright
\(c_L\) is the mean realized censoring proportion, Abias denotes absolute bias, MCSD the Monte Carlo standard deviation, ASE the average estimated standard error, and ECP the empirical coverage probability of nominal \(95\%\) Wald intervals.
\par\smallskip
For a CHW or RCW row, VR (\%) is \(100\{1-(\mathrm{MCSD}_{\mathrm{CTW}}/\mathrm{MCSD}_{\star})^2\}\), where \(\star\) denotes that row's method. Positive values favor CTW. IAT assigns response \(1/2\) to every unresolved observed pair and gives every pair unit weight.
\end{minipage}
\end{table}

\begin{table}[!htbp]
\caption{Finite-sample performance of complete history weighting (CHW), resolved comparison weighting (RCW), conditional tie weighting (CTW), and indeterminate as tie (IAT) for the binary covariate coefficient \(\beta_{L2}\) under the probit link with \(n=1000\). Death is prioritized over one nonfatal event. Results are based on 2,000 replications for each Gumbel dependence parameter \(\alpha\) and restriction time \(L\).}
\label{tab:sim-probit-n1000-binary}
\centering
\begingroup
\renewcommand{\baselinestretch}{1}
\fontsize{10}{12}\selectfont
\renewcommand{\arraystretch}{1}
\setlength{\tabcolsep}{5pt}
\begin{tabular*}{\textwidth}{@{\extracolsep{\fill}}ccclrrrrrr@{}}
\toprule
$\alpha$ & $L$ & $c_L$ & Method & Truth & Abias & MCSD & ASE & ECP & VR (\%)\\
\midrule
\multirow{4}{*}{1.25} & \multirow{4}{*}{2.0} & \multirow{4}{*}{0.356} & CHW & \multirow{4}{*}{1.394} & 0.006 & 0.109 & 0.111 & 0.949 & 13.7\\
 &  &  & RCW &  & 0.006 & 0.108 & 0.108 & 0.946 & 11.3\\
 &  &  & CTW &  & 0.006 & 0.102 & 0.102 & 0.954 & \textemdash\\
 &  &  & IAT &  & 0.904 & 0.051 & 0.052 & 0.000 & \textemdash\\
\addlinespace[3pt]
\multirow{4}{*}{1.25} & \multirow{4}{*}{2.8} & \multirow{4}{*}{0.530} & CHW & \multirow{4}{*}{1.354} & 0.001 & 0.120 & 0.123 & 0.957 & 24.6\\
 &  &  & RCW &  & 0.001 & 0.116 & 0.117 & 0.953 & 18.9\\
 &  &  & CTW &  & 0.003 & 0.104 & 0.105 & 0.954 & \textemdash\\
 &  &  & IAT &  & 1.076 & 0.044 & 0.044 & 0.000 & \textemdash\\
\addlinespace[3pt]
\multirow{4}{*}{1.25} & \multirow{4}{*}{4.5} & \multirow{4}{*}{0.699} & CHW & \multirow{4}{*}{1.274} & 0.006 & 0.164 & 0.162 & 0.946 & 49.4\\
 &  &  & RCW &  & 0.007 & 0.146 & 0.144 & 0.942 & 36.7\\
 &  &  & CTW &  & 0.000 & 0.116 & 0.115 & 0.947 & \textemdash\\
 &  &  & IAT &  & 1.100 & 0.041 & 0.041 & 0.000 & \textemdash\\
\addlinespace[6pt]
\multirow{4}{*}{2} & \multirow{4}{*}{2.0} & \multirow{4}{*}{0.356} & CHW & \multirow{4}{*}{1.491} & 0.005 & 0.116 & 0.115 & 0.945 & 9.1\\
 &  &  & RCW &  & 0.005 & 0.113 & 0.113 & 0.951 & 5.1\\
 &  &  & CTW &  & 0.005 & 0.110 & 0.109 & 0.941 & \textemdash\\
 &  &  & IAT &  & 0.982 & 0.053 & 0.053 & 0.000 & \textemdash\\
\addlinespace[3pt]
\multirow{4}{*}{2} & \multirow{4}{*}{2.8} & \multirow{4}{*}{0.530} & CHW & \multirow{4}{*}{1.442} & 0.006 & 0.126 & 0.126 & 0.946 & 20.7\\
 &  &  & RCW &  & 0.006 & 0.120 & 0.120 & 0.949 & 11.6\\
 &  &  & CTW &  & 0.009 & 0.112 & 0.111 & 0.942 & \textemdash\\
 &  &  & IAT &  & 1.155 & 0.045 & 0.045 & 0.000 & \textemdash\\
\addlinespace[3pt]
\multirow{4}{*}{2} & \multirow{4}{*}{4.5} & \multirow{4}{*}{0.699} & CHW & \multirow{4}{*}{1.350} & 0.010 & 0.166 & 0.163 & 0.950 & 43.5\\
 &  &  & RCW &  & 0.008 & 0.148 & 0.143 & 0.950 & 28.8\\
 &  &  & CTW &  & 0.014 & 0.125 & 0.122 & 0.936 & \textemdash\\
 &  &  & IAT &  & 1.173 & 0.042 & 0.041 & 0.000 & \textemdash\\
\addlinespace[6pt]
\multirow{4}{*}{4} & \multirow{4}{*}{2.0} & \multirow{4}{*}{0.356} & CHW & \multirow{4}{*}{1.555} & 0.007 & 0.122 & 0.119 & 0.950 & 8.6\\
 &  &  & RCW &  & 0.007 & 0.120 & 0.117 & 0.944 & 5.5\\
 &  &  & CTW &  & 0.008 & 0.116 & 0.114 & 0.945 & \textemdash\\
 &  &  & IAT &  & 1.032 & 0.054 & 0.053 & 0.000 & \textemdash\\
\addlinespace[3pt]
\multirow{4}{*}{4} & \multirow{4}{*}{2.8} & \multirow{4}{*}{0.530} & CHW & \multirow{4}{*}{1.506} & 0.009 & 0.132 & 0.129 & 0.944 & 18.2\\
 &  &  & RCW &  & 0.008 & 0.127 & 0.123 & 0.943 & 11.1\\
 &  &  & CTW &  & 0.010 & 0.120 & 0.116 & 0.940 & \textemdash\\
 &  &  & IAT &  & 1.213 & 0.045 & 0.045 & 0.000 & \textemdash\\
\addlinespace[3pt]
\multirow{4}{*}{4} & \multirow{4}{*}{4.5} & \multirow{4}{*}{0.698} & CHW & \multirow{4}{*}{1.417} & 0.000 & 0.175 & 0.165 & 0.938 & 41.2\\
 &  &  & RCW &  & 0.001 & 0.153 & 0.144 & 0.942 & 23.0\\
 &  &  & CTW &  & 0.013 & 0.134 & 0.127 & 0.935 & \textemdash\\
 &  &  & IAT &  & 1.238 & 0.042 & 0.042 & 0.000 & \textemdash\\
\bottomrule
\end{tabular*}
\endgroup
\par\smallskip
\begin{minipage}{\textwidth}
\renewcommand{\baselinestretch}{1}
\fontsize{9}{11}\selectfont\raggedright
\(c_L\) is the mean realized censoring proportion, Abias denotes absolute bias, MCSD the Monte Carlo standard deviation, ASE the average estimated standard error, and ECP the empirical coverage probability of nominal \(95\%\) Wald intervals.
\par\smallskip
For a CHW or RCW row, VR (\%) is \(100\{1-(\mathrm{MCSD}_{\mathrm{CTW}}/\mathrm{MCSD}_{\star})^2\}\), where \(\star\) denotes that row's method. Positive values favor CTW. IAT assigns response \(1/2\) to every unresolved observed pair and gives every pair unit weight.
\end{minipage}
\end{table}
\FloatBarrier

\subsubsection{Identity link}

Tables \ref{tab:sim-identity-n200} and \ref{tab:sim-identity-n1000} give the remaining identity-link results. CTW had small absolute bias and lower MCSD than CHW and RCW, with the corresponding variance reductions reported in the VR column.

\begin{table}[!htbp]
\caption{Finite-sample performance of complete history weighting (CHW), resolved comparison weighting (RCW), conditional tie weighting (CTW), and indeterminate as tie (IAT) for the binary covariate coefficient \(\beta_{L,B}\) under the identity link with \(n=200\). Death is prioritized over one nonfatal event. For each Gumbel dependence parameter \(\alpha\) and restriction time \(L\), 2,000 datasets were generated.}
\label{tab:sim-identity-n200}
\centering
\begingroup
\renewcommand{\baselinestretch}{1}
\fontsize{10}{10.8}\selectfont
\renewcommand{\arraystretch}{1}
\setlength{\tabcolsep}{5pt}
\begin{tabular*}{\textwidth}{@{\extracolsep{\fill}}ccclrrrrrr@{}}
\toprule
$\alpha$ & $L$ & $c_L$ & Method & Truth & Abias & MCSD & ASE & ECP & VR (\%)\\
\midrule
\multirow{4}{*}{1.25} & \multirow{4}{*}{2.0} & \multirow{4}{*}{0.355} & CHW & \multirow{4}{*}{0.617} & 0.004 & 0.112 & 0.110 & 0.929 & 23.9\\
 &  &  & RCW &  & 0.003 & 0.108 & 0.108 & 0.931 & 19.4\\
 &  &  & CTW &  & 0.000 & 0.097 & 0.098 & 0.930 & \textemdash\\
 &  &  & IAT &  & 0.373 & 0.073 & 0.075 & 0.000 & \textemdash\\
\addlinespace[3pt]
\multirow{4}{*}{1.25} & \multirow{4}{*}{2.8} & \multirow{4}{*}{0.530} & CHW & \multirow{4}{*}{0.605} & 0.011 & 0.136 & 0.129 & 0.930 & 41.1\\
 &  &  & RCW &  & 0.007 & 0.128 & 0.122 & 0.926 & 33.7\\
 &  &  & CTW &  & 0.001 & 0.104 & 0.103 & 0.925 & \textemdash\\
 &  &  & IAT &  & 0.472 & 0.064 & 0.065 & 0.000 & \textemdash\\
\addlinespace[3pt]
\multirow{4}{*}{1.25} & \multirow{4}{*}{4.5} & \multirow{4}{*}{0.699} & CHW & \multirow{4}{*}{0.579} & 0.052 & 0.252 & 0.193 & 0.889 & 77.1\\
 &  &  & RCW &  & 0.025 & 0.189 & 0.164 & 0.888 & 59.0\\
 &  &  & CTW &  & 0.000 & 0.121 & 0.114 & 0.910 & \textemdash\\
 &  &  & IAT &  & 0.502 & 0.060 & 0.060 & 0.000 & \textemdash\\
\addlinespace[6pt]
\multirow{4}{*}{2} & \multirow{4}{*}{2.0} & \multirow{4}{*}{0.356} & CHW & \multirow{4}{*}{0.647} & 0.006 & 0.107 & 0.107 & 0.936 & 22.5\\
 &  &  & RCW &  & 0.005 & 0.103 & 0.104 & 0.937 & 16.3\\
 &  &  & CTW &  & 0.002 & 0.094 & 0.096 & 0.929 & \textemdash\\
 &  &  & IAT &  & 0.396 & 0.074 & 0.076 & 0.000 & \textemdash\\
\addlinespace[3pt]
\multirow{4}{*}{2} & \multirow{4}{*}{2.8} & \multirow{4}{*}{0.530} & CHW & \multirow{4}{*}{0.633} & 0.009 & 0.127 & 0.125 & 0.921 & 39.1\\
 &  &  & RCW &  & 0.005 & 0.117 & 0.117 & 0.931 & 27.7\\
 &  &  & CTW &  & 0.001 & 0.099 & 0.101 & 0.925 & \textemdash\\
 &  &  & IAT &  & 0.498 & 0.064 & 0.066 & 0.000 & \textemdash\\
\addlinespace[3pt]
\multirow{4}{*}{2} & \multirow{4}{*}{4.5} & \multirow{4}{*}{0.698} & CHW & \multirow{4}{*}{0.604} & 0.039 & 0.234 & 0.182 & 0.892 & 75.6\\
 &  &  & RCW &  & 0.018 & 0.172 & 0.154 & 0.898 & 55.0\\
 &  &  & CTW &  & 0.005 & 0.116 & 0.112 & 0.913 & \textemdash\\
 &  &  & IAT &  & 0.528 & 0.059 & 0.061 & 0.000 & \textemdash\\
\addlinespace[6pt]
\multirow{4}{*}{4} & \multirow{4}{*}{2.0} & \multirow{4}{*}{0.355} & CHW & \multirow{4}{*}{0.663} & 0.002 & 0.106 & 0.103 & 0.920 & 16.5\\
 &  &  & RCW &  & 0.002 & 0.103 & 0.100 & 0.925 & 11.5\\
 &  &  & CTW &  & 0.001 & 0.096 & 0.095 & 0.916 & \textemdash\\
 &  &  & IAT &  & 0.402 & 0.077 & 0.075 & 0.001 & \textemdash\\
\addlinespace[3pt]
\multirow{4}{*}{4} & \multirow{4}{*}{2.8} & \multirow{4}{*}{0.529} & CHW & \multirow{4}{*}{0.652} & 0.007 & 0.127 & 0.120 & 0.921 & 34.5\\
 &  &  & RCW &  & 0.005 & 0.117 & 0.113 & 0.924 & 23.4\\
 &  &  & CTW &  & 0.001 & 0.103 & 0.099 & 0.915 & \textemdash\\
 &  &  & IAT &  & 0.511 & 0.067 & 0.066 & 0.000 & \textemdash\\
\addlinespace[3pt]
\multirow{4}{*}{4} & \multirow{4}{*}{4.5} & \multirow{4}{*}{0.698} & CHW & \multirow{4}{*}{0.626} & 0.043 & 0.240 & 0.175 & 0.886 & 75.8\\
 &  &  & RCW &  & 0.023 & 0.172 & 0.149 & 0.898 & 52.7\\
 &  &  & CTW &  & 0.003 & 0.118 & 0.111 & 0.897 & \textemdash\\
 &  &  & IAT &  & 0.548 & 0.061 & 0.060 & 0.000 & \textemdash\\
\bottomrule
\end{tabular*}
\endgroup
\par\smallskip
\begin{minipage}{\textwidth}
\renewcommand{\baselinestretch}{1}
\fontsize{9}{11}\selectfont\raggedright
The coefficient \(\beta_{L,B}\) corresponds to the binary-covariate contrast \(B_i-B_j\) and is on the marginal net benefit scale. \(c_L\) is the mean realized censoring proportion, Abias denotes absolute bias, MCSD the Monte Carlo standard deviation, ASE the average estimated standard error, and ECP the empirical coverage probability of nominal \(95\%\) Wald intervals.
\par\smallskip
For a CHW or RCW row, VR (\%) is \(100\{1-(\mathrm{MCSD}_{\mathrm{CTW}}/\mathrm{MCSD}_{\star})^2\}\), where \(\star\) denotes that row's method. Positive values favor CTW. IAT assigns response \(1/2\) to every unresolved observed pair and gives every pair unit weight.
\end{minipage}
\end{table}
\begin{table}[!htbp]
\caption{Finite-sample performance of complete history weighting (CHW), resolved comparison weighting (RCW), conditional tie weighting (CTW), and indeterminate as tie (IAT) for the binary covariate coefficient \(\beta_{L,B}\) under the identity link with \(n=1000\). Death is prioritized over one nonfatal event. Results are based on 2,000 replications for each Gumbel dependence parameter \(\alpha\) and restriction time \(L\).}
\label{tab:sim-identity-n1000}
\centering
\begingroup
\renewcommand{\baselinestretch}{1}
\fontsize{10}{12}\selectfont
\renewcommand{\arraystretch}{1}
\setlength{\tabcolsep}{5pt}
\begin{tabular*}{\textwidth}{@{\extracolsep{\fill}}ccclrrrrrr@{}}
\toprule
$\alpha$ & $L$ & $c_L$ & Method & Truth & Abias & MCSD & ASE & ECP & VR (\%)\\
\midrule
\multirow{4}{*}{1.25} & \multirow{4}{*}{2.0} & \multirow{4}{*}{0.356} & CHW & \multirow{4}{*}{0.617} & 0.000 & 0.048 & 0.049 & 0.944 & 20.8\\
 &  &  & RCW &  & 0.000 & 0.047 & 0.048 & 0.948 & 17.0\\
 &  &  & CTW &  & 0.000 & 0.043 & 0.044 & 0.953 & \textemdash\\
 &  &  & IAT &  & 0.373 & 0.033 & 0.034 & 0.000 & \textemdash\\
\addlinespace[3pt]
\multirow{4}{*}{1.25} & \multirow{4}{*}{2.8} & \multirow{4}{*}{0.530} & CHW & \multirow{4}{*}{0.605} & 0.002 & 0.056 & 0.057 & 0.945 & 34.1\\
 &  &  & RCW &  & 0.002 & 0.054 & 0.055 & 0.948 & 27.8\\
 &  &  & CTW &  & 0.001 & 0.046 & 0.047 & 0.956 & \textemdash\\
 &  &  & IAT &  & 0.472 & 0.028 & 0.029 & 0.000 & \textemdash\\
\addlinespace[3pt]
\multirow{4}{*}{1.25} & \multirow{4}{*}{4.5} & \multirow{4}{*}{0.699} & CHW & \multirow{4}{*}{0.579} & 0.009 & 0.085 & 0.082 & 0.944 & 59.4\\
 &  &  & RCW &  & 0.007 & 0.076 & 0.074 & 0.935 & 48.9\\
 &  &  & CTW &  & 0.003 & 0.054 & 0.054 & 0.941 & \textemdash\\
 &  &  & IAT &  & 0.503 & 0.027 & 0.027 & 0.000 & \textemdash\\
\addlinespace[6pt]
\multirow{4}{*}{2} & \multirow{4}{*}{2.0} & \multirow{4}{*}{0.356} & CHW & \multirow{4}{*}{0.647} & 0.000 & 0.046 & 0.047 & 0.948 & 13.4\\
 &  &  & RCW &  & 0.000 & 0.045 & 0.046 & 0.947 & 9.6\\
 &  &  & CTW &  & 0.000 & 0.043 & 0.043 & 0.949 & \textemdash\\
 &  &  & IAT &  & 0.393 & 0.034 & 0.034 & 0.000 & \textemdash\\
\addlinespace[3pt]
\multirow{4}{*}{2} & \multirow{4}{*}{2.8} & \multirow{4}{*}{0.530} & CHW & \multirow{4}{*}{0.633} & 0.001 & 0.053 & 0.055 & 0.947 & 28.2\\
 &  &  & RCW &  & 0.001 & 0.051 & 0.052 & 0.950 & 20.7\\
 &  &  & CTW &  & 0.001 & 0.045 & 0.046 & 0.948 & \textemdash\\
 &  &  & IAT &  & 0.496 & 0.029 & 0.029 & 0.000 & \textemdash\\
\addlinespace[3pt]
\multirow{4}{*}{2} & \multirow{4}{*}{4.5} & \multirow{4}{*}{0.699} & CHW & \multirow{4}{*}{0.604} & 0.003 & 0.079 & 0.078 & 0.939 & 54.7\\
 &  &  & RCW &  & 0.001 & 0.070 & 0.069 & 0.937 & 42.8\\
 &  &  & CTW &  & 0.002 & 0.053 & 0.053 & 0.943 & \textemdash\\
 &  &  & IAT &  & 0.527 & 0.027 & 0.027 & 0.000 & \textemdash\\
\addlinespace[6pt]
\multirow{4}{*}{4} & \multirow{4}{*}{2.0} & \multirow{4}{*}{0.356} & CHW & \multirow{4}{*}{0.663} & 0.001 & 0.047 & 0.046 & 0.941 & 15.1\\
 &  &  & RCW &  & 0.001 & 0.046 & 0.045 & 0.945 & 11.1\\
 &  &  & CTW &  & 0.001 & 0.043 & 0.043 & 0.941 & \textemdash\\
 &  &  & IAT &  & 0.403 & 0.035 & 0.034 & 0.000 & \textemdash\\
\addlinespace[3pt]
\multirow{4}{*}{4} & \multirow{4}{*}{2.8} & \multirow{4}{*}{0.530} & CHW & \multirow{4}{*}{0.652} & 0.001 & 0.054 & 0.053 & 0.946 & 27.1\\
 &  &  & RCW &  & 0.001 & 0.051 & 0.050 & 0.942 & 19.6\\
 &  &  & CTW &  & 0.000 & 0.046 & 0.045 & 0.940 & \textemdash\\
 &  &  & IAT &  & 0.512 & 0.030 & 0.030 & 0.000 & \textemdash\\
\addlinespace[3pt]
\multirow{4}{*}{4} & \multirow{4}{*}{4.5} & \multirow{4}{*}{0.698} & CHW & \multirow{4}{*}{0.626} & 0.008 & 0.080 & 0.076 & 0.941 & 51.9\\
 &  &  & RCW &  & 0.005 & 0.070 & 0.066 & 0.934 & 37.1\\
 &  &  & CTW &  & 0.000 & 0.055 & 0.053 & 0.920 & \textemdash\\
 &  &  & IAT &  & 0.548 & 0.028 & 0.027 & 0.000 & \textemdash\\
\bottomrule
\end{tabular*}
\endgroup
\par\smallskip
\begin{minipage}{\textwidth}
\renewcommand{\baselinestretch}{1}
\fontsize{9}{11}\selectfont\raggedright
The coefficient \(\beta_{L,B}\) corresponds to the binary-covariate contrast \(B_i-B_j\) and is on the marginal net benefit scale. \(c_L\) is the mean realized censoring proportion, Abias denotes absolute bias, MCSD the Monte Carlo standard deviation, ASE the average estimated standard error, and ECP the empirical coverage probability of nominal \(95\%\) Wald intervals.
\par\smallskip
For a CHW or RCW row, VR (\%) is \(100\{1-(\mathrm{MCSD}_{\mathrm{CTW}}/\mathrm{MCSD}_{\star})^2\}\), where \(\star\) denotes that row's method. Positive values favor CTW. IAT assigns response \(1/2\) to every unresolved observed pair and gives every pair unit weight.
\end{minipage}
\end{table}
\FloatBarrier

\subsection{Sensitivity to the fitted copula}
\label{supp:copula-sensitivity}

We examined the effect of the fitted copula on CTW at \(n=200\) and \(n=500\). Event histories were generated from the Gumbel model at the same three dependence parameters and restriction times as in Section \ref{sec:simulation} of main text, with 2,000 replications per configuration. We fitted Gumbel, Clayton, and Frank copulas to the same records, using the same parametric forms for the two marginal functions. The marginal parameters were estimated jointly with the dependence parameter for each copula. The censoring model, covariates, and link-specific targets were kept the same as the main text. Each likelihood accounted for death without a preceding nonfatal event and for death ending nonfatal follow-up. The CHW and RCW estimates were the same in all three comparisons because neither method uses a joint event model.

The tables report signed bias, ECP, and the percentage variance reductions defined in Section \ref{sec:simulation}. Here \(\mathrm{VR}_{\mathrm{CHW}}\) and \(\mathrm{VR}_{\mathrm{RCW}}\) identify the comparator for each fitted CTW copula.

The fitted copula had little effect on bias or coverage in these settings. Relative to the Gumbel fit, the largest absolute changes in bias and ECP were \(0.009\) and \(0.003\), respectively. CTW had smaller empirical variance than CHW and RCW under every fitted copula in all settings. Across both sample sizes and all links, variance reductions relative to CHW ranged from \(9.1\%\) to \(77.1\%\) for Gumbel, from \(9.4\%\) to \(77.2\%\) for Clayton, and from \(9.4\%\) to \(77.3\%\) for Frank. The corresponding ranges relative to RCW were \(5.3\%\)--\(59.0\%\), \(5.7\%\)--\(59.3\%\), and \(5.6\%\)--\(59.3\%\). Clayton and Frank gave slightly smaller variances than Gumbel, with reductions of \(0.1\%\)--\(3.3\%\) and \(0.2\%\)--\(2.5\%\), respectively.

Tables \ref{tab:copula-logit-bias-ecp} and \ref{tab:copula-logit-bias-ecp-binary} give the logit results, including the variance reductions for each coefficient and configuration. Bias was positive for both coefficients under all three fitted copulas. The maximum absolute biases at \(n=200\) were \(0.093\), \(0.092\), and \(0.093\) for Gumbel, Clayton, and Frank, respectively, decreasing to \(0.045\), \(0.046\), and \(0.047\) at \(n=500\). At the larger sample size, ECP ranged from \(0.933\) to \(0.950\) for Gumbel, from \(0.934\) to \(0.951\) for Clayton, and from \(0.933\) to \(0.951\) for Frank.

\begin{table}[!htbp]
\caption{Signed bias, empirical coverage probability (ECP) of nominal 95\% Wald confidence intervals, and percentage reduction in empirical variance for conditional tie weighting (CTW) under the logit link for the continuous covariate coefficient $\beta_{L1}$. Data were generated with a Gumbel copula at the indicated $\alpha$ and restriction time $L$, with 2,000 generated samples per configuration.}
\label{tab:copula-logit-bias-ecp}
\centering
\begingroup
\renewcommand{\baselinestretch}{1}
\fontsize{10}{12}\selectfont
\setlength{\tabcolsep}{4pt}
\begin{tabular}{@{}cclrrrr@{\hspace{1em}}rrrr@{}}
\toprule
 & & & \multicolumn{4}{c}{$n=200$} & \multicolumn{4}{c}{$n=500$}\\
\cmidrule(lr){4-7}\cmidrule(lr){8-11}
$\alpha$ & $L$ & Copula & Bias & ECP & $\mathrm{VR}_{\mathrm{CHW}}$ & $\mathrm{VR}_{\mathrm{RCW}}$ & Bias & ECP & $\mathrm{VR}_{\mathrm{CHW}}$ & $\mathrm{VR}_{\mathrm{RCW}}$\\
\midrule
\multirow{9}{*}{1.25} & \multirow{3}{*}{2.0} & Gumbel & 0.016 & 0.939 & 23.9 & 20.8 & 0.011 & 0.948 & 22.8 & 20.0\\
 &  & Clayton & 0.015 & 0.938 & 24.1 & 21.0 & 0.010 & 0.948 & 23.0 & 20.2\\
 &  & Frank & 0.016 & 0.939 & 24.1 & 21.0 & 0.011 & 0.948 & 23.0 & 20.2\\
\addlinespace[1.8pt]
 & \multirow{3}{*}{2.8} & Gumbel & 0.015 & 0.935 & 37.6 & 32.5 & 0.010 & 0.944 & 36.5 & 32.5\\
 &  & Clayton & 0.013 & 0.936 & 38.0 & 32.9 & 0.007 & 0.943 & 37.0 & 33.0\\
 &  & Frank & 0.014 & 0.935 & 37.9 & 32.9 & 0.009 & 0.944 & 36.9 & 32.9\\
\addlinespace[1.8pt]
 & \multirow{3}{*}{4.5} & Gumbel & 0.017 & 0.940 & 67.2 & 55.2 & 0.010 & 0.934 & 60.4 & 51.4\\
 &  & Clayton & 0.012 & 0.940 & 67.7 & 55.9 & 0.005 & 0.936 & 61.0 & 52.2\\
 &  & Frank & 0.016 & 0.940 & 67.6 & 55.7 & 0.008 & 0.933 & 60.9 & 52.0\\
\addlinespace[3pt]
\multirow{9}{*}{2} & \multirow{3}{*}{2.0} & Gumbel & 0.014 & 0.950 & 22.3 & 18.1 & 0.007 & 0.939 & 18.6 & 15.0\\
 &  & Clayton & 0.013 & 0.950 & 22.7 & 18.5 & 0.006 & 0.940 & 19.1 & 15.5\\
 &  & Frank & 0.014 & 0.950 & 22.6 & 18.4 & 0.006 & 0.940 & 19.1 & 15.4\\
\addlinespace[1.8pt]
 & \multirow{3}{*}{2.8} & Gumbel & 0.021 & 0.950 & 37.7 & 30.4 & 0.009 & 0.950 & 35.1 & 27.9\\
 &  & Clayton & 0.019 & 0.949 & 38.5 & 31.3 & 0.007 & 0.951 & 36.2 & 29.1\\
 &  & Frank & 0.020 & 0.950 & 38.3 & 31.1 & 0.008 & 0.951 & 36.0 & 28.9\\
\addlinespace[1.8pt]
 & \multirow{3}{*}{4.5} & Gumbel & 0.026 & 0.949 & 67.5 & 56.4 & 0.012 & 0.947 & 60.2 & 48.9\\
 &  & Clayton & 0.021 & 0.950 & 68.4 & 57.6 & 0.007 & 0.946 & 61.5 & 50.6\\
 &  & Frank & 0.024 & 0.949 & 68.2 & 57.3 & 0.010 & 0.946 & 61.3 & 50.2\\
\addlinespace[3pt]
\multirow{9}{*}{4} & \multirow{3}{*}{2.0} & Gumbel & 0.012 & 0.945 & 17.4 & 15.2 & 0.005 & 0.944 & 16.2 & 13.3\\
 &  & Clayton & 0.011 & 0.944 & 17.8 & 15.5 & 0.004 & 0.943 & 16.5 & 13.6\\
 &  & Frank & 0.012 & 0.945 & 17.7 & 15.5 & 0.004 & 0.944 & 16.5 & 13.6\\
\addlinespace[1.8pt]
 & \multirow{3}{*}{2.8} & Gumbel & 0.017 & 0.943 & 33.3 & 27.6 & 0.011 & 0.947 & 30.0 & 23.9\\
 &  & Clayton & 0.017 & 0.942 & 34.0 & 28.4 & 0.010 & 0.945 & 30.8 & 24.7\\
 &  & Frank & 0.017 & 0.942 & 33.9 & 28.2 & 0.010 & 0.946 & 30.6 & 24.6\\
\addlinespace[1.8pt]
 & \multirow{3}{*}{4.5} & Gumbel & 0.025 & 0.940 & 63.1 & 47.8 & 0.012 & 0.936 & 55.3 & 43.5\\
 &  & Clayton & 0.026 & 0.938 & 63.7 & 48.7 & 0.013 & 0.934 & 56.2 & 44.6\\
 &  & Frank & 0.026 & 0.940 & 63.6 & 48.5 & 0.013 & 0.934 & 56.0 & 44.4\\
\bottomrule
\end{tabular}
\endgroup
\par\smallskip
\begin{minipage}{0.98\linewidth}
\renewcommand{\baselinestretch}{1}
\fontsize{9}{11}\selectfont\raggedright
For \(\star\in\{\mathrm{CHW},\mathrm{RCW}\}\), \(\mathrm{VR}_{\star}=100\{1-(\mathrm{MCSD}_{\mathrm{CTW}}/\mathrm{MCSD}_{\star})^2\}\) is the percentage reduction in empirical variance with the indicated CTW copula fit relative to \(\star\). Positive values favor CTW. 
\end{minipage}
\end{table}

\begin{table}[!htbp]
\caption{Signed bias, empirical coverage probability (ECP) of nominal 95\% Wald confidence intervals, and percentage reduction in empirical variance for conditional tie weighting (CTW) under the logit link for the binary covariate coefficient $\beta_{L2}$. Data were generated with a Gumbel copula at the indicated $\alpha$ and restriction time $L$, with 2,000 generated samples per configuration. }
\label{tab:copula-logit-bias-ecp-binary}
\centering
\begingroup
\renewcommand{\baselinestretch}{1}
\fontsize{10}{12}\selectfont
\setlength{\tabcolsep}{4pt}
\begin{tabular}{@{}cclrrrr@{\hspace{1em}}rrrr@{}}
\toprule
 & & & \multicolumn{4}{c}{$n=200$} & \multicolumn{4}{c}{$n=500$}\\
\cmidrule(lr){4-7}\cmidrule(lr){8-11}
$\alpha$ & $L$ & Copula & Bias & ECP & $\mathrm{VR}_{\mathrm{CHW}}$ & $\mathrm{VR}_{\mathrm{RCW}}$ & Bias & ECP & $\mathrm{VR}_{\mathrm{CHW}}$ & $\mathrm{VR}_{\mathrm{RCW}}$\\
\midrule
\multirow{9}{*}{1.25} & \multirow{3}{*}{2.0} & Gumbel & 0.051 & 0.937 & 19.5 & 16.4 & 0.022 & 0.947 & 18.0 & 13.6\\
 &  & Clayton & 0.050 & 0.937 & 19.7 & 16.5 & 0.021 & 0.947 & 18.1 & 13.8\\
 &  & Frank & 0.051 & 0.937 & 19.7 & 16.5 & 0.022 & 0.947 & 18.1 & 13.7\\
\addlinespace[1.8pt]
 & \multirow{3}{*}{2.8} & Gumbel & 0.049 & 0.930 & 32.6 & 26.8 & 0.019 & 0.944 & 31.2 & 23.7\\
 &  & Clayton & 0.045 & 0.930 & 33.0 & 27.2 & 0.015 & 0.944 & 31.6 & 24.1\\
 &  & Frank & 0.047 & 0.930 & 32.9 & 27.2 & 0.017 & 0.944 & 31.6 & 24.1\\
\addlinespace[1.8pt]
 & \multirow{3}{*}{4.5} & Gumbel & 0.060 & 0.920 & 67.1 & 47.5 & 0.021 & 0.933 & 52.7 & 40.5\\
 &  & Clayton & 0.052 & 0.921 & 67.5 & 48.2 & 0.012 & 0.934 & 53.3 & 41.3\\
 &  & Frank & 0.058 & 0.921 & 67.4 & 48.0 & 0.019 & 0.934 & 53.2 & 41.1\\
\addlinespace[3pt]
\multirow{9}{*}{2} & \multirow{3}{*}{2.0} & Gumbel & 0.045 & 0.946 & 17.7 & 13.6 & 0.023 & 0.941 & 12.8 & 9.1\\
 &  & Clayton & 0.044 & 0.946 & 18.0 & 13.9 & 0.022 & 0.940 & 13.3 & 9.6\\
 &  & Frank & 0.045 & 0.946 & 18.0 & 13.8 & 0.022 & 0.941 & 13.2 & 9.5\\
\addlinespace[1.8pt]
 & \multirow{3}{*}{2.8} & Gumbel & 0.062 & 0.943 & 31.3 & 22.0 & 0.030 & 0.941 & 26.2 & 17.6\\
 &  & Clayton & 0.058 & 0.944 & 31.9 & 22.7 & 0.026 & 0.941 & 27.1 & 18.6\\
 &  & Frank & 0.060 & 0.945 & 31.8 & 22.6 & 0.029 & 0.940 & 27.0 & 18.5\\
\addlinespace[1.8pt]
 & \multirow{3}{*}{4.5} & Gumbel & 0.089 & 0.927 & 64.8 & 43.5 & 0.042 & 0.937 & 46.2 & 30.8\\
 &  & Clayton & 0.080 & 0.928 & 65.5 & 44.5 & 0.034 & 0.937 & 47.5 & 32.5\\
 &  & Frank & 0.085 & 0.929 & 65.3 & 44.3 & 0.040 & 0.937 & 47.2 & 32.1\\
\addlinespace[3pt]
\multirow{9}{*}{4} & \multirow{3}{*}{2.0} & Gumbel & 0.063 & 0.938 & 10.8 & 7.4 & 0.035 & 0.942 & 10.2 & 6.5\\
 &  & Clayton & 0.062 & 0.938 & 11.1 & 7.6 & 0.034 & 0.941 & 10.5 & 6.8\\
 &  & Frank & 0.062 & 0.938 & 11.1 & 7.6 & 0.035 & 0.941 & 10.5 & 6.8\\
\addlinespace[1.8pt]
 & \multirow{3}{*}{2.8} & Gumbel & 0.073 & 0.932 & 23.6 & 14.5 & 0.038 & 0.944 & 21.0 & 13.0\\
 &  & Clayton & 0.071 & 0.931 & 24.3 & 15.3 & 0.037 & 0.943 & 21.8 & 13.9\\
 &  & Frank & 0.072 & 0.931 & 24.2 & 15.2 & 0.038 & 0.944 & 21.7 & 13.7\\
\addlinespace[1.8pt]
 & \multirow{3}{*}{4.5} & Gumbel & 0.093 & 0.927 & 61.1 & 35.3 & 0.045 & 0.938 & 45.3 & 24.7\\
 &  & Clayton & 0.092 & 0.927 & 61.9 & 36.5 & 0.046 & 0.935 & 46.3 & 26.1\\
 &  & Frank & 0.093 & 0.926 & 61.8 & 36.3 & 0.047 & 0.935 & 46.1 & 25.8\\
\bottomrule
\end{tabular}
\endgroup
\par\smallskip
\begin{minipage}{0.98\linewidth}
\renewcommand{\baselinestretch}{1}
\fontsize{9}{11}\selectfont\raggedright
For \(\star\in\{\mathrm{CHW},\mathrm{RCW}\}\), \(\mathrm{VR}_{\star}=100\{1-(\mathrm{MCSD}_{\mathrm{CTW}}/\mathrm{MCSD}_{\star})^2\}\) is the percentage reduction in empirical variance with the indicated CTW copula fit relative to \(\star\). Positive values favor CTW. 
\end{minipage}
\end{table}
\FloatBarrier

The probit results in Tables \ref{tab:copula-probit-bias-ecp} and~\ref{tab:copula-probit-bias-ecp-binary} showed the same pattern. The maximum absolute biases for Gumbel, Clayton, and Frank were \(0.052\), \(0.051\), and \(0.052\) at \(n=200\), compared with \(0.025\), \(0.026\), and \(0.027\) at \(n=500\). At \(n=500\), ECP ranged from \(0.929\) to \(0.946\) for Gumbel and from \(0.929\) to \(0.945\) for both alternatives. CTW had lower empirical variance than CHW and RCW under each fitted copula.

\begin{table}[!htbp]
\caption{Signed bias, empirical coverage probability (ECP) of nominal 95\% Wald confidence intervals, and percentage reduction in empirical variance for conditional tie weighting (CTW) under the probit link for the continuous covariate coefficient $\beta_{L1}$. Data were generated with a Gumbel copula at the indicated $\alpha$ and restriction time $L$, with 2,000 generated samples per configuration.}
\label{tab:copula-probit-bias-ecp}
\centering
\begingroup
\renewcommand{\baselinestretch}{1}
\fontsize{10}{12}\selectfont
\setlength{\tabcolsep}{4pt}
\begin{tabular}{@{}cclrrrr@{\hspace{1em}}rrrr@{}}
\toprule
 & & & \multicolumn{4}{c}{$n=200$} & \multicolumn{4}{c}{$n=500$}\\
\cmidrule(lr){4-7}\cmidrule(lr){8-11}
$\alpha$ & $L$ & Copula & Bias & ECP & $\mathrm{VR}_{\mathrm{CHW}}$ & $\mathrm{VR}_{\mathrm{RCW}}$ & Bias & ECP & $\mathrm{VR}_{\mathrm{CHW}}$ & $\mathrm{VR}_{\mathrm{RCW}}$\\
\midrule
\multirow{9}{*}{1.25} & \multirow{3}{*}{2.0} & Gumbel & 0.010 & 0.934 & 23.5 & 20.4 & 0.007 & 0.943 & 22.3 & 19.8\\
 &  & Clayton & 0.010 & 0.934 & 23.6 & 20.5 & 0.007 & 0.944 & 22.5 & 20.0\\
 &  & Frank & 0.010 & 0.934 & 23.6 & 20.5 & 0.007 & 0.943 & 22.5 & 19.9\\
\addlinespace[1.8pt]
 & \multirow{3}{*}{2.8} & Gumbel & 0.010 & 0.935 & 37.0 & 31.8 & 0.006 & 0.941 & 35.7 & 31.8\\
 &  & Clayton & 0.009 & 0.935 & 37.4 & 32.2 & 0.005 & 0.940 & 36.2 & 32.3\\
 &  & Frank & 0.009 & 0.935 & 37.3 & 32.1 & 0.006 & 0.940 & 36.1 & 32.3\\
\addlinespace[1.8pt]
 & \multirow{3}{*}{4.5} & Gumbel & 0.011 & 0.934 & 66.5 & 54.5 & 0.006 & 0.933 & 59.8 & 50.8\\
 &  & Clayton & 0.008 & 0.935 & 67.0 & 55.1 & 0.003 & 0.932 & 60.5 & 51.6\\
 &  & Frank & 0.010 & 0.934 & 66.9 & 55.0 & 0.005 & 0.932 & 60.3 & 51.4\\
\addlinespace[3pt]
\multirow{9}{*}{2} & \multirow{3}{*}{2.0} & Gumbel & 0.009 & 0.948 & 21.5 & 17.2 & 0.004 & 0.938 & 17.9 & 14.3\\
 &  & Clayton & 0.009 & 0.948 & 21.8 & 17.6 & 0.004 & 0.937 & 18.3 & 14.8\\
 &  & Frank & 0.009 & 0.948 & 21.8 & 17.5 & 0.004 & 0.938 & 18.3 & 14.7\\
\addlinespace[1.8pt]
 & \multirow{3}{*}{2.8} & Gumbel & 0.013 & 0.946 & 36.9 & 29.6 & 0.006 & 0.943 & 34.5 & 27.4\\
 &  & Clayton & 0.012 & 0.945 & 37.7 & 30.5 & 0.005 & 0.945 & 35.5 & 28.5\\
 &  & Frank & 0.012 & 0.945 & 37.5 & 30.3 & 0.005 & 0.943 & 35.3 & 28.3\\
\addlinespace[1.8pt]
 & \multirow{3}{*}{4.5} & Gumbel & 0.015 & 0.946 & 67.0 & 56.0 & 0.007 & 0.941 & 59.8 & 48.5\\
 &  & Clayton & 0.013 & 0.947 & 67.9 & 57.2 & 0.004 & 0.942 & 61.1 & 50.1\\
 &  & Frank & 0.014 & 0.947 & 67.7 & 56.9 & 0.006 & 0.942 & 60.8 & 49.7\\
\addlinespace[3pt]
\multirow{9}{*}{4} & \multirow{3}{*}{2.0} & Gumbel & 0.008 & 0.940 & 16.9 & 14.7 & 0.003 & 0.946 & 15.5 & 12.7\\
 &  & Clayton & 0.008 & 0.941 & 17.2 & 15.0 & 0.003 & 0.945 & 15.8 & 13.0\\
 &  & Frank & 0.008 & 0.941 & 17.2 & 14.9 & 0.003 & 0.945 & 15.8 & 13.0\\
\addlinespace[1.8pt]
 & \multirow{3}{*}{2.8} & Gumbel & 0.012 & 0.943 & 32.7 & 27.0 & 0.007 & 0.944 & 29.0 & 23.2\\
 &  & Clayton & 0.011 & 0.941 & 33.4 & 27.7 & 0.006 & 0.945 & 29.8 & 24.0\\
 &  & Frank & 0.011 & 0.942 & 33.2 & 27.6 & 0.007 & 0.944 & 29.7 & 23.8\\
\addlinespace[1.8pt]
 & \multirow{3}{*}{4.5} & Gumbel & 0.016 & 0.937 & 62.4 & 47.1 & 0.007 & 0.934 & 54.5 & 43.0\\
 &  & Clayton & 0.016 & 0.937 & 63.0 & 48.1 & 0.008 & 0.932 & 55.4 & 44.1\\
 &  & Frank & 0.016 & 0.937 & 62.9 & 47.9 & 0.008 & 0.933 & 55.2 & 43.9\\
\bottomrule
\end{tabular}
\endgroup
\par\smallskip
\begin{minipage}{0.98\linewidth}
\renewcommand{\baselinestretch}{1}
\fontsize{9}{11}\selectfont\raggedright
For \(\star\in\{\mathrm{CHW},\mathrm{RCW}\}\), \(\mathrm{VR}_{\star}=100\{1-(\mathrm{MCSD}_{\mathrm{CTW}}/\mathrm{MCSD}_{\star})^2\}\) is the percentage reduction in empirical variance with the indicated CTW copula fit relative to \(\star\). Positive values favor CTW. 
\end{minipage}
\end{table}

\begin{table}[!htbp]
\caption{Signed bias, empirical coverage probability (ECP) of nominal 95\% Wald confidence intervals, and percentage reduction in empirical variance for conditional tie weighting (CTW) under the probit link for the binary covariate coefficient $\beta_{L2}$. Data were generated with a Gumbel copula at the indicated $\alpha$ and restriction time $L$, with 2,000 generated samples per configuration.}
\label{tab:copula-probit-bias-ecp-binary}
\centering
\begingroup
\renewcommand{\baselinestretch}{1}
\fontsize{10}{12}\selectfont
\setlength{\tabcolsep}{4pt}
\begin{tabular}{@{}cclrrrr@{\hspace{1em}}rrrr@{}}
\toprule
 & & & \multicolumn{4}{c}{$n=200$} & \multicolumn{4}{c}{$n=500$}\\
\cmidrule(lr){4-7}\cmidrule(lr){8-11}
$\alpha$ & $L$ & Copula & Bias & ECP & $\mathrm{VR}_{\mathrm{CHW}}$ & $\mathrm{VR}_{\mathrm{RCW}}$ & Bias & ECP & $\mathrm{VR}_{\mathrm{CHW}}$ & $\mathrm{VR}_{\mathrm{RCW}}$\\
\midrule
\multirow{9}{*}{1.25} & \multirow{3}{*}{2.0} & Gumbel & 0.029 & 0.932 & 18.7 & 15.3 & 0.013 & 0.943 & 16.9 & 12.6\\
 &  & Clayton & 0.029 & 0.932 & 18.8 & 15.5 & 0.013 & 0.942 & 17.1 & 12.7\\
 &  & Frank & 0.029 & 0.932 & 18.8 & 15.4 & 0.013 & 0.943 & 17.1 & 12.7\\
\addlinespace[1.8pt]
 & \multirow{3}{*}{2.8} & Gumbel & 0.028 & 0.927 & 31.3 & 25.1 & 0.011 & 0.942 & 29.7 & 22.0\\
 &  & Clayton & 0.026 & 0.928 & 31.7 & 25.4 & 0.009 & 0.942 & 30.1 & 22.4\\
 &  & Frank & 0.027 & 0.927 & 31.6 & 25.4 & 0.010 & 0.942 & 30.0 & 22.4\\
\addlinespace[1.8pt]
 & \multirow{3}{*}{4.5} & Gumbel & 0.033 & 0.919 & 66.5 & 45.7 & 0.012 & 0.929 & 51.1 & 38.1\\
 &  & Clayton & 0.029 & 0.919 & 66.9 & 46.3 & 0.007 & 0.929 & 51.7 & 38.9\\
 &  & Frank & 0.032 & 0.919 & 66.8 & 46.2 & 0.011 & 0.929 & 51.6 & 38.7\\
\addlinespace[3pt]
\multirow{9}{*}{2} & \multirow{3}{*}{2.0} & Gumbel & 0.025 & 0.941 & 16.2 & 11.8 & 0.013 & 0.939 & 11.6 & 7.9\\
 &  & Clayton & 0.025 & 0.940 & 16.4 & 12.1 & 0.012 & 0.940 & 12.0 & 8.3\\
 &  & Frank & 0.025 & 0.941 & 16.4 & 12.1 & 0.012 & 0.940 & 11.9 & 8.3\\
\addlinespace[1.8pt]
 & \multirow{3}{*}{2.8} & Gumbel & 0.034 & 0.941 & 29.4 & 19.6 & 0.017 & 0.943 & 24.8 & 15.9\\
 &  & Clayton & 0.032 & 0.942 & 30.0 & 20.2 & 0.015 & 0.944 & 25.7 & 16.9\\
 &  & Frank & 0.033 & 0.943 & 29.9 & 20.1 & 0.016 & 0.943 & 25.5 & 16.7\\
\addlinespace[1.8pt]
 & \multirow{3}{*}{4.5} & Gumbel & 0.049 & 0.927 & 63.5 & 40.4 & 0.023 & 0.934 & 44.8 & 28.5\\
 &  & Clayton & 0.044 & 0.930 & 64.2 & 41.4 & 0.019 & 0.935 & 46.0 & 30.1\\
 &  & Frank & 0.047 & 0.929 & 64.0 & 41.2 & 0.022 & 0.934 & 45.8 & 29.8\\
\addlinespace[3pt]
\multirow{9}{*}{4} & \multirow{3}{*}{2.0} & Gumbel & 0.035 & 0.935 & 9.5 & 5.9 & 0.020 & 0.940 & 9.1 & 5.3\\
 &  & Clayton & 0.035 & 0.935 & 9.8 & 6.2 & 0.019 & 0.940 & 9.4 & 5.7\\
 &  & Frank & 0.035 & 0.935 & 9.7 & 6.1 & 0.019 & 0.940 & 9.4 & 5.6\\
\addlinespace[1.8pt]
 & \multirow{3}{*}{2.8} & Gumbel & 0.042 & 0.930 & 22.0 & 12.5 & 0.022 & 0.942 & 19.3 & 11.1\\
 &  & Clayton & 0.041 & 0.929 & 22.6 & 13.2 & 0.022 & 0.944 & 20.1 & 11.9\\
 &  & Frank & 0.041 & 0.929 & 22.5 & 13.1 & 0.022 & 0.943 & 19.9 & 11.8\\
\addlinespace[1.8pt]
 & \multirow{3}{*}{4.5} & Gumbel & 0.052 & 0.922 & 60.4 & 32.6 & 0.025 & 0.935 & 43.7 & 22.4\\
 &  & Clayton & 0.051 & 0.923 & 61.0 & 33.7 & 0.026 & 0.935 & 44.7 & 23.7\\
 &  & Frank & 0.052 & 0.923 & 60.9 & 33.6 & 0.027 & 0.935 & 44.5 & 23.4\\
\bottomrule
\end{tabular}
\endgroup
\par\smallskip
\begin{minipage}{0.98\linewidth}
\renewcommand{\baselinestretch}{1}
\fontsize{9}{11}\selectfont\raggedright
For \(\star\in\{\mathrm{CHW},\mathrm{RCW}\}\), \(\mathrm{VR}_{\star}=100\{1-(\mathrm{MCSD}_{\mathrm{CTW}}/\mathrm{MCSD}_{\star})^2\}\) is the percentage reduction in empirical variance with the indicated CTW copula fit relative to \(\star\), computed from unrounded MCSDs. Positive values favor CTW. 
\end{minipage}
\end{table}
\FloatBarrier

For the identity link, Table \ref{tab:copula-identity-bias-ecp} shows small bias under all three copulas and gives the variance reductions relative to CHW and RCW. Maximum absolute bias was \(0.005\), \(0.004\), and \(0.005\) for Gumbel, Clayton, and Frank at \(n=200\), and \(0.003\) for each at \(n=500\). ECP also changed little across fitted copulas. The largest variance reductions relative to RCW were \(59.0\%\), \(59.3\%\), and \(59.3\%\) for Gumbel, Clayton, and Frank, respectively, at \(n=200\), \(\alpha=1.25\), and \(L=4.5\).

\begin{table}[!htbp]
\caption{Signed bias, empirical coverage probability (ECP) of nominal 95\% Wald confidence intervals, and percentage reduction in empirical variance for conditional tie weighting (CTW) under the identity link for the binary covariate coefficient $\beta_{L,B}$. Data were generated with a Gumbel copula at the indicated $\alpha$ and restriction time $L$, with 2,000 generated samples per configuration. }
\label{tab:copula-identity-bias-ecp}
\centering
\begingroup
\renewcommand{\baselinestretch}{1}
\fontsize{10}{12}\selectfont
\setlength{\tabcolsep}{4pt}
\begin{tabular}{@{}cclrrrr@{\hspace{1em}}rrrr@{}}
\toprule
 & & & \multicolumn{4}{c}{$n=200$} & \multicolumn{4}{c}{$n=500$}\\
\cmidrule(lr){4-7}\cmidrule(lr){8-11}
$\alpha$ & $L$ & Copula & Bias & ECP & $\mathrm{VR}_{\mathrm{CHW}}$ & $\mathrm{VR}_{\mathrm{RCW}}$ & Bias & ECP & $\mathrm{VR}_{\mathrm{CHW}}$ & $\mathrm{VR}_{\mathrm{RCW}}$\\
\midrule
\multirow{9}{*}{1.25} & \multirow{3}{*}{2.0} & Gumbel & 0.000 & 0.930 & 23.9 & 19.4 & -0.001 & 0.938 & 19.6 & 15.7\\
 &  & Clayton & 0.000 & 0.930 & 23.9 & 19.5 & -0.001 & 0.938 & 19.7 & 15.8\\
 &  & Frank & 0.000 & 0.930 & 24.0 & 19.5 & -0.001 & 0.938 & 19.7 & 15.8\\
\addlinespace[1.8pt]
 & \multirow{3}{*}{2.8} & Gumbel & -0.001 & 0.925 & 41.1 & 33.7 & -0.002 & 0.944 & 35.8 & 28.5\\
 &  & Clayton & -0.001 & 0.925 & 41.3 & 33.9 & -0.002 & 0.944 & 36.1 & 28.8\\
 &  & Frank & -0.001 & 0.924 & 41.4 & 34.0 & -0.002 & 0.944 & 36.2 & 28.9\\
\addlinespace[1.8pt]
 & \multirow{3}{*}{4.5} & Gumbel & 0.000 & 0.910 & 77.1 & 59.0 & -0.002 & 0.926 & 59.2 & 47.2\\
 &  & Clayton & -0.001 & 0.912 & 77.2 & 59.3 & -0.003 & 0.927 & 59.5 & 47.7\\
 &  & Frank & 0.000 & 0.911 & 77.3 & 59.3 & -0.002 & 0.926 & 59.6 & 47.8\\
\addlinespace[3pt]
\multirow{9}{*}{2} & \multirow{3}{*}{2.0} & Gumbel & -0.002 & 0.929 & 22.5 & 16.3 & 0.000 & 0.936 & 16.7 & 12.4\\
 &  & Clayton & -0.002 & 0.930 & 22.7 & 16.5 & 0.000 & 0.937 & 17.0 & 12.8\\
 &  & Frank & -0.002 & 0.929 & 22.7 & 16.5 & 0.000 & 0.937 & 17.0 & 12.8\\
\addlinespace[1.8pt]
 & \multirow{3}{*}{2.8} & Gumbel & 0.001 & 0.925 & 39.1 & 27.7 & 0.001 & 0.934 & 33.6 & 25.0\\
 &  & Clayton & 0.000 & 0.927 & 39.6 & 28.2 & 0.000 & 0.935 & 34.2 & 25.7\\
 &  & Frank & 0.000 & 0.926 & 39.6 & 28.3 & 0.001 & 0.935 & 34.3 & 25.8\\
\addlinespace[1.8pt]
 & \multirow{3}{*}{4.5} & Gumbel & 0.005 & 0.913 & 75.6 & 55.0 & 0.003 & 0.919 & 55.5 & 42.2\\
 &  & Clayton & 0.004 & 0.912 & 76.0 & 55.7 & 0.002 & 0.920 & 56.4 & 43.3\\
 &  & Frank & 0.005 & 0.913 & 76.0 & 55.6 & 0.002 & 0.919 & 56.4 & 43.3\\
\addlinespace[3pt]
\multirow{9}{*}{4} & \multirow{3}{*}{2.0} & Gumbel & 0.001 & 0.916 & 16.5 & 11.5 & 0.003 & 0.940 & 15.0 & 11.1\\
 &  & Clayton & 0.001 & 0.916 & 16.8 & 11.7 & 0.003 & 0.940 & 15.4 & 11.4\\
 &  & Frank & 0.001 & 0.916 & 16.7 & 11.7 & 0.003 & 0.940 & 15.4 & 11.4\\
\addlinespace[1.8pt]
 & \multirow{3}{*}{2.8} & Gumbel & 0.001 & 0.915 & 34.5 & 23.4 & 0.002 & 0.936 & 31.8 & 22.9\\
 &  & Clayton & 0.001 & 0.915 & 35.2 & 24.1 & 0.002 & 0.935 & 32.5 & 23.7\\
 &  & Frank & 0.001 & 0.916 & 35.1 & 24.1 & 0.002 & 0.935 & 32.4 & 23.6\\
\addlinespace[1.8pt]
 & \multirow{3}{*}{4.5} & Gumbel & 0.003 & 0.897 & 75.8 & 52.7 & 0.002 & 0.922 & 57.8 & 40.0\\
 &  & Clayton & 0.002 & 0.899 & 76.3 & 53.7 & 0.002 & 0.923 & 58.7 & 41.3\\
 &  & Frank & 0.003 & 0.898 & 76.3 & 53.6 & 0.003 & 0.921 & 58.6 & 41.2\\
\bottomrule
\end{tabular}
\endgroup
\par\smallskip
\begin{minipage}{0.98\linewidth}
\renewcommand{\baselinestretch}{1}
\fontsize{9}{11}\selectfont\raggedright
For \(\star\in\{\mathrm{CHW},\mathrm{RCW}\}\), \(\mathrm{VR}_{\star}=100\{1-(\mathrm{MCSD}_{\mathrm{CTW}}/\mathrm{MCSD}_{\star})^2\}\) is the percentage reduction in empirical variance with the indicated CTW copula fit relative to \(\star\). Positive values favor CTW.
\end{minipage}
\end{table}
\FloatBarrier

\clearpage
\section{Additional HF-ACTION analyses}
\label{supp:hfaction}

The analysis uses the cohort, eight baseline covariates, and censoring model in Section \ref{sec:hfaction} of main text. CHW, RCW, and CTW use the same observed death and first-hospitalization records and the same PIM covariates.

\subsection{Joint model for death and hospitalization}
\label{supp:hfaction-model}

The Gumbel model in \eqref{supp:eq:ctw-gumbel} describes first hospitalization before death and death without a prior hospitalization, as specified in \eqref{supp:eq:ctw-event-distribution}. Let \(\bm x\) contain the eight covariates in the joint model. We use \(\Lambda_0(d\mid\bm x)=\left\{\sum_{r=1}^4\lambda_{Dr}\ell_r(d)\right\}\exp(\bm\gamma_0^\top\bm x)\) and \(\Lambda_1(t\mid\bm x)=\kappa_Tt^{\nu_T}\exp(\bm\gamma_1^\top\bm x)\), where \(\ell_r(d)\) is the time accumulated by \(d\) in the intervals \([0,1)\), \([1,2)\), \([2,3)\), and \([3,\infty)\), measured in years. The knots are fixed in advance. The 23 parameters comprise four baseline death rates, eight death coefficients, a scale and shape for first hospitalization, eight hospitalization coefficients, and the Gumbel dependence parameter. When fitting the joint model, we center the covariates and divide them by their sample standard deviations. The reported PIM coefficients and covariance matrices are transformed back to the original covariate units, including age per year and CPX duration per minute.

For a nonfatal event recorded in this interval and observed follow-up through \(u>\varepsilon\), the likelihood contribution is
\begin{equation*}
\begin{cases}
\dot{\mathcal S}_0(u,\varepsilon\mid\bm x)-\dot{\mathcal S}_0(u,0\mid\bm x),
&\text{death recorded at }u,\\
\mathcal S(u,0\mid\bm x)-\mathcal S(u,\varepsilon\mid\bm x),
&\text{censored alive at }u.
\end{cases}
\end{equation*}
For such an individual censored alive at \(u<L\), the conditional tie probability is \(\omega_i^{(1)}(L)=\{\mathcal S(L,0\mid\bm X_i)-\mathcal S(L,\varepsilon\mid\bm X_i)\}/\{\mathcal S(u,0\mid\bm X_i)-\mathcal S(u,\varepsilon\mid\bm X_i)\}\). These expressions integrate the likelihood contributions over the first day.

The pair censoring probability is capped by its maximum with \(0.0001\), while \(\omega_i^{(1)}(L)\omega_j^{(1)}(L)\) equals its fitted value from the joint model. The fitted Gumbel dependence parameter was \(\widehat\theta=1.166\).

\clearpage
\subsection{Additional link and precision analyses}
\label{supp:hfaction-results}

\begin{figure}[H]
\centering
\includegraphics[width=0.98\textwidth,height=0.72\textheight,keepaspectratio]{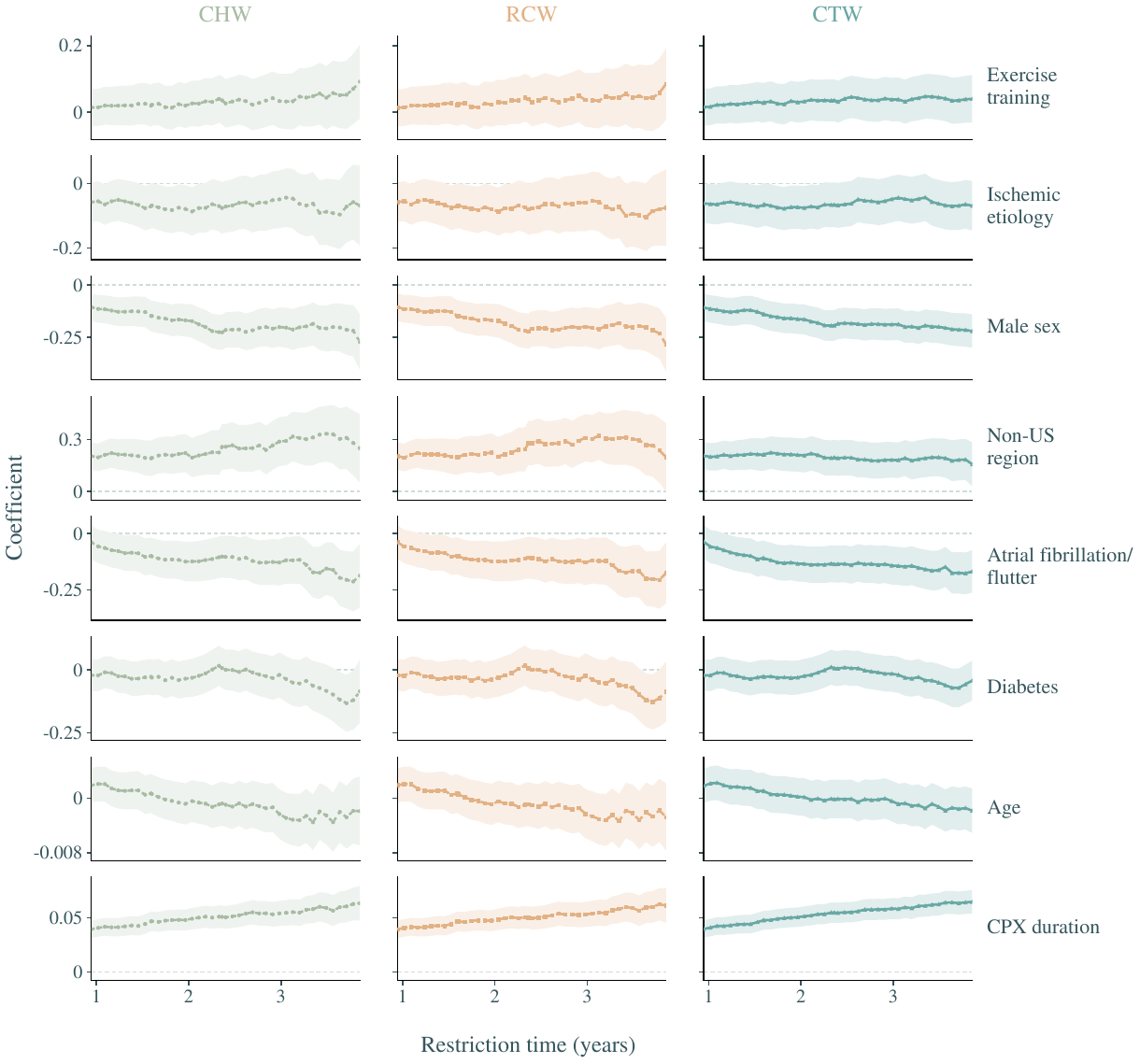}
\caption{HF-ACTION probit coefficient estimates and pointwise \(95\%\) confidence bands over \(L=0.945\)--\(3.855\) years. Columns identify methods, and vertical scales are shared within covariate rows.}
\label{fig:hfaction-probit-trajectories}
\end{figure}

Figure \ref{fig:hfaction-probit-trajectories} and Table \ref{tab:hfaction-probit-final} give the probit results. The directions of the principal covariate associations and their patterns over restriction time were similar to those under the logit link. At \(L=3.855\) years, the treatment coefficients were \(0.092\), \(0.085\), and \(0.040\) for CHW, RCW, and CTW, respectively, and all three confidence intervals included zero. Male sex and atrial fibrillation or flutter were unfavorably associated with the prioritized outcome, whereas longer CPX duration was favorable.

\begin{table}[!htbp]
\caption{HF-ACTION probit coefficient estimates from complete history weighting (CHW), resolved comparison weighting (RCW), and conditional tie weighting (CTW) at \(L=3.855\) years. The analysis includes \(2{,}109\) participants and prioritizes death over first hospitalization.}
\label{tab:hfaction-probit-final}
\centering
\begingroup
\renewcommand{\baselinestretch}{1}
\fontsize{10}{12}\selectfont
\setlength{\tabcolsep}{6pt}
\renewcommand{\arraystretch}{1.10}
\begin{tabular*}{\textwidth}{@{\extracolsep{\fill}}llrrr@{}}
\toprule
\textbf{Covariate} & \textbf{Method} & \textbf{Estimate} & \textbf{SE} & \textbf{95\% CI}\\
\midrule
Exercise training & CHW & $0.092$ & $0.057$ & $(-0.019,\,0.203)$\\
 & RCW & $0.085$ & $0.055$ & $(-0.022,\,0.193)$\\
 & CTW & $0.040$ & $0.036$ & $(-0.031,\,0.112)$\\
\addlinespace[4pt]
Ischemic etiology & CHW & $-0.068$ & $0.063$ & $(-0.193,\,0.056)$\\
 & RCW & $-0.075$ & $0.061$ & $(-0.194,\,0.044)$\\
 & CTW & $-0.069$ & $0.039$ & $(-0.146,\,0.008)$\\
\addlinespace[4pt]
Male sex & CHW & $-0.272$ & $0.067$ & $(-0.403,\,-0.140)$\\
 & RCW & $-0.283$ & $0.065$ & $(-0.410,\,-0.156)$\\
 & CTW & $-0.220$ & $0.041$ & $(-0.299,\,-0.140)$\\
\addlinespace[4pt]
Non-US region & CHW & $0.249$ & $0.101$ & $(0.052,\,0.446)$\\
 & RCW & $0.197$ & $0.101$ & $(0.000,\,0.394)$\\
 & CTW & $0.156$ & $0.065$ & $(0.029,\,0.283)$\\
\addlinespace[4pt]
Atrial fibrillation/flutter & CHW & $-0.187$ & $0.072$ & $(-0.329,\,-0.046)$\\
 & RCW & $-0.176$ & $0.069$ & $(-0.310,\,-0.041)$\\
 & CTW & $-0.169$ & $0.048$ & $(-0.262,\,-0.075)$\\
\addlinespace[4pt]
Diabetes & CHW & $-0.085$ & $0.064$ & $(-0.210,\,0.040)$\\
 & RCW & $-0.086$ & $0.061$ & $(-0.205,\,0.033)$\\
 & CTW & $-0.042$ & $0.040$ & $(-0.121,\,0.037)$\\
\addlinespace[4pt]
Age, per year & CHW & $-0.002$ & $0.003$ & $(-0.007,\,0.003)$\\
 & RCW & $-0.003$ & $0.003$ & $(-0.008,\,0.002)$\\
 & CTW & $-0.002$ & $0.002$ & $(-0.005,\,0.001)$\\
\addlinespace[4pt]
CPX duration, per minute & CHW & $0.063$ & $0.008$ & $(0.047,\,0.080)$\\
 & RCW & $0.061$ & $0.008$ & $(0.045,\,0.077)$\\
 & CTW & $0.065$ & $0.006$ & $(0.054,\,0.076)$\\
\bottomrule
\end{tabular*}
\endgroup
\par\smallskip
\begin{minipage}{\textwidth}
\renewcommand{\baselinestretch}{1}
\fontsize{9}{11}\selectfont\raggedright
Coefficients are on the probit scale. All listed covariates enter the PIM and the Cox censoring model. SE denotes the estimated standard error, and CI denotes a pointwise Wald confidence interval.
\end{minipage}
\end{table}
\FloatBarrier

\begin{figure}[H]
\centering
\includegraphics[width=0.98\textwidth,height=0.72\textheight,keepaspectratio]{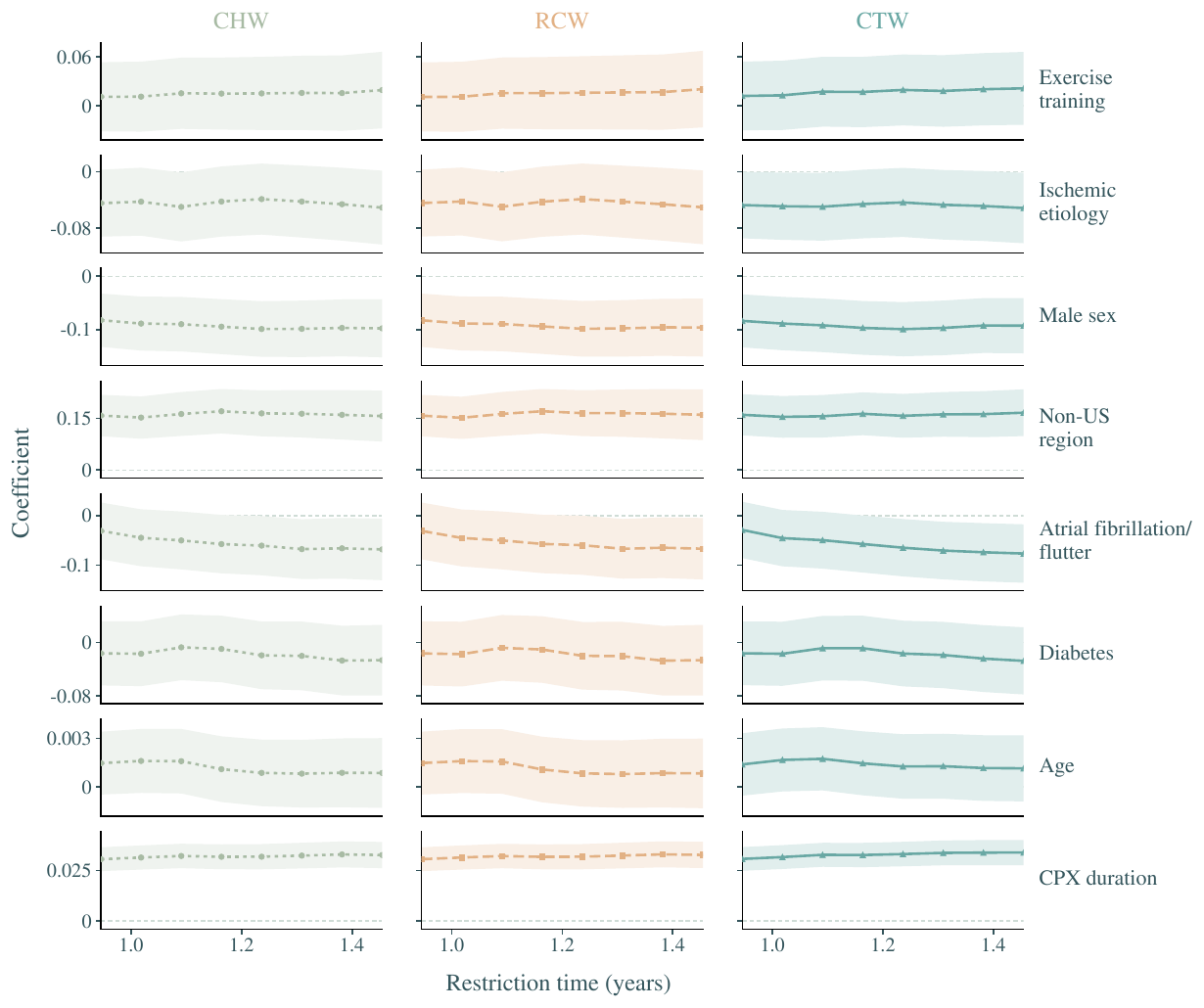}
\caption{HF-ACTION centered identity coefficient estimates and pointwise \(95\%\) confidence bands over \(L=0.945\)--\(1.455\) years. Columns identify methods, and vertical scales are shared within covariate rows. Coefficients are on the conditional net benefit scale.}
\label{fig:hfaction-identity-trajectories}
\end{figure}

\clearpage
Figure \ref{fig:hfaction-identity-trajectories} shows the centered identity estimates over \(L=0.945\)--\(1.455\) years. Table \ref{tab:hfaction-identity-converged} reports the estimates at \(L=1.455\) years. The treatment coefficients were \(0.019\), \(0.021\), and \(0.021\) for CHW, RCW, and CTW, respectively. The CTW \(95\%\) confidence interval was \((-0.023,0.066)\). Treatment and age intervals included zero throughout this range. At the final restriction time, the confidence interval for ischemic etiology narrowly excluded zero under CTW but included zero under CHW and RCW.

\begin{table}[!htbp]
\caption{HF-ACTION identity coefficient estimates from complete history weighting (CHW), resolved comparison weighting (RCW), and conditional tie weighting (CTW) at \(L=1.455\) years. The analysis includes \(2{,}109\) participants and prioritizes death over first hospitalization.}
\label{tab:hfaction-identity-converged}
\centering
\begingroup
\renewcommand{\baselinestretch}{1}
\fontsize{10}{12}\selectfont
\setlength{\tabcolsep}{6pt}
\renewcommand{\arraystretch}{1.10}
\begin{tabular*}{\textwidth}{@{\extracolsep{\fill}}llrrr@{}}
\toprule
\textbf{Covariate} & \textbf{Method} & \textbf{Estimate} & \textbf{SE} & \textbf{95\% CI}\\
\midrule
Exercise training & CHW & $0.019$ & $0.024$ & $(-0.028,\,0.066)$\\
 & RCW & $0.021$ & $0.024$ & $(-0.027,\,0.068)$\\
 & CTW & $0.021$ & $0.023$ & $(-0.023,\,0.066)$\\
\addlinespace[4pt]
Ischemic etiology & CHW & $-0.051$ & $0.027$ & $(-0.103,\,0.002)$\\
 & RCW & $-0.050$ & $0.027$ & $(-0.103,\,0.002)$\\
 & CTW & $-0.051$ & $0.025$ & $(-0.101,\,-0.001)$\\
\addlinespace[4pt]
Male sex & CHW & $-0.098$ & $0.028$ & $(-0.152,\,-0.043)$\\
 & RCW & $-0.096$ & $0.028$ & $(-0.151,\,-0.042)$\\
 & CTW & $-0.093$ & $0.026$ & $(-0.145,\,-0.041)$\\
\addlinespace[4pt]
Non-US region & CHW & $0.156$ & $0.038$ & $(0.083,\,0.230)$\\
 & RCW & $0.160$ & $0.037$ & $(0.087,\,0.234)$\\
 & CTW & $0.166$ & $0.035$ & $(0.098,\,0.234)$\\
\addlinespace[4pt]
Atrial fibrillation/flutter & CHW & $-0.068$ & $0.032$ & $(-0.130,\,-0.006)$\\
 & RCW & $-0.067$ & $0.032$ & $(-0.129,\,-0.005)$\\
 & CTW & $-0.077$ & $0.030$ & $(-0.135,\,-0.018)$\\
\addlinespace[4pt]
Diabetes & CHW & $-0.026$ & $0.027$ & $(-0.079,\,0.026)$\\
 & RCW & $-0.026$ & $0.027$ & $(-0.079,\,0.026)$\\
 & CTW & $-0.027$ & $0.026$ & $(-0.078,\,0.023)$\\
\addlinespace[4pt]
Age, per year & CHW & $0.001$ & $0.001$ & $(-0.001,\,0.003)$\\
 & RCW & $0.001$ & $0.001$ & $(-0.001,\,0.003)$\\
 & CTW & $0.001$ & $0.001$ & $(-0.001,\,0.003)$\\
\addlinespace[4pt]
CPX duration, per minute & CHW & $0.033$ & $0.003$ & $(0.026,\,0.039)$\\
 & RCW & $0.033$ & $0.003$ & $(0.026,\,0.039)$\\
 & CTW & $0.034$ & $0.003$ & $(0.028,\,0.040)$\\
\bottomrule
\end{tabular*}
\endgroup
\par\smallskip
\begin{minipage}{\textwidth}
\renewcommand{\baselinestretch}{1}
\fontsize{9}{11}\selectfont\raggedright
Coefficients are on the conditional net benefit scale. All listed covariates enter the PIM and the Cox censoring model. SE denotes the estimated standard error, and CI denotes a pointwise Wald confidence interval. Age and CPX duration are expressed per year and per minute, respectively.
\end{minipage}
\end{table}
\FloatBarrier

Table \ref{tab:hfaction-efficiency-two-times} compares CTW with CHW and RCW at two restriction times for each link. The comparisons use \(L=1.018\) and \(3.855\) years for logit and probit, and \(L=1.018\) and \(1.455\) years for the centered identity link. Variance reductions were small near one year and larger at the later time. The probit comparisons closely followed the logit results, while the identity-link reductions at \(L=1.455\) were smaller. The percentages use the estimated coefficient variances.

\begin{table}[!htbp]
\caption{Percentage reduction in estimated coefficient variance with conditional tie weighting (CTW) relative to complete history weighting (CHW) and resolved comparison weighting (RCW) in HF-ACTION. Results are shown at two restriction times for each link, with death prioritized over first hospitalization.}
\label{tab:hfaction-efficiency-two-times}
\centering
\begingroup
\renewcommand{\baselinestretch}{1}
\fontsize{10}{12}\selectfont
\setlength{\tabcolsep}{6pt}
\renewcommand{\arraystretch}{1.10}
\begin{tabularx}{\textwidth}{@{}>{\raggedright\arraybackslash}X*{4}{>{\raggedleft\arraybackslash}p{0.13\textwidth}}@{}}
\toprule
\multicolumn{5}{@{}l}{\textbf{Logit link}}\\
 & \multicolumn{2}{c}{$L=1.018$ years} & \multicolumn{2}{c}{$L=3.855$ years}\\
\cmidrule(lr){2-3}\cmidrule(lr){4-5}
\textbf{Covariate} & \textbf{CHW} & \textbf{RCW} & \textbf{CHW} & \textbf{RCW}\\
\midrule
Exercise training & 2.0 & 2.0 & 59.0 & 56.0\\
Ischemic etiology & 1.9 & 1.9 & 62.2 & 58.8\\
Male sex & 2.0 & 2.0 & 63.9 & 61.0\\
Non-US region & 1.9 & 2.0 & 59.0 & 58.8\\
Atrial fibrillation/flutter & 1.9 & 1.9 & 56.0 & 51.4\\
Diabetes & 1.9 & 1.9 & 60.5 & 56.4\\
Age, per year & 2.9 & 2.8 & 58.3 & 55.5\\
CPX duration, per minute & 1.8 & 1.8 & 56.3 & 54.5\\
\midrule
\multicolumn{5}{@{}l}{\textbf{Probit link}}\\
 & \multicolumn{2}{c}{$L=1.018$ years} & \multicolumn{2}{c}{$L=3.855$ years}\\
\cmidrule(lr){2-3}\cmidrule(lr){4-5}
\textbf{Covariate} & \textbf{CHW} & \textbf{RCW} & \textbf{CHW} & \textbf{RCW}\\
\midrule
Exercise training & 2.1 & 2.1 & 58.7 & 55.8\\
Ischemic etiology & 1.9 & 1.9 & 61.8 & 58.4\\
Male sex & 2.0 & 2.1 & 63.6 & 60.7\\
Non-US region & 2.0 & 2.1 & 58.6 & 58.6\\
Atrial fibrillation/flutter & 1.9 & 1.9 & 55.8 & 51.2\\
Diabetes & 1.9 & 2.0 & 60.2 & 56.2\\
Age, per year & 2.9 & 2.8 & 58.4 & 55.8\\
CPX duration, per minute & 1.8 & 1.8 & 55.3 & 53.7\\
\midrule
\multicolumn{5}{@{}l}{\textbf{Centered identity link}}\\
 & \multicolumn{2}{c}{$L=1.018$ years} & \multicolumn{2}{c}{$L=1.455$ years}\\
\cmidrule(lr){2-3}\cmidrule(lr){4-5}
\textbf{Covariate} & \textbf{CHW} & \textbf{RCW} & \textbf{CHW} & \textbf{RCW}\\
\midrule
Exercise training & 2.3 & 2.2 & 9.6 & 9.5\\
Ischemic etiology & 2.1 & 2.1 & 9.1 & 8.9\\
Male sex & 2.2 & 2.2 & 8.7 & 8.9\\
Non-US region & 2.2 & 2.4 & 15.8 & 15.1\\
Atrial fibrillation/flutter & 2.1 & 2.1 & 10.0 & 9.6\\
Diabetes & 2.1 & 2.1 & 9.6 & 9.3\\
Age, per year & 3.1 & 3.0 & 10.2 & 9.9\\
CPX duration, per minute & 2.1 & 2.1 & 7.4 & 7.5\\
\bottomrule
\end{tabularx}
\endgroup
\par\smallskip
\begin{minipage}{\textwidth}
\renewcommand{\baselinestretch}{1}
\fontsize{9}{11}\selectfont\raggedright
Entries are \(100\{1-\widehat{\operatorname{Var}}(\widehat\beta_{Lk}^{\mathrm{CTW}})/\widehat{\operatorname{Var}}(\widehat\beta_{Lk}^{\star})\}\) for comparator \(\star\in\{\mathrm{CHW},\mathrm{RCW}\}\), where \(k\) indexes the covariate in each row. Positive values indicate greater estimated precision with CTW. 
\end{minipage}
\end{table}
\FloatBarrier

\section{Tutorial for the \texttt{gpim} package}
\label{supp:gpim-tutorial}

The \texttt{gpim} package implements the CHW, RCW, and CTW estimators with logit, probit, and identity links. The source is available at \url{https://github.com/fancy575/gpim}. We illustrate RCW and CTW using the package's synthetic \texttt{hfaction} data, which contain 407 event records from 300 individuals. The package example uses synthetic records, and the analysis in the main paper uses the HF-ACTION trial data.

\subsection{Installation and data structure}
\label{supp:gpim-installation}

The following commands install the package from GitHub and load the example data. Installation from source requires the appropriate R compilation tools.
\begingroup
\begin{verbatim}
install.packages("remotes")
remotes::install_github(
  "fancy575/gpim",
  upgrade = "never", build_vignettes = FALSE,
  INSTALL_opts = "--preclean"
)
library(gpim)
options(width = 100)
data("hfaction", package = "gpim")
\end{verbatim}
\endgroup

Each individual has one death or censoring record and an additional record if first hospitalization was observed. The columns \texttt{id}, \texttt{time}, and \texttt{status} identify the individual, observed time, and event type. Time is measured in years, with \texttt{status = 0} denoting censoring, \texttt{status = 1} death, and \texttt{status = 2} first hospitalization. Baseline covariates are repeated across an individual's records. The default hierarchy compares death before hospitalization, and later event times are favorable. An absent hospitalization record means that hospitalization was not observed before death or censoring.

\subsection{Fitting and summarizing the model}
\label{supp:gpim-fitting}

We fit RCW and CTW at restriction times \(L=0.5,1,3\) years. The outcome model includes treatment, age per ten years, sex, and left ventricular ejection fraction per five percentage points. The Cox censoring model is specified separately through \texttt{censor\_formula}. For CTW, \texttt{pim\_joint\_model} specifies a Gumbel copula with Weibull proportional hazards margins. By default, both margins use the covariates in the outcome model. Setting \texttt{weight\_floor = 0} leaves the inverse censoring weights untruncated.
\begingroup
\begin{verbatim}
joint <- pim_joint_model(
  copula = "gumbel",
  margins = c("weibull", "weibull")
)

fit <- gpim(
  HierSurv(time, status)  
    treatment + I(age / 10) + sex + I(lvef / 5),
  data = hfaction,
  id = "id",
  L = c(0.5, 1, 3),
  method = c("RCW", "CTW"),
  link = "logit",
  censor_formula =   treatment + I(age / 10) + sex,
  joint_model = joint,
  control = pim_control(weight_floor = 0)
)
\end{verbatim}
\endgroup

The \texttt{summary} method returns coefficient estimates, standard errors, and pointwise \(95\%\) Wald confidence intervals. Standard errors account for overlapping pairs and estimation of the censoring distribution, together with estimation of the joint event model for CTW. The following code displays the treatment results separately for each method, with three decimal places.
\begingroup
\begin{verbatim}
for (m in c("RCW", "CTW")) {
  trt <- subset(
    summary(fit, method = m)$coefficients,
    term == "treatmentExercise",
    select = c(L, estimate, std.error, conf.low, conf.high)
  )
  trt[] <- lapply(trt, formatC, format = "f", digits = 3)
  cat("\n", m, "\n", sep = "")
  print(trt, row.names = FALSE)
}
\end{verbatim}
\begin{verbatim}
RCW
     L estimate std.error conf.low conf.high
 0.500    0.127     0.060    0.008     0.245
 1.000    0.168     0.102   -0.032     0.368
 3.000    0.615     0.176    0.271     0.960

CTW
     L estimate std.error conf.low conf.high
 0.500    0.124     0.060    0.007     0.241
 1.000    0.141     0.099   -0.053     0.336
 3.000    0.562     0.159    0.250     0.874
\end{verbatim}
\endgroup

At \(L=3\) years, exponentiating the treatment coefficients and confidence limits gives conditional win odds of \(1.851\) (\(95\%\) CI \(1.311,2.611\)) for RCW and \(1.754\) (\(95\%\) CI \(1.284,2.396\)) for CTW, comparing exercise with usual care at equal values of the remaining covariates. CTW has a smaller estimated standard error at each horizon in this synthetic example.

The arguments \texttt{link = "probit"} and \texttt{link = "identity"} select the other supported links. Identity coefficients describe conditional net benefit. Alternative CTW joint models can be specified through the \texttt{copula} and \texttt{margins} arguments of \texttt{pim\_joint\_model}, including Cox margins with unspecified baseline hazards. The standard methods \texttt{coef}, \texttt{vcov}, and \texttt{confint} extract coefficients, covariance matrices, and confidence intervals. With several restriction times, \texttt{plot(fit)} displays the coefficient estimates and pointwise intervals.

\end{document}